\documentclass[10pt,twocolumn,amsmath,amssymb,aps,prb]{revtex4-2}

\usepackage{graphicx}
\usepackage{dcolumn}
\usepackage{bm}
\usepackage[cp1250]{inputenc}
\usepackage[usenames,dvipsnames]{color}
\usepackage{tabularx}

\usepackage{hyperref}

\usepackage{cancel}
\usepackage{color}

\newcommand{\rmi}{{\rm i}}
\newcommand{\rmd}{{\rm d}}
\newcommand{\phan}{{\phantom{+}}}
\newcommand{\phani}{{\phantom{i}}}
\newcommand{\EulerGamma}{\gamma_{\rm E}}
\newcommand{\rhotilde}{\tilde{\rho}}

\newcommand{\kreis}[1]{\,\unitlength1ex\begin{picture}(2,2)%
\put(0.75,0.75){\circle{1.35}}\put(0.75,0.75){\makebox(0,0){\tiny#1}\;}\end{picture}}
\newcommand{\xBkreis}[1]{x_B^{\hspace*{-0.7mm}\kreis{#1}}\hspace*{-1mm}}
\newcommand{\xBtildekreis}[1]{\tilde{x}_B^{\hspace*{-0.7mm}\kreis{#1}}\hspace*{-1mm}}

\newcommand{\SMsubsecDoubleIntegral}{SM-VI~B~1}
\newcommand{\SMsubsecTripleIntegral}{SM-VI~B~2}
\newcommand{\SMSMsubsubsecderivationellBT}{SM~V~A}
\newcommand{\SMFIKsecondorder}{S342}
\newcommand{\SMSMsubsubsecderivationlb}{SM~V~B}
\newcommand{\SMsmxBtermderivation}{SM~II}
\newcommand{\SMsmxAtermderivation}{SM-III}
\newcommand{\SMFreeEnergyIsingKondo}{SM~I}

\renewcommand{\today}{Resubmitted version of March 25, 2024}

\begin{document}

\title{Second Wilson number from third-order perturbation theory\\ for the
symmetric single-impurity Kondo model at low temperatures}

\author{Kevin Bauerbach}
\author{Florian Gebhard$^1$}
\email{florian.gebhard@physik.uni-marburg.de}
\affiliation{$^1$Fachbereich Physik, Philipps-Universit\"at Marburg,
  35032 Marburg, Germany}
  
\date{\today}

\begin{abstract}
    We determine the impurity-induced free energy and the impurity-induced zero-field susceptibility of the symmetric single-impurity Kondo model 
    from weak-coupling perturbation theory up to third order in the Kondo coupling at low temperatures and small magnetic fields. 
    We reproduce the analytical structure of the zero-field magnetic susceptibility as obtained from Wilson's renormalization group method. This permits us 
    to obtain analytically the first two Wilson numbers. 
\end{abstract}

\maketitle

\section{Overview}
\label{sec:Intro}
\label{sec:overview}

\subsection{Introduction}

The Kondo model represents one of the fundamental models in theoretical many-particle physics. It is the basis of a theoretical description
phenomena due to magnetic impurities in metals, e.g., the increase of the electrical resistance at low temperatures, typically below the so-called Kondo temperature $T_{\rm K}$~\cite{deHaas}. 
The underlying fundamental interaction is the antiferromagnetic $s$-$d$ spin-exchange interaction 
between a localized electron and the conduction electron spin polarization at the magnetic impurity~\cite{Zener}.
At low temperatures, $T\ll T_{\rm K}$, the conduction-electron cloud completely screens the spin-1/2 impurity, i.e., the conduction-electron Kondo cloud 
and the impurity spin form the so-called Kondo singlet state. Recently, the Kondo cloud was observed experimentally for the first time in a quantum dot~\cite{Borzenets}. 

After the establishment of the model Hamiltonian by Kasuya in 1956~\cite{Kasuya} and the first theoretical explanation of the electrical resistance minimum by Kondo in 1964~\cite{Kondo1964}, 
many important contributions to the solution of the Kondo problem were published in the 1970s and 1980s. These include, in particular, 
Wilson's renormalization group (RG)~\cite{RevModPhys.47.773} treatment, the numerical renormalization group (NRG)~\cite{RevModPhys.47.773, PhysRevB.21.1003, PhysRevB.21.1044}, 
and also the Bethe Ansatz~\cite{TW,AFL} treatments. A detailed historical overview of the physics of the Kondo model in the 20th century can be found in Hewson's book~\cite{Hewson}. 
A more recent study~\cite{Barczaetal} treats the ground-state properties of the symmetric single-impurity Kondo model in detail, and compares  
various analytical methods, e.g., the Gutzwiller and Yosida variational wave functions, and numerical approaches such as the density-matrix renormalization group (DMRG) and 
the NRG methods. Also in the year 2020, Bauerbach et al.~\cite{IsingKondo} introduced a simplification of the Kondo model, the so-called Ising-Kondo model, 
in which the spin-exchange term of the full Kondo Hamiltonian is neglected. Thereby, the complexity of the Kondo model is reduced to that of potential scattering of spinless fermions, 
which can be solved analytically using the equation of motion method.

The Kondo singlet is discernible in the magnetic susceptibility at zero field. The contribution of a free spin-1/2 to the zero-field magnetic susceptibility
is given by Curie's law,
\begin{equation}
    \frac{\chi_0^{\rm ii, free}(T)}{(g_e\mu_{\rm B})^2}=\frac{1}{4T} \; ,
\end{equation}
where $g_e\approx 2$ is the electronic $g$-factor, and $\mu_{\rm B}$ is the Bohr magneton, and the superscript `ii' denotes the
impurity-induced contribution to the magnetic susceptibility.
Apparently, $\chi_0^{\rm free}(T)$ diverges for $T\to 0$ because the spin can freely orient itself in the magnetic field.
In the spin-1/2 Kondo model, the impurity spin is locked into the Kondo singlet at zero temperature with a small but finite excitation energy of the order
of $T_{\rm K}$. It is one of the central achievements of Wilson's renormalization group method
to show that the impurity contribution to the zero-field magnetic susceptibility is finite, and given by~\cite{RevModPhys.47.773}
\begin{equation}
    \frac{\chi_0^{\rm ii}(T=0)}{(g_e\mu_{\rm B})^2}= \frac{c_0}{\sqrt{\pi e}}\frac{D}{\tilde{D}(j)}
    \frac{\exp(1/j)}{\sqrt{j}} \exp\left[-\sum_{r=1}^{\infty}\alpha_r j^r \right] \; ,
    \label{eq:zerofieldintro}
\end{equation}
where $j>0$ is the antiferromagnetic Kondo coupling parameter, $c_0$ and $\alpha_r$ are undetermined constants, 
and $\tilde{D}(j)/D$ is the effective bandwidth in units of the bare bandwidth~$D$.
For small couplings, $j \ll 1$, the numerical renormalization group can be used to calculate the unknown parameters with high accuracy, see, e.g., Ref.~\cite{Barczaetal}
for a recent study. The constant
\begin{equation}
    w= \frac{4c_0}{\sqrt{\pi e}}= \frac{e^{\gamma_{\rm E}+1/4}}{\pi^{3/2}} \approx 0.410705 \; ,
    \label{eq:wilsonnumberintro}
\end{equation}
where $\gamma_{\rm E}\approx 0.577216$ is Euler's number, is often called the ``Wilson number"~\cite{Hewson}.

More generally, the bandwidth renormalization factor has a Taylor series expansion,
\begin{equation}\label{tildeDExpansionintro}
\tilde{D}(j)=D\Bigl(c_0-c_1 \frac{j}{2}+\mathcal{O}[j^2]\Bigr)\;,
\end{equation}
where we denote $c_0$ and $c_1$ as the first and second Wilson numbers, respectively.
In this work, we determine the first two Wilson numbers analytically.

\subsection{Summary of results}
\label{subsec:summaryofresults}

Wilson's renormalization group gives an analytic expression for the impurity-induced zero-field magnetic susceptibility
at low temperatures~$T$ in terms of a series expansion in the Kondo coupling, see p.~830 in Ref.~\cite{RevModPhys.47.773},  ($k\equiv k_{\rm B}\equiv 1$)
\begin{eqnarray}
\chi_0^{\rm ii,W}(T)&=&\frac{1}{4T}\biggl[1-j+j^2\ln\left(\frac{T}{\tilde{D}(j)}\right)\nonumber\\
&&\qquad -j^3\left(\ln^2\left(\frac{T}{\tilde{D}(j)}\right)+\frac{1}{2}\ln\left(\frac{T}{\tilde{D}(j)}\right)\right)\nonumber \\
&& \qquad + \mathcal{O}[j^4]\biggr] \; .
\label{eq:WilsonsChi0intro}
\end{eqnarray}
In this work, we calculate the free energy of the symmetric single-impurity Kondo model up to third
order in~$j$ for small temperatures~$T$ and even smaller magnetic fields, $B\ll T\ll D\equiv 1$.
The expansion up to second order in~$B$ permits us to reproduce the analytic structure given in Eq.~(\ref{eq:WilsonsChi0intro}).
This is one of the major achievements of this work. 

Starting from this comparison we 
derive the  effective bandwidth $\tilde{D}(j)$ up to and including first order in~$j$ with the result
\begin{eqnarray}
    c_0&=&\frac{e^{3/4+\EulerGamma}}{4\pi}\approx 0.300049\;,\nonumber \\
    c_1&=&\frac{e^{3/4+\EulerGamma}}{32\pi}\left(-13+\pi^2-8\ln 2\right)\approx -0.325387
    \label{eq:bothWilsonnumbers}
\end{eqnarray}
for the first two Wilson numbers. 
While $c_0$ is known from the literature, see, e.g., Refs.~\cite{AFL,PhysRevB.101.075132},
the second Wilson number $c_1$ is a new result.

\subsection{Outline} 
\label{subsec:outline}

Our work is organized as follows. In Sect.~\ref{sec:aSIKM}, we introduce the Hamiltonian of the ${\rm XXZ}$ anisotropic single-impurity Kondo model and define the free energy  
and the magnetic susceptibility. 
In Sect.~\ref{thePT}, we address perturbation theory in the Kondo coupling to calculate the free energy up to and including third order. 
In Sect.~\ref{zerofieldsuscep}, we determine the zero-field susceptibility and compare our result with that of Wilson's renormalization group
to identify the first two Wilson numbers. Short final remarks in Sect.~\ref{sec:conclusions} close our presentation.

The calculations to third order are lengthy and cumbersome. To avoid the interruption of the flow of reading, 
we discuss strategies for the evaluation of the terms to second and third order in the Appendix, and merely quote the results in the main text.  
All details of the analytical calculations can be found in the 49~pages of the Supplemental Material.
We note in passing that we heavily rely on {\sc Mathematica}~\cite{Mathematica12} to derive our results.

\section{Anisotropic single-impurity Kondo model}
\label{sec:aSIKM}

In this section we define the model Hamiltonian of the ${\rm XXZ}$ anisotropic single-impurity Kondo model and introduce the physical quantities of interest in this study.

\subsection{Model Hamiltonian}
\label{subsec:Hamiltonian}

The Hamiltonian of the anisotropic single-impurity Kondo model
reads~\cite{Hewson,PhysRev.149.491,Solyom}
\begin{equation}\label{eq:generalHamiltonian}
\hat{H}_{\rm K}=\hat{H}_0 +\hat{V}_{\rm sd} 
\end{equation}
with
\begin{equation}
\hat{H}_0=\hat{T}+\hat{H}_{\rm m}\;,
\end{equation}
where $\hat{T}$ is the kinetic energy of the host electrons,
$\hat{V}_{\rm sd}$ is their interaction with the impurity spin at the lattice origin, and
$\hat{H}_{\rm m}$ describes the electrons' interaction with an external magnetic field.

\subsubsection{Host electrons}
\label{subsec:hostelectronmodelDOS}

The kinetic energy of the host electrons is given by
\begin{equation}
  \hat{T}= \sum_{\sigma} \hat{T}_{\sigma} \; , \;
  \hat{T}_{\sigma}=
  \sum_{i,j} t_{i,j} \hat{c}_{i,\sigma}^+\hat{c}_{j,\sigma}^{\vphantom{+}}
    \; .
\end{equation}
Here, $\hat{c}_{i,\sigma}^+$ ($\hat{c}_{i,\sigma}^{\vphantom{+}}$) creates
(annihilates) an electron with spin $\sigma=\uparrow,\downarrow$ on lattice site~$i$,
where $t_{i,j}=t_{j,i}^*$ are the matrix elements for the tunneling
of an electron from site~$j$ to site~$i$ on a lattice with $L$~sites ($L$ is even).
Assuming translational invariance, $t_{i,j}=t(i-j)$, the 
kinetic energy is diagonal in momentum space~\cite{PhysRevB.101.075132},
\begin{equation}
  \hat{T}_{\sigma}=
  \sum_{k} \epsilon(k)\hat{a}_{k,\sigma}^+\hat{a}_{k,\sigma}^{\vphantom{+}} \; ,
\end{equation}
where
\begin{eqnarray}
  \hat{a}^+_{k,\sigma}&=&\frac{1}{\sqrt{L}}\sum_r e^{\rmi k r}\hat{c}^+_{r,\sigma}\;,
  \nonumber \\
\hat{c}^{+}_{r,\sigma}
  &=&\frac{1}{\sqrt{L}}\sum_k e^{-\rmi k r}\hat{a}^+_{k\sigma}
\end{eqnarray}
with $k$ from the first Brillouin zone.
The corresponding density of states of the host electrons is given by
\begin{equation}
\rho_0(\omega)=\frac{1}{L}\sum_k\delta(\omega-\epsilon(k))\;.
\end{equation}
We assume a bipartite lattice so that
there exists half a reciprocal lattice vector~$Q$ for which
$\epsilon(Q-k)=-\epsilon(k)$ for all~$k$ from the first Brillouin zone. 
As a consequence, the density of states of the host electrons is symmetric,
$\rho_0(-\omega)=\rho_0(\omega)$.

In the following, we set half the bandwidth~$W$ as our energy unit, i.e., $W=2$,
so that $\rho_0(|\omega|>1)=0$.
To be definite, we choose to work with a constant density of states,
\begin{equation}
  \rho_0(|\omega|\leq 1)=\frac{1}{2}\;.
  \label{rhoconst}
\end{equation}
We shall consider
the case where the paramagnetic host electron system is half-filled, i.e., 
the electron numbers fulfill
$N_{\uparrow}=N_{\downarrow}=L/2$ and $N=N_{\uparrow}+N_{\downarrow}=L$; 
the thermodynamic limit, $N,L\to \infty$ with $n=N/L=1$
fixed, is implicit. In the supplemental material we outline a strategy to generalize the results 
to a general density of states that is finite at the Fermi energy.

\subsubsection{Kondo interaction}
\label{subsec:KondoInteraction}

In the (anisotropic) Kondo model, the host electrons interact locally 
with the impurity spin at the origin,
\begin{eqnarray}\label{DefinitionKondoInteraction}
\hat{V}_{\rm sd}&=& \hat{V}_{\perp} + \hat{V}_z \;, \nonumber \\
\hat{V}_\perp&=&J_{\perp} \left(\hat{s}^x_0\hat{S}^x+\hat{s}^y_0\hat{S}^y\right)
\nonumber\\
&=&\frac{J_{\perp}}{2} \left( \hat{s}_0^+\hat{S}^- + \hat{s}_0^-\hat{S}^+\right)
\nonumber \\
&=&\frac{J_\perp}{2L}\sum_{k,k'}\left(\hat{a}^+_{k,\uparrow}\hat{a}^\phan_{k',\downarrow}\hat{d}^+_\Downarrow\hat{d}^\phani_\Uparrow+\hat{a}^+_{k,\downarrow}\hat{a}^\phan_{k',\uparrow}\hat{d}^+_\Uparrow\hat{d}^\phani_\Downarrow\right)\;,\nonumber\\
\hat{V}_{\rm z}&=&J_z \hat{s}^z_0
\hat{S}^z \nonumber \\
&=& \frac{J_z}{4L}\sum_{k,k'}\left(\hat{n}^d_\Uparrow-\hat{n}^d_\Downarrow\right)\left(\hat{a}^+_{k,\uparrow}\hat{a}^\phan_{k',\uparrow}-\hat{a}^+_{k,\downarrow}\hat{a}^\phan_{k',\downarrow}\right),
\label{eq:Vdefintion}
\end{eqnarray}
where $\hat{V_\perp}$ describes the spin-flip component of the Hamiltonian and $\hat{V}_z$ is the $z$ component. The operators $\hat{d}^+_s$ ($\hat{d}^{\vphantom{+}}_s$) create (annihilate)
an impurity electron with spin $s=\;\Uparrow,\Downarrow$.
In Eq.~(\ref{eq:Vdefintion}) it is implicitly understood that
the impurity is either filled with an electron with
spin $\Uparrow$ or with spin $\Downarrow$. The individual operators are defined as
\begin{equation}\label{SsndsOperators}
\hat{S}^+=\hat{d}^+_\Uparrow\hat{d}^\phani_\Downarrow\;, \quad \hat{S}^-=\hat{d}^+_\Downarrow\hat{d}^\phani_\Uparrow\;, \quad
\hat{n}^d_{s}=\hat{d}^+_s\hat{d}^\phani_s\;,
\end{equation}
and
\begin{equation}
\hat{s}_0^+=\frac{1}{L}\sum_{k,k'}\hat{a}^+_{k,\uparrow}\hat{a}^\phan_{k',\downarrow}\;, \quad\hat{s}^-_0=\frac{1}{L}\sum_{k,k'}\hat{a}^+_{k,\downarrow}\hat{a}^\phan_{k',\uparrow}\; .
\end{equation}
Note that in the isotropic single-impurity Kondo model we have $J_{\perp}= J_z\equiv J_{\rm K}$. Furthermore, for $J_\perp=0$, $J_z\neq 0$, we obtain the Ising-Kondo model~\cite{IsingKondo}.

\subsubsection{External magnetic field}
\label{subsec:ExternalMagneticField}

To investigate the magnetic properties of the Kondo model, we couple the electrons to a global external magnetic field
\begin{equation}
  \hat{H}_{\rm m}=
  -B\left(\hat{n}^d_{\Uparrow}-\hat{n}^d_{\Downarrow}\right)
  -B\sum_{i}\left(\hat{n}_{i,\uparrow}
  -\hat{n}_{i,\downarrow}\right)\end{equation}
with the local density operators
$\hat{n}^d_s$ 
and
$\hat{n}_{i,\sigma}=\hat{c}^+_{i,\sigma}\hat{c}^{\vphantom{+}}_{i,\sigma}$.
Here, the magnetic energy reads
\begin{equation}
B=g_e\mu_{\rm B}\mathcal{H}/2 >0\;,
\end{equation}
where ${\cal H}$ is the external field,
$g_e\approx 2$ is the electrons' gyromagnetic factor, and $\mu_{\rm B}$
is Bohr's magneton. 

\subsubsection{Particle-hole transformation}
\label{subsec:ParticleHoleTrafo}

In the definition of the particle-hole transformation we include a spin-flip operation,
\begin{eqnarray}
\widetilde{\tau}_S:\;  
& &
\hat{a}_{k,\sigma}^{\vphantom{+}} \mapsto \hat{a}_{Q-k,\bar{\sigma}}^+ \; , 
\hat{d}_s^{\vphantom{+}} \mapsto \hat{d}_{\bar{s}}^+\; ,
\nonumber \\
&& \hat{a}_{k,\sigma}^+ \mapsto \hat{a}_{Q-k,\bar{\sigma}}^{\vphantom{+}} \; ,
\hat{d}_s^+ \mapsto \hat{d}_{\bar{s}}^{\vphantom{+}} \; ,
\label{eq:phtrafo}
\end{eqnarray}
where $\bar{\uparrow}=\downarrow$ ($\bar{\Uparrow}=\Downarrow$)
and $\bar{\downarrow}=\uparrow$ ($\bar{\Downarrow}=\Uparrow$)
denotes the flipped spin. The transformation~$\widetilde{\tau}_S$ implies $\hat{c}_{0,\sigma}^+
\mapsto\hat{c}_{0,\bar{\sigma}}^{\vphantom{+}}$.

It is not difficult to show that the chemical potential $\mu(T)=0$ guarantees a half-filled host-electron system for all temperatures $T$, see also Ref.~\cite{Barczaetal}. 
To see this, we consider the average particle number
\begin{equation}
\langle\hat{N}\rangle(\mu)=\frac{1}{Z(\mu)}{\rm Tr}\left[e^{-\beta(\hat{H}-\mu\hat{N})}\hat{N}\right]
\end{equation}
with the grand canonical partition function
\begin{equation}
Z(\mu)={\rm Tr}\left[e^{-\beta(\hat{H}-\mu\hat{N})}\right]\;.
\end{equation}
The particle-hole transformation~$\widetilde{\tau}_S$, Eq.~(\ref{eq:phtrafo}), leaves the Hamiltonian invariant, $\tilde{\tau}_S:\;\hat{H}_{\rm K}\mapsto \hat{H}_{\rm K}$, but it affects the particle number operator, $\tilde{\tau}_S:\;\hat{N}\mapsto 2L-\hat{N}$. Thus,
\begin{equation}
Z(\mu)={\rm Tr}\left[e^{-\beta(\hat{H}-\mu(2L-\hat{N}))}\right]=e^{2L\beta\mu}Z(-\mu)
\end{equation}
and
\begin{eqnarray}
\langle\hat{N}\rangle(\mu) &=& e^{-2 L\beta \mu}\frac{1}{Z(-\mu)}{\rm Tr}\left[e^{-\beta[\hat{H}-\mu(2L-\hat{N})]}(2L-\hat{N})\right]\nonumber\\
&=&2L-\frac{1}{Z(-\mu)}{\rm Tr}\left[e^{-\beta(\hat{H}+\mu\hat{N})}\hat{N}\right]\nonumber\\
&=&2L-\langle\hat{N}\rangle(-\mu)\;.
\end{eqnarray}
This implies that $\langle\hat{N}\rangle(\mu=0)=L$ for all $T$.

\subsection{Impurity-induced zero-field susceptibility}
\label{def:impurity-induced-free-energy}
In our study, we are interested in the impurity contribution to the zero-field magnetic susceptibility, which can be determined from the impurity contribution to the free energy.

\subsubsection{Impurity-induced free energy}
The free energy of the anisotropic single-impurity Kondo model is defined by ($\beta=1/T$, $k_{\rm B}\equiv 1$)
\begin{equation}\label{def:freeenergy}
    \mathcal{F}^{\rm aSIKM}(B,T,J_\perp,J_z)=-T\ln Z=-T\ln{\rm Tr}\left[e^{-\beta\hat{H}_{\rm K}}\right]\;,
\end{equation}
where $Z$ is the partition function. In general, the free energy consists of a contribution of the host electrons and of an impurity-induced part,
\begin{equation}
\mathcal{F}^{\rm aSIKM}(B,T,J_\perp,J_z)=\mathcal{F}^{\rm h}(B,T)+\mathcal{F}^{\rm ii}(B,T,J_\perp,J_z)\;,
\end{equation} 
where
\begin{equation}
    \mathcal{F}^{\rm h}(B,T)=-T\ln{\rm Tr}\left[e^{-\beta\hat{H}_0}\right]\;.
\end{equation}
Note that
\begin{eqnarray}\label{H0PartitionFunction}
Z_0&=&{\rm Tr}\left[e^{-\beta\hat{H}_0}\right]\nonumber\\
&=&{\rm Tr}_F\left[e^{-\beta\left(\hat{T}-B\left(\hat{N}_\uparrow-\hat{N}_\downarrow\right)\right)}\right]2\cosh(B/T)\;.
\end{eqnarray}
Here, ${\rm Tr}_F$ means that we perform the trace only over the bath states.

We focus on the impurity-induced contribution to the free energy, whereby we drop the index `${\rm ii}$' in the following to simplify the notation,
\begin{equation}
\mathcal{F}^{\rm ii}(B,T,J_\perp,J_z)\equiv \mathcal{F}(B,T,J_\perp,J_z)\;.
\end{equation}

\subsubsection{Impurity-induced zero-field susceptibility}
\label{def:zero-field-suscep}
The impurity-induced zero-field susceptibility is defined as the second derivative of the impurity-induced free energy~\cite{PhysRevB.101.075132}
\begin{eqnarray}\label{defzerofield}
\frac{\chi_0^{\rm ii}(B,T,J_\perp,J_z)}{(g_e\mu_{\rm B})^2}=-\frac{1}{4}\frac{\partial^2}{\partial B^2}\mathcal{F}(B,T,J_\perp,J_z)\;.
\end{eqnarray}
For the zero-field susceptibility, we need to know the free energy up to and including second order in $B$.

\section{Perturbation theory at finite temperatures}
\label{thePT}
In this section, we derive the impurity-induced free energy of the anisotropic single-impurity Kondo model  
from weak-coupling perturbation theory up to and including third order in the Kondo coupling 
at low temperatures and weak magnetic fields, $B\ll T\ll1$.

\subsection{Formal expansion}
\label{subsec:FormalExpansion}

We define
\begin{equation}\label{TauDependenceOfOperators}
\hat{A}(\tau)=e^{\tau\hat{H}_0}\hat{A}e^{-\tau\hat{H}_0} \; .
\end{equation}
For our purposes, where $\hat{A}$ does not change the total magnetization, we have
\begin{equation}
\hat{A}(\tau)=e^{\tau\hat{T}}\hat{A}e^{-\tau\hat{T}}
\end{equation} 
because $[\hat{A},\hat{n}^d_\Uparrow-\hat{n}^d_\Downarrow+\hat{N}_\uparrow-\hat{N}_\downarrow]_-=0$. Moreover, we define
\begin{equation}\label{DefinitionLangleRangle}
\langle\hat{A}\rangle=\frac{1}{Z_0}{\rm Tr}\left(e^{-\beta\hat{H}_0}\hat{A}\right)\;.
\end{equation}
At small couplings we find the partition function
\begin{equation}
Z=Z_0(1-V_1+V_2-V_3)
\label{eq:DefinitionPartitionFunctionAtSmallCouplings}
\end{equation}
so that the impurity-induced contribution to the free energy can be written as a power series in the coupling,
\begin{eqnarray}\label{Formal-Expansion}
\mathcal{F}&=&\mathcal{F}^{(0)}+\mathcal{F}^{(1)}+\mathcal{F}^{(2)}+\mathcal{F}^{(3)}
\end{eqnarray}
with
\begin{eqnarray}\label{F0-F3-Definitions}
\mathcal{F}^{(0)}&=&-T\ln(2\cosh(\beta B))\;,\nonumber\\
\mathcal{F}^{(1)}&=&TV_1\;,\nonumber\\
\mathcal{F}^{(2)}&=&-T V_2+\frac{T}{2}V_1^2\;,\nonumber\\
\mathcal{F}^{(3)}&=&TV_3-TV_1 V_2+\frac{T}{3}V_1^3\; , 
\end{eqnarray}
up to and including the third order in the coupling constant. The individual terms read
\begin{eqnarray}\label{V1-V2-V3-Definition}
V_1&=&\int_0^\beta\rmd\tau\,\langle\hat{V}(\tau)\rangle\;,\nonumber\\
V_2&=&\int_0^\beta\rmd\tau_2\int_0^{\tau_2}\rmd\tau_1\,\langle\hat{V}(\tau_2)\hat{V}(\tau_1)\rangle\;,\nonumber\\
V_3&=&\int_0^\beta\rmd\tau_3\,\int_0^{\tau_3}\rmd \tau_2\,\int_0^{\tau_2}\rmd \tau_1\,\langle\hat{V}(\tau_3)\hat{V}(\tau_2)\hat{V}(\tau_1)\rangle\;.\nonumber\\
\end{eqnarray}
In the following, we separately discuss the individual orders of the perturbation theory.

\subsection{Leading order term}
\label{subsec:formalleadingorderterm}

{}From Eq.~(\ref{F0-F3-Definitions}) we find the leading order term
\begin{eqnarray}
\label{FleadingOrder}
\mathcal{F}^{(0)}&=&-T\ln(2\cosh(B/T))\nonumber\\
&=&-T\ln 2-\frac{B^2}{2T}+\mathcal{O}[B^3]
\end{eqnarray}
for $B\ll T\ll1$.

\subsection{First order term}
\label{FirstOrderTermGeneral}

The integrand in the first order term from Eq.~(\ref{V1-V2-V3-Definition}) can be written as
\begin{eqnarray}
\langle\hat{V}(\tau)\rangle&=&\frac{1}{Z_0}{\rm Tr}\left[e^{-\beta\hat{H}_0}e^{\tau\hat{T}}\hat{V}e^{-\tau\hat{T}}\right]\nonumber\\
&=&\frac{1}{Z_0}{\rm Tr}\left[e^{-\beta\hat{H}_0}\hat{V}\right]=\langle\hat{V}\rangle\;,
\end{eqnarray}
where we used the fact that the trace is cyclic and that $[\hat{H}_0,\hat{T}]=0$. Moreover, ${\rm Tr}$ is the trace over states with definite spin orientation into the $z$-direction. 
This implies that we need an even number of spin-flip operators within the trace, see Eq.~(\ref{SsndsOperators}), to obtain a finite result.
For example, for an odd number of spin-flip operators, we see for an expectation value of impurity operators
\begin{equation}
\langle\Uparrow|\hat{S}^+|\Uparrow\rangle=\langle\Downarrow|\hat{S}^-|\Downarrow\rangle=0\; .
\end{equation}
Hence, 
\begin{equation}
\langle\hat{V}(\tau)\rangle=\langle\hat{V}\rangle=\langle\hat{V}_z\rangle \; .
\end{equation}
With 
\begin{equation}
    \hat{M}=\frac{1}{L}\sum_{k,k'}\left(\hat{a}^+_{k,\uparrow}\hat{a}^\phan_{k',\uparrow}-\hat{a}^+_{k,\downarrow}\hat{a}^\phan_{k,\downarrow}\right)\;,
\end{equation}
we find
\begin{eqnarray}
V_1&=&\int_0^\beta\rmd\tau\,\langle\hat{V}(\tau)\rangle\nonumber\\
&=&\frac{1}{T}\langle\hat{V}_z\rangle\nonumber\\
&=&\frac{J_z}{4T}\frac{1}{Z_0}{\rm Tr}\left[e^{-\beta\hat{H}_0}\left(\hat{n}^d_\Uparrow-\hat{n}^d_\Downarrow\right)\hat{M}\right]\nonumber\\
&=&\frac{J_z}{4T}\frac{2\sinh(\beta B)}{2\cosh(\beta B)}\langle\hat{M}\rangle\nonumber\\
&=&\frac{J_z}{4T}\tanh(\beta B)M_0(B,T)\; ,
\end{eqnarray}
where we defined
\begin{eqnarray}\label{M0definition}
M_0(B,T)&=&\langle\hat{M}\rangle=\frac{1}{L}\left(N_\uparrow(B,T)-N_\downarrow(B,T)\right)\nonumber\\
&=&\int_{-1}^1\rmd \epsilon\,\rho_0(\epsilon)\left(\frac{1}{1+e^{\beta(\epsilon-B)}}-\frac{1}{1+e^{\beta(\epsilon+B)}}\right)\nonumber\\
\end{eqnarray}
because
\begin{eqnarray}
n_{k,\uparrow}\equiv\langle\hat{n}_{k,\uparrow}\rangle&=&\frac{1}{1+e^{\beta(\epsilon(k)-B)}}\;,\nonumber\\
n_{k,\downarrow}\equiv\langle\hat{n}_{k,\downarrow}\rangle&=&\frac{1}{1+e^{\beta(\epsilon(k)+B)}}
\end{eqnarray}
hold for the non-interacting host electrons.

At low temperatures and small fields, $B\ll T\ll1$, we have
\begin{eqnarray}\label{M0forsmallB}
\tanh(B/T)&\approx& \frac{B}{T}\;,\nonumber\\
M_0(B,T)&\approx& \frac{2B}{T}\int_{-\infty}^{\infty}\rmd \epsilon\,\frac{\rho_0(\epsilon)}{\left[2\cosh\left(\epsilon/(2T)\right)\right]^2}\nonumber\\
&\approx& \frac{2B}{T}\rho_0(0)T=2B\rho_0(0)\;.
\end{eqnarray}
We define $j_z=J_z\rho_0(0)$ and use Eq.~(\ref{F0-F3-Definitions}) to find the first order contribution
\begin{equation}\label{FirstOrderTerm}
V_1\approx\frac{B^2}{2T^2}j_z\;,\quad \mathcal{F}^{(1)}=TV_1\approx \frac{B^2}{2T}j_z\;.
\end{equation}

\subsection{Second-order term}
\label{subsec:formalsecondorderterm}

According to Eq.~(\ref{F0-F3-Definitions}), the second-order term reads
\begin{equation}
\mathcal{F}^{(2)}=-T V_2+\frac{T}{2}V_1^2\;,
\end{equation}
with $V_1$ and $V_2$ from Eq.~(\ref{V1-V2-V3-Definition}). With our results of the first order, see Eq.~(\ref{FirstOrderTerm}), we find
\begin{equation}
\frac{T}{2}V_1^2=\frac{T}{2}\frac{B^4}{4T^4}j_z^2=\frac{B^4}{8T^3}j_z^2\sim \mathcal{O}[B^4]\;,
\end{equation}
so we can ignore this contribution at small fields. 
We are left with
\begin{eqnarray}\label{secondorderfreeenergy}
\mathcal{F}^{(2)}&=&-T V_2+\mathcal{O}[B^3].
\end{eqnarray}
With Eq.~(\ref{V1-V2-V3-Definition}), we have
\begin{eqnarray}
V_2&=&\int_0^\beta\rmd \lambda_2\,\int_0^{\lambda_2}\rmd \lambda_1\,\left\langle\hat{V}(\lambda_2)\hat{V}(\lambda_1)\right\rangle\nonumber\\
&=&\int_0^\beta\rmd\lambda_2\,\int_0^{\lambda_2}\rmd\lambda_1\,\frac{1}{Z_0}\nonumber\\
&&\times{\rm Tr}\left[e^{-\beta\hat{H}_0}e^{\lambda_2\hat{T}}\hat{V}e^{(-\lambda_2+\lambda_1)\hat{T}}\hat{V}e^{-\lambda_1\hat{T}}\right]\nonumber\\
&=&\int_0^\beta\rmd\lambda_2\,\int_0^{\lambda_2}\rmd \lambda_1\,\frac{1}{Z_0}\nonumber\\
&&\times{\rm Tr}\left[e^{-\beta\hat{H}_0}\hat{V}e^{(\lambda_1-\lambda_2)\hat{T}} \hat{V}e^{-(\lambda_1-\lambda_2)\hat{T}}\right]\;,
\end{eqnarray}
where made use of the fact that the trace is cyclic. The substitution $\lambda=\lambda_2-\lambda_1$ then leads to
\begin{eqnarray}
V_2&=&\int_0^\beta\rmd \lambda\,\left\langle\hat{V}\hat{V}(-\lambda)\right\rangle\int_\lambda^\beta\rmd\lambda_2\nonumber\\
&=&\int_0^\beta\rmd\lambda\,(\beta-\lambda)\left\langle\hat{V}\hat{V}(-\lambda)\right\rangle\;.
\end{eqnarray}
Now we use the fact that the trace runs over definite spin states, see~section~\ref{FirstOrderTermGeneral},
\begin{eqnarray}\label{TheTwoPartsOfTheSecondOrder}
V_2&=&\int_0^\beta\rmd\lambda\,(\beta-\lambda)\left\langle\hat{V}_z\hat{V}_z(-\lambda)\right\rangle\nonumber\\
&&\quad+\int_0^\beta\rmd\lambda\,(\beta-\lambda)\left\langle\hat{V}_\perp\hat{V}_\perp(-\lambda)\right\rangle\nonumber\\
&=&V_2^z+V_2^\perp\;.
\end{eqnarray}

The calculations of $V_2^z$ and $V_2^\perp$ are rather lengthy, and thus moved to the Appendix. In Appendix~\ref{appendix:secondorderterms} we show that
\begin{eqnarray}\label{V2secondorder}
V_2^z&=&-\frac{j_z^2}{8T}\left[-2\ln 2+\frac{\pi^2}{3}T^2+B^2-\frac{B^2}{T}\right]\;,\nonumber\\
V_2^\perp&=&-\frac{j_\perp^2}{4T}\biggl[-2\ln 2 +\frac{\pi^2T^2}{3}\nonumber\\
&&\qquad+B^2\left(1-\frac{2}{T}\left(-1-\EulerGamma+\ln(2\pi T)\right)\right)\biggr]\;.\nonumber\\
\end{eqnarray}
Note that $V_2^z$ is the second-order Ising-Kondo contribution to the single-impurity Kondo model, see Ref.~\cite{IsingKondo}. 
In Sect.~\SMFreeEnergyIsingKondo\ of the Supplemental Material, we explicitly calculate the exact result of the free energy of the Ising-Kondo model 
for a constant density of states, see Eq.~(\ref{rhoconst}). 
In this way, we can compare the results from perturbation theory of Appendix~\ref{appendix:secondorderterms} with those from the Taylor series of the exact result. 
Evidently, the two results agree.

Hence, with Eq.~(\ref{secondorderfreeenergy}) the second order of the impurity-induced free energy reads
\begin{eqnarray}
\label{F2ndorder}
\mathcal{F}^{(2)}&=&\frac{1}{8}\left(-2\ln 2+\frac{\pi^2T^2}{3}\right)j_z^2+\frac{B^2}{8}j_z^2-\frac{B^2}{8T}j_z^2\nonumber\\
&&+\frac{1}{4}\left(-2\ln 2+\frac{\pi^2T^2}{3}\right)j_\perp^2+\frac{B^2}{4}j_\perp^2\nonumber\\
&&+\frac{B^2}{2T}j_\perp^2\left(1+\EulerGamma-\ln(2\pi T)\right)
\end{eqnarray}
with $j_\perp=J_\perp\rho_0(0)$. This result agrees with our previous calculation~\cite{PhysRevB.101.075132}.

\subsection{Third-order term}
\label{formalthirdorderterm}

According to Eq.~(\ref{F0-F3-Definitions}), the third order is given by
\begin{equation}
\mathcal{F}^{(3)}=TV_3-TV_1V_2+\frac{T}{3}V_1^3\;,
\end{equation}
where $V_1$ and $V_2$ were determined in eqs.~(\ref{FirstOrderTerm}),~(\ref{TheTwoPartsOfTheSecondOrder}) and~(\ref{V2secondorder}). 
Since $V_1\propto B^2/T^2$, we see that $(T/3) V_1^3\sim B^6/T^5$ which is negligible for $B\ll T\ll1$. In this limit, the remaining contribution is
\begin{equation}
\label{thirdorder-definition-up-to-Bsquared}
\mathcal{F}^{(3)}= TV_3-TV_1V_2(B=0)\;,
\end{equation}
where $T V_1 V_2(B=0)$ is known, and we may focus on $V_3$. 
In the second term we only need $V_2(B=0)$ because $V_1$ is already second order in~$B$, see Eq.~(\ref{FirstOrderTerm}). From Eq.~(\ref{V1-V2-V3-Definition}) we find
\begin{eqnarray}\label{RemainingpartsofV3}
V_3&=&\int_0^\beta\rmd\lambda_3\,\int_0^{\lambda_3}\rmd\lambda_2\,\int_0^{\lambda_2}\rmd\lambda_1\,\frac{1}{Z_0}\nonumber\\
&&\times{\rm Tr}\left[e^{-\beta\hat{H}_0}\hat{V}(\lambda_3)\hat{V}(\lambda_2)\hat{V}(\lambda_1)\right]\nonumber\\
&=&\int_0^\beta\rmd\lambda_3\,\int_0^{\lambda_3}\rmd\lambda_2\,\int_0^{\lambda_2}\rmd\tau_1\,\frac{1}{Z_0}\nonumber\\
&&\times{\rm Tr}\bigg[
e^{-\beta\hat{H}_0}\left(\hat{V}_z(\lambda_3)+\hat{V}_\perp(\lambda_3)\right)\nonumber\\
&& \qquad\times \left(\hat{V}_z(\lambda_2)+\hat{V}_\perp(\lambda_2)\right)\left(\hat{V}_z(\lambda_1)+\hat{V}_\perp(\lambda_1)\right)
\bigg]\nonumber\\
&=&V_3^z+V_3^a+V_3^b+V_3^c
\end{eqnarray}
with the four contributions
\begin{eqnarray}\label{V3z-V3a-V3b-V3c-Definition}
V_3^z&=&\int_0^\beta\rmd\lambda_3\,\int_0^{\lambda_3}\rmd\lambda_2\,\int_0^{\lambda_2}\rmd\lambda_1\,\frac{1}{Z_0}\nonumber\\
&&\times{\rm Tr}\left[e^{-\beta\hat{H}_0} \hat{V}_z(\lambda_3)\hat{V}_z(\lambda_2)\hat{V}_z(\lambda_1)\right]\;,\\[3pt]
%
V_3^a&=&\int_0^\beta\rmd\lambda_3\,\int_0^{\lambda_3}\rmd\lambda_2\,\int_0^{\lambda_2}\rmd\lambda_1\,\frac{1}{Z_0}\nonumber\\
&&\times{\rm Tr}\left[e^{-\beta\hat{H}_0} \hat{V}_z(\lambda_3)\hat{V}_\perp(\lambda_2)\hat{V}_\perp(\lambda_1)\right]\;,\\
V_3^b&=&\int_0^\beta\rmd\lambda_3\,\int_0^{\lambda_3}\rmd\lambda_2\,\int_0^{\lambda_2}\rmd\lambda_1\,\frac{1}{Z_0}\nonumber\\
&&\times{\rm Tr}\left[e^{-\beta\hat{H}_0}\hat{V}_\perp(\lambda_3)\hat{V}_z(\lambda_2)\hat{V}_\perp(\lambda_1) \right]\; ,\\
V_3^c&=&\int_0^\beta\rmd\lambda_3\,\int_0^{\lambda_3}\rmd\lambda_2\,\int_0^{\lambda_2}\rmd\lambda_1\,\frac{1}{Z_0}\nonumber\\
&&\times{\rm Tr}\left[e^{-\beta\hat{H}_0} \hat{V}_\perp(\lambda_3)\hat{V}_\perp(\lambda_2)\hat{V}_z(\lambda_1)\right]\;.
\end{eqnarray}
Again, we used the fact that the trace runs over definite spin states. Therefore, we need an even number of spin-flip operations within the trace.

The Ising-Kondo contribution $V_z^{(3)}$ is known from the exact result, see Sect.~\SMFreeEnergyIsingKondo\ of the Supplemental Material, ($V_1^z\equiv V_1$)
\begin{equation}
\label{3rdorderIsing-Kondo-Vz}
V_3^z-V_1V_2^z+\frac{V_1^3}{3}=-\frac{\pi^2}{96}\frac{B^2}{T^2}j_z^3
\end{equation}
and thus
\begin{equation}\label{IsingKondoThirdOrder}
V_3^z=B^2j_z^3\left(\frac{\ln 2}{8 T^3}-\frac{\pi^2}{96 T^2}-\frac{\pi^2}{48 T}\right)+\mathcal{O}[B^3]
\end{equation}
with $j_z=J_z\rho_0(0)$.

The calculation of the terms $V_3^i$ in Eq.~(\ref{RemainingpartsofV3}), $i=a,b,c$, are very lengthy, 
so we defer their analysis to Appendix~\ref{appendix:thirdorderterms}, and simply state the result of the third-order contribution to the free energy,
\begin{eqnarray}
\label{FreeEnergyThirdOrderResult}
\mathcal{F}^{(3)}&=&j_\perp^2 j_z \biggl[-\frac{12\ln(2)^2+\pi^2}{16}+\frac{\pi^2 T}{4}\nonumber\\
&&\qquad\quad+\frac{\pi^2 T^2}{4}(-1+\ln 2)+\mathcal{O}[B^0 T^3]\biggr]\nonumber\\
&&+B^2j_\perp^2 j_z\left(\frac{3}{4}\left(-2+\ln 2\right)+\mathcal{O}[T]\right)\nonumber\\
&&-\frac{B^2}{T}j_z^3\frac{\pi^2}{96}+\mathcal{O}[j_z^3 B^2 T]\nonumber\\
&&+\frac{B^2}{2T}j_\perp^2j_z\biggl[1+\EulerGamma+\EulerGamma^2-\frac{\pi^2}{24}+\frac{\ln 2}{2}\nonumber\\
&&\qquad\qquad\quad+\ln^2(2)+ 2\ln 2\ln \pi+\ln^2(\pi)\nonumber\\
&&\qquad\qquad\quad -\ln(2\pi T)(1+2\EulerGamma)+2\ln(2\pi)\ln T\nonumber\\
&&\qquad\qquad\quad+\ln^2(T)
\biggr]+\mathcal{O}[B^3]\:.
\end{eqnarray}

\section{Zero-field susceptibility}
\label{zerofieldsuscep}

In this section, we determine the zero-field susceptibility and compare our result with the series expansion of the zero-field susceptibility determined by Wilson using his renormalization group~\cite{RevModPhys.47.773}. From now on we consider the isotropic single-impurity Kondo model,
\begin{equation}
    J_z\equiv J_{\perp}\equiv J_{\rm K}\quad\Leftrightarrow\quad j_z=j_\perp=j\;,
\end{equation}
where $j=J_{\rm K}\rho_0(0)$.

\subsection{Calculation}

Using the definition of the zero-field susceptibility~(\ref{defzerofield}) and eqs.~(\ref{Formal-Expansion}),~(\ref{FleadingOrder}),~(\ref{FirstOrderTerm}),~(\ref{F2ndorder}) 
and~(\ref{FreeEnergyThirdOrderResult}), we find
\begin{eqnarray}
\frac{\chi_0^{\rm ii}}{(g_e\mu_{\rm B})^2}&=&\chi_0^{(0)}+\chi_0^{(1)}+\chi_0^{(2)}+\chi_0^{(3)} +\mathcal{O}[j^4]
\end{eqnarray}
with [$j_\perp=j_z=j=J_{\rm K}\rho_0(0)$]
\begin{eqnarray}\label{chi0terms}
\chi_0^{(0)}&=&\frac{1}{4 T}+\mathcal{O}[T^0,j^0]\;,\nonumber\\
\chi_0^{(1)}&=&-\frac{1}{4 T} j+\mathcal{O}[T^0,j^1]\;,\nonumber\\
\chi_0^{(2)}&=&\frac{1}{4 T} j^2\frac{1}{4}\left(-3-4\EulerGamma+4\ln(2\pi T)\right)+\mathcal{O}[T^0,j^2]\;,\nonumber\\
&=&\frac{1}{4 T} j^2\left(-\frac{3}{4}-\EulerGamma+\ln(2\pi T)\right)+\mathcal{O}[T^0,j^2]\;,\nonumber\\
\chi_0^{(3)}&=&\frac{1}{4T} j^3
\biggl[
-1-\EulerGamma-\EulerGamma^2+\frac{\pi^2}{16}-\frac{\ln 2}{2}-\ln^2(2)\nonumber\\
&&\quad\qquad -2\ln 2\ln \pi-\ln^2(\pi)+\ln(2\pi T)(1+2\EulerGamma)\nonumber\\
&&\quad\qquad -2\ln(2\pi)\ln T-\ln^2(T)
\biggr]+\mathcal{O}[T^0,j^3]\;.\nonumber \\
\end{eqnarray}

\subsection{Zero-field susceptibility from Wilson's renormalization group}

Wilson used his renormalization group to derive the zero-field susceptibility 
in terms of a series expansion, see p.~830 in Ref.~\cite{RevModPhys.47.773},  ($k=k_{\rm B}=1$)
\begin{eqnarray}\label{WilsonsChi0}
\chi_{0,{\rm W}}^{\rm ii}&=&\frac{1}{4T}\biggl[1+2J_{\rm W}\rho_0(0)+4(J_{\rm W}\rho_0(0))^2\ln\left(\frac{T}{\tilde{D}}\right)\nonumber\\
&&\qquad +(J_{\rm W}\rho_0(0))^3\left(8\ln^2\left(\frac{T}{\tilde{D}}\right)+4\ln\left(\frac{T}{\tilde{D}}\right)\right)
\biggr]\nonumber\\
\end{eqnarray}
with Wilson's coupling constant $J_{\rm W}$ and
\begin{equation}\label{tildeDExpansion}
\tilde{D}\equiv \tilde{D}(J_{\rm W}\rho_0(0))=D(c_0+c_1 J_{\rm W}\rho_0(0)+\mathcal{O}[j^2])\;,
\end{equation}
where $D$ is the band width; $D\equiv2$ in our work. 
Our goal is to determine analytical expressions for the first and second Wilson numbers $c_0$ and $c_1$.

\subsection{Comparison}

Now we are in the position to compare the individual orders in the coupling~$j$ between our zero-field susceptibility $\chi_0^{\rm ii}$ from Eq.~(\ref{chi0terms}) 
with those of Wilson's $\chi_{0,{\rm W}}^{\rm ii}$ from Eq.~(\ref{WilsonsChi0}). The leading orders, describing a free spin, naturally agree. From first order we identify
\begin{equation}
2J_{\rm W}\rho_0(0)\equiv - j=- J_{\rm K}\rho_0(0)\;.
\end{equation}
Therefore, we can rewrite Eq.~(\ref{WilsonsChi0}) in terms of our coupling parameter~$j$ in the form
\begin{eqnarray}\label{chi0withj}
\chi_{0,{\rm W}}^{\rm ii}&=&\frac{1}{4T}\biggl[1-j+j^2\ln\left(\frac{T}{\tilde{D}}\right)\nonumber\\
&&\qquad-j^3\left(\ln^2\left(\frac{T}{\tilde{D}}\right)+\frac{1}{2}\ln\left(\frac{T}{\tilde{D}}\right)\right)
\biggr]\;.
\end{eqnarray}
Furthermore, using Eq.~(\ref{tildeDExpansion}), we find that Wilson's constant term to second order can be written as
\begin{eqnarray}
-j^2\ln \tilde{D}&=&-j^2\ln\left(2 c_0-jc_1+\mathcal{O}[j^2]\right)\nonumber\\
&=&-j^2\left[\ln(2c_0)+\ln\left(1-\frac{jc_1}{2c_0}+\mathcal{O}[j^2]\right)\right]\nonumber\\
&=&-j^2\left[\ln(2c_0)-j\frac{c_1}{2c_0}+\mathcal{O}[j^2]\right]\nonumber\\
&=&-j^2\ln(2c_0)+j^3\frac{c_1}{2c_0}+\mathcal{O}[j^4]\;,
\end{eqnarray}
where we used that $j\ll 1$ in the next to last step. Thus, we have 
\begin{equation}
-\ln \tilde{D}=-\ln(2c_0)+j\frac{c_1}{2c_0}+\mathcal{O}[j^2]\;.
\end{equation}
Using this expression, we obtain for Eq.~(\ref{chi0withj}) up to and including the third order in $j$
\begin{equation}\label{chi0WilsonForComparison}
\chi_{0,{\rm W}}^{\rm ii}=\chi_{0,{\rm W}}^{(0)}+\chi_{0,{\rm W}}^{(1)}+\chi_{0,{\rm W}}^{(2)}+\chi_{0,{\rm W}}^{(3)}+\mathcal{O}[j^4]
\end{equation}
with
\begin{eqnarray}
\chi_{0,{\rm W}}^{(0)}&=&\frac{1}{4 T}\;,\nonumber\\
\chi_{0,{\rm W}}^{(1)}&=&-\frac{1}{4T} j\;,\nonumber\\
\chi_{0,{\rm W}}^{(2)}&=&\frac{1}{4T}j^2\left( \ln T- \ln(2 c_0)\right)\; , \nonumber \\
\chi_{0,{\rm W}}^{(3)}&=&-\frac{1}{4T}j^3\bigg(-\frac{c_1}{2c_0}+
\ln^2(T)-2\ln T\ln(2c_0)\nonumber\\
&&\qquad\qquad+\ln^2(2c_0)
+\frac{\ln T}{2}-\frac{\ln(2c_0)}{2}\bigg)
\,.
\end{eqnarray}
The leading and first-order terms are seen to agree explicitly, as discussed before. From the comparison of eqs.~(\ref{chi0terms}) and~(\ref{chi0WilsonForComparison})
to second order we can determine the first Wilson number~$c_0$,
\begin{eqnarray}
\chi_0^{(2)} &=& \frac{1}{4 T} j^2\frac{1}{4}\left(-3-4\EulerGamma+4\ln(2\pi T)\right)\nonumber\\
\chi_{0,{\rm W}}^{(2)}&=&\frac{1}{4T} j^2\left(\ln T- \ln(2 c_0)\right)
\end{eqnarray}
which leads to
%
%
\begin{eqnarray}
\ln(2c_0)&=&\frac{3}{4}+\EulerGamma-\ln(2\pi)\approx -0.510661 \; , \nonumber \\
c_0&=&\frac{e^{3/4+\EulerGamma}}{4\pi}\approx 0.300049\;.
\end{eqnarray}
This result was obtained earlier, e.g., in Refs.~\cite{AFL,PhysRevB.101.075132}.
Frequently, the number $w= 4c_0/(\sqrt{\pi e}) \approx 0.410705$ is denoted `the' Wilson number, see Eq.~(\ref{eq:wilsonnumberintro}).

Finally, from third order,
\begin{eqnarray}
\chi_0^{(3)} &=& \frac{1}{4T} j^3
\bigg(
-1-\EulerGamma-\EulerGamma^2+\frac{\pi^2}{16}-\frac{\ln 2}{2}-\ln^2(2)\nonumber\\
&&\qquad\quad -2\ln 2\ln \pi-\ln^2(\pi)+\ln(2\pi T)[1+2\EulerGamma]\nonumber\\
&&\qquad\quad -2\ln(2\pi)\ln T-\ln^2(T)
\bigg)\nonumber \; , \\
\chi_{0,{\rm W}}^{(3)}&=&-\frac{1}{4T}j^3\bigg(-\frac{c_1}{2c_0}+
\ln^2(T)-2\ln T \ln(2c_0)\nonumber\\
&&\qquad\qquad+\ln^2(2c_0)
+\frac{\ln T}{2}-\frac{\ln(2c_0)}{2}\bigg)\;,
\end{eqnarray}
we obtain the second Wilson number
\begin{eqnarray}
\frac{c_1}{2 c_0}&=&\frac{-13+\pi^2}{16}-\frac{\ln 2}{2}\approx -0.542223 \; , \\
c_1&=&\frac{e^{3/4+\EulerGamma}}{32\pi}\left(-13+\pi^2-8\ln 2\right)\approx -0.325387\;.\nonumber
\end{eqnarray}

\section{Final remarks}
\label{sec:conclusions}

As seen from the analytical structure of the perturbation expansion in Eq.~(\ref{WilsonsChi0}),
the theory cannot be applied at very low temperatures because of the logarithmic divergences.
Eq.~(\ref{eq:zerofieldintro}) shows that the 
zero-field susceptibility at zero temperature to first order in~$j$
also requires the coefficient $\alpha_1$ that is not within reach using 
our perturbation theory approach. It was determined numerically by Wilson~\cite{RevModPhys.47.773}, and also in Ref.~\cite{Barczaetal}.
Unfortunately, we do not have a recipe to determine $\alpha_1$ analytically. Our attempts to use perturbation theory
at zero temperature and finite magnetic field failed because we were not able to sum the contributions 
proportional to $j\ln(B)$.

We also tried to make contact with the Bethe Ansatz solution at finite temperatures. 
The leading-order corrections were extracted from the Takahashi equations~\cite{RevModPhys.55.331,TsvelickWiegmann},
see Refs.~\cite{Hewson,PhysRevB.101.075132} for a short review. Bortz and Kl\"umper~\cite{BortzKluemper} succeeded to derive
the free energy at finite fields from two coupled integral equations that can be treated numerically, and confirmed the earlier results to second order
in the Kondo coupling. Unfortunately, we did not succeed to derive a systematic series expansion of the Kl\"umper equations 
for weak coupling so that we cannot compare Bethe Ansatz results with those from our third-order perturbation theory. 
Such a comparison would permit to identify the parameters in the Bethe Ansatz treatment with those of the lattice model.

\bibliography{kondo}

\begin{thebibliography}{20}%
\makeatletter
\providecommand \@ifxundefined [1]{%
 \@ifx{#1\undefined}
}%
\providecommand \@ifnum [1]{%
 \ifnum #1\expandafter \@firstoftwo
 \else \expandafter \@secondoftwo
 \fi
}%
\providecommand \@ifx [1]{%
 \ifx #1\expandafter \@firstoftwo
 \else \expandafter \@secondoftwo
 \fi
}%
\providecommand \natexlab [1]{#1}%
\providecommand \enquote  [1]{``#1''}%
\providecommand \bibnamefont  [1]{#1}%
\providecommand \bibfnamefont [1]{#1}%
\providecommand \citenamefont [1]{#1}%
\providecommand \href@noop [0]{\@secondoftwo}%
\providecommand \href [0]{\begingroup \@sanitize@url \@href}%
\providecommand \@href[1]{\@@startlink{#1}\@@href}%
\providecommand \@@href[1]{\endgroup#1\@@endlink}%
\providecommand \@sanitize@url [0]{\catcode `\\12\catcode `\$12\catcode
  `\&12\catcode `\#12\catcode `\^12\catcode `\_12\catcode `\%12\relax}%
\providecommand \@@startlink[1]{}%
\providecommand \@@endlink[0]{}%
\providecommand \url  [0]{\begingroup\@sanitize@url \@url }%
\providecommand \@url [1]{\endgroup\@href {#1}{\urlprefix }}%
\providecommand \urlprefix  [0]{URL }%
\providecommand \Eprint [0]{\href }%
\providecommand \doibase [0]{https://doi.org/}%
\providecommand \selectlanguage [0]{\@gobble}%
\providecommand \bibinfo  [0]{\@secondoftwo}%
\providecommand \bibfield  [0]{\@secondoftwo}%
\providecommand \translation [1]{[#1]}%
\providecommand \BibitemOpen [0]{}%
\providecommand \bibitemStop [0]{}%
\providecommand \bibitemNoStop [0]{.\EOS\space}%
\providecommand \EOS [0]{\spacefactor3000\relax}%
\providecommand \BibitemShut  [1]{\csname bibitem#1\endcsname}%
\let\auto@bib@innerbib\@empty
\bibitem [{\citenamefont {de~Haas}\ \emph {et~al.}(1934)\citenamefont
  {de~Haas}, \citenamefont {de~Boer},\ and\ \citenamefont {van~den
  Berg}}]{deHaas}%
  \BibitemOpen
  \bibfield  {author} {\bibinfo {author} {\bibfnamefont {W.~J.}\ \bibnamefont
  {de~Haas}}, \bibinfo {author} {\bibfnamefont {J.}~\bibnamefont {de~Boer}},\
  and\ \bibinfo {author} {\bibfnamefont {G.~J.}\ \bibnamefont {van~den Berg}},\
  }\bibfield  {title} {\bibinfo {title} {The electrical resistance of gold,
  copper and lead at low temperatures},\ }\href@noop {} {\bibfield  {journal}
  {\bibinfo  {journal} {Physica}\ }\textbf {\bibinfo {volume} {1}},\ \bibinfo
  {pages} {1115} (\bibinfo {year} {1934})}\BibitemShut {NoStop}%
\bibitem [{\citenamefont {Zener}(1951)}]{Zener}%
  \BibitemOpen
  \bibfield  {author} {\bibinfo {author} {\bibfnamefont {C.}~\bibnamefont
  {Zener}},\ }\bibfield  {title} {\bibinfo {title} {Interaction between the $d$
  shells in the transition metals},\ }\href@noop {} {\bibfield  {journal}
  {\bibinfo  {journal} {Phys.~Rev.}\ }\textbf {\bibinfo {volume} {81}},\
  \bibinfo {pages} {440} (\bibinfo {year} {1951})}\BibitemShut {NoStop}%
\bibitem [{\citenamefont {Borzenets}\ \emph {et~al.}(2020)\citenamefont
  {Borzenets}, \citenamefont {Shim}, \citenamefont {Chen}, \citenamefont
  {Ludwig}, \citenamefont {Wieck}, \citenamefont {Tarucha},\ and\ \citenamefont
  {Yamamoto}}]{Borzenets}%
  \BibitemOpen
  \bibfield  {author} {\bibinfo {author} {\bibfnamefont {I.~V.}\ \bibnamefont
  {Borzenets}}, \bibinfo {author} {\bibfnamefont {J.}~\bibnamefont {Shim}},
  \bibinfo {author} {\bibfnamefont {J.~C.~H.}\ \bibnamefont {Chen}}, \bibinfo
  {author} {\bibfnamefont {A.}~\bibnamefont {Ludwig}}, \bibinfo {author}
  {\bibfnamefont {A.~D.}\ \bibnamefont {Wieck}}, \bibinfo {author}
  {\bibfnamefont {S.}~\bibnamefont {Tarucha}},\ and\ \bibinfo {author}
  {\bibfnamefont {M.}~\bibnamefont {Yamamoto}},\ }\bibfield  {title} {\bibinfo
  {title} {{Observation of the Kondo screening cloud}},\ }\href@noop {}
  {\bibfield  {journal} {\bibinfo  {journal} {Nature}\ }\textbf {\bibinfo
  {volume} {579}},\ \bibinfo {pages} {210} (\bibinfo {year}
  {2020})}\BibitemShut {NoStop}%
\bibitem [{\citenamefont {Kasuya}(1956)}]{Kasuya}%
  \BibitemOpen
  \bibfield  {author} {\bibinfo {author} {\bibfnamefont {T.}~\bibnamefont
  {Kasuya}},\ }\bibfield  {title} {\bibinfo {title} {{A Theory of Metallic
  Ferro- and Antiferromagnetism on Zener's Model}},\ }\href@noop {} {\bibfield
  {journal} {\bibinfo  {journal} {Progress of Theoretical Physics}\ }\textbf
  {\bibinfo {volume} {16}},\ \bibinfo {pages} {45} (\bibinfo {year}
  {1956})}\BibitemShut {NoStop}%
\bibitem [{\citenamefont {Kondo}(1964)}]{Kondo1964}%
  \BibitemOpen
  \bibfield  {author} {\bibinfo {author} {\bibfnamefont {J.}~\bibnamefont
  {Kondo}},\ }\bibfield  {title} {\bibinfo {title} {{Resistance Minimum in
  Dilute Magnetic Alloys}},\ }\href@noop {} {\bibfield  {journal} {\bibinfo
  {journal} {Progress of Theoretical Physics}\ }\textbf {\bibinfo {volume}
  {32}},\ \bibinfo {pages} {37} (\bibinfo {year} {1964})}\BibitemShut {NoStop}%
\bibitem [{\citenamefont {Wilson}(1975)}]{RevModPhys.47.773}%
  \BibitemOpen
  \bibfield  {author} {\bibinfo {author} {\bibfnamefont {K.~G.}\ \bibnamefont
  {Wilson}},\ }\bibfield  {title} {\bibinfo {title} {The renormalization group:
  {C}ritical phenomena and the {K}ondo problem},\ }\href@noop {} {\bibfield
  {journal} {\bibinfo  {journal} {Rev. Mod. Phys.}\ }\textbf {\bibinfo {volume}
  {47}},\ \bibinfo {pages} {773} (\bibinfo {year} {1975})}\BibitemShut
  {NoStop}%
\bibitem [{\citenamefont {Krishna-murthy}\ \emph
  {et~al.}(1980{\natexlab{a}})\citenamefont {Krishna-murthy}, \citenamefont
  {Wilkins},\ and\ \citenamefont {Wilson}}]{PhysRevB.21.1003}%
  \BibitemOpen
  \bibfield  {author} {\bibinfo {author} {\bibfnamefont {H.~R.}\ \bibnamefont
  {Krishna-murthy}}, \bibinfo {author} {\bibfnamefont {J.~W.}\ \bibnamefont
  {Wilkins}},\ and\ \bibinfo {author} {\bibfnamefont {K.~G.}\ \bibnamefont
  {Wilson}},\ }\bibfield  {title} {\bibinfo {title} {{Renormalization-group
  approach to the Anderson model of dilute magnetic alloys. I. Static
  properties for the symmetric case}},\ }\href@noop {} {\bibfield  {journal}
  {\bibinfo  {journal} {Phys. Rev. B}\ }\textbf {\bibinfo {volume} {21}},\
  \bibinfo {pages} {1003} (\bibinfo {year} {1980}{\natexlab{a}})}\BibitemShut
  {NoStop}%
\bibitem [{\citenamefont {Krishna-murthy}\ \emph
  {et~al.}(1980{\natexlab{b}})\citenamefont {Krishna-murthy}, \citenamefont
  {Wilkins},\ and\ \citenamefont {Wilson}}]{PhysRevB.21.1044}%
  \BibitemOpen
  \bibfield  {author} {\bibinfo {author} {\bibfnamefont {H.~R.}\ \bibnamefont
  {Krishna-murthy}}, \bibinfo {author} {\bibfnamefont {J.~W.}\ \bibnamefont
  {Wilkins}},\ and\ \bibinfo {author} {\bibfnamefont {K.~G.}\ \bibnamefont
  {Wilson}},\ }\bibfield  {title} {\bibinfo {title} {{Renormalization-group
  approach to the Anderson model of dilute magnetic alloys. II. Static
  properties for the asymmetric case}},\ }\href@noop {} {\bibfield  {journal}
  {\bibinfo  {journal} {Phys. Rev. B}\ }\textbf {\bibinfo {volume} {21}},\
  \bibinfo {pages} {1044} (\bibinfo {year} {1980}{\natexlab{b}})}\BibitemShut
  {NoStop}%
\bibitem [{\citenamefont {Tsvelick}\ and\ \citenamefont
  {Wiegmann}(1983{\natexlab{a}})}]{TW}%
  \BibitemOpen
  \bibfield  {author} {\bibinfo {author} {\bibfnamefont {A.}~\bibnamefont
  {Tsvelick}}\ and\ \bibinfo {author} {\bibfnamefont {P.}~\bibnamefont
  {Wiegmann}},\ }\bibfield  {title} {\bibinfo {title} {Exact results in the
  theory of magnetic alloys},\ }\href@noop {} {\bibfield  {journal} {\bibinfo
  {journal} {Advances in Physics}\ }\textbf {\bibinfo {volume} {32}},\ \bibinfo
  {pages} {453} (\bibinfo {year} {1983}{\natexlab{a}})}\BibitemShut {NoStop}%
\bibitem [{\citenamefont {Andrei}\ \emph
  {et~al.}(1983{\natexlab{a}})\citenamefont {Andrei}, \citenamefont {Furuya},\
  and\ \citenamefont {Lowenstein}}]{AFL}%
  \BibitemOpen
  \bibfield  {author} {\bibinfo {author} {\bibfnamefont {N.}~\bibnamefont
  {Andrei}}, \bibinfo {author} {\bibfnamefont {K.}~\bibnamefont {Furuya}},\
  and\ \bibinfo {author} {\bibfnamefont {J.~H.}\ \bibnamefont {Lowenstein}},\
  }\bibfield  {title} {\bibinfo {title} {{Solution of the Kondo problem}},\
  }\href@noop {} {\bibfield  {journal} {\bibinfo  {journal} {Rev. Mod. Phys.}\
  }\textbf {\bibinfo {volume} {55}},\ \bibinfo {pages} {331} (\bibinfo {year}
  {1983}{\natexlab{a}})}\BibitemShut {NoStop}%
\bibitem [{\citenamefont {Hewson}(1993)}]{Hewson}%
  \BibitemOpen
  \bibfield  {author} {\bibinfo {author} {\bibfnamefont {A.}~\bibnamefont
  {Hewson}},\ }\href@noop {} {\emph {\bibinfo {title} {{\sl The Kondo Problem
  to Heavy Fermions}}}}\ (\bibinfo  {publisher} {Cambridge University Press},\
  \bibinfo {address} {Cambridge},\ \bibinfo {year} {1993})\BibitemShut
  {NoStop}%
\bibitem [{\citenamefont {Barcza}\ \emph {et~al.}(2019)\citenamefont {Barcza},
  \citenamefont {Gebhard}, \citenamefont {Linneweber},\ and\ \citenamefont
  {Legeza}}]{Barczaetal}%
  \BibitemOpen
  \bibfield  {author} {\bibinfo {author} {\bibfnamefont {G.}~\bibnamefont
  {Barcza}}, \bibinfo {author} {\bibfnamefont {F.}~\bibnamefont {Gebhard}},
  \bibinfo {author} {\bibfnamefont {T.}~\bibnamefont {Linneweber}},\ and\
  \bibinfo {author} {\bibfnamefont {{\"O}.}~\bibnamefont {Legeza}},\ }\bibfield
   {title} {\bibinfo {title} {{Ground-state properties of the symmetric
  single-impurity Anderson model on a ring from density-matrix renormalization
  group, Hartree-Fock, and Gutzwiller theory}},\ }\href@noop {} {\bibfield
  {journal} {\bibinfo  {journal} {Phys. Rev. B}\ }\textbf {\bibinfo {volume}
  {99}},\ \bibinfo {pages} {165130} (\bibinfo {year} {2019})}\BibitemShut
  {NoStop}%
\bibitem [{\citenamefont {Bauerbach}\ \emph {et~al.}(2020)\citenamefont
  {Bauerbach}, \citenamefont {Mahmoud},\ and\ \citenamefont
  {Gebhard}}]{IsingKondo}%
  \BibitemOpen
  \bibfield  {author} {\bibinfo {author} {\bibfnamefont {K.}~\bibnamefont
  {Bauerbach}}, \bibinfo {author} {\bibfnamefont {Z.~M.}\ \bibnamefont
  {Mahmoud}},\ and\ \bibinfo {author} {\bibfnamefont {F.}~\bibnamefont
  {Gebhard}},\ }\bibfield  {title} {\bibinfo {title} {Thermodynamics and
  screening in the {I}sing-{K}ondo model},\ }\href@noop {} {\bibfield
  {journal} {\bibinfo  {journal} {phys. stat. sol. (b)}\ }\textbf {\bibinfo
  {volume} {258}},\ \bibinfo {pages} {2000367} (\bibinfo {year}
  {2020})}\BibitemShut {NoStop}%
\bibitem [{\citenamefont {Barcza}\ \emph {et~al.}(2020)\citenamefont {Barcza},
  \citenamefont {Bauerbach}, \citenamefont {Eickhoff}, \citenamefont {Anders},
  \citenamefont {Gebhard},\ and\ \citenamefont {Legeza}}]{PhysRevB.101.075132}%
  \BibitemOpen
  \bibfield  {author} {\bibinfo {author} {\bibfnamefont {G.}~\bibnamefont
  {Barcza}}, \bibinfo {author} {\bibfnamefont {K.}~\bibnamefont {Bauerbach}},
  \bibinfo {author} {\bibfnamefont {F.}~\bibnamefont {Eickhoff}}, \bibinfo
  {author} {\bibfnamefont {F.~B.}\ \bibnamefont {Anders}}, \bibinfo {author}
  {\bibfnamefont {F.}~\bibnamefont {Gebhard}},\ and\ \bibinfo {author}
  {\bibfnamefont {{\"O}.}~\bibnamefont {Legeza}},\ }\bibfield  {title}
  {\bibinfo {title} {Symmetric single-impurity {K}ondo model on a tight-binding
  chain: Comparison of analytical and numerical ground-state approaches},\
  }\href@noop {} {\bibfield  {journal} {\bibinfo  {journal} {Phys. Rev. B}\
  }\textbf {\bibinfo {volume} {101}},\ \bibinfo {pages} {075132} (\bibinfo
  {year} {2020})}\BibitemShut {NoStop}%
\bibitem [{\citenamefont {{Wolf\-ram Research{,} Inc.}}(2019)}]{Mathematica12}%
  \BibitemOpen
  \bibfield  {author} {\bibinfo {author} {\bibnamefont {{Wolf\-ram Research{,}
  Inc.}}},\ }\href@noop {} {\emph {\bibinfo {title} {{\sc Mathematica},
  ver.~12}}}\ (\bibinfo  {publisher} {{Wolf\-ram Research{,} Inc.}},\ \bibinfo
  {address} {Champaign, IL},\ \bibinfo {year} {2019})\BibitemShut {NoStop}%
\bibitem [{\citenamefont {Schrieffer}\ and\ \citenamefont
  {Wolff}(1966)}]{PhysRev.149.491}%
  \BibitemOpen
  \bibfield  {author} {\bibinfo {author} {\bibfnamefont {J.~R.}\ \bibnamefont
  {Schrieffer}}\ and\ \bibinfo {author} {\bibfnamefont {P.~A.}\ \bibnamefont
  {Wolff}},\ }\bibfield  {title} {\bibinfo {title} {Relation between the
  {A}nderson and {K}ondo models},\ }\href@noop {} {\bibfield  {journal}
  {\bibinfo  {journal} {Phys. Rev.}\ }\textbf {\bibinfo {volume} {149}},\
  \bibinfo {pages} {491} (\bibinfo {year} {1966})}\BibitemShut {NoStop}%
\bibitem [{\citenamefont {S{\'o}lyom}(2009)}]{Solyom}%
  \BibitemOpen
  \bibfield  {author} {\bibinfo {author} {\bibfnamefont {J.}~\bibnamefont
  {S{\'o}lyom}},\ }\href@noop {} {\emph {\bibinfo {title} {{\sl Fundamentals of
  the Physics of Solids}}}},\ Vol.\ \bibinfo {volume} {1-3}\ (\bibinfo
  {publisher} {Sprin\-ger, Berlin},\ \bibinfo {year} {2009})\BibitemShut
  {NoStop}%
\bibitem [{\citenamefont {Andrei}\ \emph
  {et~al.}(1983{\natexlab{b}})\citenamefont {Andrei}, \citenamefont {Furuya},\
  and\ \citenamefont {Lowenstein}}]{RevModPhys.55.331}%
  \BibitemOpen
  \bibfield  {author} {\bibinfo {author} {\bibfnamefont {N.}~\bibnamefont
  {Andrei}}, \bibinfo {author} {\bibfnamefont {K.}~\bibnamefont {Furuya}},\
  and\ \bibinfo {author} {\bibfnamefont {J.~H.}\ \bibnamefont {Lowenstein}},\
  }\bibfield  {title} {\bibinfo {title} {{Solution of the Kondo problem}},\
  }\href@noop {} {\bibfield  {journal} {\bibinfo  {journal} {Rev. Mod. Phys.}\
  }\textbf {\bibinfo {volume} {55}},\ \bibinfo {pages} {331} (\bibinfo {year}
  {1983}{\natexlab{b}})}\BibitemShut {NoStop}%
\bibitem [{\citenamefont {Tsvelick}\ and\ \citenamefont
  {Wiegmann}(1983{\natexlab{b}})}]{TsvelickWiegmann}%
  \BibitemOpen
  \bibfield  {author} {\bibinfo {author} {\bibfnamefont {A.}~\bibnamefont
  {Tsvelick}}\ and\ \bibinfo {author} {\bibfnamefont {P.}~\bibnamefont
  {Wiegmann}},\ }\bibfield  {title} {\bibinfo {title} {Exact results in the
  theory of magnetic alloys},\ }\href@noop {} {\bibfield  {journal} {\bibinfo
  {journal} {Advances in Physics}\ }\textbf {\bibinfo {volume} {32}},\ \bibinfo
  {pages} {453} (\bibinfo {year} {1983}{\natexlab{b}})}\BibitemShut {NoStop}%
\bibitem [{\citenamefont {Bortz}\ and\ \citenamefont
  {Kl\"{u}mper}(2004)}]{BortzKluemper}%
  \BibitemOpen
  \bibfield  {author} {\bibinfo {author} {\bibfnamefont {M.}~\bibnamefont
  {Bortz}}\ and\ \bibinfo {author} {\bibfnamefont {A.}~\bibnamefont
  {Kl\"{u}mper}},\ }\bibfield  {title} {\bibinfo {title} {Lattice path integral
  approach to the one-dimensional {K}ondo model},\ }\href@noop {} {\bibfield
  {journal} {\bibinfo  {journal} {J.\ Phys.\ A}\ }\textbf {\bibinfo {volume}
  {37}},\ \bibinfo {pages} {6413} (\bibinfo {year} {2004})}\BibitemShut
  {NoStop}%
\end{thebibliography}%


\begin{thebibliography}{3}%
\makeatletter
\providecommand \@ifxundefined [1]{%
 \@ifx{#1\undefined}
}%
\providecommand \@ifnum [1]{%
 \ifnum #1\expandafter \@firstoftwo
 \else \expandafter \@secondoftwo
 \fi
}%
\providecommand \@ifx [1]{%
 \ifx #1\expandafter \@firstoftwo
 \else \expandafter \@secondoftwo
 \fi
}%
\providecommand \natexlab [1]{#1}%
\providecommand \enquote  [1]{``#1''}%
\providecommand \bibnamefont  [1]{#1}%
\providecommand \bibfnamefont [1]{#1}%
\providecommand \citenamefont [1]{#1}%
\providecommand \href@noop [0]{\@secondoftwo}%
\providecommand \href [0]{\begingroup \@sanitize@url \@href}%
\providecommand \@href[1]{\@@startlink{#1}\@@href}%
\providecommand \@@href[1]{\endgroup#1\@@endlink}%
\providecommand \@sanitize@url [0]{\catcode `\\12\catcode `\$12\catcode
  `\&12\catcode `\#12\catcode `\^12\catcode `\_12\catcode `\%12\relax}%
\providecommand \@@startlink[1]{}%
\providecommand \@@endlink[0]{}%
\providecommand \url  [0]{\begingroup\@sanitize@url \@url }%
\providecommand \@url [1]{\endgroup\@href {#1}{\urlprefix }}%
\providecommand \urlprefix  [0]{URL }%
\providecommand \Eprint [0]{\href }%
\providecommand \doibase [0]{https://doi.org/}%
\providecommand \selectlanguage [0]{\@gobble}%
\providecommand \bibinfo  [0]{\@secondoftwo}%
\providecommand \bibfield  [0]{\@secondoftwo}%
\providecommand \translation [1]{[#1]}%
\providecommand \BibitemOpen [0]{}%
\providecommand \bibitemStop [0]{}%
\providecommand \bibitemNoStop [0]{.\EOS\space}%
\providecommand \EOS [0]{\spacefactor3000\relax}%
\providecommand \BibitemShut  [1]{\csname bibitem#1\endcsname}%
\let\auto@bib@innerbib\@empty
\bibitem [{\citenamefont {Bauerbach}\ \emph {et~al.}(2020)\citenamefont
  {Bauerbach}, \citenamefont {Mahmoud},\ and\ \citenamefont
  {Gebhard}}]{IsingKondo}%
  \BibitemOpen
  \bibfield  {author} {\bibinfo {author} {\bibfnamefont {K.}~\bibnamefont
  {Bauerbach}}, \bibinfo {author} {\bibfnamefont {Z.~M.}\ \bibnamefont
  {Mahmoud}},\ and\ \bibinfo {author} {\bibfnamefont {F.}~\bibnamefont
  {Gebhard}},\ }\bibfield  {title} {\bibinfo {title} {Thermodynamics and
  screening in the {I}sing-{K}ondo model},\ }\href@noop {} {\bibfield
  {journal} {\bibinfo  {journal} {phys. stat. sol. (b)}\ }\textbf {\bibinfo
  {volume} {258}},\ \bibinfo {pages} {2000367} (\bibinfo {year}
  {2020})}\BibitemShut {NoStop}%
\bibitem [{\citenamefont {{Wolf\-ram Research{,} Inc.}}(2019)}]{Mathematica12}%
  \BibitemOpen
  \bibfield  {author} {\bibinfo {author} {\bibnamefont {{Wolf\-ram Research{,}
  Inc.}}},\ }\href@noop {} {\emph {\bibinfo {title} {{\sc Mathematica},
  ver.~12}}}\ (\bibinfo  {publisher} {{Wolf\-ram Research{,} Inc.}},\ \bibinfo
  {address} {Champaign, IL},\ \bibinfo {year} {2019})\BibitemShut {NoStop}%
\bibitem [{\citenamefont {Barcza}\ \emph {et~al.}(2020)\citenamefont {Barcza},
  \citenamefont {Bauerbach}, \citenamefont {Eickhoff}, \citenamefont {Anders},
  \citenamefont {Gebhard},\ and\ \citenamefont {Legeza}}]{PhysRevB.101.075132}%
  \BibitemOpen
  \bibfield  {author} {\bibinfo {author} {\bibfnamefont {G.}~\bibnamefont
  {Barcza}}, \bibinfo {author} {\bibfnamefont {K.}~\bibnamefont {Bauerbach}},
  \bibinfo {author} {\bibfnamefont {F.}~\bibnamefont {Eickhoff}}, \bibinfo
  {author} {\bibfnamefont {F.~B.}\ \bibnamefont {Anders}}, \bibinfo {author}
  {\bibfnamefont {F.}~\bibnamefont {Gebhard}},\ and\ \bibinfo {author}
  {\bibfnamefont {{\"O}.}~\bibnamefont {Legeza}},\ }\bibfield  {title}
  {\bibinfo {title} {Symmetric single-impurity {K}ondo model on a tight-binding
  chain: Comparison of analytical and numerical ground-state approaches},\
  }\href@noop {} {\bibfield  {journal} {\bibinfo  {journal} {Phys. Rev. B}\
  }\textbf {\bibinfo {volume} {101}},\ \bibinfo {pages} {075132} (\bibinfo
  {year} {2020})}\BibitemShut {NoStop}%
\end{thebibliography}%

\appendix
\setcounter{equation}{0}
\renewcommand{\theequation}{\Alph{section}.\arabic{equation}}
\section*{Appendix}

In Appendix~\ref{appendix:secondorderterms} we derive the second-order contributions to the free energy of the Kondo model, 
followed by the third-order contributions in Appendix~\ref{appendix:thirdorderterms}. 
In addition to the analytical results, we also present a numerical approach for the validation of the third-order analytical results.

\section{Derivation of the second-order contributions}
\label{appendix:secondorderterms}
In this section we derive the two contributions to second order of the impurity-induced free energy from Sect.~\ref{subsec:formalsecondorderterm}, namely, 
the Ising-Kondo contribution and the spin-flip contribution.

\subsection{Second-order Ising-Kondo contribution}
\label{appendix:secondorderIsingKondo}
To derive the second-order Ising-Kondo contribution, we start from Eq.~(\ref{TheTwoPartsOfTheSecondOrder})
\begin{equation}
V_2^z=\int_0^\beta\rmd\lambda_2\,\int_0^{\lambda_2}\rmd\lambda_1\,\left\langle\hat{V}_z(\lambda_2)\hat{V}_z(\lambda_1)\right\rangle\;,
\label{appeq:V2zdef}
\end{equation}
see also eqs.~(\ref{DefinitionKondoInteraction}),~(\ref{TauDependenceOfOperators}), and~(\ref{DefinitionLangleRangle}). Recall that 
the free Hamiltonian for the Kondo model in a magnetic field is defined by ($\sigma=\uparrow,\downarrow$)
\begin{equation}
\hat{H}_0=\sum_{k,\sigma}\left(\epsilon(k)-\sigma B\right)\hat{a}^+_{k,\sigma}\hat{a}^\phan_{k,\sigma}-B\left(\hat{n}^d_\Uparrow-\hat{n}^d_\Downarrow\right)\;.
\end{equation}
In momentum space, the Ising-Kondo coupling reads
\begin{eqnarray}
\hat{V}_z(\tau)&=&\frac{J_z}{4}\left(\hat{n}^d_\Uparrow-\hat{n}^d_\Downarrow\right)\\
&&\times\frac{1}{L}\sum_{k,k'}e^{(\epsilon(k)-\epsilon(k'))\lambda}\left(\hat{a}^+_{k,\uparrow}\hat{a}^\phan_{k',\uparrow}-\hat{a}^+_{k,\downarrow}\hat{a}^\phan_{k',\downarrow}\right)
\nonumber \;,
\end{eqnarray}
where $\hat{n}^d_\Uparrow-\hat{n}^d_\Downarrow$ does not change the impurity-spin~$s=\Uparrow,\Downarrow$, 
just as $\hat{c}^+_{0,\uparrow}\hat{c}^\phan_{0,\uparrow}-\hat{c}^+_{0,\downarrow}\hat{c}^\phan_{0,\downarrow}$ does not change the $\uparrow$ and $\downarrow$ spins in the host-electron bath.

We use
\begin{equation}
\left(\hat{n}^d_\Uparrow-\hat{n}^d_\Downarrow\right)\left(\hat{n}^d_\Uparrow-\hat{n}^d_\Downarrow\right)=\hat{n}^d_\Uparrow+\hat{n}^d_\Downarrow-2\hat{n}^d_\Uparrow\hat{n}^d_\Downarrow=1\;,
\end{equation}
since $\hat{n}^d_\Uparrow+\hat{n}^d_\Downarrow=1$ and $\hat{n}^d_\Uparrow\hat{n}^d_\Downarrow=0$. This identity is a direct consequence of the single-impurity model. We find
\begin{eqnarray}
\left\langle\hat{V}_z(\lambda_2)\hat{V}_z(\lambda_1)\right\rangle
&=&\left(\frac{J_z}{4L}\right)^2\sum_{k,k',p,p'}\left\langle\hat{M}_{k,k'}\hat{M}_{p,p'}\right\rangle\nonumber\\
&&\qquad\times\,e^{(\epsilon(k)-\epsilon(k'))\lambda_2}e^{(\epsilon(p)-\epsilon(p'))\lambda_1}\nonumber\;,\\
\end{eqnarray}
where
\begin{equation}
    \hat{M}_{k,k'}=\hat{a}^+_{k,\uparrow}\hat{a}^\phan_{k',\uparrow}-\hat{a}^+_{k,\downarrow}\hat{a}^\phan_{k',\downarrow}\;.
\end{equation}
With 
\begin{equation}
n_{k,\uparrow}=\frac{1}{1+e^{\beta(\epsilon(k)-B)}}\;,\qquad n_{k,\downarrow}=\frac{1}{1+e^{\beta(\epsilon(k)+B)}}
\end{equation}
we further obtain [$\hat{V}_1\hat{V}_2\equiv \hat{V}_z(\lambda_2)\hat{V}_z(\lambda_1)$]
\begin{eqnarray}
\langle\hat{V}_1\hat{V}_2\rangle&=&\left(\frac{J_z}{4L}\right)^2\sum_{k,p}\left(n_{k,\uparrow}-n_{k,\downarrow}\right)\left(n_{p,\uparrow}-n_{p,\downarrow}\right)\nonumber\\
&&+\left(\frac{J_z}{4L}\right)^2\sum_{k,p}e^{(\epsilon(k)-\epsilon(p))(\lambda_2-\lambda_1)}\nonumber\\
&&\qquad\qquad\times\left(n_{k,\uparrow}(1-n_{p,\uparrow})+n_{k,\downarrow}(1-n_{p,\downarrow})\right)\nonumber\\
&=&\left(\frac{J_z}{4}\right)^2\left(M_0(B)\right)^2\nonumber\\
&&+\left(\frac{J_z}{4}\right)^2\Bigl(f(\lambda_2-\lambda_1,B)f(\lambda_2-\lambda_1,-B)\nonumber\\
&&\qquad\qquad+f(\lambda_2-\lambda_1,-B)f(\lambda_2-\lambda_1,B)\Bigr)\nonumber\\
\end{eqnarray}
with $M_0$ from Eq.~(\ref{M0definition}) and
\begin{equation}\label{flambdaBdefinition}
f(\lambda,B)=\frac{1}{L}\sum_kn_{k,\uparrow}e^{\epsilon(k)\lambda}=\int_{-\infty}^\infty\rmd \epsilon\,\rho_0(\epsilon)\frac{e^{\epsilon\lambda}}{1+e^{\beta(\epsilon-B)}}\;.
\end{equation}
Note that
\begin{eqnarray}
1-n_{p,\uparrow}&=&\frac{1+e^{\beta(\epsilon(p)-B)}-1}{1+e^{\beta(\epsilon(p)-B)}}=\frac{1}{1+e^{-\beta(\epsilon(p)-B)}}\nonumber\\
\end{eqnarray}
and therefore [$\rho_0(-\epsilon)=\rho_0(\epsilon)$]
\begin{eqnarray}
\frac{1}{L}\sum_pe^{-\epsilon(p)\lambda} (1-n_{p,\uparrow})&=&\int_{-\infty}^\infty\rmd\epsilon\,\rho_0(\epsilon)\frac{e^{-\beta\lambda}}{1+e^{-\beta(\epsilon-B)}}\nonumber\\
&=&\int_{-\infty}^\infty\rmd\epsilon\,\rho_0(\epsilon)\frac{e^{\epsilon\lambda}}{1+e^{\beta(\epsilon+B)}}\nonumber\\
&=&f(\lambda,-B)\;,\nonumber\\
\frac{1}{L}\sum_ke^{\epsilon(k)\lambda}n_{k,\downarrow}&=&\int_{-\infty}^\infty\rmd\epsilon\,\rho_0(\epsilon)\frac{e^{\epsilon\lambda}}{1+e^{\beta(\epsilon-B)}}\nonumber\\
&=&f(\lambda,-B)\;,\nonumber\\
\frac{1}{L}\sum_pe^{-\epsilon(p)\lambda}(1-n_{p,\downarrow})&=&f(\lambda,B)\;.
\end{eqnarray}

In the next step, we use the substitution $\lambda=\lambda_2-\lambda_1$, so that for a general function $h(x)$ we can reduces the double integral
\begin{eqnarray}
\int_0^\beta\rmd\lambda_2\,\int_0^{\lambda_2}\rmd\lambda_1\,h(\lambda_2-\lambda_1)&=&\int_0^\beta\rmd\lambda_2\,\int_0^{\lambda_2}\rmd\lambda\,h(\lambda)\nonumber\\
&=&\int_0^\beta\rmd\lambda\,h(\lambda)\int_{\lambda}^\beta\rmd\lambda_2\nonumber\\
&=&\int_0^\beta\rmd\lambda\,(\beta-\lambda)h(\lambda)\;,\nonumber\\
\end{eqnarray}
into a single integral. 
For a more detailed derivation, see Sect.~\SMsubsecDoubleIntegral\ in the Supplemental Material. In Eq.~(\ref{appeq:V2zdef}) this results in
\begin{eqnarray}\label{ForSecondOrderIsingKondo}
V_2^z&=&\int_0^\beta\rmd\lambda_2\int_0^{\lambda_2}\rmd\lambda_1\,\left\langle\hat{V}_z(\lambda_2)\hat{V}_z(\lambda_1)\right\rangle\nonumber\\
&=&\frac{\beta^2}{2}\left(\frac{J_z}{4}M_0(B)\right)^2+2\left(\frac{J_z}{4}\right)^2 \ell(B,T)\;\;
\end{eqnarray}
with
\begin{equation}\label{ellBTdef}
    \ell(B,T)=\int_0^\beta\rmd\lambda\,(\beta-\lambda)f(\lambda,B)f(\lambda,-B)\; .
\end{equation}
Apparently, the problem is reduced to the calculation of the integral $\ell(B,T)$. 
The derivation of this integral is lengthy and can be found in the Supplemental Material, Sect.~\SMSMsubsubsecderivationellBT. 
At $B\ll T\ll 1$ the result is
\begin{eqnarray}\label{ellBT-definition}
\ell(B,T)&=&-\beta\left(-2\ln 2+\frac{T^2\pi^2}{3}+B^2+\frac{\pi^2}{3}B^2T^2\right)\nonumber\\
&&+\mathcal{O}[B^3]\;.
\end{eqnarray}
Finally, with eqs.~(\ref{ForSecondOrderIsingKondo}) and~(\ref{M0forsmallB}), i.e., $M_0(B\ll1)\approx 2B\rho_0(0)$, we obtain [$j_z=J_z\rho_0(0)$]
\begin{eqnarray}
V_2^z=-\frac{1}{T}\left(-2\ln 2+\frac{\pi^2}{3}T^2+B^2-\frac{B^2}{T}\right)\left(\frac{j_z^2}{8}\right)\;,\nonumber\\
\end{eqnarray}
in agreement with the analytical solution in Ref.~\cite{IsingKondo}, see also Eq.~(\SMFIKsecondorder) of the Supplemental Material.

\subsection{Second-order spin-flip contribution}
\label{appendix:secondorderSpinFlip}

Now we calculate the spin-flip contribution to second order. Using Eq.~(\ref{DefinitionKondoInteraction}), (\ref{TauDependenceOfOperators}), and~(\ref{TheTwoPartsOfTheSecondOrder}), we 
evaluate
\begin{equation}
V_2^\perp(\lambda)=\frac{J_\perp^2}{4}\Big\langle \hat{s}_0^+\hat{S}^-\hat{s}_0^-(-\lambda)\hat{S}^+ 
+\hat{s}_0^-\hat{S}^+\hat{s}_0^+(-\lambda)\hat{S}^- \Big\rangle\;,
\end{equation}
because $\hat{\mathbf{S}}(\lambda)=\hat{\mathbf{S}}$ for the local spin. Then, [$\beta=1/T$]
\begin{eqnarray}
V_2^\perp(\lambda)&=&\frac{J_\perp^2}{4}\frac{1}{2\cosh(\beta B)}\nonumber\\
&&\times\bigg(\left\langle\Downarrow|e^{\beta B(\hat{n}^d_\Uparrow-\hat{n}^d_\Downarrow)}|\Downarrow\right\rangle\left\langle\hat{s}^+_0\hat{s}_0^-(-\lambda)\right\rangle_F \nonumber\\
&&\quad + \left\langle\Uparrow|e^{\beta B(\hat{n}^d_\Uparrow-\hat{n}^d_\Downarrow)}|\Uparrow\right\rangle\left\langle\hat{s}^-_0\hat{s}_0^+(-\lambda)\right\rangle_F\bigg)\nonumber\\
&=&\left(\frac{J_\perp}{2L}\right)^2\frac{1}{2\cosh(\beta B)}\sum_{k,k'}\sum_{p,p'}\nonumber\\
&&\qquad\times\bigg(e^{-\beta B}
\left\langle\hat{a}^+_{k,\uparrow}\hat{a}^\phan_{k',\downarrow}e^{-\lambda\hat{T}}\hat{a}^+_{p,\downarrow}\hat{a}^\phan_{p',\uparrow}e^{\lambda\hat{T}}\right\rangle\nonumber\\
&&\qquad\quad+e^{\beta B}\left\langle\hat{a}^+_{k,\downarrow}\hat{a}^\phan_{k',\uparrow}e^{-\lambda\hat{T}}\hat{a}^+_{p,\uparrow}\hat{a}^\phan_{p',\downarrow}e^{\lambda\hat{T}}\right\rangle
\bigg)\;.\nonumber\\
\end{eqnarray}
The application of Wick's theorem then leads to
\begin{eqnarray}
V_2^\perp(\lambda)&=&\left(\frac{J_\perp}{2L}\right)^2\frac{1}{2\cosh(\beta B)}\nonumber\\
&&\times\sum_{k,k'}\bigg(
e^{-\beta B}e^{\lambda(\epsilon(k)-\epsilon(k'))}n_{k,\uparrow}\left(1-n_{k',\downarrow}\right)\nonumber\\
&&\qquad\quad +e^{\beta B}e^{\lambda(\epsilon(k)-\epsilon(k'))}n_{k,\downarrow}(1-n_{k',\uparrow})
\bigg).\nonumber\\
\end{eqnarray}
We introduce the function $\tilde{f}(\lambda,B)$, see also Eq.~(\ref{flambdaBdefinition}),
\begin{equation}\label{ftildelambdaBdefinition}
    \tilde{f}(\lambda,B)=\frac{f(\lambda,B)}{\rho_0(0)}=\int_{-1}^1\rmd\epsilon\frac{\tilde\rho(\epsilon)e^{\lambda\epsilon}}{1+e^{\beta(\epsilon-B)}}
\end{equation}
with $\rhotilde(\epsilon)=\rho_0(\epsilon)/\rho_0(0)$. Thus,
\begin{eqnarray}
\frac{1}{L}\sum_ke^{\lambda\epsilon(k)}n_{k,\uparrow}=\rho_0(0)\tilde{f}(\lambda,B)\;.
\end{eqnarray}
Furthermore, with $\rho_0(\epsilon)=\rho_0(-\epsilon)$ we find
\begin{eqnarray}
p(\lambda,B)&=&\frac{1}{L}\sum_ke^{-\lambda\epsilon(k')}(1-n_{k',\downarrow})\nonumber\\
&=&\int_{-\infty}^\infty\rmd\epsilon\,\rho_0(\epsilon)e^{-\lambda\epsilon}\left(1-\frac{1}{1+e^{\beta(\epsilon+B)}}\right)\nonumber\\
&=&\int_{-\infty}^\infty\rmd\epsilon\,\rho_0(\epsilon)\frac{e^{-\lambda\epsilon}}{1+e^{-\beta(\epsilon+B)}}\nonumber\\
&=&\int_{-\infty}^\infty\rmd\epsilon\,\rho_0(\epsilon)\frac{e^{\lambda\epsilon}}{1+e^{\beta(\epsilon-B)}}\nonumber\\
&=&\rho_0(0)\tilde{f}(\lambda,B)\;.
\end{eqnarray}
Thus, [$j_\perp=J_\perp\rho_0(0)$]
\begin{eqnarray}
V_2^\perp(\lambda)&=&\left(\frac{j_\perp}{2}\right)^2\frac{1}{2\cosh(\beta B)}\nonumber\\
&&\quad\times\left(e^{-\beta B}\tilde{f}^2(\lambda,B)+e^{\beta B}\tilde{f}^2(\lambda,-B)\right)\;.\nonumber\\
\end{eqnarray}
We note that
\begin{eqnarray}
\tilde{f}(\beta-\lambda,B)&=&\int_{-1}^1\rmd\epsilon\,\rhotilde(\epsilon)\frac{e^{(\beta-\lambda)\epsilon}}{1+e^{\beta(\epsilon-B)}}\nonumber\\
&=&\int_{-1}^1\rmd\epsilon\,\rhotilde(\epsilon)\frac{e^{\lambda \epsilon}e^{-\beta\epsilon}e^{\beta B}e^{-\beta B}}{1+e^{-\beta}(\epsilon+B)}\nonumber\\
&=&e^{\beta B}\tilde{f}(\lambda,-B)\;,
\end{eqnarray}
where we used $\rhotilde(\epsilon)=\rhotilde(-\epsilon)$. Then,
\begin{eqnarray}
V_2^\perp&=&\int_0^{\beta/2}\rmd\lambda\,(\beta-\lambda)\frac{j_\perp^2}{4}\frac{1}{2\cosh(\beta B)}\nonumber\\
&&\quad\times\left(e^{-\beta B}\tilde{f}^2(\lambda,B)+e^{\beta B}\tilde{f}^2(\lambda,-B)\right)\nonumber\\
&&+\int_0^{\beta/2}\rmd\mu\,\mu\frac{j_\perp^2}{4}\frac{1}{2\cosh(\beta B)}\Bigl(e^{-\beta B}e^{2\beta B}\tilde{f}^2(\mu,-B)\nonumber\\
&&\qquad\qquad\qquad+e^{\beta B}e^{-2\beta B}\tilde{f}^2(\mu,B)\Bigr)\;,\nonumber\\
\end{eqnarray}
where we substituted $\mu=\beta-\lambda$. Thus,
\begin{eqnarray}
V_2^\perp&=&\beta\int_0^{\beta/2}\rmd \lambda\,\left(\frac{j_\perp}{2}\right)^2\frac{1}{2\cosh(\beta B)}\\
&&\times\left(e^{-\beta B}\tilde{f}^2(\lambda,B)+e^{\beta B}\tilde{f}^2(\lambda,-B)\right)\,.\nonumber
\end{eqnarray}
We note
\begin{equation}
\int_0^{\beta/2}\rmd\lambda\,e^{\lambda(\epsilon_1+\epsilon_2)}=\frac{e^{\beta/2(\epsilon_1+\epsilon_2)}-1}{\epsilon_1+\epsilon_2}\;,
\end{equation}
which leads to
\begin{eqnarray}
    V_2^\perp=V_{2,\alpha}^\perp + V_{2,\beta}^\perp
\end{eqnarray}
with
\begin{eqnarray}
V_{2,\alpha}^\perp&=&\beta\left(\frac{j_\perp}{2}\right)^2\frac{1}{2\cosh(\beta B)}
\nonumber\\
&&\times\biggl(e^{-\beta B}\int_{-1}^1\rmd\epsilon_1\,\frac{\rhotilde(\epsilon_1)}{1+e^{\beta(\epsilon_1-B)}}\nonumber\\
&&\quad\qquad\times \int_{-1}^1\rmd\epsilon_2\,\frac{\rhotilde(\epsilon_2)}{1+e^{\beta(\epsilon_2-B)}}\left(\frac{e^{\beta/2(\epsilon_1+\epsilon_2)}}{\epsilon_1+\epsilon_2}\right)\nonumber\\
&&+\left[B\to(-B)\right]
\bigg)\;.
\end{eqnarray}
The second term in $V_{2,\alpha}^\perp$, where we simply have to change the sign of $B$, is the negative of the first term. This can be seen by performing the substitution $\epsilon_1\to-\epsilon_2$ and $\epsilon_2\to-\epsilon_1$ in the first term, so that the term vanishes altogether,
\begin{equation}
V_{2,\alpha}^\perp=0\;.
\end{equation}
Thus,
\begin{eqnarray}
V_2^\perp&=&V_{2,\beta}^\perp\nonumber\\
&=&(-\beta)\left(\frac{j_\perp}{2}\right)^2\frac{1}{2\cosh(\beta B)}\nonumber\\
&&\times\bigg(\int_{-1}^1\rmd\epsilon_1\,\frac{\rhotilde(\epsilon_1)}{1+e^{\beta(\epsilon_1-B)}}\nonumber\\
&&\qquad\times\int_{-1}^1\rmd\epsilon_2\,\frac{\rhotilde(\epsilon_2)}{1+e^{\beta(\epsilon_2-B)}}\frac{e^{-\beta B}}{\epsilon_1+\epsilon_2}\nonumber \\
&&\qquad\quad+\left[B\to(-B)\right]\bigg)\;.
\end{eqnarray}
We define
\begin{equation}\label{lBTdef}
l(B,T)=\int_{-1}^{1}\frac{\rmd\epsilon_1}{1+e^{\beta(\epsilon_1-B)}}\int_{-1}^1\frac{\rmd\epsilon_2}{1+e^{\beta(\epsilon_2-B)}}\frac{1}{\epsilon_1+\epsilon_2}
\end{equation}
for a constant density of states, $\rhotilde(\epsilon)=1$. Then, for a constant density of states,
\begin{eqnarray}\label{VvonlBl-B}
V_2^\perp&=&-\beta\left(\frac{j_\perp}{2}\right)^2\frac{1}{2\cosh(\beta B)}\nonumber\\
&&\times\left(e^{-\beta B}l(B,T)+e^{\beta B}l(-B,T)\right)\;.
\end{eqnarray}
The evaluation of $l(B)$ is lengthy and can be found in the Supplemental Material, see Sect.~\SMSMsubsubsecderivationlb. The result is 
\begin{equation}
l(B)=-2\ln 2 +\frac{\pi^2T^2}{3}+2B\left(-1-\EulerGamma+\ln(2\pi T)\right)+B^2\;.
\end{equation}
Finally, we arrive at
\begin{eqnarray}
V_2^\perp&=&-\beta\left(\frac{j_\perp}{2}\right)^2\biggl[-2\ln 2+\frac{\pi^2T^2}{3}\\
&&+B^2\left[1-\frac{2}{T}\left(-1-\EulerGamma+\ln(2\pi T)\right)\right)\biggr]\;.\nonumber
\end{eqnarray}

\section{Derivation of the third-order contributions}
\setcounter{equation}{0}
\label{appendix:thirdorderterms}
In this section, we derive the three contributions, $V_3^a$, $V_3^b$ and $V_3^c$, in third order of the impurity-induced free energy, see Eq.~(\ref{RemainingpartsofV3}). 
The Ising-Kondo contribution $V_3^z$ is already known, see Eq.~(\ref{IsingKondoThirdOrder}).

\subsection{Term \texorpdfstring{$\mathbf{V_3^a}$}{V3a} from third order}
\label{appendix:thirdorder-V3a}

\subsubsection{Evaluation of the trace}

With $\hat{n}^{kk'}_\sigma=\hat{a}^+_{k,\sigma}\hat{a}^\phan_{k',\sigma}$ we find from eqs.~(\ref{RemainingpartsofV3}) and~(\ref{V3z-V3a-V3b-V3c-Definition})
\begin{eqnarray}
V_a^{(3)}&=&\left(\frac{J_\perp}{2}\right)^2\frac{J_z}{4}\int_0^\beta\rmd\lambda_3\,\int_0^{\lambda_3}\rmd\lambda_2\,\int_0^{\lambda_2}\rmd\lambda_1\,w_3^a\;,\nonumber\\
\end{eqnarray}
where
\begin{eqnarray}
w_3^a&=&\frac{1}{L}\sum_{k,k'}\frac{1}{Z_0}\nonumber\\
&&\quad\times{\rm Tr}\biggl[
e^{-\beta\hat{H}_0}\left(\hat{n}^d_\Uparrow-\hat{n}^d_\Downarrow\right)\left(\hat{n}^{kk'}_\uparrow(\lambda_3)-\hat{n}^{kk'}_\downarrow(\lambda_3)\right)\nonumber\\
&&\quad\times\left(\hat{s}_0^+(\lambda_2)\hat{S}^-+\hat{s}_0^-(\lambda_2)\hat{S}^+\right)\nonumber\\
&&\quad\times\left(\hat{s}_0^+(\lambda_1)\hat{S}^-+\hat{s}_0^-(\lambda_1)\hat{S}^+\right)
\biggr]\;.
\end{eqnarray}
Note that 
\begin{eqnarray}\label{Spm-s0pm-Operators}
\hat{S}^+\hat{S}^+&=&0\;,\qquad \hat{S}^-\hat{S}^-=0\;,\nonumber\\[1pt]
\hat{S}^+\hat{S}^-&=&\hat{n}^d_\Uparrow\;,\qquad\hat{S}^-\hat{S}^+=\hat{n}^d_\Downarrow\;,\nonumber\\[1pt]
(\hat{n}^d_s)^2&=&\hat{n}^d_s\;,\qquad\hat{n}^d_\Uparrow\hat{n}^d_\Downarrow=\hat{n}^d_\Downarrow\hat{n}^d_\Uparrow=0
\end{eqnarray}
in the fermionic single-impurity model. Using these properties we find
\begin{eqnarray}
w_3^a&=&\frac{1}{Z_0}\frac{1}{L}\sum_{k,k'}\nonumber\\
&&{\rm Tr}\biggl[e^{-\beta\hat{H}_0}\Big(
\hat{n}^d_\Uparrow\mathcal{N}_1^{k,k'}\hat{n}^d_\Uparrow -\hat{n}^d_\Uparrow\mathcal{N}_2^{k,k'}\hat{n}^d_\Uparrow\nonumber\\
&&\qquad\qquad-\hat{n}^d_\Downarrow\mathcal{N}_3^{k,k'}\hat{n}^d_\Downarrow +\hat{n}^d_\Downarrow\mathcal{N}_4^{k,k'}\hat{n}^d_\Downarrow
\Big)
\biggr] \; .\nonumber \\
\end{eqnarray}
We perform the trace both over the impurity spin and the fermionic bath, [$\bar{w}_3^a=2Z_0 L\cosh(\beta B)w^a_3$]
\begin{eqnarray}
\bar{w}_3^a&=&\sum_{k,k'}
{\left\langle\Uparrow|e^{\beta B(\hat{n}^d_\Uparrow-\hat{n}^d_\Downarrow)}\hat{n}^d_\Uparrow|\Uparrow\right\rangle}\Big(\mathcal{N}^{k,k'}_1-\mathcal{N}^{k,k'}_2\Big)\nonumber\\
&&+\sum_{k,k'}\left\langle\Downarrow|e^{\beta B(\hat{n}^d_\Uparrow-\hat{n}^d_\Downarrow)}\hat{n}^d_\Downarrow|\Downarrow\right\rangle\left(-\mathcal{N}^{k,k'}_3+\mathcal{N}^{k,k'}_4\right)\;,\nonumber\\
\end{eqnarray}
where the fermionic traces are given by
\begin{eqnarray}
\mathcal{N}^{k,k'}_1&=&\left\langle\hat{n}^{kk'}_\uparrow(\lambda_3)\hat{s}^-_0(\lambda_2)\hat{s}^+_0(\lambda_1)\right\rangle_F\;,\nonumber\\[2pt]
\mathcal{N}^{k,k'}_2&=&\left\langle\hat{n}^{kk'}_\downarrow(\lambda_3)\hat{s}^-_0(\lambda_2)\hat{s}^+_0(\lambda_1)\right\rangle_F\;,\nonumber\\[2pt]
\mathcal{N}^{k,k'}_3&=&\left\langle\hat{n}^{kk'}_\uparrow(\lambda_3)\hat{s}^+_0(\lambda_2)\hat{s}^-_0(\lambda_1)\right\rangle_F\;,\nonumber\\[2pt]
\mathcal{N}^{k,k'}_4&=&\left\langle\hat{n}^{kk'}_\downarrow(\lambda_3)\hat{s}^+_0(\lambda_2)\hat{s}^-_0(\lambda_1)\right\rangle_F\;.
\end{eqnarray}
The evaluation of the impurity-spin operators leaves us with the traces over the fermionic bath,
\begin{eqnarray}
2\cosh(\beta B)L^3w_3^a&=&\sum_{k,k'}
e^{\beta B}\left(\mathcal{N}^{k,k'}_1-\mathcal{N}^{k,k'}_2\right)\nonumber\\
&&+\sum_{k,k'}e^{-\beta B}\left(-\mathcal{N}^{k,k'}_3+\mathcal{N}^{k,k'}_4\right)
\;.\nonumber\\
\end{eqnarray}
To evaluate the expectation values $\mathcal{N}_i^{k,k'}$ ($i=1,\ldots,4$) we use Wick's theorem,
\begin{widetext}
\begin{eqnarray}
\sum_{k,k'}\mathcal{N}_1^{k,k'}&=&\sum_{\substack{k,k',p\\p',m,m'}}\left\langle e^{\lambda_3\hat{T}}\hat{a}^+_{k,\uparrow}\hat{a}^\phan_{k',\uparrow}e^{-\lambda_3\hat{T}}e^{\lambda_2\hat{T}}\hat{a}^+_{p,\downarrow}\hat{a}^\phan_{p',\uparrow}e^{-\lambda_2\hat{T}}e^{\lambda_1\hat{T}}\hat{a}^+_{m,\uparrow}\hat{a}^\phan_{m',\downarrow}e^{\lambda_1\hat{T}} \right\rangle_F\nonumber\\
&=&\sum_{k,p,p'}n_{k,\uparrow}(1-n_{p',\uparrow})n_{p,\downarrow}e^{\lambda_2(\epsilon(p)-\epsilon(p'))}e^{\lambda_1(\epsilon(p')-\epsilon(p))}\nonumber\\
&&-\sum_{k,k',p}n_{k,\uparrow}(1-n_{k',\uparrow})n_{p,\downarrow}e^{\lambda_3\left(\epsilon(k)-\epsilon(k')\right)}e^{\lambda_2(\epsilon(p)-\epsilon(k))}e^{\lambda_1(\epsilon(k')-\epsilon(p))}
\;,\\[12pt]
\sum_{k,k'}\mathcal{N}_2^{k,k'}&=&\sum_{\substack{k,k',p\\p',m,m'}}\left\langle 
e^{\lambda_3\hat{T}}\hat{a}^+_{k,\downarrow}\hat{a}^\phan_{k',\downarrow} e^{-\lambda_3\hat{T}}e^{\lambda_2\hat{T}}\hat{a}^+_{p,\downarrow}\hat{a}^\phan_{p',\uparrow}e^{-\lambda_2\hat{T}}e^{\lambda_1\hat{T}}\hat{a}^+_{m,\uparrow}\hat{a}^\phan_{m',\downarrow}e^{-\lambda_1\hat{T}}
\right\rangle_F\nonumber\\
&=&-\sum_{k,p,p'}n_{k,\downarrow}n_{p,\downarrow}(1-n_{p',\uparrow})e^{\lambda_2(\epsilon(p)-\epsilon(p'))}e^{\lambda_1(\epsilon(p')-\epsilon(p))}\nonumber\\
&&-\sum_{k,k',p'}n_{k,\downarrow}(1-n_{k',\downarrow})(1-n_{p',\uparrow})e^{\lambda_3(\epsilon(k)-\epsilon(k'))}e^{\lambda_2(\epsilon(k')-\epsilon(p'))}e^{\lambda_1(\epsilon(p')-\epsilon(k))}
\; ,\\[12pt]
\sum_{k,k'}\mathcal{N}_3^{k,k'}&=&\sum_{\substack{k,k',p\\p',m,m'}}\left\langle 
e^{\lambda_3\hat{T}}\hat{a}^+_{k,\uparrow}\hat{a}^\phan_{k',\uparrow} e^{-\lambda_3\hat{T}}e^{\lambda_2\hat{T}}\hat{a}^+_{p,\uparrow}\hat{a}^\phan_{p',\downarrow}e^{-\lambda_2\hat{T}}e^{\lambda_1\hat{T}}\hat{a}^+_{m,\downarrow}\hat{a}^\phan_{m',\uparrow}e^{-\lambda_1\hat{T}}
\right\rangle_F\nonumber\\
&=&-\sum_{k,p,p'}n_{k,\uparrow}n_{p,\uparrow}(1-n_{p',\downarrow})e^{\lambda_2(\epsilon(p)-\epsilon(p'))}e^{\lambda_1(\epsilon(p')-\epsilon(p))}\nonumber\\
&&-\sum_{k,k',p'}n_{k,\uparrow}(1-n_{k',\uparrow})(1-n_{p',\downarrow})e^{\lambda_3(\epsilon(k)-\epsilon(k'))}e^{\lambda_2(\epsilon(k')-\epsilon(p'))}e^{\lambda_1(\epsilon(p')-\epsilon(k))}
\;,
\end{eqnarray}
and
\begin{eqnarray}
\sum_{k,k'}\mathcal{N}_4^{k,k'}&=&\sum_{\substack{k,k',p\\p',m,m'}}\left\langle 
e^{\lambda_3\hat{T}}\hat{a}^+_{k,\downarrow}\hat{a}^\phan_{k',\downarrow} e^{-\lambda_3\hat{T}}e^{\lambda_2\hat{T}}\hat{a}^+_{p,\uparrow}\hat{a}^\phan_{p',\downarrow}e^{-\lambda_2\hat{T}}e^{\lambda_1\hat{T}}\hat{a}^+_{m,\downarrow}\hat{a}^\phan_{m',\uparrow}e^{-\lambda_1\hat{T}}
\right\rangle_F\nonumber\\
&=&\sum_{k,p,p'}n_{k,\downarrow}(1-n_{p',\downarrow})n_{p,\uparrow}e^{\lambda_2(\epsilon(p)-\epsilon(p'))}e^{\lambda_1(\epsilon(p')-\epsilon(p))}\nonumber\\
&&-\sum_{k,k',p}n_{k,\downarrow}(1-n_{k',\downarrow})n_{p,\uparrow}e^{\lambda_3(\epsilon(k)-\epsilon(k'))}e^{\lambda_2(\epsilon(p)-\epsilon(k))}e^{\lambda_1(\epsilon(k')-\epsilon(p))}\;.
\end{eqnarray}
\end{widetext}

\subsubsection{Further simplifications}

To make progress, we define 
\begin{equation}\label{Wpm-Definition}
W^\pm=\frac{e^{\mp \beta B}}{2\cosh(\beta B)} \; ,
\end{equation}
and combine the next to last terms of $\mathcal{N}^{k,k'}_1$ and $\mathcal{N}^{k,k'}_2$, as well as the next to last terms of $\mathcal{N}^{k,k'}_3$ and $\mathcal{N}^{k,k'}_4$, which leads to
\begin{eqnarray}\label{V3aAfterEvaluationOfTheTrace}
L^3w_3^a&=&W^-\big(A_1-A_2-A_3\big) - W^+\big(A_4+A_5+A_6\big)\nonumber\\
\end{eqnarray}
with
\begin{eqnarray}
    A_1&=&\sum_{k,p,p'}(n_{k,\uparrow}-n_{k,\downarrow})(1-n_{p',\uparrow})n_{p,\downarrow}\nonumber\\
    &&\qquad\times e^{\epsilon(p)(\lambda_2-\lambda_1)}e^{-\epsilon(p')(\lambda_2-\lambda_1)}\;,
    \end{eqnarray}
    \begin{eqnarray}
     A_2&=&\sum_{k,k',p}n_{k,\uparrow}(1-n_{k',\uparrow})n_{p,\downarrow}e^{\epsilon(k)(\lambda_3-\lambda_2)}\nonumber\\
    &&\qquad\times e^{-\epsilon(k')(\lambda_3-\lambda_1)}e^{\epsilon(p)(\lambda_2-\lambda_1)}\;,
    \end{eqnarray}
    \begin{eqnarray}
    A_3&=&-\sum_{k,k',p'}n_{k,\downarrow}(1-n_{k',\downarrow})(1-n_{p',\uparrow})\nonumber\\
    &&\qquad\quad\times e^{\epsilon(k)(\lambda_3-\lambda_1)}e^{-\epsilon(k')(\lambda_3-\lambda_2)}e^{-\epsilon(p')(\lambda_2-\lambda_1)}\;,\nonumber \\
    \end{eqnarray}
    \begin{eqnarray}
    A_4&=&\sum_{k,p,p'}(n_{k,\uparrow}-n_{k,\downarrow})n_{p,\uparrow}(1-n_{p',\downarrow})\nonumber\\
        &&\qquad \times e^{\epsilon(p)(\lambda_2-\lambda_1)}e^{-\epsilon(p')(\lambda_2-\lambda_1)}\;,
        \end{eqnarray}
        \begin{eqnarray}
    A_5&=&\sum_{k,k',p'}n_{k,\uparrow}(1-n_{k',\uparrow})(1-n_{p',\downarrow})\nonumber\\
    &&\qquad\times e^{\epsilon(k)(\lambda_3-\lambda_1)}e^{-\epsilon(k')(\lambda_3-\lambda_2)}e^{-\epsilon(p')(\lambda_2-\lambda_1)}\;,\nonumber \\
\end{eqnarray}
and
\begin{eqnarray}
    A_6&=&\sum_{k,k',p}n_{k,\downarrow}(1-n_{k',\downarrow})n_{p,\uparrow}e^{\epsilon(k)(\lambda_3-\lambda_2)}\nonumber\\
    &&\qquad\times e^{-\epsilon(k')(\lambda_3-\lambda_1)}e^{\epsilon(p)(\lambda_2-\lambda_1)}\;.
\end{eqnarray}
Again, we use the function $\tilde{f}(\lambda,B)$, see Eq.~(\ref{ftildelambdaBdefinition}), and find
\begin{eqnarray}\label{f-tauB-function-properties}
\frac{1}{L}\sum_ke^{\lambda\epsilon(k)}n_{k,\uparrow}&=&\frac{1}{L}\sum_{k'}e^{-\lambda\epsilon(k')}(1-n_{k',\downarrow})\nonumber\\
&=&\rho_0(0)\tilde{f}(\lambda,B)\;,\nonumber\\[3pt]
\frac{1}{L}\sum_ke^{\lambda\epsilon(k)}n_{k,\downarrow}&=&\frac{1}{L}\sum_{k'}e^{-\lambda\epsilon(k')}(1-n_{k',\uparrow})\nonumber\\
&=&\rho_0(0)\tilde{f}(\lambda,-B)\;.
\end{eqnarray}
Now we can rewrite the first term of Eq.~(\ref{V3aAfterEvaluationOfTheTrace}) as
\begin{eqnarray}
{\rm 1}^{\rm st}&=&\frac{W^-}{L^3}\sum_{k,p,p'}(n_{k,\uparrow}-n_{k,\downarrow})(1-n_{p',\uparrow})\nonumber \\
&& \qquad \quad \times n_{p,\downarrow}e^{(\epsilon(p)-\epsilon(p'))(\lambda_2-\lambda_1)}\nonumber\\
&=&W^-M_0(B)\rho_0^2(0)\tilde{f}^2(\lambda_2-\lambda_1,-B)
\end{eqnarray}
with $M_0(B,T)$ from Eq.~(\ref{M0definition}). Treating the other terms accordingly, 
we arrive at
\begin{eqnarray}\label{Result-of-wa3}
w_3^a&=&W^-M_0(B)\rho_0^2(0)\tilde{f}^2(\lambda_2-\lambda_1,-B)\nonumber\\
&&-W^+M_0(B)\rho_0^2(0)\tilde{f}^2(\lambda_2-\lambda_1,B)\nonumber\\
&&-2W^-\rho_0^3(0)\tilde{f}(\lambda_3-\lambda_2,B)\tilde{f}(\lambda_3-\lambda_1,-B)\nonumber\\
&&\quad \times \tilde{f}(\lambda_2-\lambda_1,-B)\nonumber\\
&&-2W^+\rho_0^3(0)\tilde{f}(\lambda_3-\lambda_2,-B)\tilde{f}(\lambda_3-\lambda_1,B)\nonumber\\
&&\quad \times \tilde{f}(\lambda_2-\lambda_1,B)\
\end{eqnarray}
for the first contribution. 

\subsection{Term \texorpdfstring{$\mathbf{V_3^b}$}{V3b} from third order}
\label{appendix:thirdorder-V3b}


\subsubsection{Evaluation of the trace}

We start from Eq.~(\ref{RemainingpartsofV3}) and~(\ref{V3z-V3a-V3b-V3c-Definition})
\begin{eqnarray}
V_3^b&=&\left(\frac{J_\perp}{2}\right)^2\frac{J_z}{4}\int_0^\beta\rmd\lambda_3\,\int_0^{\lambda_3}\rmd\lambda_2\,\int_0^{\lambda_2}\rmd\lambda_1\,w_3^b\;,\nonumber\\
\end{eqnarray}
where
\begin{eqnarray}
w_3^b&=&\frac{1}{L}\sum_{k,k'}\frac{1}{Z_0}
{\rm Tr}\bigg(
e^{-\beta\hat{H}_0}\left(\hat{s}_0^+(\lambda_3)\hat{S}^-+\hat{s}_0^-(\lambda_3)\hat{S}^+\right)\nonumber\\
&&\qquad\qquad\times\left(\hat{n}^d_\Uparrow-\hat{n}^d_\Downarrow\right)\left(\hat{n}^{kk'}_\uparrow(\lambda_2)-\hat{n}^{kk'}_\downarrow(\lambda_2)\right)\nonumber\\
&&\qquad\qquad\times\left(\hat{s}_0^+(\lambda_1)\hat{S}^-+\hat{s}_0^-(\lambda_1)\hat{S}^+\right)
\bigg)\;.
\end{eqnarray}
Note that 
\begin{eqnarray}
\hat{n}^d_\Uparrow\hat{S}^+&=&\hat{S}^+\;,\nonumber\\
\hat{n}^d_\Downarrow\hat{S}^-&=&\hat{S}^-\;,\nonumber\\
\hat{n}^d_\Uparrow\hat{S}^-&=&\hat{n}^d_\Downarrow\hat{S}^+=0\;,\nonumber\\
\hat{S}^+\hat{S}^-&=&\hat{n}^d_\Uparrow\;,\nonumber\\
\hat{S}^-\hat{S}^+&=&\hat{n}^d_\Downarrow\;,
\end{eqnarray}
in the spin-1/2 single-impurity model. With these properties we find
\begin{eqnarray}
w_3^b&=&\frac{1}{Z_0}\frac{1}{L}\sum_{k,k'}
{\rm Tr}\biggl[e^{-\beta\hat{H}_0}\Big(\hat{n}^d_\Downarrow\mathcal{N}_7^{k,k'}\hat{n}^d_\Downarrow-\hat{n}^d_\Downarrow\mathcal{N}_8^{k,k'}\hat{n}^d_\Downarrow\nonumber\\
&&\qquad\qquad\qquad -\hat{n}^d_\Uparrow\mathcal{N}_6^{k,k'}\hat{n}^d_\Uparrow+\hat{n}^d_\Uparrow\mathcal{N}_5^{k,k'}\hat{n}^d_\Uparrow\Big)
\biggr]\nonumber \\
\end{eqnarray}
and therefore [$\bar{w}_3^b=2Z_0 L\cosh(\beta B)w^b_3$]
\begin{eqnarray}
\bar{w}_3^b&=&\sum_{k,k'}
{\left\langle\Uparrow|e^{\beta B(\hat{n}^d_\Uparrow-\hat{n}^d_\Downarrow)}\hat{n}^d_\Uparrow|\Uparrow\right\rangle}\left(\mathcal{N}^{k,k'}_5-\mathcal{N}^{k,k'}_6\right)\nonumber\\
&&-\sum_{k,k'}\left\langle\Downarrow|e^{\beta B(\hat{n}^d_\Uparrow-\hat{n}^d_\Downarrow)}\hat{n}^d_\Downarrow|\Downarrow\right\rangle\left(-\mathcal{N}^{k,k'}_7+\mathcal{N}^{k,k'}_8
\right)\nonumber\\
\end{eqnarray}
holds with
\begin{eqnarray}
    \mathcal{N}^{k,k'}_5&=&\left\langle \hat{s}^-_0(\lambda_3)\hat{n}^{kk'}_\downarrow(\lambda_2)\hat{s}^+_0(\lambda_1) \right\rangle_F\;,\nonumber\\[2pt]
    \mathcal{N}^{k,k'}_6&=&\left\langle \hat{s}^-_0(\lambda_3)\hat{n}^{kk'}_\uparrow(\lambda_2)\hat{s}^+_0(\lambda_1) \right\rangle_F\;,\nonumber\\[2pt]
    \mathcal{N}^{k,k'}_7&=&\left\langle \hat{s}^+_0(\lambda_3)\hat{n}^{kk'}_\uparrow(\lambda_2)\hat{s}^-_0(\lambda_1) \right\rangle_F\;,\nonumber\\[2pt]
    \mathcal{N}^{k,k'}_8&=&\left\langle \hat{s}^+_0(\lambda_3)\hat{n}^{kk'}_\downarrow(\lambda_2)\hat{s}^-_0(\lambda_1) \right\rangle_F\;.
\end{eqnarray}
The evaluation of the impurity-spin operators leaves us with the traces over the fermionic bath,
\begin{eqnarray}
2\cosh(\beta B)L^3w_3^b&=&\sum_{k,k'}e^{\beta B}\left(\mathcal{N}^{k,k'}_5-\mathcal{N}^{k,k'}_6\right)\nonumber\\
&&-\sum_{k,k'}e^{-\beta B}\left(-\mathcal{N}^{k,k'}_7+\mathcal{N}^{k,k'}_8\right)\;.\nonumber\\
\end{eqnarray}
To evaluate the expectation values $\mathcal{N}_i^{k,k'}$ ($i=5,\ldots,8$) we again use Wick's theorem, 
\begin{widetext}
\begin{eqnarray}
    \sum_{k,k'} \mathcal{N}^{k,k'}_5&=&\sum_{\substack{k,k',p\\p',m,m'}} \left\langle e^{\lambda_3\hat{T}}\hat{a}^+_{k,\downarrow}\hat{a}^\phan_{k',\uparrow}e^{-\lambda_3\hat{T}}e^{\lambda_2\hat{T}}\hat{a}^+_{p,\downarrow}\hat{a}^\phan_{p',\downarrow}e^{-\lambda_2\hat{T}}e^{\lambda_1\hat{T}}\hat{a}^+_{m,\uparrow}\hat{a}^\phan_{m',\downarrow}e^{\lambda_1\hat{T}} \right\rangle_F\nonumber\\
    &=&-\sum_{k,k',p}n_{k,\downarrow}n_{p,\downarrow}(1-n_{k',\uparrow})e^{-\lambda_3(\epsilon(k')-\epsilon(k))}e^{-\lambda_2(\epsilon(k)-\epsilon(p))}e^{-\lambda_1(\epsilon(p)-\epsilon(k'))}\nonumber\\
    &&+\sum_{k,k',p}n_{k,\downarrow}n_{p,\downarrow}(1-n_{k',\uparrow})e^{-\lambda_3(\epsilon(k')-\epsilon(k))}e^{-\lambda_1(\epsilon(k)-\epsilon(k'))}\;,\\
    \sum_{k,k'} \mathcal{N}^{k,k'}_6&=&\sum_{\substack{k,k',p\\p',m,m'}}\left\langle e^{\lambda_3\hat{T}}\hat{a}^+_{k,\downarrow}\hat{a}^\phan_{k',\uparrow}e^{-\lambda_3\hat{T}}e^{\lambda_2\hat{T}}\hat{a}^+_{p,\uparrow}\hat{a}^\phan_{p',\uparrow}e^{-\lambda_2\hat{T}}e^{\lambda_1\hat{T}}\hat{a}^+_{m,\uparrow}\hat{a}^\phan_{m',\downarrow}e^{\lambda_1\hat{T}} \right\rangle_F\nonumber\\
    &=&-\sum_{k,k',p'}(1-n_{k',\uparrow})(1-n_{p',\uparrow})n_{k,\downarrow}e^{-\lambda_3(\epsilon(k')-\epsilon(k))}e^{-\lambda_2(\epsilon(p')-\epsilon(k'))}e^{-\lambda_1(\epsilon(k)-\epsilon(p'))}\nonumber\\
    &&-\sum_{k,k',p'}(1-n_{k',\uparrow})n_{p,\uparrow}n_{k,\downarrow}e^{-\lambda_3(\epsilon(k')-\epsilon(k))}e^{-\lambda_1(\epsilon(k)-\epsilon(k'))}\; ,\\
    \sum_{k,k'} \mathcal{N}^{k,k'}_7&=&\sum_{\substack{k,k',p\\p',m,m'}}\left\langle e^{\lambda_3\hat{T}}\hat{a}^+_{k,\uparrow}\hat{a}^\phan_{k',\downarrow}e^{-\lambda_3\hat{T}}e^{\lambda_2\hat{T}}\hat{a}^+_{p,\uparrow}\hat{a}^\phan_{p',\uparrow}e^{-\lambda_2\hat{T}}e^{\lambda_1\hat{T}}\hat{a}^+_{m,\downarrow}\hat{a}^\phan_{m',\uparrow}e^{\lambda_1\hat{T}} \right\rangle_F\nonumber\\
    &=&-\sum_{k,k',p}n_{k,\uparrow}n_{p,\uparrow}(1-n_{k',\downarrow})e^{-\lambda_3(\epsilon(k')-\epsilon(k))}e^{-\lambda_2(\epsilon(k)-\epsilon(p))}e^{-\lambda_1(\epsilon(p)-\epsilon(k'))}\nonumber\\
    &&+\sum_{k,k',p}n_{k,\uparrow}n_{p,\uparrow}(1-n_{k',\downarrow})e^{-\lambda_3(\epsilon(k')-\epsilon(k))}e^{-\lambda_1(\epsilon(k)-\epsilon(k'))}\;,
 \end{eqnarray}
and
    \begin{eqnarray}
    \sum_{k,k'} \mathcal{N}^{k,k'}_8&=&\sum_{\substack{k,k',p\\p',m,m'}}\left\langle e^{\lambda_3\hat{T}}\hat{a}^+_{k,\uparrow}\hat{a}^\phan_{k',\downarrow}e^{-\lambda_3\hat{T}}e^{\lambda_2\hat{T}}\hat{a}^+_{p,\downarrow}\hat{a}^\phan_{p',\downarrow}e^{-\lambda_2\hat{T}}e^{\lambda_1\hat{T}}\hat{a}^+_{m,\downarrow}\hat{a}^\phan_{m',\uparrow}e^{\lambda_1\hat{T}} \right\rangle_F\nonumber\\
    &=&-W^{+}\sum_{k,k',p'}(1-n_{k',\downarrow})(1-n_{p',\downarrow})n_{k,\uparrow}e^{-\lambda_3(\epsilon(k')-\epsilon(k))}e^{-\lambda_2(\epsilon(p')-\epsilon(k'))}e^{-\lambda_1(\epsilon(k)-\epsilon(p'))}\nonumber\\
&&-W^{+}\sum_{k,k',p}(1-n_{k',\downarrow})n_{p,\downarrow}n_{k,\uparrow}e^{-\lambda_3(\epsilon(k')-\epsilon(k))}e^{-\lambda_1(\epsilon(k)-\epsilon(k'))}\;.
\end{eqnarray}
\end{widetext}

\subsubsection{Further simplifications}

We combine the last terms of $\mathcal{N}^{k,k'}_5$ and $\mathcal{N}^{k,k'}_6$, and of $\mathcal{N}^{k,k'}_7$ and $\mathcal{N}^{k,k'}_8$. With Eq.~(\ref{Wpm-Definition}), we then find
\begin{eqnarray}\label{V3bAfterEvaluationOfTheTrace}
    L^3w_3^b&=&-W^{-}\left(A_7+A_8+A_9\right)\nonumber\\
    &&+W^+\left(-A_{10}+A_{11}-A_{12}\right)\;,
\end{eqnarray}
where
\begin{eqnarray}\label{A7A8vonV3b}
    A_7&=&\sum_{k,k',p}n_{k,\downarrow}n_{p,\downarrow}(1-n_{k',\uparrow})\nonumber\\
    &&\qquad\times e^{-\epsilon(k)(\lambda_2-\lambda_3)}e^{\epsilon(k')(\lambda_1-\lambda_3)}e^{-\epsilon(p)(\lambda_1-\lambda_2)}\;,\nonumber \\
    \end{eqnarray}
    \begin{eqnarray}
    A_8&=&\sum_{k,k',p}n_{k,\downarrow}(n_{p,\uparrow}-n_{p,\downarrow})(1-n_{k',\uparrow})\nonumber\\
    &&\qquad\times e^{-\epsilon(k)(\lambda_1-\lambda_3)}e^{\epsilon(k')(\lambda_1-\lambda_3)}\;,
    \end{eqnarray}
    \begin{eqnarray}
    A_9&=&\sum_{k,k',p'}(1-n_{k',\uparrow})(1-n_{p',\uparrow})n_{k,\downarrow}e^{-\epsilon(k)(\lambda_1-\lambda_3)}\nonumber \\
        &&\qquad\times e^{\epsilon(k')(\lambda_2-\lambda_3)}e^{\epsilon(p')(\lambda_1-\lambda_2)}\;,
    \end{eqnarray}
    \begin{eqnarray}
    A_{10}&=&\sum_{k,k',p}n_{k,\uparrow}n_{p,\uparrow}(1-n_{k',\downarrow})\nonumber\\
    &&\qquad\times e^{-\epsilon(k)(\lambda_2-\lambda_3)}e^{\epsilon(k')(\lambda_1-\lambda_3)}e^{-\epsilon(p)(\lambda_1-\lambda_2)}\;,\nonumber \\
    \end{eqnarray}
    \begin{eqnarray}
    A_{11}&=&\sum_{k,k',p}n_{k,\uparrow}(n_{p,\uparrow}-n_{p,\downarrow})(1-n_{k',\downarrow})\nonumber\\
    &&\qquad\times e^{-\epsilon(k)(\lambda_1-\lambda_3)}e^{\epsilon(k')(\lambda_1-\lambda_3)}\;,
    \end{eqnarray}
\begin{eqnarray}    A_{12}&=&\sum_{k,k',p'}(1-n_{k',\downarrow})(1-n_{p',\downarrow})n_{k,\uparrow}\nonumber\\
    &&\qquad\times e^{-\epsilon(k)(\lambda_1-\lambda_3)}e^{\epsilon(k')(\lambda_2-\lambda_3)}e^{\epsilon(p')(\lambda_1-\lambda_2)}\;.\nonumber \\
\end{eqnarray}
With Eq.~(\ref{f-tauB-function-properties}), and~(\ref{ftildelambdaBdefinition}) we  arrive at
\begin{eqnarray}\label{Result-of-wb3}
w_3^b &=&\rho_0^2(0)W^+M_0(B)\tilde{f}^2(\lambda_3-\lambda_1,B)\nonumber\\
&&-\rho_0^2(0)W^-M_0(B)\tilde{f}^2(\lambda_3-\lambda_1,-B)\nonumber\\
&&-2\rho_0^3(0) W^+\tilde{f}(\lambda_3-\lambda_2,B)\tilde{f}(\lambda_2-\lambda_1,B)\nonumber\\
&&\qquad\qquad  \times \tilde{f}(\lambda_3-\lambda_1,B)\nonumber\\
&&-2\rho_0^3(0)W^-\tilde{f}(\lambda_3-\lambda_2,-B)\tilde{f}(\lambda_2-\lambda_1,-B)\nonumber\\
&&\qquad\qquad\times \tilde{f}(\lambda_3-\lambda_1,-B)
\end{eqnarray}
for the second contribution. 

\subsection{Term \texorpdfstring{$\mathbf{V_3^c}$}{V3c} from third order}
\label{appendix:thirdorder-V3c}

We can obtain the term $V_3^c$ from the result for $V_3^a$, as we will show next. We start with its definition in Eq.~(\ref{RemainingpartsofV3}),
\begin{eqnarray}
V_3^c&=&\int_0^\beta\rmd\lambda_3\,\int_0^{\lambda_3}\rmd\lambda_2\,\int_0^{\lambda_2}\rmd\lambda_1\,\nonumber\\
&&\quad\times \frac{1}{Z_0}{\rm Tr}\left(
e^{-\beta\hat{H}_0}\hat{V}_\perp(\lambda_3)\hat{V}_\perp(\lambda_2)\hat{V}_z(\lambda_1)
\right) \nonumber\\
&=&\int_0^\beta\rmd\lambda_3\,\int_0^{\lambda_3}\rmd\lambda_2\,\int_0^{\lambda_2}\rmd\lambda_1\,\nonumber\\
&&\quad\times \frac{1}{Z_0}{\rm Tr}\left(
V_z^+(\lambda_1)\hat{V}_\perp^+(\lambda_2)\hat{V}_\perp^+(\lambda_3)e^{-\beta\hat{H}_0}
\right)\;.\nonumber\\
\end{eqnarray}
Note that for observables $\hat{A}^+(\lambda)=\hat{A}(-\lambda)$\;.
Thus,
\begin{eqnarray}\label{eq:V3c-rewritten}
V_3^c&=&\int_0^\beta\rmd\lambda_3\,\int_0^{\lambda_3}\rmd\lambda_2\,\int_0^{\lambda_2}\rmd\lambda_1\,\nonumber\\
&&\times \frac{1}{Z_0}{\rm Tr}\left(e^{-\beta\hat{H}_0}
\hat{V}_z(-\lambda_1)\hat{V}_\perp(-\lambda_2)\hat{V}_\perp(-\lambda_3)\right)\;,\nonumber\\
\end{eqnarray}
where we used that the trace is cyclic. Equation~(\ref{eq:V3c-rewritten}) is known from $V_3^a$ 
when we replace $\lambda_3\to -\lambda_1$, $\lambda_2\to-\lambda_2$ and $\lambda_1\to -\lambda_3$. Then, 
we find for the third contribution
\begin{eqnarray}\label{Result-of-wc3}
V_3^c&=&\int_0^\beta\rmd\lambda_3\,\int_0^{\lambda_3}\rmd\lambda_2\,\int_0^{\lambda_2}\rmd\lambda_1\, \nonumber\\
&& \times \Big( W^- M_0(B)\rho_0^2(0)\tilde{f}^2(\lambda_3-\lambda_2,-B)\nonumber\\
&&\quad -W^+M_0(B)\rho_0^2(0)\tilde{f}^2(\lambda_3-\lambda_2,B)\nonumber\\
&&\quad-2 W^-\rho_0^3(0)\tilde{f}(\lambda_3-\lambda_2,-B)\tilde{f}(\lambda_3-\lambda_1,-B)\nonumber\\
&&\quad \qquad \times \tilde{f}(\lambda_2-\lambda_1,B)\nonumber\\
&&\quad -2 W^+ \rho_0^3(0)\tilde{f}(\lambda_3-\lambda_2,B)\tilde{f}(\lambda_3-\lambda_1,B)\nonumber\\
&&\quad\qquad\times \tilde{f}(\lambda_2-\lambda_1,-B)\Big)\; .
\end{eqnarray}

\subsection{Rearrangement of the third-order contributions}
\label{section:Rearrangement}

Looking more closely at the three terms $w_3^a$, $w_3^b$ and $w_3^c$, see eqs.~(\ref{Result-of-wa3}),~(\ref{Result-of-wb3}), and~(\ref{Result-of-wc3}), 
we can combine those terms that are proportional to $M_0(B)$, and regroup the remaining terms. Sorting the term $V_3$ thus leads to 
\begin{eqnarray}
 TV_3-TV_3^z &=& F_3^\alpha+F_3^\beta\;,\nonumber\\
 F_3^\alpha&=&\frac{1}{16}j_\perp^2 j_z x_A\;,\nonumber \\
 F_3^\beta&=&\frac{1}{8}j_\perp^2j_z x_B
\end{eqnarray}
with
\begin{widetext}
\begin{eqnarray}\label{xA-Definition-third-order}
x_A&=&T\int_0^\beta\rmd\lambda_3\,\int_0^{\lambda_3}\rmd\lambda_2\,\int_0^{\lambda_2}\rmd\lambda_1\,\nonumber\\
&&\times \Big(
-W^+\tilde{M}_0(B)\tilde{f}^2(\lambda_2-\lambda_1,B)+ W^- \tilde{M}_0(B)\tilde{f}^2(\lambda_2-\lambda_1,-B+ W^+ \tilde{M}_0(B)\tilde{f}^2(\lambda_3-\lambda_1,B)\nonumber\\
&&\quad - W^- \tilde{M}_0(B)\tilde{f}^2(\lambda_3-\lambda_1,-B)- W^+ \tilde{M}_0(B)\tilde{f}^2(\lambda_3-\lambda_2,B)+ W^- \tilde{M}_0(B)\tilde{f}^2(\lambda_3-\lambda_2,-B)\Big)
\end{eqnarray}
and
\begin{eqnarray}\label{xB-Definition-third-order}
x_B&=&-T\int_0^\beta\rmd\lambda_3\,\int_0^{\lambda_3}\rmd\lambda_2\,\int_0^{\lambda_2}\rmd\lambda_1\nonumber\\
&&\times\Big(
W^+ \tilde{f}(\lambda_3-\lambda_2,-B)\tilde{f}(\lambda_3-\lambda_1,B)\tilde{f}(\lambda_2-\lambda_1,B)+W^- \tilde{f}(\lambda_3-\lambda_2,B)\tilde{f}(\lambda_3-\lambda_1,-B) \tilde{f}(\lambda_2-\lambda_1,-B)\nonumber\\
&&\quad+W^+ \tilde{f}(\lambda_3-\lambda_2,B)\tilde{f}(\lambda_3-\lambda_1,B) \tilde{f}(\lambda_2-\lambda_1,B)+W^- \tilde{f}(\lambda_3-\lambda_2,-B)\tilde{f}(\lambda_3-\lambda_1,-B) \tilde{f}(\lambda_2-\lambda_1,-B)\nonumber\\
&&\quad +W^+ \tilde{f}(\lambda_3-\lambda_2,B)\tilde{f}(\lambda_3-\lambda_1,B) \tilde{f}(\lambda_2-\lambda_1,-B)+W^- \tilde{f}(\lambda_3-\lambda_2,-B)\tilde{f}(\lambda_3-\lambda_1,-B) \tilde{f}(\lambda_2-\lambda_1,B)
\Big)\;,\nonumber\\
\end{eqnarray}
where $j_\perp=\rho_0(0)J_\perp$, $j_z=\rho_0(0)J_z$ and $\tilde{M}_0(B)=M_0(B)/\rho_0(0)$, $M_0(B)$ from Eq.~(\ref{M0definition}), and $\tilde{f}(\lambda,B)$ from Eq.~(\ref{ftildelambdaBdefinition}). Note that $\tilde{M}_0(B)\approx 2 B$ for $B\ll1$. Our remaining task is the evaluation of the two terms $x_A$ and $x_B$.
\end{widetext}

\subsection{Term \texorpdfstring{$\mathbf{x_B}$}{xB}}
\label{chapter:the-term-xB}

The calculation of the term $x_B$ is extremely lengthy, and is thus deferred to the Supplemental Material, Sect.~\SMsmxBtermderivation. Here, we give only the analytical result for $0\ll B\ll T\ll1 $. In order for this work to be self-contained, we provide an additional way to warrant $x_B$ numerically.

\subsubsection{Result for \texorpdfstring{$x_B$}{xB}}
The analytical expression of the term $x_B$ at low temperatures and small magnetic fields reads
\begin{equation}\label{SummaryXB}
x_B=x_B^{(0)}+B^2 x_B^{(2)}
\end{equation}
with 
\begin{equation}
x_B^{(0)}=-\frac{\pi^2}{2}-6\ln^2(2)+2\pi^2 T+2\pi^2 T^2(\ln 2 -1)
\end{equation}
and
\begin{eqnarray}
x_B^{(2)}&=&\frac{1}{T}\bigg(
8+8\EulerGamma+4\EulerGamma^2-\frac{\pi^2}{6}+4\ln^2(2)\nonumber\\
&&\quad+8\ln 2 \ln\pi+4\ln^2(\pi)-8\ln(2\pi T)(1+\EulerGamma)\nonumber\\
&&\quad+8\ln(2\pi)\ln T +4\ln^2(T)\bigg)\;.\nonumber\\
&&+(-8+6\ln2) +\mathcal{O}[B^3]\;.
\end{eqnarray}
It is of utmost importance to consider the order of the respective corrections because the separation in $x_A$ and $x_B$ could lead to terms which do not exist in the sum $x_A+x_B$.

\subsubsection{Numerical check for \texorpdfstring{$x_B$}{xB}}
\label{section:numericalcheck-xB}

The numerical evaluation of the $x_B$ term requires a specific procedure to obtain stable results. 
We start from the definition~(\ref{xB-Definition-third-order}) and use the identity
\begin{eqnarray}
N(\beta)&=&\int_0^\beta\rmd\lambda_3\,\int_0^{\lambda_3}\rmd\lambda_2\,\int_0^{\lambda_2}\rmd\lambda_1\nonumber \\
&&\quad\times F(\lambda_3-\lambda_2)G(\lambda_2-\lambda_1)H(\lambda_3-\lambda_1) \nonumber\\
&=&\int_0^\beta\rmd u\,(\beta-u)H(u)\int_0^u\rmd v\, F(v)G(u-v)\;,\nonumber\\
\end{eqnarray}
see Sect.~\SMsubsecTripleIntegral\ of the Supplemental Material, to write
\begin{eqnarray}
    x_B&=&-T\int_0^\beta \rmd u\,(\beta- u)\left(W^+ \mathcal{L}_1 + W^-\mathcal{L}_2\right)
\end{eqnarray}
with 
\begin{eqnarray}
    \mathcal{L}_1&=&\tilde{f}(u,B)\int_0^u\rmd v\,\tilde{f}(v,-B)\tilde{f}(u-v,B)\nonumber\\
    &&+\tilde{f}(u,B)\int_0^u\rmd v\,\tilde{f}(v,B)\tilde{f}(u-v,B)\nonumber\\
    &&+\tilde{f}(u,B)\int_0^u\rmd v\,\tilde{f}(v,B)\tilde{f}(u-v,-B)
\end{eqnarray}
and 
\begin{eqnarray}
    \mathcal{L}_2&=&\tilde{f}(u,-B)\int_0^u\rmd v\,\tilde{f}(v,B)\tilde{f}(u-v,-B)\nonumber\\
    &&+\tilde{f}(u,-B)\int_0^u\rmd v\,\tilde{f}(v,-B)\tilde{f}(u-v,-B)\nonumber\\
    &&+\tilde{f}(u,-B)\int_0^u\rmd v\,\tilde{f}(v,-B)\tilde{f}(u-v,B)\;.\nonumber\\
\end{eqnarray}
We define
\begin{eqnarray}
x_B(B,T,\tau_1,\tau_2,\tau_3)&=&-T\int_0^\beta\rmd u\,(\beta-u)\tilde{f}(u,\tau_1 B)\nonumber\\
&&\quad\times\!\int_0^u \rmd v\, \tilde{f}(v,\tau_2 B)\tilde{f}(u-v,\tau_3 B)\nonumber\\
\end{eqnarray}
so that
\begin{eqnarray}\label{connection-xB-xBtau}
x_B(B,T)&=&W^+ x_B(B,T,1,-1,1)\nonumber\\
&&+W^- x_B(B,T,-1,1,-1)\nonumber\\
&&+W^+ x_B(B,T,1,1,1)\nonumber\\
&&+W^- x_B(B,T,-1,-1,-1) \nonumber\\
&&+ W^+ x_B(B,T,1,1,-1)\nonumber\\
&&+W^- x_B(B,T,-1,-1,1)\;,
\end{eqnarray}
where ($b=B/T\ll 1$)
\begin{eqnarray}
\tilde{f}(\lambda,B)&=&\int_{-1}^1\rmd \epsilon\,\frac{\rhotilde(\epsilon)e^{\lambda\epsilon}}{1+e^{\beta(\epsilon-B)}}=\int_{-1}^1\rmd \epsilon\,\frac{e^{\lambda\epsilon}}{1+e^{\beta(\epsilon-B)}}\;,\nonumber\\
\tilde{F}(x,b)&=&\frac{1}{T}\int_{-1}^1\rmd \epsilon\,\frac{\rhotilde(\epsilon/T)e^{x\epsilon/T}}{1+e^{\epsilon/T-b}}=\int_{-1/T}^{1/T}\rmd s\,\frac{e^{xs}}{1+e^{s-b}}\;,\nonumber\\
\end{eqnarray}
and we used that $\rhotilde(\epsilon)=1$ for our constant density of states.
Moreover, with the substitutions $u=\beta x$ and $v=\beta y$, we obtain 
\begin{eqnarray}\label{xB-numerical-integration-fig}
x_B(B,T,\tau_1,\tau_2,\tau_3)&=&-T\int_0^1\rmd x\,(1-x)\tilde{F}(x,\tau_1 b)\nonumber\\
&&\times\int_0^x\rmd y\,\tilde{F}(y,\tau_2 b)\tilde{F}(x-y,\tau_3 b)\;.\nonumber\\
\end{eqnarray}
The double-integral must be split in order to obtain numerically tractable integrands.

\begin{figure}[t]
    \includegraphics[width=8.5cm]{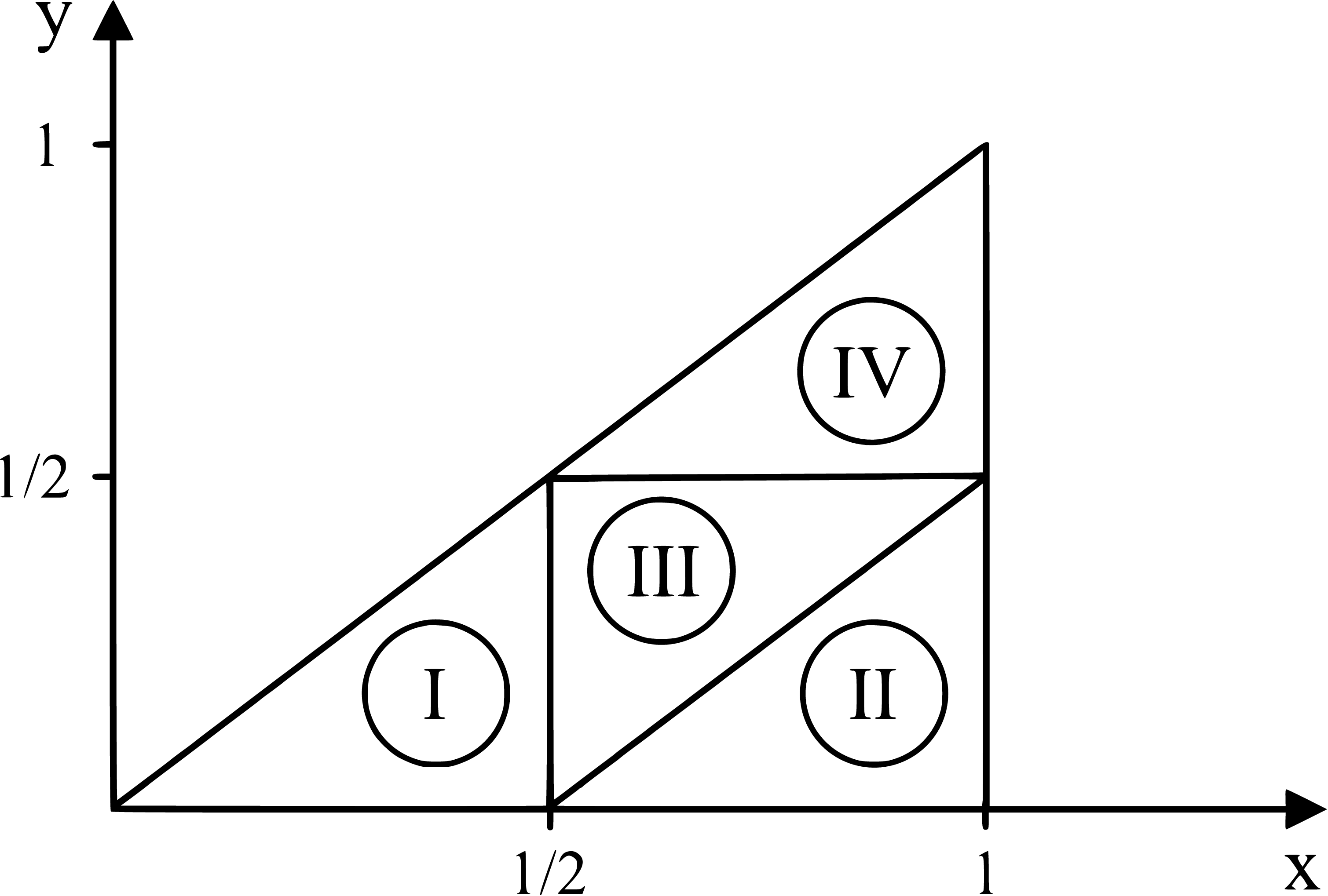}
 \caption{The four integration regions in Eq.~(\ref{xB-numerical-integration-fig}) are \newline
\begin{tabular}{@{}ll@{}}
I: & $0\leq x\leq 1/2$ and $0\leq y\leq x$; \\[2pt]
II: & $1/2\leq x\leq 1$ and $0\leq y\leq x-1/2$; \\[2pt]
III: &$1/2\leq x\leq 1$ and $x-1/2\leq y\leq 1/2$; \\[2pt]
IV: &  $1/2\leq x \leq 1$ and $1/2\leq y\leq x$.
\end{tabular}
\label{fig:xB-integration-regions}}
\end{figure}

To this end we note that
\begin{eqnarray}
\tilde{F}(1-x,\tau b)&=&\int_{-1/T}^{1/T}\rmd s\,\frac{e^{(1-x)s}}{1+e^{s-\tau b}}\nonumber\\
&=&\int_{-1/T}^{1/T}\rmd s\,\frac{e^{(-1+x)s}}{1+e^{-s-\tau b}}\frac{e^{\tau b}}{e^{\tau b}}\nonumber\\
&=&e^{\tau b}\tilde{F}(x,-\tau b)\;.
\end{eqnarray}
This permits a splitting of the integration region.
The four proper integration regions in Eq.~(\ref{xB-numerical-integration-fig}) are visualized in figure~\ref{fig:xB-integration-regions}.  
The individual integration regions contribute [$\tilde{x}=1-x$, $x_B^i=x_B^i(B,T,\tau_1,\tau_2,\tau_3)$ with $i={\rm I},{\rm II},{\rm III},{\rm IV}$]
\begin{eqnarray}
x_B^{\rm I}&=&-T\int_0^{1/2}\rmd x\,(1-x)\tilde{F}(x,\tau_1 b)\nonumber\\
&&\times\int_0^x\rmd y\,\tilde{F}(y,\tau_2 b)\tilde{F}(x-y,\tau_3 b)
\end{eqnarray}
and
\begin{eqnarray}
x_B^{\rm II}&=&-T\int_0^{1/2}\rmd \tilde{x}\, \tilde{x}e^{\tau_1 b}\tilde{F}(\tilde{x},-\tau_1 b)\nonumber\\
&&\times\int_0^{1/2-\tilde{x}}\rmd y \tilde{F}(y,\tau_2 b)\tilde{F}(1-\tilde{x}-y,\tau_3 b)\nonumber\\
&=&-T\int_0^{1/2}\rmd x\, xe^{\tau_1 b}\tilde{F}(x,-\tau_1b)\nonumber\\
&&\times\int_0^{1/2-x}\rmd y\, \tilde{F}(y,\tau_2 b)e^{\tau_3 b}\tilde{F}(x+y,-\tau_3 b)\nonumber\\
\end{eqnarray}
because $0\leq x+y\leq 1/2$. Next, we have [$\tilde{x}=1-x$]
\begin{eqnarray}
x_B^{\rm III}&=&-T\int_0^{1/2}\rmd \tilde{x}\,\tilde{x}e^{\tau_2 b}\tilde{F}(\tilde{x},-\tau_1 b)\nonumber\\
&&\times\int_{1/2-\tilde{x}}^{1/2}\rmd y\, \tilde{F}(y,\tau_2 b)\tilde{F}(1-\tilde{x}-y,\tau_3 b)\nonumber\\
\end{eqnarray}
because $1/2-\tilde{x} \leq y \leq 1/2$, $1/2\leq \tilde{x}+y\leq 1/2+\tilde{x}$, $-1/2\geq - (x+y) \geq -1/2-\tilde{x}$ and $1/2\geq 1-(\tilde{x}+y)\geq 1/2 -\tilde{x}\geq 0$. Finally, [$\tilde{x}=1-x$]
\begin{eqnarray}
x_B^{\rm IV}&=&-T\int_0^{1/2}\rmd \tilde{x}\,\tilde{x}e^{\tau_1 b}\tilde{F}(\tilde{x},-\tau_1 b)\nonumber\\
&&\times\int_{1/2}^{1-\tilde{x}}\rmd y\, \tilde{F}(y,\tau_2b)\tilde{F}(1-\tilde{x}-y,\tau_3 b)\nonumber\\
&=&-T\int_0^{1/2}\rmd x\,xe^{\tau_1 b}\tilde{F}(x,-\tau_1 b)\nonumber\\
&&\times\int_{x}^{1/2}\rmd \tilde{y} e^{\tau_2 b}\tilde{F}(\tilde{y},-\tau_2 b)\tilde{F}(\tilde{y}-x,\tau_3 b)\nonumber\\
\end{eqnarray}
because $0\leq \tilde{y}-x\leq 1/2$ with $\tilde{y}=1-y$.

In $x_B^{\rm I}$ to $x_B^{\rm IV}$, all arguments of the functions $F$ are between zero and one half. We rearrange the terms in the following way, where $\xBkreis{j}=\xBkreis{j}(B,T,\tau_1,\tau_2,\tau_3)$, $j=1,2,3$:
\begin{itemize}
	\item[i)] We take the first term in $x_B^{\rm I}$ and define
	\begin{eqnarray}\label{xBkreis1def}
	\xBkreis{1}&=&-T\int_0^{1/2}\rmd x\, \tilde{F}(x,\tau_1 b)\nonumber\\
 &&\times\int_0^x\rmd y\, \tilde{F}(y,\tau_2 b)\tilde{F}(x-y,\tau_3 b)\;.
	\end{eqnarray}
	\item[ii)] We substitute $\tilde{y}=1/2-y$ in $x_B^{\rm III}$ and combine it with the second term in $x_B^{\rm I}$  to
	\begin{eqnarray}\label{xBkreis2def}
	\xBkreis{2}&=&-T\int_0^{1/2}\rmd x\,\int_0^x\rmd y\nonumber\\
 &&\times \bigg(
	-x \tilde{F}(x,\tau_1 b) \tilde{F}(y,\tau_2 b) \tilde{F}(x-y,\tau_3 b)\nonumber\\
	&&\qquad + xe^{\tau_1 b} \tilde{F}(x,-\tau_1 b)\tilde{F}(1/2-y,\tau_2 b)\nonumber\\
 &&\qquad\quad\times \tilde{F}(1/2-x+y,\tau_3 b)
	\bigg)\;.
	\end{eqnarray}
	\item[iii)] We substitute $\tilde{y}=1/2-y$ in $x_B^{\rm IV}$ and combine it with $x_B^{\rm II}$ to 
	\begin{eqnarray}\label{xBkreis3def}
	\xBkreis{3}&=&-T\int_0^{1/2}\rmd x\, \int_0^{1/2-x}\rmd y\,\nonumber\\
	&&\times \bigg(
	x e^{\tau_1 b}\tilde{F}(x,-\tau_1b)\tilde{F}(y,\tau_2 b) \nonumber\\
 &&\qquad \quad\times e^{\tau_3 b}\tilde{F}(x+y,-\tau_3b)\nonumber\\
	&& \qquad + xe^{\tau_1 b}\tilde{F}(x,-\tau_1 b)e^{\tau_2 b}\tilde{F}(1/2-y,-\tau_2 b)\nonumber\\
 &&\qquad\qquad\times \tilde{F}(1/2-x+y,\tau_3 b)
	\bigg)\;.
	\end{eqnarray}
\end{itemize}
Our considerations therefore imply 
\begin{equation}\label{connection-xBtau-and-xB123tau}
x_B(B,T,\tau_1,\tau_2,\tau_3)=\xBkreis{1}+\xBkreis{2}+\xBkreis{3}\;.
\end{equation}

Next, we expand $x_B(B,T,\tau_1,\tau_2,\tau_3)$  for small fields and temperatures, $B\ll T\ll1$, which is equivalent to an expansion for small $b=B/T \ll1$ . 
Therefore, we may approximate
\begin{eqnarray}
\tilde{F}(\lambda,\tau b)&=&\int_{-1/T}^{1/T}\rmd s\, \frac{e^{\lambda s}}{1+e^{s-\tau b}}\nonumber\\
&=&\tilde{F}_0(\lambda)+\tau b\tilde{F}_1(\lambda)+b^2\tilde{F}_2(\lambda)
\end{eqnarray}
with 
\begin{eqnarray}
\tilde{F}_0(\lambda)&=&\frac{\pi \lambda}{\sin(\pi\lambda)}-\frac{e^{-\lambda/T}}{\lambda}\;,\nonumber\\
\tilde{F}_1(\lambda)&=&\frac{\pi\lambda}{\sin(\pi\lambda)}\;,\nonumber\\
\tilde{F}_2(\lambda)&=&\frac{\lambda}{2}\frac{\pi\lambda}{\sin(\pi\lambda)}
\end{eqnarray}
for $0\leq \lambda\leq 1/2$ and $T\to 0$. These expressions are valid for the constant density of states.

Using Eq.~(\ref{connection-xB-xBtau}) and~(\ref{connection-xBtau-and-xB123tau}), we are able to give expressions that permits a reliable numerical evaluation of $x_B$. 
For the individual orders in $b$, these expressions are very lengthy. Therefore, we first define
\begin{eqnarray}
    \xBtildekreis{j}(B,T)&=&W^+ \xBkreis{j}(B,T,1,-1,1)\nonumber\\
&&+W^- \xBkreis{j}(B,T,-1,1,-1)\nonumber\\
&&+W^+ \xBkreis{j}(B,T,1,1,1)\nonumber\\
&&+W^- \xBkreis{j}(B,T,-1,-1,-1) \nonumber\\
&&+ W^+ \xBkreis{j}(B,T,1,1,-1)\nonumber\\
&&+W^- \xBkreis{j}(B,T,-1,-1,1)\;,
\end{eqnarray}
see Eq.~(\ref{connection-xBtau-and-xB123tau}), and expand ($b=B/T\ll1 $)
\begin{equation}
    \xBtildekreis{j}(B,T)=\xBtildekreis{j}^{(0)}(T)+b^2\xBtildekreis{j}^{(2)}(T)\;.
\end{equation}
As always, there is no linear order in $b$. Moreover, with eqs.~(\ref{xBkreis1def}--\ref{xBkreis3def}), we have 
\begin{eqnarray}
    \xBtildekreis{1}^{(n)}(T)&=&-T\int_0^{1/2}\rmd x\,\int_0^x\rmd y\,\tilde{\mathcal{X}}_n(T)\;,\nonumber\\
    \xBtildekreis{2}^{(n)}(T)&=&-T\int_0^{1/2}\rmd x\,\int_0^x\rmd y\,\tilde{\mathcal{Y}}_n(T)\;,\nonumber\\
    \xBtildekreis{3}^{(n)}(T)&=&-T\int_0^{1/2}\rmd x\,\int_0^{1/2-x}\rmd y\,\tilde{\mathcal{Z}}_n(T)\;.\nonumber\\
    \label{appeq:tildexBnumerics}
\end{eqnarray}
With the help of~\textsc{Mathematica}~\cite{Mathematica12}, we find for the second-order terms,
\begin{eqnarray}
    \tilde{\mathcal{X}}_2&=&-3\tilde{F}_0(x-y)\tilde{F}_0(y)\tilde{F}_1(x)
    -\tilde{F}_0(x)\tilde{F}_0(y)\tilde{F}_1(x-y)\nonumber \\
&&+\tilde{F}_0(y)\tilde{F}_1(x)\tilde{F}_1(x-y)
 -\tilde{F}_0(x)\tilde{F}_0(x-y)\tilde{F}_1(y)\nonumber\\
&&+\tilde{F}_0(x-y)\tilde{F}_1(x)\tilde{F}_1(y)
-\tilde{F}_0(x)\tilde{F}_1(x-y)\tilde{F}_1(y)\nonumber\\
&&+3 \tilde{F}_0(x-y)\tilde{F}_0(y)\tilde{F}_2(x)
+3 \tilde{F}_0(x)\tilde{F}_0(y)\tilde{F}_2(x-y)\nonumber\\
    &&+3 \tilde{F}_0(x)\tilde{F}_0(x-y)\tilde{F}_2(y)\; ,
\end{eqnarray}
\begin{eqnarray}
    \frac{\tilde{\mathcal{Y}}_2}{x}&=&-(3/2) \tilde{F}_0(x) \tilde{F}_0(1/2-y)\tilde{F}_0(1/2-x)\nonumber\\
    &&+3 \tilde{F}_0(x-y)\tilde{F}_0(y)\tilde{F}_1(x)\nonumber\\
    &&- \tilde{F}_0(1/2-x+y)\tilde{F}_1(x)\tilde{F}_1(1/2-y)\nonumber\\
    &&+ \tilde{F}_0(x)\tilde{F}_0(y)\tilde{F}_1(x-y)
    -\tilde{F}_0(y)\tilde{F}_1(x)\tilde{F}_1(x-y)\nonumber\\
    &&+ \tilde{F}_0(x)\tilde{F}_0(x-y)\tilde{F}_1(y)
    - \tilde{F}_0(x-y)\tilde{F}_1(x)\tilde{F}_1(y)\nonumber\\
    &&+ \tilde{F}_0(x)\tilde{F}_1(x-y)\tilde{F}_1(y)\nonumber\\
    &&- \tilde{F}_0(1/2-y)\tilde{F}_1(x)\tilde{F}_1(1/2-x+y)\nonumber\\
    &&- \tilde{F}_0(x) \tilde{F}_1(1/2-y)\tilde{F}_1(1/2-x+y)\nonumber\\
    &&-3 \tilde{F}_0(x-y)\tilde{F}_0(y)\tilde{F}_2(x)\nonumber\\
    &&+3 \tilde{F}_0(1/2-y)\tilde{F}_0(1/2-x+y)\tilde{F}_2(x)\nonumber\\
    &&+3 \tilde{F}_0(x)\tilde{F}_0(1/2-x+y)\tilde{F}_2(1/2-y)\nonumber\\
    &&-3 \tilde{F}_0(x)\tilde{F}_0(y)\tilde{F}_2(x-y)
    -3 \tilde{F}_0(x)\tilde{F}_0(x-y)\tilde{F}_2(y)\nonumber\\
    &&+3 \tilde{F}_0(x)\tilde{F}_0(1/2-y)\tilde{F}_2(1/2-x+y)\;,
\end{eqnarray}
and
\begin{eqnarray}
    \frac{\tilde{\mathcal{Z}}_2}{x}&=&-\tilde{F}_0(1/2-y)\tilde{F}_0(1/2-x-y)\tilde{F}_1(x)\\
    &&-\tilde{F}_0(y)\tilde{F}_0(x+y)\tilde{F}_1(x)\nonumber\\
    &&-3 \tilde{F}_0(x)\tilde{F}_0(1/2-x-y)\tilde{F}_1(1/2-y)\nonumber\\
    &&+ \tilde{F}_0(1/2-x-y)\tilde{F}_1(x)\tilde{F}_1(1/2-y)\nonumber\\
    &&-\tilde{F}_0(x)\tilde{F}_0(1/2-y)\tilde{F}_1(1/2-x-y)\nonumber\\
    &&- \tilde{F}_0(1/2-y)\tilde{F}_1(x)\tilde{F}_1(1/2-x-y)\nonumber\\
    &&+ \tilde{F}_0(x)\tilde{F}_1(1/2-y)\tilde{F}_1(1/2-x-y)\nonumber\\
    &&- \tilde{F}_0(x)\tilde{F}_0(x+y)\tilde{F}_1(y)
    -\tilde{F}_0(x+y)\tilde{F}_1(x)\tilde{F}_1(y)\nonumber\\
    &&-3 \tilde{F}_0(x)\tilde{F}_0(y)\tilde{F}_1(x+y)\nonumber\\
    &&+ \tilde{F}_0(y)\tilde{F}_1(x)\tilde{F}_1(x+y)
    + \tilde{F}_0(x)\tilde{F}_1(y)\tilde{F}_1(x+y)\nonumber\\
    &&+3  \tilde{F}_0(1/2-y)\tilde{F}_0(1/2-x-y)\tilde{F}_2(x)\nonumber\\
    &&+3  \tilde{F}_0(y) \tilde{F}_0(x+y)\tilde{F}_2(x)\nonumber \\    
    &&    +3  \tilde{F}_0(x) \tilde{F}_0(1/2-x-y)\tilde{F}_2(1/2-y)\nonumber\\
    &&+3 \tilde{F}_0(x) \tilde{F}_0(1/2-y) \tilde{F}_2(1/2-x-y)\nonumber\\
    &&+3 \tilde{F}_0(x) \tilde{F}_0(x+y) \tilde{F}_2(y)
    +3 \tilde{F}_0(x) \tilde{F}_0(y) \tilde{F}_2(x+y)\nonumber
\end{eqnarray}
while the leading-order terms read
\begin{eqnarray}
    \tilde{\mathcal{X}}_0&=&3 \tilde{F}_0(x)\tilde{F}_0(x-y)\tilde{F}_0(y)\;,\nonumber\\
    \tilde{\mathcal{Y}}_0&=&3x\Big(-\tilde{F}_0(x)\tilde{F}_0(x-y)\tilde{F}_0(y)\nonumber\\
    &&\qquad  +\tilde{F}_0(x)\tilde{F}_0(1/2-y)\tilde{F}_0(1/2-x+y)\Big)\;,\nonumber\\
    \tilde{\mathcal{Z}}_0&=&3x\tilde{F}_0(x)\Big(
    \tilde{F}_0(1/2-y)\tilde{F}_0(1/2-x-y)\nonumber\\
    &&\qquad \qquad+\tilde{F}_0(y)\tilde{F}_0(x+y)
    \Big)\;.
\end{eqnarray}
The numerical integration is stable down to $T=0.001$. The results agrees with the analytic expression~(\ref{SummaryXB}). 

\subsection{Term \texorpdfstring{$\mathbf{x_A}$}{xA}}
\label{chapter:the-term-xA}

Analogous to the previous section, the derivation of an analytic expression for the term $x_A$ is very lengthy. It can be found in the Supplemental Material, Sect.~\SMsmxAtermderivation. 
We again give only the analytical result for $0\ll B\ll T\ll 1$ as well as a variant for the numerical evaluation of $x_A$.

\subsubsection{Result for \texorpdfstring{$x_A$}{xA}}
We summarize the term $x_A$
\begin{eqnarray}
\label{SummaryXA}
x_A(B,T)&=&B^2\bigg(4\ln 2\frac{1}{T^2}\nonumber\\
&&\qquad+\frac{1}{T}\Big(8\big(\ln(2\pi T)-1-\EulerGamma\big)+4\ln2\Big)\nonumber\\
&&\qquad -\frac{2}{3}(12+\pi^2)+\mathcal{O}[T]\bigg)+\mathcal{O}[B^3]\;.\nonumber \\
\end{eqnarray}
Obviously, $x_A$ has no contribution to leading order in~$B$. 
Furthermore, the correction terms for $x_A$ and $x_B$ are identical, of the order $T$, as is mandatory.

\subsubsection{Numerical check for \texorpdfstring{$x_A$}{xA}}
\label{section:numericalcheck-xA}
In the present representation, a numerically stable evaluation of the $x_A$ term is not feasible. Therefore, 
we start again from Eq.~(\ref{xA-Definition-third-order}) [$j_\perp=\rho_0(0)J_\perp$, $j_z=\rho_0(0)J_z$, $\beta=1/T$]
\begin{eqnarray}\label{x3Aeq}
x_A&=&T\tilde{M}_0(B)\int_0^\beta\rmd\lambda_3\,\int_0^{\lambda_3}\rmd\lambda_2\,\int_0^{\lambda_2}\rmd\lambda_1\,\nonumber\\
&&\times \Big(  W^+ \tilde{f}^2(\lambda_3-\lambda_1,B)- W^- \tilde{f}^2(\lambda_3-\lambda_1,-B)\nonumber\\
&&\quad -W^+\tilde{f}^2(\lambda_2-\lambda_1,B)+ W^-\tilde{f}^2(\lambda_2-\lambda_1,-B)\nonumber\\
&& \quad - W^+\tilde{f}^2(\lambda_3-\lambda_2,B)+ W^- \tilde{f}^2(\lambda_3-\lambda_2,-B)\Big)\nonumber\\
\end{eqnarray}
with
\begin{eqnarray}
W^\pm&=&\frac{e^{\mp\beta B}}{2\cosh(\beta B)}\;,\nonumber\\
\tilde{f}(\lambda,B)&=&\int_{-1}^1\rmd\epsilon\,\frac{\rhotilde(\epsilon)e^{\lambda\epsilon}}{1+e^{\beta(\epsilon-B)}}\nonumber\\
&=&\int_{-1}^1\rmd\epsilon\,\frac{e^{\lambda\epsilon}}{1+e^{\beta(\epsilon-B)}}\;,\nonumber\\
\end{eqnarray}
where we used that
\begin{equation}
    \rhotilde(\epsilon)=\rho_0(\epsilon)/\rho_0(0)=1
\end{equation}
for a constant density of states. Note that
\begin{eqnarray}
\tilde{M}_0(B)=\frac{M_0(B)}{\rho_0(0)}\approx 2 B
\end{eqnarray}
for $B\ll1$, see Eq.~(\ref{M0forsmallB}).
We define $(\tau=\pm 1)$
\begin{eqnarray}
x_A^{\alpha,\tau}&=&T\int_0^\beta\rmd\lambda_3\,\int_0^{\lambda_3}\rmd\lambda_2\,\int_0^{\lambda_2}\rmd\lambda_1\,\tilde{f}^2(\lambda_3-\lambda_1,\tau B)\;,\nonumber\\
x_A^{\beta,\tau}&=&T\int_0^\beta\rmd\lambda_3\,\int_0^{\lambda_3}\rmd\lambda_2\,\int_0^{\lambda_2}\rmd\lambda_1\,\tilde{f}^2(\lambda_2-\lambda_1,\tau B)\;,\nonumber\\
x_A^{\gamma,\tau}&=&T\int_0^\beta\rmd\lambda_3\,\int_0^{\lambda_3}\rmd\lambda_2\,\int_0^{\lambda_2}\rmd\lambda_1\,\tilde{f}^2(\lambda_3-\lambda_2,\tau B)\;,
\nonumber\\
\end{eqnarray}
to find
\begin{eqnarray}\label{x_AapproxOrigin}
x_A&\approx& 2 B\Big(W^+x_A^{\alpha,+}-W^-x_A^{\alpha,-}-W^+x_A^{\beta,+}\nonumber\\
&&\qquad+ W^-x_A^{\beta,-}-W^+x_A^{\gamma,+}+W^- x_A^{\gamma,-}\Big)\;.\nonumber \\
\end{eqnarray}
Since $M_0(B)\sim B$ we only need to expand the remaining terms up to and including the linear order in $B$.

To make progress, we define
\begin{eqnarray}
g_A^\alpha&=&\int_0^\beta\rmd\lambda_3\,\int_0^{\lambda_3}\rmd\lambda_2\,\int_0^{\lambda_2}\rmd\lambda_1\,e^{(\epsilon_1+\epsilon_2)(\lambda_3-\lambda_1)}\nonumber\\
g_A^\beta&=&\int_0^\beta\rmd\lambda_3\,\int_0^{\lambda_3}\rmd\lambda_2\,\int_0^{\lambda_2}\rmd\lambda_1\,e^{(\epsilon_1+\epsilon_2)(\lambda_2-\lambda_1)}\nonumber\\
g_A^\gamma&=&\int_0^\beta\rmd\lambda_3\,\int_0^{\lambda_3}\rmd\lambda_2\,\int_0^{\lambda_2}\rmd\lambda_1\,e^{(\epsilon_1+\epsilon_2)(\lambda_3-\lambda_2)}\;.\nonumber\\
\end{eqnarray}
This gives ($j=\alpha,\beta,\gamma$)
\begin{eqnarray}\label{xAjtau-Def}
x_A^{j,\tau}&=&T\int_{-1}^{1}\rmd\epsilon_1\,\int_{-1}^1\rmd\epsilon_2\,\nonumber\\
&&\times \frac{\rhotilde(\epsilon_1)}{1+e^{\beta(\epsilon_1-\tau B)}}\frac{\rhotilde(\epsilon_2)}{1+e^{\beta(\epsilon_2-\tau B)}} g_A^j\;.
\end{eqnarray}
Note that
\begin{eqnarray}
Q(\lambda_3)&=&\int_0^{\lambda_3}\rmd\lambda_2\,\int_0^{\lambda_2}\rmd\lambda_1\,e^{(\epsilon_1+\epsilon_2)(\lambda_3-\lambda_2)}\nonumber\\
&=&\int_0^{\lambda_3}\rmd\lambda_2\,e^{(\epsilon_1+\epsilon_2)(\lambda_3-\lambda_2)}\lambda_2\nonumber\\
&=&-\frac{\lambda_3}{\epsilon_1+\epsilon_2}-\frac{1}{(\epsilon_1+\epsilon_2)^2}+\frac{e^{(\epsilon_1+\epsilon_2)\lambda_3}}{(\epsilon_1+\epsilon_2)^2}\nonumber\\
&=&\int_0^{\lambda_3}\rmd\lambda_2\,\left[-\frac{1}{\epsilon_1+\epsilon_2}+\frac{e^{(\epsilon_1+\epsilon_2)\lambda_2}}{\epsilon_1+\epsilon_2}\right]\nonumber\\
&=&\int_0^{\lambda_3}\rmd\lambda_2\,\int_0^{\lambda_2}\rmd\lambda_1\,e^{(\epsilon_1+\epsilon_2)(\lambda_2-\lambda_1)}\;,
\end{eqnarray}
which implies 
\begin{equation}
g_A^\beta\equiv g_A^\gamma
\end{equation}
and thus
\begin{equation}
x_A^{\beta,\tau}\equiv x_A^{\gamma,\tau}\;.
\end{equation}
We define
\begin{eqnarray}\label{xAalpha-and-xAbeta-Definition}
x_A^\alpha &=& W^+ x_A^{\alpha,+}- W^- x_A^{\alpha,-}\;,\nonumber\\
x_A^\beta&=&-2W^+x_A^{\beta,+}+2 W^-x_A^{\beta,-}\;,
\end{eqnarray}
which shows that
\begin{equation}\label{xAapprox2BxAalpha}
x_A\approx 2 B(x_A^\alpha+x_A^\beta)
\end{equation}
to leading order in~$B$.
Lastly, we consider $[b=B/T]$
\begin{eqnarray}\label{Ftildedef}
\tilde{F}(x,b)=\frac{1}{T}\int_{-1}^1\rmd\epsilon\,\frac{\rhotilde(\epsilon/T)e^{x\epsilon/T}}{1+e^{\epsilon/T-b}}=\int_{-1/T}^{1/T}\rmd u\,\frac{e^{xu}}{1+e^{u-b}}\nonumber\\
\end{eqnarray}
for $0\leq x\leq 1$. Again, we used that $\rhotilde(\epsilon)=1$ for a constant density of states. Then, we have
\begin{eqnarray}\label{eq:ftilde1-x}
\tilde{F}(1-x,b)&=&\int_{-1/T}^{1/T}\rmd u\,\frac{e^u e^{-ux}}{1+e^{u-b}}\nonumber\\
&=&\int_{-1/T}^{1/T}\rmd u\,\frac{e^{ux}}{e^u+e^{-b}}=e^b \tilde{F}(x,-b)\;,\nonumber\\
\end{eqnarray}
where we substituted $u\to- u$ in the last step. We discuss the two contributions to $x_A$ separately.

\paragraph{Term \texorpdfstring{${x_A^{\alpha}}$}{xAalpha}:}
\label{xAalphaterm-vanishes}

By definition,
\begin{eqnarray}
x_A^{\alpha,\tau}&=&T\int_0^\beta\rmd\lambda_3\,\int_0^{\lambda_3}\rmd\lambda_2\,\int_0^{\lambda_2}\rmd\lambda_1\,\tilde{f}^2(\lambda_3-\lambda_1,\tau B)\nonumber\\
&=&T\int_0^\beta\rmd\lambda_3\,\int_0^{\lambda_3}\rmd\lambda_1\,\int_{\lambda_1}^{\lambda_3}\rmd\lambda_2\, \tilde{f}^2(\lambda_3-\lambda_1,\tau B)\;.\nonumber\\
\end{eqnarray}
We substitute $u=\lambda_3-\lambda_1$ with $0\leq \lambda_1=\lambda_3-u\leq\lambda_3$, which implies $0\leq u\leq \lambda_3$. Thus,
\begin{eqnarray}
x_A^{\alpha,\tau}&=&T\int_0^\beta\rmd u\,u \tilde{f}^2(u,\tau B)\int_u^\beta\rmd\lambda_3\nonumber\\
&=&T\int_0^\beta\rmd u\, u\tilde{f}^2(u,\tau B)(\beta-u)\;.
\end{eqnarray}
Now we substitute $u=\beta x$, 
\begin{eqnarray}
x_A^{\alpha,\tau}&=&T\beta^3\int_0^1\rmd x\,x(1-x)\tilde{f}^2(\beta x,\tau B)\nonumber\\
&=&\int_0^1\rmd x\, x(1-x)\left(\tilde{F}(x,\tau B)\right)^2\nonumber\\
&=&\int_0^{1/2}\rmd x\,x(1-x)\left(\tilde{F}(x,\tau B)\right)^2\nonumber\\
&&+\int_0^{1/2}\rmd y\,(1-y)y\left(\tilde{F}(1-y,\tau B)\right)^2\;,\nonumber\\
\end{eqnarray}
where $y=1-x$. Note that ($b=B/T=\beta B$)
\begin{eqnarray}
\left[\tilde{F}(1-y,\tau B)\right]^2=e^{2\tau b}\left[\tilde{F}(y,-\tau b)\right]^2\;,
\end{eqnarray}
see Eq.~(\ref{eq:ftilde1-x}).

Using the definition of $x_A^\alpha$ in Eq.~(\ref{xAalpha-and-xAbeta-Definition}),
\begin{eqnarray}
x_A^\alpha &=& W^+ x_A^{\alpha,+}- W^- x_A^{\alpha,-}\;,
\end{eqnarray}
we obtain $[\bar{x}_A^{\alpha}=2\cosh(b)x_A^\alpha]$
\begin{eqnarray}
\bar{x}_A^{\alpha}&=&e^{-b}\int_0^{1/2}\rmd x\,x(1-x)\tilde{F}^2(x,b)\nonumber\\
&&+e^{-b}\int_0^{1/2}\rmd x\,x(1-x)\tilde{F}^2(x,-b)e^{2 b}\nonumber\\
&&-e^b\int_0^{1/2}\rmd x\, x(1-x)\tilde{F}^2(x,-b)\nonumber\\
&&-e^b\int_0^{1/2}\rmd x\,x(1-x)\tilde{F}^2(x,b)e^{-2b} \nonumber\\
&=&e^{b}\int_0^{1/2}\rmd x\,x(1-x)\Big(\tilde{F}^2(x,-b)-\tilde{F}^2(x,-b)\Big)\nonumber\\
&&+e^{-b}\int_0^{1/2}\rmd x\,x(1-x)\Big(\tilde{F}^2(x,b)-\tilde{F}^2(x,b)\Big)\nonumber\\
&=& 0\;.
\end{eqnarray}
For convenience, here is the result once again explicitly,
\begin{eqnarray}
0&=&T\int_0^\beta\rmd\lambda_3\,\int_0^{\lambda_3}\rmd\lambda_2\,\int_0^{\lambda_2}\rmd \lambda_1\,\nonumber\\
&&\times\left(W^+f^2(\lambda_3-\lambda_1,B)-W^-f^2(\lambda_3-\lambda_1,-B)\right)\;,\nonumber\\
\label{appeq:xAalphaiszero}
\end{eqnarray}
which agrees with our analytic calculations to $\mathcal{O}[T]$.

\paragraph{Term \texorpdfstring{$x_A^{\beta}$}{xAbeta}:}
\label{xAbetaterm}

By definition, 
\begin{eqnarray}
x_A^{\beta,\tau}&=&T\int_0^\beta\rmd\lambda_3\,\int_0^{\lambda_3}\rmd\lambda_2\,\int_0^{\lambda_2}\rmd\lambda_1\,\tilde{f}^2(\lambda_2-\lambda_1,\tau B)\nonumber\\
&=&T\int_0^\beta\rmd\lambda_2\,\int_0^{\lambda_2}\rmd\lambda_1\,\int_{\lambda_2}^\beta\rmd\lambda_3\,\tilde{f}^2(\lambda_2-\lambda_1,\tau B)\nonumber\\
&=&T\int_0^\beta\rmd\lambda_2\,\int_0^{\lambda_2}\rmd\lambda_1\,(\beta-\lambda_2)\,\tilde{f}^2(\lambda_2-\lambda_1,\tau B)\;.\nonumber\\
\end{eqnarray}
We substitute $u=\lambda_2-\lambda_1$, which implies $0\leq\lambda_1=\lambda_2-u$, $u\geq 0$, $u\leq \lambda_2$ and thus $0\leq \lambda_2\leq\beta$, $0\leq u\leq \lambda_2$,
\begin{eqnarray}
x_A^{\beta,\tau}&=&T\int_0^\beta\rmd\lambda_2\,\int_0^{\lambda_2}\rmd u\,(\beta-\lambda_2)\tilde{f}^2(u,\tau B)\nonumber\\
&=&T\int_0^\beta\rmd u\,\frac{1}{2}(\beta-u)^2 \tilde{f}^2(u,\tau B)\nonumber\\
&=&T\beta^3\int_0^1\rmd x\,\frac{1}{2}(1-x)^2\tilde{f}^2(\beta x,\tau B)\nonumber\\
&=&\int_0^1\rmd x\,\frac{1}{2}(1-x)^2\tilde{F}^2(x,\tau b)
\end{eqnarray}
with $\tilde{F}$ from Eq.~(\ref{Ftildedef}) and $b=B/T$. Moreover, with Eq.~(\ref{eq:ftilde1-x}), we find
\begin{eqnarray}
x_A^{\beta,\tau}&=&\int_0^{1/2}\rmd x\,\frac{1}{2}(1-x)^2\tilde{F}^2(x,\tau b)\nonumber\\
&&+\int_0^{1/2}\rmd x\,\frac{x^2}{2}e^{2\tau b}\tilde{F}^2(x,-\tau b)\;.
\end{eqnarray}
Now,
\begin{eqnarray}
x_A^\beta&=&-2W^+x_A^{\beta,+}+2W^-x_A^{\beta,-}\nonumber\\
\end{eqnarray}
and thus
\begin{eqnarray}
2\cosh(b)x_A^\beta&=&-e^{-b}\int_0^{1/2}\rmd x\,(1-x)^2\tilde{F}^2(x,b)\nonumber\\
&&-e^{-b}\int_0^{1/2}\rmd x\,x^2 e^{2 b}\tilde{F}^2(x,-b)\nonumber\\
&&+e^b\int_0^{1/2}\rmd x\,(1-x)^2\tilde{F}^2(x,-b)\nonumber\\
&&+e^b\int_0^{1/2}\rmd x\,x^2e^{-2b}\tilde{F}^2(x,b)\nonumber\\
&=&e^{-b}\int_0^{1/2}\rmd x\,(2x-1)\tilde{F}^2(x,b)\nonumber\\
&&-e^b\int_0^{1/2}\rmd x\,(2x-1)\tilde{F}^2(x,-b)\;.\nonumber\\
\end{eqnarray}
Furthermore, for small $b\ll1$ we find up to and including the first order
\begin{eqnarray}
\tilde{F}(x,b)&=&\tilde{F}_0(x)+b\tilde{F}_1(x)\;,\nonumber\\
\tilde{F}_0(x)&=&\int_{-1/T}^{1/T}\rmd u\,\frac{e^{xu}}{1+e^u}\;,\nonumber\\
\tilde{F}_1(x)&=&\int_{-1/T}^{1/T}\rmd u\,\frac{e^{xu}e^u}{(1+e^u)^2}
\end{eqnarray}
and we can write
\begin{eqnarray}\label{xAbetanum}
x_A^\beta&\approx&\frac{B}{T}\bigg(\int_0^{1/2}\rmd x\,(1-2x)\tilde{F}^2_0(x)\nonumber\\
&&\qquad -2\int_0^{1/2}\rmd x\,(1-2 x)\tilde{F}_0(x)\tilde{F}_1(x)\bigg)\;.\nonumber \\
\end{eqnarray}
At low temperatures, $T\ll1$, and $0\leq x\leq 1/2$, we find
\begin{eqnarray}
\tilde{F}_0(x)&\approx&\frac{\pi}{\sin(\pi x)}-\frac{e^{-x/T}}{x}\;,\nonumber\\
\tilde{F}_1(x)&=&\frac{\pi x}{\sin(\pi x)}\;.
\end{eqnarray}
Equation~(\ref{xAbetanum}) can be readily evaluated numerically and agrees with our analytical result up to $\mathcal{O}[T]$.
With eqs.~(\ref{xAapprox2BxAalpha}) and~(\ref{appeq:xAalphaiszero})  we have
\begin{equation}
    x_A=2B x^\beta_A\;.
\end{equation}

\subsection{Final result of the third-order contribution}
\label{FinalResultThirdOrder}

With Eq.~(\ref{thirdorder-definition-up-to-Bsquared}) and section~\ref{section:Rearrangement} we have
\begin{eqnarray}
\mathcal{F}^{(3)}&\approx& TV_3-TV_1V_2(B=0)\nonumber\\
&=&TV_3^z+F_3^\alpha+F_3^\beta-T V_1 V_2(B=0)\nonumber \\
\end{eqnarray}
with [$j_z=J_z\rho_0(0)$] 
\begin{equation}
    TV_3^z= B^2 j_z^3\left(\frac{\ln 2}{8 T^2}-\frac{\pi^2}{96T}-\frac{\pi^2}{48}\right)+\mathcal{O}[B^3]
\end{equation}
from Eq.~(\ref{IsingKondoThirdOrder}) and [$j_\perp=J_\perp\rho_0(0)$]
\begin{eqnarray}
F_3^\alpha&=&\frac{1}{16}j_\perp^2j_z x_A\;,\nonumber\\
F_3^\beta&=&\frac{1}{8}j_\perp^2 j_z x_B
\end{eqnarray}
where [$x_A=x_A(B,T)$, $x_B=x_B(B,T)$]
\begin{eqnarray}
x_A&=&B^2\bigg(4\ln 2\frac{1}{T^2}\nonumber\\
&&\qquad +\frac{1}{T}\Big(8\big(\ln(2\pi T)-1-\EulerGamma\big)+4\ln2\Big)\nonumber\\
&&\qquad -\frac{2}{3}(12+\pi^2)+\mathcal{O}[T]\bigg)+\mathcal{O}[B^3]\;,\nonumber\\
x_B&=&-\frac{\pi^2}{2}-6\ln^2(2)+2\pi^2 T\nonumber\\
&&+2\pi^2 T^2(\ln 2 -1)+\mathcal{O}[B^0 T^3,B^0 T^3\ln T]\nonumber\\
&&+\frac{B^2}{T}\bigg(
8+8\EulerGamma+4\EulerGamma^2-\frac{\pi^2}{6}+4\ln^2(2)\nonumber\\
&&+8\ln 2 \ln\pi+4\ln^2(\pi)-8\ln(2\pi T)(1+\EulerGamma)\nonumber\\
&&+8\ln(2\pi)\ln T +4\ln^2(T)\bigg)\nonumber\\
&&+B^2(-8+6\ln2+\mathcal{O}[T]) +\mathcal{O}[B^3]
\end{eqnarray}
from Eq.~(\ref{SummaryXB}) and~(\ref{SummaryXA}). Furthermore,
\begin{eqnarray}
    TV_1V_2(B=0)&=&j_z^3 B^2\left(\frac{\pi^2}{48}+\frac{\ln 2}{4 T^2}\right)\nonumber\\
    &&+ j_\perp^2 j_z B^2\left(-\frac{\pi^2}{24}+\frac{\ln 2}{4 T^2}\right)\nonumber\\
\end{eqnarray}
from our previous results, see Eq.~(\ref{FirstOrderTerm}) and~(\ref{F2ndorder}).

Finally, we obtain the third-order contribution,
\begin{eqnarray}
\label{FreeEnergyThirdOrder}
\mathcal{F}^{(3)}&=&j_\perp^2 j_z \bigg(-\frac{12\ln(2)^2+\pi^2}{16}+\frac{\pi^2 T}{4}\nonumber\\
&&+\frac{\pi^2 T^2}{4}(-1+\ln 2)+\mathcal{O}[B^0 T^3\bigg)\nonumber\\
&&+B^2j_\perp^2 j_z\left(\frac{3}{4}\left(-2+\ln 2\right)+\mathcal{O}[T]\right)\nonumber\\
&&-\frac{B^2}{T}j_z^3\frac{\pi^2}{96}+\mathcal{O}[j_z^3 B^2 T]\nonumber\\
&&+\frac{B^2}{2T}j_\perp^2j_z\bigg(1+\EulerGamma+\EulerGamma^2-\frac{\pi^2}{24}+\frac{\ln 2}{2}\nonumber\\
&&\quad+\ln^2(2)+ 2\ln 2\ln \pi+\ln^2(\pi)\nonumber\\
&&\quad -\ln(2\pi T)(1+2\EulerGamma)+2\ln(2\pi)\ln T\nonumber\\
&&\quad+\ln^2(T)
\bigg)+\mathcal{O}[B^3]\:.
\end{eqnarray}

\section{Outline of the Supplemental Material}

\renewcommand{\thesubsection}{SM-\Roman{subsection}}
In this section, we summarize the content of the Supplemental Material.
Note that the seven sections comprise of 49~text pages.

\subsection{Free energy of the Ising-Kondo model for a constant density of states}
Following Bauerbach et al.~\cite{IsingKondo}, we derive the free energy of the Ising-Kondo model for a constant density of states using the equation-of-motion method. 
Since we have explicitly performed the second-order perturbation theory for the Ising-Kondo model, see Appendix~\ref{appendix:secondorderIsingKondo}, 
we can directly compare the Taylor series of the exact result with perturbation theory. 

\subsection{Analytical calculation of an approximation of the term \texorpdfstring{$\mathbf{x_B}$}{xB}}

We explicitly derive an analytic approximation of the term $x_B$ from Sect.~\ref{chapter:the-term-xB} in the limit $0\ll B\ll T\ll 1$ up to and including second order in $B$.

\subsection{Analytical calculation of an approximation of the term \texorpdfstring{$\mathbf{x_A}$}{xA}}

We explicitly derive an analytic approximation of the term $x_A$ from Sect.~\ref{chapter:the-term-xA} in the limit $0\ll B\ll T\ll 1$ up to and including second order in $B$.

\mbox{}\par\mbox{}

\subsection{Properties of the functions \texorpdfstring{$\mathbf{H_T(\boldsymbol{\omega},\boldsymbol{\tau} B)}$}{HT-Func} and \texorpdfstring{$\mathbf{\overline{R}(x)}$}{Rbar(x)}}

The functions $H_T(\omega,\tau B)$ and $\overline{R}(x)$ are of paramount importance to the whole perturbation theory. In this section, we derive an analytical expression 
for the function $H_T$ at low temperatures and we discuss its properties in detail. This includes various integral representations and series expansions at low temperatures. 
We also show how to calculate the function $H_T$ for a general density of states. This crucial step will be helpful for future work where
a general density of state is used.

\subsection{Derivation of the integrals}

We explicitly derive analytical expressions at low temperatures of all integrals occurring in this study. 

\subsection{Mathematical details}

In this section we list other important mathematical properties that are particularly useful in the context of this study. 
This includes specific substitutions applied in second and third order to compute multi-dimensional integrals. Moreover, we discuss various properties of Fermi functions and other
useful functions.

\subsection{Integral tables}

For a better overview, we list two integral tables, for the integrals $\mathcal{I}_i$ and for the integrals $\alpha_j$ that occur in Sect.~SM-V.

\end{document}


\title{Supplemental Material:\\
Second Wilson number from third-order perturbation theory\\ for the
symmetric single-impurity Kondo model at low temperatures}

\author{Kevin Bauerbach}
\author{Florian Gebhard$^1$}
\email{florian.gebhard@physik.uni-marburg.de}
\affiliation{$^1$Fachbereich Physik, Philipps-Universit\"at Marburg,
  35032 Marburg, Germany}
  
\date{November 16, 2023}

\begin{abstract}
    The Supplemental Material contains seven parts. In Sect.~\ref{FreeEnergyIsingKondo}, 
    we follow Bauerbach et al.~\cite{IsingKondo} and determine the free energy 
    of the single-impurity Ising-Kondo model for a constant density of states using the equation of motion method. 
    In Sects.~\ref{sm:xBterm-derivation} and~\ref{sm:xAterm-derivation} 
    we derive analytic expressions of the two terms $x_B$ and $x_A$, respectively, 
    that determine the free energy to third order perturbation theory. Within these 
    derivations, the functions $H_T$ and $\Rbar$ frequently occur, and we devote
    Sect.~\ref{chapter:properties-of-the-Rbar-function} to their analysis. 
    Sect.~\ref{subsec:DerviationOfAllIntegrals} contains the explicit derivation of the 
    low-temperature asymptotics of all integrals occurring in this work. 
    Moreover, in Sect.~\ref{mathematicaldetails} we list other important mathematical 
    properties that are particularly useful in the context of this study. The final section~\ref{Integraltables} 
    provides two integral tables for a better overview.
\end{abstract}

\maketitle

\section{Free energy of the Ising-Kondo model}
\label{FreeEnergyIsingKondo}
In this section, we derive the free energy of the Ising-Kondo model for a constant density of states at $0 < B\ll T\ll1 $. Using the Taylor series of the exact result, we can verify the second order perturbation theory $V_2^z$ of the Ising-Kondo model, 
see section~\mainsubsecformalsecondorderterm 
of the main paper, and explicitly compute the third-order contribution.

\subsection{Ising-Kondo model}
The Ising-Kondo model arises from the Kondo model, 
see Eq.~(\maineqgeneralHamiltonian) 
and especially~(\mainDefinitionKondoInteraction) 
of the main paper, for a vanishing spin-flip component in the interaction term,
\begin{equation}
\hat{H}_{\rm IK}=H_{\rm K}(J_\perp\equiv 0)\;.
\end{equation}

The interaction term of the single-impurity Ising-Kondo model reads
\begin{eqnarray}
\hat{V}&=&\hat{V}_{\rm z}=J_z \hat{s}^z_0
\hat{S}^z \nonumber \\
&=& \frac{J_z}{4L}\sum_{k,k'}\left(\hat{n}^d_\Uparrow-\hat{n}^d_\Downarrow\right)\left(\hat{a}^+_{k,\uparrow}\hat{a}^\phan_{k',\uparrow}-\hat{a}^+_{k,\downarrow}\hat{a}^\phan_{k',\downarrow}\right).\;
\end{eqnarray}

The density of states of the host electrons is given by
\begin{equation}
\rho_0(\omega)=\frac{1}{L}\sum_k\delta(\omega-\epsilon(k))\;,
\end{equation}
and we choose to work with a constant density of states,
\begin{equation}
\rho_0(\omega)=1/2\;,\quad\textnormal{for }|\omega|\leq 1
\end{equation}
with bandwidth $W=2$ so that half the bandwidth is our unit of energy. The Hilbert transform 
of the density of states $\rho_0(\omega)$ provides the real part $\Lambda_0(\omega)$
of the local host-electron Green function~$g_0(\omega)$, ($\eta=0^+$)
\begin{equation}
  g_0(\omega) = \frac{1}{L}\sum_k \frac{1}{\omega-\epsilon(k)+\rmi \eta}
  \equiv \Lambda_0(\omega) -\rmi \pi \rho_0(\omega)
  \label{eq:g0def}
  \end{equation}
with
\begin{equation}
\Lambda_0(\omega)=\int_{-1}^1\rmd\epsilon\,\frac{\rho_0(\epsilon)}{\omega-\epsilon}\;.
\label{Hilberttransformed}
\end{equation}
For $|\omega|<1$, this is a principal-value integral. 
The corresponding Hilbert transform is
\begin{eqnarray}\label{HilberttransformedConstDos}
  \Lambda_0(|\omega|>1)&=&
  \frac{1}{2}\ln\left(\frac{\omega+1}{\omega-1}\right)\;,
\nonumber
  \\
  \Lambda_0(|\omega|<1)&=&
  \frac{1}{2}\ln\left( \frac{1+\omega}{1-\omega}\right)\;.
\end{eqnarray}

In the following we shall consider particle-hole symmetry and the case of half band-filling, where the number of host electrons equals the number of sites,
$N=N_{\uparrow}+N_{\downarrow}=L$;
the thermodynamic limit, $N,L\to \infty$ with $n=N/L=1$ fixed, is implicit. 

\subsection{General considerations}
In Sect.~3.3.1 of Bauerbach et al~\cite{IsingKondo}, we derived the impurity-induced free energy as
\begin{eqnarray}
F_{\rm IK}^{\rm i}(B,T,V)&=& \overline{F}_s(B,T,V)\nonumber\\
&&-T\ln\Big(2\cosh\left(B^{\rm eff}(B,T,V)/T\right)\Big)\;,\nonumber\\
\overline{F}_s(B,T,V)&=&\frac{1}{2}\Big(F_{\rm sf}(B,T,V)+F_{\rm sf}(-B,T,-V)\nonumber\\
&&+F_{\rm sf}(B,T,-V)+F_{\rm sf}(-B,T,V)\Big)\;,\nonumber\\
\overline{F}_a(B,T,V)&=&\frac{1}{2}\Big(F_{\rm sf}(B,T,V)+F_{\rm sf}(-B,T,-V)\nonumber\\
&&-F_{\rm sf}(B,T,-V)-F_{\rm sf}(-B,T,V)\Big)\;,\nonumber\\
\end{eqnarray}
where $V\equiv J_z/4>0$ and
\begin{eqnarray}
F_{\rm sf}&=&-T\int_{-\infty}^\infty\rmd \omega \,D_0(\omega,V)\ln\left(1+e^{-\beta(\omega-B)}\right)\;,\nonumber\\
B^{\rm eff}&=&B-\overline{F}_a(B,T,V)
\end{eqnarray}
with $F_{\rm sf}=F_{\rm sf}(B,T,V)$ and $B^{\rm eff}=B^{\rm eff}(B,T,V)$.
Here, ($\eta=0^+$)
\begin{eqnarray}
D_0(\omega,V)&=&-\frac{1}{\pi}\frac{\partial}{\partial\omega}{\rm Im}\big[\ln(1-Vg_0(\omega))\big]\;,\nonumber\\
g_0(\omega)&=&\int_{-1}^1\rmd\epsilon\,\frac{\rho_0(\epsilon)}{\omega-\epsilon+\rmi\eta}\;,
\end{eqnarray}
where $D_0(\omega,V)$ is the single-particle density of states and $g_0(\omega)$ the local Green function of non-interacting fermions,
see Eq.~(\ref{eq:g0def}).

\subsection{Terms of order \texorpdfstring{$\mathbf{B^2}$}{B2}}
\label{FIKsecondorder}
To isolate the terms of order $B^2$, we write [$\overline{F}_s=\overline{F}_s(B,T,V)$, $\overline{F}_a=\overline{F}_a(B,T,V)$]
\begin{eqnarray}
F_{\rm sf}&=&F_{\rm sf}(T,V)-B\int_{-\infty}^\infty\rmd \omega\,\frac{D_0(\omega,V)}{1+e^{\beta\omega}}\nonumber\\
&&-\frac{B^2}{2 T}\int_{-\infty}^\infty\rmd\omega\,\frac{D_0(\omega,V)}{(1+e^{\beta\omega})^2}e^{\beta\omega}+\mathcal{O}[B^3]\;,\nonumber\\[6pt]
\overline{F}_s&=&F_{\rm sf}(T,V)+F_{\rm sf}(T,-V)-\frac{B^2}{2T}\nonumber\\
&&\times\int_{-\infty}^\infty\rmd\omega\,\left(D_0(\omega,V)+D_0(\omega,-V)\right)\frac{e^{\beta\omega}}{(1+e^{\beta\omega})^2}\nonumber\\
&&+\mathcal{O}[B^4]\;,\nonumber\\[6pt]
\overline{F}_a&=&-B\int_{-\infty}^\infty\rmd\omega\,\frac{D_0(\omega,V)-D_0(\omega,-V)}{1+e^{\beta\omega}}+\mathcal{O}[B^3]\;.\nonumber\\
\end{eqnarray}
For $\overline{F}_s$ we thus find for the term of order $B^2$
\begin{eqnarray}\label{Dvonomega-gleich0}
\overline{F}_s^{(2)}&=& -\frac{B^2}{2 T} \int_{-\infty}^{\infty}
\rmd T x \,\Bigl(\frac{e^x}{(1+e^x)^2}\nonumber\\
&&\hphantom{-\frac{B^2}{2 T} \int_{-\infty}^{\infty}}
\times \left(D_0(xT,V)+D_0(xT,-V)\right)\Bigr)\nonumber\\
&=&-\frac{B^2}{2}\left(\int_{-\infty}^\infty\rmd x\,\frac{e^x}{(1+e^x)^2}\right)\nonumber\\
&&\times\left(D_0(0,V)+D_0(0,-V)\right)+\mathcal{O}[B^2T^2]\;.
\end{eqnarray}
For small frequencies we approximate
\begin{equation}
g_0(\omega)=\Lambda_0(\omega)-\rmi\pi \rho_0(\omega)\nonumber\\
\approx \omega\Lambda^\prime_0(0)-\rmi\pi\rho_0(0)+\mathcal{O}[\omega^2]\;,
\end{equation}
and find [$\zeta(\omega,V)=1-V\omega\Lambda_0^\prime(0)+\rmi\pi V\rho_0(0)$]
\begin{eqnarray}\label{lambda0primeoccur}
D_0(\omega=0,V)&=&-\frac{1}{\pi}\frac{\partial}{\partial\omega}{\rm Im}\left(\ln\zeta(\omega,V)\right)\Big|_{\omega=0}\nonumber\\
&=&\frac{1}{\pi}{\rm Im}\left(\frac{V\Lambda_0^\prime(0)}{1+\rmi\pi V\rho_0(0)}\right)\nonumber\\
&=&-\frac{V^2\Lambda_0^\prime(0)\rho_0(0)}{1+(\pi V\rho_0(0))^2}\nonumber\\
&=&D_0(\omega=0,-V)\;.
\end{eqnarray}
Note that
\begin{eqnarray}\label{derivativeofhilberttransformed}
\Lambda_0^\prime(0)=2\rho_0(0)-\int_{-1}^1\rmd\epsilon\,\frac{\rho_0(\epsilon)-\rho_0(0)}{\epsilon^2}=1
\end{eqnarray}
for a constant density of states, see~\ref{appendix:Lambda0PrimeVon0}.
Thus,
\begin{eqnarray}
\overline{F}_s^{(2)}&=&B^2\frac{V^2\Lambda_0^\prime(0)\rho_0(0)}{1+(\pi V\rho_0(0))^2}+\mathcal{O}[B^2 T^2]\;.
\end{eqnarray}
For a constant density of states we thus have
\begin{equation}
  \overline{F}_s^{(2)}=B^2\frac{2V^2}{4+(\pi V)^2}\;.  
\end{equation}
We thus find for the symmetric part of $\overline{F}_s$ 
\begin{eqnarray}
\overline{F}_s(B,T,V)&\approx &\overline{F}_s(T,V)\nonumber\\
&&+B^2\frac{V^2\Lambda_0^\prime(0)\rho_0(0)}{1+(\pi V\rho_0(0))^2}+\mathcal{O}[B^2T^2]\;,\nonumber\\
\overline{F}_s(T,V)&=&F_{\rm sf}(T,V)+F_{\rm sf}(T,-V)
\end{eqnarray}
up to and including order~$B^2$.
Next, with
\begin{equation}
    \Delta D_0(\omega,V)=D_0(\omega,V)-D_0(\omega,-V)\;,
\end{equation}
we have
\begin{eqnarray}
\overline{F}_a(B,T,V)&=&-B\bigg(\int_{-\infty}^0\rmd\omega\,\Delta D_0(\omega,V)\nonumber\\
&&+2\int_0^\infty\frac{\rmd\omega}{1+e^{\beta\omega}}\Delta D_0(\omega,V)\bigg)+\mathcal{O}[B^3]\;,\nonumber\\
\end{eqnarray}
where we used that $D_0(\omega,V)=D_0(-\omega,-V)$ in the second term. The second term is $\mathcal{O}[T^2]$ because $D_0(0,V)=D_0(0,-V)$,
\begin{eqnarray}
D_0(xT,V)-D_0(xT,-V)=\mathcal{O}[xT]\;.
\end{eqnarray}
Thus,
\begin{eqnarray}
\overline{F}_a(B,T,V)&=&-B\int_{-\infty}^0\rmd\omega\,\left(-\frac{1}{\pi}\right)\nonumber\\
&&\times\,{\rm Im}\left(\frac{\partial}{\partial\omega}\ln\left(\frac{1-Vg_0(\omega)}{1+Vg_0(\omega)}\right)\right)\nonumber\\
&&+\mathcal{O}[BT^2,B^3]\nonumber\\
&=&\frac{B}{\pi}{\rm Im}\left(\ln\left(\frac{1+\rmi\pi V\rho_0(0)}{1-\rmi\pi V\rho_0(0)}\right)\right)\nonumber\\
&&+\mathcal{O}[BT^2,B^3]\nonumber\\
&=&\frac{2B}{\pi}{\arctan}\left(\pi V\rho_0(0)\right)+\mathcal{O}[BT^2,B^3]\;.\nonumber\\
\end{eqnarray}
Therefore,
\begin{eqnarray}
B^{\rm eff}(B,T,V)&=&B\left(1-\alpha(V)\right)\;,\nonumber\\
\alpha(V)&=&\frac{2}{\pi}{\rm arctan}\left(\pi  V\rho_0(0)\right)
\end{eqnarray}
for $T\ll1 $, and thus the impurity-contribution to the free energy of the Ising-Kondo model
at small fields and low temperatures reads
\begin{eqnarray}
F_{\rm IK}^{\rm i}(B,T,V)&\approx &\overline{F}_s(T,V)+B^2\frac{V^2\Lambda_0^\prime(0)\rho_0(0)}{1+(\pi V\rho_0(0))^2}\nonumber\\
&&+\mathcal{O }[B^2T^2]\nonumber\\
&&-T\ln 2-\frac{B^2}{2 T}(1-\alpha(V))^2\nonumber\\
&&+\mathcal{O}[B^2 T,B^4]
\end{eqnarray}
for a general density of states.
The $B^2$-terms become [$j_z=J_z\rho_0(0)$]
\begin{eqnarray}
F_{\rm IK}^{{\rm i},{(2)}}&=&-\frac{B^2}{2T}\Bigl[\left(1-\frac{2}{\pi}{\rm arctan}\left(\frac{\pi j_z}{4}\right)\right)^2\nonumber\\
&&-4T\left(\frac{j_z}{4}\right)^2\frac{\tilde{\Lambda}_0^\prime(0)}{1+(\pi j_z/4)^2}\Bigr]
\end{eqnarray}
with $\tilde{\Lambda}_0^\prime(0)$ from Eq.~(\ref{derivativeofhilberttransformed}).

\subsection{Terms of order \texorpdfstring{$\mathbf{B^0}$}{B0}}

It remains to analyze the terms for $B=0$
\begin{eqnarray}
\overline{F}_s(T,V)&=&-T\int_{-\infty}^\infty\rmd \omega\,D_0(\omega,V)\ln(1+e^{-\beta\omega})\nonumber\\
&&-T\int_{-\infty}^\infty\rmd\omega\, D_0(\omega,-V)\ln(1+e^{-\beta\omega})\;,\nonumber\\
D_0(\omega,V)&=&-\frac{1}{\pi}\frac{\partial}{\partial\omega}{\rm Im}\left(\ln\left(1-Vg_0(\omega)\right)\right)\;,\nonumber\\
g_0(\omega)&=&\int_{-1}^1\rmd\epsilon\,\frac{\rho_0(\epsilon)}{\omega-\epsilon+\rmi\eta}
\end{eqnarray}
with $\eta=0^+$. For the constant density of states we have
\begin{eqnarray}
g_0(\omega)&=&\Lambda_0(\omega)-\rmi\pi \rho_0(\omega)
\end{eqnarray}
with $\rho_0(\omega)=1/2$ for $|\omega|\leq 1$ and $\Lambda_0(\omega)$ from Eq.~(\ref{Hilberttransformed}).

\subsubsection{Pole contributions}
For $V>0$, there is an anti-bound state,
\begin{eqnarray}
1-V\Lambda_0(\omega_{\rm ab})&=&0\;,\nonumber\\
\omega_{\rm ab}&=&{\rm coth}\left(1/V\right)>1\;,
\end{eqnarray}
and, for $V<0$, there is a bound state,
\begin{eqnarray}
1+|V|\Lambda_0(\omega_{\rm b})&=&0\;,\nonumber\\
\omega_{\rm b}&=&-{\rm coth}(1/|V|)<-1\;.
\end{eqnarray}
Thus, for a constant density of states we find
\begin{eqnarray}
D_0^{\rm c}(\omega,V)&=&\delta(\omega-\omega_{\rm ab})\theta_{\rm H}(V)+\delta(\omega-\omega_{\rm b})\theta_{\rm H}(-V)\nonumber\\
&&+D_0^{{\rm c},{\rm band}}(\omega,V)\;,
\end{eqnarray}
where $\theta_{\rm H}$ is the Heaviside step function.

\subsubsection{Band contribution}
{}From Eq.~(155) of Bauerbach et al.\ we have
\begin{eqnarray}
D_0^{{\rm c},{\rm band}}(\omega,V)&=&-\frac{1}{\pi}\frac{\partial}{\partial\omega}{\rm Cot}^{-1}\left[\frac{1-V\Lambda_0(\omega)}{\pi V\rho_0(0)}\right]\nonumber\\
&&+\frac{1}{\pi}\frac{\partial}{\partial\omega}{\rm Cot}^{-1}\left[\frac{1+V\Lambda_0(\omega)}{\pi V\rho_0(0)}\right]
\end{eqnarray}
with $V=J_z/4>0$. Note that
\begin{equation}
{\rm Cot}^{-1}(x)=\pi\theta_{\rm H}(-x)+{\rm cot}^{-1}(x)
\end{equation}
is continuous and differentiable across $x=0$. For $\overline{F}_s(T,V)$ at small temperatures we write
\begin{eqnarray}
\overline{F}_s(T,V)&=&\int_{-\infty}^0\rmd\omega\,\omega\Delta D_0(\omega,V)\nonumber\\
&&-2 T\int_0^\infty\rmd\omega\,\ln(1+e^{-\beta\omega})\Delta D_0(\omega,V)\nonumber\\
&=&e_0(V)+\delta \overline{F}_s(T,V)\;,
\end{eqnarray}
where we used $\ln(1+e^{-\beta\omega})=-\beta\omega+\ln(1+e^{\beta\omega})$ and $D_0(\omega,V)=D_0(-\omega,-V)$ in the second step. At low temperatures, the second term is 
\begin{eqnarray}
\delta\overline{F}_s(T,V)&=&-2T^2\int_0^\infty\rmd x\,\ln(1+e^{-x})\Delta D_0(xT,V)\nonumber\\
&\approx&-4 T^2D_0(0,V)\int_0^\infty\rmd x\,\ln(1+e^x)+\mathcal{O}[T^4]\nonumber\\
&=&-\frac{\pi^2 T^2}{3}D_0(0,V)
\end{eqnarray}
with 
\begin{eqnarray}
D_0(0,V)&=&-\frac{V^2\Lambda_0^\prime(0)\rho_0(0)}{1+(\pi V\rho_0(0))^2}\;,
\end{eqnarray}
see~Eq.~(\ref{Dvonomega-gleich0}).

It remains to calculate the ground-state energy. 

\subsubsection{Ground-state energy}

The ground-state energy reads
\begin{eqnarray}
e_0(V)&=&\int_{-\infty}^0\rmd\omega\,\omega\left(D_0(\omega,V)+D_0(\omega,-V)\right)\nonumber\\
&=&\omega_{\rm b}+\int_{-1}^0\rmd\omega\,\omega D_0^{{\rm c},{\rm band}}(\omega,V)\nonumber\\
&=&\omega_{\rm b}+\left.\vphantom{A^B_C}\omega\mathcal{D}^{{\rm c},{\rm band}}(\omega,V)\right|_{-1}^0\nonumber \\
&& -\int_{-1}^0\rmd\omega\,\mathcal{D}_0^{{\rm c},{\rm band}}(\omega,V)
\end{eqnarray}
with
\begin{eqnarray}
\mathcal{D}_0^{{\rm c},{\rm band}}(\omega,V)&=&-\frac{1}{\pi}{\rm Cot}^{-1}\left(\frac{1-V\Lambda_0(\omega)}{\pi V\rho_0(0)}\right)\nonumber\\
&&+\frac{1}{\pi}{\rm Cot}^{-1}\left(\frac{1+V\Lambda_0(\omega)}{\pi V\rho_0(0)}\right)\;,\nonumber\\
\mathcal{D}_0(-1,V)&=&-\frac{1}{\pi}{\rm Cot}^{-1}(\infty)+\frac{1}{\pi}{\rm Cot}^{-1}(-\infty)=1\;.\nonumber\\
\end{eqnarray}
Thus,
\begin{eqnarray}
e_0(V)&=&\omega_{\rm b}+\left(0-(-1)1\right)-\int_{-1}^0\rmd\omega\,\mathcal{D}_0^{{\rm c},{\rm band}}(\omega,V)\;,\nonumber\\
&=&\left(1-\coth(1/V)\right)-\int_{-1}^0\rmd\omega\,\mathcal{D}_0^{{\rm c},{\rm band}}(\omega,V)\;.\nonumber\\
\end{eqnarray}
The formula is applicable for all $V$. For $V\ll1 $, and ignoring exponentially small corrections $e^{-1/V}$, we find the asymptotic series expansion
\begin{eqnarray}
e_0(V\ll1)&\approx &a_2 V^2 +a_4 V^4 + \mathcal{O}[V^6]\; , \nonumber \\
a_2&=& -\int_{-1}^0\rmd\omega\,\frac{1}{2}\ln\left(\frac{1+\omega}{1-\omega}\right) = -\ln2 \; , \nonumber\\
a_4&=& -\frac{V^4}{8}\int_{-1}^0\rmd\omega
\biggl[-\pi^2\ln\left(\frac{1+\omega}{1-\omega}\right)\nonumber \\
&&  \hphantom{-\frac{V^4}{8}\int_{-1}^0\rmd\omega\biggl[}
+\ln^3\left(\frac{1+\omega}{1-\omega}\right)\biggr]\nonumber\\
&=&\frac{1}{8}\left(2\pi^2\ln 2-9\zeta(3)\right)
\end{eqnarray}
or 
\begin{eqnarray}
e_0(j_z)&=&-\frac{\ln 2}{4}j_z^2+\frac{1}{128}j_z^4\left(2\pi^2\ln2-9\zeta(3)\right)\,.\nonumber\\
\end{eqnarray}
This result can also be obtained directly using 
\begin{equation}
\mathcal{D}_0^{\rm asymp}(\omega,V)=-\frac{1}{\pi}{\rm Im}\left[\ln\left(1-V^2\left(\Lambda_0(\omega)-\frac{\rmi\pi}{2}\right)^2\right)\right]
\end{equation}
and performing the expansion in $V^2$.

\subsection{Summary}

In the limit $0<B\ll T\ll1$, we have 
\begin{eqnarray}
F_{\rm IK}^{\rm i}(B,J_z,T)&\approx & -T\ln2+\overline{F}_s(T,J_z)\nonumber\\
&&-\frac{B^2}{2 T}\biggl[\left(1-\frac{2}{\pi}{\rm arctan}\left(\frac{\pi j_z}{4}\right)\right)^2\nonumber\\
&&-T\left(\frac{j_z}{4}\right)^2\frac{2\tilde{\Lambda}_0^\prime(0)}{1+(\pi j_z/4)^2}\biggr]
\end{eqnarray}
with
\begin{equation}
\tilde{\Lambda}_0^\prime(0)=1
\end{equation}
for the constant density of states, see~\ref{appendix:Lambda0PrimeVon0};
corrections are of the order [$B^2T,B^4$]. Furthermore,
\begin{eqnarray}
\overline{F}_s(T,J_z)&=&e_0(J_z)-\frac{\pi^2 T^2}{3}D_0(0,J_z)+\mathcal{O}[T^4]\;,\nonumber\\
D_0(0,J_z)&=&-\left(\frac{j_z}{4}\right)^2\tilde{\Lambda}_0^\prime(0)\frac{1}{1+\left(\pi j_z/4\right)^2}\;.
\end{eqnarray}
The ground-state energy $e_0(J_z)$ must be calculated individually for each density of states. With
\begin{equation}
    g_0(\omega)=\Lambda_0(\omega)-\rmi\pi\rho_0(\omega)
\end{equation}
we have
\begin{eqnarray}
e_0(J_z)&=&\int_{-1}^0\rmd\omega\,\frac{1}{\pi}{\rm Im}\left[]
\ln\left(1-\left(\frac{J_z}{4}\right)^2\Xi(\omega)\right)\right]\nonumber\\
&=&-\left(\frac{J_z}{4}\right)^2\int_{-1}^0\rmd\omega\,\frac{1}{\pi}{\rm Im}\left(g_0(\omega)\right)^2\nonumber\\
&&-\frac{1}{2}\left(\frac{J_z}{4}\right)^4\int_{-1}^0\rmd\omega\,\frac{1}{\pi}{\rm Im}\left(g_0(\omega)\right)+\mathcal{O}[j_z^6]\nonumber\\
&=&\frac{j_z^2}{8}\int_{-1}^0\rmd\omega\,\tilde{\Lambda}_0(\omega)\tilde{\rho}_0(\omega)\nonumber\\
&&-\frac{j_z^4}{128}\int_{-1}^0\rmd\omega\,\left(\pi^2\tilde{\Lambda}_0(\omega)\tilde{\rho}_0^3(\omega)-\tilde{\Lambda}_0^3(\omega)\tilde{\rho}_0(\omega)\right)\nonumber\\
&&+\mathcal{O}[j_z^6]\;,
\end{eqnarray}
where
\begin{equation}
\tilde{\Lambda}_0(\omega)=\frac{\Lambda_0(\omega)}{\rho_0(0)}\;,\tilde{\rho}_0(0)=\frac{\rho_0(\omega)}{\rho_0(0)}\;,
j_z=J_z\rho_0(0)\;.
\end{equation}
For a constant density of states,
\begin{eqnarray}
\tilde{\rho}_0(\omega)&=&1\;,\nonumber\\
\tilde{\Lambda}_0(\omega)&=&\ln\left(\frac{1+\omega}{1-\omega}\right)\:,\nonumber\\
\tilde{\Lambda}_0^\prime(0)&=&2\;,\nonumber\\
e_0^{\rm const}(j_z)&=&-\frac{j_z^2}{4}\ln2+\frac{j_z^4}{128}\left(2\pi^2\ln2-9\zeta(3)\right)\; ,\nonumber\\
\end{eqnarray}
and the free energy of the Ising-Kondo model becomes, up to and including order $B^2$ and $j_z^3$,
\begin{eqnarray}\label{IsingKondoContributionFreeEnergy?}
F_{\rm IK}^i&=&-T\ln2-\frac{B^2}{2T} + \frac{B^2}{2T}j_z \nonumber\\
&&+\left(-2\ln 2+\frac{\pi^2T^2}{3}+B^2-\frac{B^2}{T}\right)\left(\frac{j_z^2}{8}\right)\nonumber\\
&&-\frac{B^2\pi^2}{96 T}j_z^3+\mathcal{O}[j_z^4]\;.
\end{eqnarray}

\section{Approximation of the term \texorpdfstring{$\mathbf{x_B}$}{xB}}
\label{sm:xBterm-derivation}
In this chapter we derive an analytic expression of the term $x_B$ occurring in 
Appendix~\mainsectionRearrangement
at $0\ll B\ll T\ll 1$. 
A numerical treatment of this term is equally possible and can be found in 
Appendix~\mainsectionnumericalcheckxB
of the main paper.

\subsection{Evaluation of the \texorpdfstring{$\mathbf{x_B}$}{xB} term}
\label{chapter:evaluation-of-the-xB-term}

We start with the formal derivation of the analytic expression of the term $x_B$. Explicit calculations follow in the subsequent subsections.

\subsubsection{Formal expansion of \texorpdfstring{${x_B}$}{xB}}

\label{chapter:XBTerm-general-considerations}
According to 
Eq.~(\mainxBDefinitionthirdorder) 
of the main paper, the term $x_B=x_B(B,T)$ is defined as ($\beta=1/T$)
\begin{eqnarray}
x_B&=&-T\int_0^\beta\rmd\lambda_3\,\int_0^{\lambda_3}\rmd\lambda_2\,\int_0^{\lambda_2}\rmd\lambda_1\nonumber\\
&&\times\Big(
W^+ \tilde{f}(\lambda_3-\lambda_2,-B)\tilde{f}(\lambda_3-\lambda_1,B)\nonumber\\
&&\times \tilde{f}(\lambda_2-\lambda_1,B)\nonumber\\
&&+W^- \tilde{f}(\lambda_3-\lambda_2,B)\tilde{f}(\lambda_3-\lambda_1,-B) \nonumber\\
&&\times \tilde{f}(\lambda_2-\lambda_1,-B)\nonumber\\
&&+W^+ \tilde{f}(\lambda_3-\lambda_2,B)\tilde{f}(\lambda_3-\lambda_1,B) \nonumber\\
&&\times \tilde{f}(\lambda_2-\lambda_1,B)\nonumber\\
&&+W^- \tilde{f}(\lambda_3-\lambda_2,-B)\tilde{f}(\lambda_3-\lambda_1,-B) \nonumber\\
&&\times \tilde{f}(\lambda_2-\lambda_1,-B)\nonumber\\
&&+W^+ \tilde{f}(\lambda_3-\lambda_2,B)\tilde{f}(\lambda_3-\lambda_1,B) \nonumber\\
&&\times \tilde{f}(\lambda_2-\lambda_1,-B)\nonumber\\
&&+W^- \tilde{f}(\lambda_3-\lambda_2,-B)\tilde{f}(\lambda_3-\lambda_1,-B) \nonumber\\
&&\times \tilde{f}(\lambda_2-\lambda_1,B)
\Big)\;.
\end{eqnarray}
Recall that 
\begin{eqnarray}
j_\perp&=&J_\perp \rho_0(0)\;,\nonumber\\
j_z&=&J_z\rho_0(0)\;,\nonumber\\
\end{eqnarray}
and
\begin{eqnarray}\label{eq:Wplusminus-and-f-definition}
\rhotilde(\epsilon)&=&\rho_0(\epsilon)/\rho_0(0)\;,\nonumber\\
\tilde{f}(\lambda,B)&=&\int_{-1}^1\rmd\epsilon\,\frac{\rhotilde(\epsilon)e^{\lambda\epsilon}}{1+e^{\beta(\epsilon-B)}}\;,\nonumber\\
W^{\pm}(B,T)&=&\frac{e^{\mp\beta B}}{2\cosh(\beta B)}
\end{eqnarray}
hold, see 
Eq.~(\mainftildelambdaBdefinition) 
and~(\mainWpmDefinition) 
of the main paper. Note that $\tilde{\rho}(\epsilon)=1$ for a constant density of states.
With the help of \textsc{Mathematica}~\cite{Mathematica12} the evaluation of the $\lambda_i$ integrals can be done analytically  [$\mathcal{G}\equiv \mathcal{G}(T,\epsilon_1,\epsilon_2,\epsilon_3)$]
\begin{eqnarray}\label{eq:tau-integrals}
\mathcal{G}&=&\int_0^\beta\rmd\lambda_3\,\int_0^{\lambda_3}\rmd\lambda_2\,\int_0^{\lambda_2}\rmd\lambda_1\,\nonumber\\
&&\times{\rm exp}\big((\lambda_3-\lambda_2)\epsilon_1+(\lambda_3-\lambda_1)\epsilon_2+(\lambda_2-\lambda_1)\epsilon_3\big)\nonumber\\
&=&\frac{\beta}{(\epsilon_1+\epsilon_2)(\epsilon_2+\epsilon_3)}+\frac{1-e^{\beta(\epsilon_2+\epsilon_3)}}{(\epsilon_2+\epsilon_3)^2(\epsilon_1-\epsilon_3)}\nonumber\\
&&+\frac{1-e^{\beta(\epsilon_1+\epsilon_2)}}{(\epsilon_1+\epsilon_2)^2(\epsilon_3-\epsilon_1)}\;.
\end{eqnarray}
We define [$F(\tau_1,\tau_2,\tau_3)\equiv F(\tau_1,\tau_2,\tau_3,B,T)$]
\begin{eqnarray}\label{Fvontau1tau2tau3Def}
F(\tau_1,\tau_2,\tau_3)&=&-T\int_{-1}^1\frac{\rmd\epsilon_1\,\rhotilde(\epsilon_1)}{1+e^{\beta(\epsilon_1-\tau_1 B)}}\nonumber\\
&&\times\int_{-1}^1\frac{\rmd\epsilon_2\,\rhotilde(\epsilon_2)}{1+e^{\beta(\epsilon_2-\tau_2B)}}\nonumber\\
&&\times \int_{-1}^1\frac{\rmd\epsilon_3\,\rhotilde(\epsilon_3)}{1+e^{\beta(\epsilon_3-\tau_3B)}}\mathcal{G}(T,\epsilon_1,\epsilon_2,\epsilon_3)\nonumber\\
\end{eqnarray}
to find
\begin{eqnarray}\label{xbftau1tau2tau3connection}
x_B&\equiv& W^+ F(-1,1,1)+W^- F(1,-1,-1)\nonumber\\
&&+W^+F(1,1,1)+W^- F(-1,-1,-1)\nonumber\\
&&+W^+F(1,1,-1)+W^- F(-1,-1,1)\;.\nonumber\\
\end{eqnarray}
Note that in Eq.~(\ref{Fvontau1tau2tau3Def}) the $\tau_i$ represent the sign of the magnetic field in the Fermi functions.

We note that $\mathcal{G}$ in Eq.~(\ref{eq:tau-integrals}) as a whole has no singularities for $T>0$. However, we intend to treat the five terms in $\mathcal{G}$ separately. Unfortunately, this is prohibited because of the singularities arising from zeros in the individual denominators. For example, the first term,
\begin{equation}
1^{\rm st}=\frac{1}{(\epsilon_2+\epsilon_3)(\epsilon_1+\epsilon_2)}\;,
\end{equation}
has first-order poles at $(\epsilon_1=-\epsilon_2,$ $\epsilon_3\neq -\epsilon_2)$ and $(\epsilon_1\neq -\epsilon_2,$ $\epsilon_3=-\epsilon_2)$ and a second-order pole at $\epsilon_1=-\epsilon_2=\epsilon_3$. Therefore, we rather address
\begin{eqnarray}
&&\mathcal{G}^{(\eta_1,\eta_2,\eta_3)}(T,\epsilon_1,\epsilon_2,\epsilon_3)\nonumber\\
&&\quad={\rm Re}\left(G(T,\epsilon_1+\rmi\eta_1, \epsilon_2+\rmi\eta_2,\epsilon_3+\rmi\eta_3)\right)\,.
\end{eqnarray}
Since $G^{(0,0,0)}(T,\epsilon_1,\epsilon_2,\epsilon_3)$ is real, we can restrict ourselves to the real part from the beginning. 

Thus, we work with
\begin{eqnarray}
&&F^{(\eta_1,\eta_2,\eta_3)}(T,\tau_1,\tau_2,\tau_3)\nonumber\\
&&\quad =-T\int_{-1}^1\frac{\rmd\epsilon_1\,\rhotilde(\epsilon_1)}{1+e^{\beta(\epsilon_1-\tau_1 B)}}\nonumber\\
&&\quad\times\int_{-1}^1\frac{\rmd\epsilon_2\,\rhotilde(\epsilon_2)}{1+e^{\beta(\epsilon_2-\tau_2 B)}}\int_{-1}^1\frac{\rmd\epsilon_3\,\rhotilde(\epsilon_3)}{1+e^{\beta(\epsilon_3-\tau_3 B)}}\nonumber\\
&&\quad\quad\quad\times\,{\rm Re}\left[\,({\rm I})+{\rm (II)}+{\rm (III)}\,\right]
\end{eqnarray}
with 
\begin{eqnarray}
{\rm (I)}&=&\frac{1}{T}\frac{1}{(\epsilon_1+\rmi\eta_1+\epsilon_2+\rmi\eta_2)(\epsilon_2+\rmi\eta_2+\epsilon_3+\rmi\eta_3)}\;,\nonumber\\
{\rm (II)}&=&\frac{1-e^{\beta(\epsilon_2+\epsilon_3)}}{(\epsilon_2+\rmi\eta_2+\epsilon_3+\rmi\eta_3)^2(\epsilon_1+\rmi\eta_1-\epsilon_3-\rmi\eta_3)}\;,\nonumber\\
{\rm (III)}&=&\frac{1-e^{\beta(\epsilon_1+\epsilon_2)}}{(\epsilon_1+\rmi\eta_1+\epsilon_2+\rmi\eta_2)^2(\epsilon_3+\rmi\eta_3-\epsilon_1-\rmi\eta_1)}\;.\nonumber\\
\end{eqnarray}
To be definite, we let $\eta_3>\eta_1\to 0^+$ and let $\eta_2>0$ to cope with the remaining singularities.
In ${\rm (II)}$ we find the contribution
\begin{eqnarray}
&&\frac{1-e^{\beta(\epsilon_1+\epsilon_2)}}{(\epsilon_2+\epsilon_3+\rmi \eta_2)^2}\left(P\left(\frac{1}{\epsilon_1-\epsilon_3}\right)+\rmi\pi\delta(\epsilon_1-\epsilon_3)\right)\nonumber\\
&&\quad=\frac{1-e^{\beta(\epsilon_2+\epsilon_3)}}{(\epsilon_2+\epsilon_3+\rmi\eta_2)^2}P\left(\frac{1}{\epsilon_1-\epsilon_3}\right)\nonumber\\
&&\quad\quad+\rmi\pi\frac{1+e^{\beta(\epsilon_1+\epsilon_2)}}{(\epsilon_1+\epsilon_2+\rmi\eta_2)^2}\delta(\epsilon_1-\epsilon_3)\;,
\end{eqnarray}
where $P$ denotes Cauchy's principle value. 
In ${\rm (III)}$ we encounter
\begin{equation}
\frac{1-e^{\beta(\epsilon_1+\epsilon_2)}}{(\epsilon_1+\epsilon_2+\rmi\eta_2)^2}P\left[\frac{1}{\epsilon_3-\epsilon_1}\right]
-\frac{\rmi\pi\delta(\epsilon_1-\epsilon_3)(1-e^{\beta(\epsilon_1+\epsilon_2)})}{(\epsilon_1+\epsilon_2+\rmi\eta_2)^2}
\end{equation}
when the proper limits for $\eta_1,\eta_2,\eta_3$ are taken.

Thus, we find
\begin{eqnarray}
&&F^{(0,\eta_2,0)}(T,\tau_1,\tau_2,\tau_3)\nonumber\\
&&\quad=-T\int_{-1}^1\frac{\rmd\epsilon_1\,\rhotilde(\epsilon_1)}{1+e^{\beta(\epsilon_1-\tau_1 B)}}\int_{-1}^1\frac{\rmd \epsilon_2\,\rhotilde(\epsilon_2)}{1+e^{\beta(\epsilon_2-\tau_2 B)}}\nonumber\\
&&\quad\qquad\times \int_{-1}^1\frac{\rmd\epsilon_3\,\rhotilde(\epsilon_3)}{1+e^{\beta(\epsilon_3-\tau_3 B)}} \nonumber\\
&&\qquad\qquad\times\,{\rm Re}\bigg[ 
\frac{1}{T}\left(P\left(\frac{1}{\epsilon_1+\epsilon_2}\right)-\rmi\pi\delta(\epsilon_1+\epsilon_2)\right)\nonumber\\
&&\qquad\qquad\hphantom{\times\,{\rm Re}\bigg[}
\left(P\left(\frac{1}{\epsilon_2+\epsilon_3}\right)-\rmi\pi\delta(\epsilon_2+\epsilon_3)\right)\nonumber\\
&&\qquad\qquad\hphantom{\times\,{\rm Re}\bigg[}
+\frac{1-e^{\beta(\epsilon_2+\epsilon_3)}}{(\epsilon_2+\epsilon_3+\rmi\eta_2)^2}P\left(\frac{1}{\epsilon_1-\epsilon_3}\right)\nonumber\\
&&\qquad\qquad\hphantom{\times\,{\rm Re}\bigg[}
+\frac{1-e^{\beta(\epsilon_1+\epsilon_2)}}{(\epsilon_2+\epsilon_1+\rmi\eta_2)^2}P\left(\frac{1}{\epsilon_3-\epsilon_1}\right)
\bigg]\;.\nonumber\\
\end{eqnarray}
We write 
\begin{eqnarray}
\left(\frac{1}{\epsilon_2+x-\rmi\eta}\right)^2&=&\left(-\frac{\partial}{\partial\epsilon_2}\right)\frac{1}{\epsilon_2+x+\rmi\eta}\nonumber\\
&=&-\frac{\partial}{\partial\epsilon_2}P\left(\frac{1}{\epsilon_2+x}\right)+\rmi\pi\frac{\partial}{\partial\epsilon_2}\delta(\epsilon_2+x).\nonumber\\
\end{eqnarray}
The imaginary part drops out when we take ${\rm Re}[\dots]$. Since we have separated the contribution of the poles, we can also calculate their contribution directly. For the constant density of states we find
\begin{eqnarray}\label{Fpolecontribution}
F^{\rm p}(\tau_1,\tau_2,\tau_3)&=&\int_{-1}^1\frac{\rmd\epsilon_1\,\rhotilde(\epsilon_1)}{1+e^{\beta(\epsilon_1-\tau_1 B)}}\int_{-1}^1\frac{\rmd\epsilon_2\,\rhotilde(\epsilon_2)}{1+e^{\beta(\epsilon_2-\tau_2 B)}}\nonumber\\
&&\times \int_{-1}^1\frac{\rmd\epsilon_3\,\rhotilde(\epsilon_3)}{1+e^{\beta(\epsilon_3-\tau_3 B)}}\nonumber\\
&&\quad \times \pi^2\delta(\epsilon_1+\epsilon_2)\delta(\epsilon_2+\epsilon_3)\nonumber\\
&=&\pi^2\int_{-1}^1\frac{\rmd\epsilon}{1+e^{\beta(\epsilon-\tau_2 B)}}\frac{1}{1+e^{-\beta(\epsilon+\tau_1B)}}\nonumber\\
&&\quad \times\frac{1}{1+e^{-\beta(\epsilon+\tau_3B)}}\nonumber\\
&=&\pi^2 T\int_{-\infty}^\infty\frac{\rmd x}{1+e^{x-b_2}}\frac{1}{1+e^{-x-b_1}}\nonumber\\
&&\quad \times \frac{1}{1+e^{-x-b_3}}\nonumber\\
&=&\frac{\pi^2 T}{2}+\frac{\pi^2}{6}B(\tau_1+2\tau_2+\tau_3)\nonumber\\
&&+\frac{\pi^2}{12 T}B^2\left(1+\tau_1\tau_2+\tau_2\tau_3+\tau_1\tau_3\right)\nonumber\\
\end{eqnarray} 
for $B\to 0$ and $T\to 0$, $B\ll T$. In the next to last step we defined $b_i=\tau_i B/T$. Note that $\tau_i^2=1$. Therefore, with Eq.~(\ref{xbftau1tau2tau3connection}) we define 
\begin{eqnarray}
x_B^{\rm p}&\equiv& W^+ \Fkrei{6}(-1,1,1)+W^- \Fkrei{6}(1,-1,-1)\nonumber\\
&&+W^+\Fkrei{6}(1,1,1)+W^- \Fkrei{6}(-1,-1,-1)\nonumber\\
&&+W^+\Fkrei{6}(1,1,-1)+W^- \Fkrei{6}(-1,-1,1)\nonumber\\
\end{eqnarray}
to find the contribution of the poles at $0\ll B\ll 1$
\begin{equation}
x_B^{\rm p}(B,T)=\frac{3\pi^2 T}{2}-\frac{\pi^2 B^2}{T}\;.
\end{equation}

For the treatment of the remaining terms we abbreviate
$f^\prime(\omega)={\partial f(\omega)}/({\partial \omega})$
and
\begin{eqnarray}
    \mathcal{Y}(\epsilon_1,\epsilon_2,\epsilon_3)&=&\frac{\rhotilde(\epsilon_1)}{1+e^{\beta(\epsilon_1-\tau_1 B)}}\frac{\rhotilde(\epsilon_2)}{1+e^{\beta(\epsilon_2-\tau_2 B)}}\nonumber\\
&&\times \frac{\rhotilde(\epsilon_3)}{1+e^{\beta(\epsilon_3-\tau_3 B)}}\nonumber\;.
\end{eqnarray}
With these definitions, we consider the six terms of $\mathcal{G}(T,\epsilon_1,\epsilon_2,\epsilon_3)$ separately -- including the pole contribution -- and find
\begin{equation}\label{Ftau1tau2tau3definition}
F(\tau_1,\tau_2,\tau_3)=\sum_{j=1}^5 \Fkrei{j}(\tau_1,\tau_2,\tau_3) + F^{\rm p}(\tau_1,\tau_2,\tau_3)
\end{equation}
with ($\beta=1/T$)
\begin{eqnarray}\label{eq:Fkreis-introduction}
\Fkrei{1}(\tau_1,\tau_2,\tau_3)&=&-T\beta\int_{-1}^1\rmd\epsilon_1\int_{-1}^1\rmd\epsilon_2\int_{-1}^1\rmd \epsilon_3\,\nonumber\\
&&\quad \times \frac{\mathcal{Y}(\epsilon_1,\epsilon_2,\epsilon_3)}{(\epsilon_1+\epsilon_2)(\epsilon_2+\epsilon_3)}\nonumber\\
&=&-\int_{-1}^1\rmd\omega\,\frac{\rhotilde(\omega)}{1+e^{\beta(\omega-\tau_2 B)}}\nonumber\\
&&\quad \times H_T(\omega,\tau_1 B)H_T(\omega,\tau_3B)\;,\\\label{eq:Fkreistwo-introduction}
\Fkrei{2}(\tau_1,\tau_2,\tau_3)&=&-T\int_{-1}^1\rmd\epsilon_1\int_{-1}^1\rmd\epsilon_2\int_{-1}^1\rmd \epsilon_3\,\nonumber\\
&&\quad \times \frac{\mathcal{Y}(\epsilon_1,\epsilon_2,\epsilon_3)}{(\epsilon_2+\epsilon_3)^2(\epsilon_1-\epsilon_3)}\nonumber\\
&=&-T\int_{-1}^1\rmd\omega\,\frac{\rhotilde(\omega)}{1+e^{\beta(\omega-\tau_3 B)}}\nonumber\\
&&\quad \times H_T(-\omega,\tau_1B)\left(-H_T^\prime(\omega,\tau_2 B)\right)\;,\nonumber\\
&&\phantom{ }\\\label{eq:Fkreisthree-introduction}
\Fkrei{3}(\tau_1,\tau_2,\tau_3)&=&T\int_{-1}^1\rmd\epsilon_1\int_{-1}^1\rmd\epsilon_2\int_{-1}^1\rmd \epsilon_3\,\nonumber\\
&&\quad \times \mathcal{Y}(\epsilon_1,\epsilon_2,\epsilon_3)\frac{e^{\beta(\epsilon_2+\epsilon_3)}}{(\epsilon_2+\epsilon_3)^2(\epsilon_1-\epsilon_3)}\nonumber\\
&=&T\int_{-1}^1\rmd\epsilon_2\frac{\rhotilde(\epsilon_2)e^{\tau_2\beta B}}{1+e^{-\beta(-\epsilon_2-\tau_2 B)}}\nonumber\\
&&\quad \times \int_{-1}^1\rmd\epsilon_3\frac{\rhotilde(\epsilon_3)e^{\tau_3 \beta B}}{1+e^{-\beta(-\epsilon_3-\tau_3 B)}}\nonumber\\
&&\quad \times\int_{-1}^1\rmd\epsilon_1\frac{\rhotilde(\epsilon_1)}{(-\epsilon_2-\epsilon_3)^2(\epsilon_1+\epsilon_3)}\nonumber\\
&&\quad \times \frac{1}{1+e^{\beta(\epsilon_1-\tau_1 B)}}\nonumber\\
&=&Te^{(\tau_2+\tau_3)\beta B}\int_{-1}^1\rmd\omega\,\frac{\rhotilde(\omega)}{1+e^{\beta(\omega+\tau_3 B)}}\nonumber\\
&&\quad \times H_T(\omega,\tau_1 B)\left(-H_T^\prime(\omega,-\tau_2 B)\right)\;,\nonumber\\
\end{eqnarray}
where we multiplied $\Fkrei{3}(\tau_1,\tau_2,\tau_3)$ with $e^{+\tau_i\beta B}e^{-\tau_i\beta B}$, $i=1,2,3$, and used $e^x/(1+e^x)=(1+e^{-x})^{-1}$. In these expressions we defined for $|\omega|<1$
\begin{eqnarray}\label{eq:HTfunction-definition}
H_T(\omega,\tau B)&=&\int_{-1}^1\rmd\epsilon\,\frac{\rhotilde(\epsilon)}{1+e^{\beta(\epsilon-\tau B)}}\frac{1}{\omega+\epsilon}\nonumber\\
&=&-\ln(1-\omega)+\ln(2\pi T)+\Rbar\left(\frac{\omega+\tau B}{T}\right)\;,\nonumber\\
\end{eqnarray}
where the second line is correct for small $T\ll1$ and a constant density of states.
$\bar{R}((\omega+\tau B)/T)$ and its expansion in $B$ is known at low temperatures and small fields, 
see section~\ref{chapter:properties-of-the-Rbar-function}. The first derivative of $H_T(\omega,\tau B)$ with respect to $\omega$ reads
\begin{eqnarray}\label{eq:HTfunctionPrime-definition}
H_T^\prime(\omega,\tau B)&=&-\int_{-1}^1\rmd \epsilon\,\frac{\rhotilde(\epsilon)}{1+e^{\beta(\epsilon-\tau B)}}\frac{1}{(\omega+\epsilon)^2}\nonumber\\
&=&\frac{1}{1-\omega}+\frac{1}{T}\Rbar^\prime\left(\frac{\omega+\tau B}{T}\right)\;.
\end{eqnarray}
To eliminate the prefactor in front of the integral in the third term, we define 
\begin{eqnarray}\label{eq:F3bardefintion}
\Fkrei{3}(\tau_1,\tau_2,\tau_3)&=&e^{(\tau_2+\tau_3)\beta B}\Fbarkrei(\tau_1,\tau_2,\tau_3) \; ,
\end{eqnarray}
and further analyze $\Fbarkrei(\tau_1,\tau_2,\tau_3)$.
Furthermore,
\begin{eqnarray}\label{Fkreis4and5}
\Fkrei{4}(\tau_1,\tau_2,\tau_3)&=&\Fkrei{2}(\tau_3,\tau_2,\tau_1)\;,\nonumber\\
\Fkrei{5}(\tau_1,\tau_2,\tau_3)&=&\Fkrei{3}(\tau_3,\tau_2,\tau_1)\;.
\end{eqnarray}

To determine the functions $\Fkrei{j}(\tau_1,\tau_2,\tau_3)$, $j=1,\dots,5$, we expand 
[$\Fkrei{j}\equiv \Fkrei{j}(\tau_1,\tau_2,\tau_3)$]
\begin{eqnarray}\label{Fkreiformalexpansion}
\Fkrei{j}&=&{\rm sgn}\big(\Fkrei{j}\,\big) F_{gh\ell;j} \; ,\nonumber\\
F_{gh\ell;j}&=&
\int_{-1}^1\rmd\omega \gkrei{j}(\omega,\tau_1 B)\hkrei{j}(\omega,\tau_2 B)\lkrei{j}(\omega,\tau_3 B)\nonumber\;,\\
\end{eqnarray}
where `${\rm sgn}$' is the sign function and
\begin{eqnarray}
\gkrei{j}(\omega,\tau B)&=&\gkreis{j}{0}(\omega)+\tau B \gkreis{j}{1}(\omega)\nonumber\\
&&+(\tau B)^2 \gkreis{j}{2}(\omega)+\mathcal{O}[B^3]\;,\nonumber\\
\hkrei{j}(\omega,\tau B)&=&\hkreis{j}{0}(\omega)+\tau B \hkreis{j}{1}(\omega)\nonumber\\
&&+(\tau B)^2 \hkreis{j}{2}(\omega)+\mathcal{O}[B^3]\;,\nonumber\\
\lkrei{j}(\omega,\tau B)&=&\lkreis{j}{0}(\omega)+\tau B \lkreis{j}{1}(\omega)\nonumber\\
&&+(\tau B)^2 \lkreis{j}{2}(\omega)+\mathcal{O}[B^3]\;.
\end{eqnarray}
Now we define [$l,m,n\in\{0,1,2\}$]
\begin{equation}\label{eq:Flmn-Definition}
\Fkreis{j}{lmn}=\int_{-1}^1\rmd \omega\, \gkreis{j}{l}(\omega)\hkreis{j}{m}(\omega)\lkreis{j}{n}(\omega)
\end{equation}
to obtain the formal expansion
\begin{eqnarray}
    \Fkrei{j}(\tau_1,\tau_2,\tau_3)&=&\Fkreis{j}{000}+B\big(\tau_1\Fkreis{j}{100}+\tau_2\Fkreis{j}{010}+\tau_3\Fkreis{j}{001}\big)\nonumber\\
    &&+B^2\big(\tau_1^2 \Fkreis{j}{200}+\tau_2^2\Fkreis{j}{020}+\tau_3^2\Fkreis{j}{002}\nonumber\\
    &&+\tau_1\tau_2\Fkreis{j}{110}+\tau_1\tau_3\Fkreis{j}{101}+\tau_2\tau_3\Fkreis{j}{011}
    \big)\;.\nonumber\\
\end{eqnarray}
Evidently,
\begin{eqnarray}\label{eq:ghl1-Definitions}
\gkrei{1}(\omega,\tau B)&=&\frac{1}{1+e^{\beta(\omega-\tau B)}}\;,\nonumber\\
\hkrei{1}(\omega,\tau B)&=&H_T(\omega,\tau B)\;,\nonumber\\
\lkrei{1}(\omega,\tau B)&=&H_T(\omega,\tau B)\;.
\end{eqnarray}
so that
\begin{eqnarray}
\gkreis{1}{0}(\omega)&=&\frac{1}{1+e^{\omega/T}}\;,\nonumber\\
\gkreis{1}{1}(\omega)&=&\frac{1}{T}\frac{e^{\omega/T}}{(1+e^{\omega/T})^2}\;,\nonumber\\
\gkreis{1}{2}(\omega)&=&\frac{1}{2T^2}\frac{e^{\omega/T}(e^{\omega/T}-1)}{(1+e^{\omega/T})^3}\;,\nonumber\\
\hkreis{1}{0}(\omega)&=&H_T^{(0)}(\omega)\nonumber\\
&=&-\ln(1-\omega)+\ln(2\pi T)+\Rbar(\omega/T)\;,\nonumber\\
\hkreis{1}{1}(\omega)&=&H_T^{(1)}(\omega)=\frac{1}{T}\Rbar^{\prime}(\omega/T)\;,\nonumber\\
\hkreis{1}{2}(\omega)&=&H_T^{(2)}(\omega)=\frac{1}{2T^2}\Rbar^{\prime\prime}(\omega/T)\;,\nonumber\\
\lkreis{1}{0}(\omega)&=&\hkreis{1}{0}(\omega)\;,\nonumber\\
\lkreis{1}{1}(\omega)&=&\hkreis{1}{1}(\omega)\;,\nonumber\\
\lkreis{1}{2}(\omega)&=&\hkreis{1}{2}(\omega)\;.
\end{eqnarray}
Furthermore,
\begin{eqnarray}\label{eq:ghl2-Definitions}
\gkrei{2}(\omega,\tau B)&=&\frac{1}{1+e^{\beta(\omega-\tau B)}}\;,\nonumber\\
\hkrei{2}(\omega,\tau B)&=&H_T(-\omega,\tau B)\;,\nonumber\\
\lkrei{2}(\omega,\tau B)&=&H_T^\prime(\omega,\tau B)\;.
\end{eqnarray}
Note that $\Rbar(x)$ and $\Rbar''(x)$ are even in $x$, $\Rbar'(x)$ is odd in $x$, thus 
\begin{eqnarray}
\gkreis{2}{0}(\omega)&=&\frac{1}{1+e^{\omega/T}}\;,\nonumber\\
\gkreis{2}{1}(\omega)&=&\frac{1}{T}\frac{e^{\omega/T}}{(1+e^{\omega/T})^2}\;,\nonumber\\
\gkreis{2}{2}(\omega)&=&\frac{1}{2T^2}\frac{e^{\omega/T}(e^{\omega/T}-1)}{(1+e^{\omega/T})^3}\;,\nonumber\\
\hkreis{2}{0}(\omega)&=&H_T^{(0)}(-\omega)=-\ln(1+\omega)+\ln(2\pi T)\nonumber\\
&&+\Rbar(\omega/T)\;,\nonumber\\
\hkreis{2}{1}(\omega)&=&H_T^{(1)}(-\omega)=-H_T^{(1)}(\omega)=-\frac{1}{T}\Rbar^{\prime}(\omega/T)\;,\nonumber\\
\hkreis{2}{2}(\omega)&=&H_T^{(2)}(-\omega)=H_T^{(2)}(\omega)=\frac{1}{2T^2}\Rbar^{\prime\prime}(\omega/T)\;,\nonumber\\
\lkreis{2}{0}(\omega)&=&(H_T^\prime)^{(0)}(\omega)=\frac{1}{1-\omega}+\frac{1}{T}\Rbar^\prime(\omega/T)\;,\nonumber\\
\lkreis{2}{1}(\omega)&=&(H_T^\prime)^{(1)}(\omega)=\frac{1}{T^2}\Rbar^{\prime\prime}(\omega/T)\;,\nonumber\\
\lkreis{2}{2}(\omega)&=&(H_T^\prime)^{(2)}(\omega)=\frac{1}{2T^3}\Rbar^{\prime\prime\prime}(\omega/T)\;.
\end{eqnarray}
For the third term, we likewise expand the function $\Fbarkrei$ from Eq.~(\ref{eq:F3bardefintion}) using the functions
\begin{eqnarray}\label{eq:ghl3-Definitions}
\gbarkrei(\omega,\tau B)&=&\frac{1}{1+e^{\beta(\omega+\tau B)}}\;,\nonumber\\
\hbarkrei(\omega,\tau B)&=&H_T(\omega,\tau B)\;,\nonumber\\
\lbarkrei(\omega,\tau B)&=&H_T^\prime(\omega,-\tau B)
\end{eqnarray}
so that we employ
\begin{eqnarray}
\gbarkreis{0}(\omega)&=&\frac{1}{1+e^{\omega/T}}\;,\nonumber\\
\gbarkreis{1}(\omega)&=&-\frac{1}{T}\frac{e^{\omega/T}}{(1+e^{\omega/T})^2}\;,\nonumber\\
\gbarkreis{2}(\omega)&=&\frac{1}{2T^2}\frac{e^{\omega/T}(e^{\omega/T}-1)}{(1+e^{\omega/T})^3}\;,\nonumber\\
\hbarkreis{0}(\omega)&=&H_T^{(0)}(\omega)\nonumber\\
&=&-\ln(1-\omega)+\ln(2\pi T)+\Rbar(\omega/T)\;,\nonumber\\
\hbarkreis{1}(\omega)&=&H_T^{(1)}(\omega)=\frac{1}{T}\Rbar^\prime(\omega/T)\;,\nonumber\\
\hbarkreis{2}(\omega)&=&H_T^{(2)}(\omega)=\frac{1}{2T^2}\Rbar^{\prime\prime}(\omega/T)\;,\nonumber\\
\lbarkreis{0}(\omega)&=&(H_T^\prime)^{(0)}(\omega)=\frac{1}{1-\omega}+\frac{1}{T}\Rbar^\prime(\omega/T)\;,\nonumber\\
\lbarkreis{1}(\omega)&=&-(H_T^\prime)^{(1)}(\omega)=-\frac{1}{T^2}\Rbar^{\prime\prime}(\omega/T)\;,\nonumber\\
\lbarkreis{2}(\omega)&=&(H_T^\prime)^{(2)}(\omega)=\frac{1}{2T^3}\Rbar^{\prime\prime\prime}(\omega/T)\;.
\end{eqnarray}
Note that, in $\lbarkreis{1}(\omega)$, the minus sign comes from the minus sign in the $H_T^\prime(\omega,-\tau B)$ term, see Eq.~(\ref{eq:ghl3-Definitions}), 
and the minus sign in $\gbarkreis{1}(\omega)$ results from the exponent $\omega+\tau B$ in $\gbarkrei(\omega,\tau B)$, see Eq.~(\ref{eq:ghl3-Definitions}).
A detailed discussion of the functions $H_T(\omega,\tau B)$ and $\Rbar(x)$ can be found in Sect.~\ref{chapter:properties-of-the-Rbar-function}.

At this point it is worthwhile to identify the symmetries of the individual terms. 
We find 
\begin{eqnarray}
    \Fkreis{1}{100}(\tau_1, \tau_2,\tau_3)&=&\Fkreis{1}{001}(\tau_1, \tau_2,\tau_3)\; , \nonumber \\
    \Fkreis{1}{200}(\tau_1, \tau_2,\tau_3)&=&\Fkreis{1}{002}(\tau_1, \tau_2,\tau_3)\;, \nonumber \\
    \Fkreis{1}{110}(\tau_1, \tau_2,\tau_3)&=&\Fkreis{1}{011}(\tau_1, \tau_2,\tau_3)\; ,
\end{eqnarray}
and
\begin{eqnarray}
\Fkrei{4}(\tau_1, \tau_2,\tau_3)&=&\Fkrei{2}(\tau_3,\tau_2,\tau_1)\;,\nonumber\\
\Fkrei{5}(\tau_1,\tau_2,\tau_3)&=&\Fkrei{3}(\tau_3,\tau_2,\tau_1)\; ,
\end{eqnarray} 
see Eq.~(\ref{Fkreis4and5}). Consequently, 
\begin{equation}\label{SymmetryOfFtau1tau2tau3}
F(\tau_1,\tau_2,\tau_3)\equiv F(\tau_3,\tau_2,\tau_1)
\end{equation}
as it must.
Because of the special function $\Fkrei{3}$, see Eq.~(\ref{eq:F3bardefintion}), 
we separately give the relations between $\Fkreis{3}{lmn}$ and $\Fbarkreis{lmn}$ for $B\ll1$,
\begin{eqnarray}
\Fkreis{3}{000}&=&\Fbarkreis{000}\;,\nonumber\\
\Fkreis{3}{100}&=&\Fbarkreis{100}\;,\nonumber\\
\Fkreis{3}{010}&=&\Fbarkreis{010}+\frac{\Fbarkreis{000}}{T}\;,\nonumber\\
\Fkreis{3}{001}&=&\Fbarkreis{001}+\frac{\Fbarkreis{000}}{T}\;,\nonumber\\
\Fkreis{3}{200}&=&\Fbarkreis{200}\;,\nonumber\\
\Fkreis{3}{020}&=&\Fbarkreis{020}+\frac{\Fbarkreis{000}}{2T^2}+\frac{\Fbarkreis{010}}{T}\;,\nonumber\\
\Fkreis{3}{002}&=&\Fbarkreis{002}+\frac{\Fbarkreis{000}}{2T^2}+\frac{\Fbarkreis{001}}{T}\;,\nonumber\\
\Fkreis{3}{110}&=&\Fbarkreis{110}+\frac{\Fbarkreis{100}}{T}\;,\nonumber\\
\Fkreis{3}{101}&=&\Fbarkreis{101}+\frac{\Fbarkreis{100}}{T}\;,\nonumber\\
\Fkreis{3}{011}&=&\Fbarkreis{011}+\frac{1}{T}\left(\Fbarkreis{010}+\Fbarkreis{001}\right)+\frac{\Fbarkreis{000}}{T^2}\;.
\end{eqnarray}

After this extensive formal expansion, it is helpful to explicitly write down the series expansion. 
Starting from the definition in Eq.~(\ref{Ftau1tau2tau3definition}), and using the properties~(\ref{eq:F3bardefintion}) and~(\ref{Fkreis4and5}), we 
analyzed the function
\begin{eqnarray}\label{Ftau1tau2tau2formaldef}
    F(\tau_1,\tau_2,\tau_3)&=&\Fkrei{1}(\tau_1,\tau_2,\tau_3)\nonumber\\
    &&+\Fkrei{2}(\tau_1,\tau_2,\tau_3)+\Fkrei{2}(\tau_3,\tau_2,\tau_1)\nonumber\\
    &&+e^{(\tau_2+\tau_3)\beta B}\Fbarkrei(\tau_1,\tau_2,\tau_3)\nonumber\\
    &&+e^{(\tau_2+\tau_1)\beta B}\Fbarkrei(\tau_1,\tau_2,\tau_3)\nonumber\\
    &&+F^{\rm p}(\tau_1,\tau_2,\tau_3)\;.
\end{eqnarray}
The functions $\Fkrei{j}$, $j=1,2,3$, are expanded separately, see Eq.~(\ref{Fkreiformalexpansion})
and the following expressions.

As before, using \textsc{Mathematica}~\cite{Mathematica12}, we find the formal expansion for $B\ll1$,
\begin{eqnarray}\label{eq:Ftau1tau2tau3smallB}
F(\tau_1,\tau_2,\tau_3)&=&F_{000}+B\Big(F_{100}(\tau_1+\tau_3)+\tau_2 F_{010}\Big)\nonumber\\
&&+B^2\Big(
 F_{200}(\tau_1^2+\tau_3^2)+\tau_2^2 F_{020}\nonumber\\
&&+ F_{110}(\tau_1\tau_2+\tau_2\tau_3)+\tau_1\tau_3 F_{101}
\Big)\nonumber\\
&&+F^{\rm p}(\tau_1,\tau_2,\tau_3)+\mathcal{O}[B^3]\;,
\end{eqnarray}
where $F^{\rm p}(\tau_1,\tau_2,\tau_3)$ is the contribution of the poles, see Eq.~(\ref{Fpolecontribution}). 
Here, we used that $F_{100}=F_{001}$, $F_{200}=F_{002}$ and $F_{110}=F_{011}$ due to Eq.~(\ref{SymmetryOfFtau1tau2tau3}). The individual terms are as follows,
\begin{eqnarray}\label{FijkFormalExpansion}
    F_{000}&=&\Fkreis{1}{000}+2(\Fkreis{2}{000}+\Fbarkreis{000})\;,\nonumber\\
    F_{100}&=&\frac{\Fbarkreis{000}}{T}+\Fkreis{1}{100}+\Fkreis{2}{100}+\Fkreis{2}{001}+\Fbarkreis{100}+\Fbarkreis{001}\;,\nonumber\\
    F_{010}&=&\frac{2\Fbarkreis{000}}{T}+\Fkreis{1}{010}+2(\Fkreis{2}{010}+\Fbarkreis{010})\;,\nonumber\\
    F_{200}&=&\frac{\Fbarkreis{000}}{2 T^2}+\frac{\Fbarkreis{001}}{T}+\Fkreis{1}{200}+\Fkreis{2}{200}+\Fkreis{2}{002}\nonumber\\
    &&+\Fbarkreis{200}+\Fbarkreis{002}\;,\nonumber\\
    F_{020}&=&\frac{\Fbarkreis{000}}{T^2}+\frac{2\Fbarkreis{010}}{T}+\Fkreis{1}{020}+2(\Fkreis{2}{020}+\Fbarkreis{020})\;,\nonumber\\
    F_{110}&=&\frac{\Fbarkreis{000}}{T^2}+\frac{1}{T}\left(\Fbarkreis{100}+\Fbarkreis{010}+\Fbarkreis{001}\right)\nonumber\\
    &&+\Fkreis{1}{110}+\Fkreis{2}{110}+\Fkreis{2}{011}+\Fbarkreis{110}+\Fbarkreis{011}\;,\nonumber\\
    F_{101}&=&\frac{2\Fbarkreis{100}}{T}+\Fkreis{1}{101}+2(\Fkreis{2}{101}+\Fbarkreis{101})\;.
\end{eqnarray}
Note that
\begin{equation}
    e^{(\tau_1+\tau_2)B/T}=1+\frac{2B}{T}+\frac{2B^2}{T^2}+\mathcal{O}[B^3]\;,
\end{equation}
see Eq.~(\ref{Ftau1tau2tau2formaldef}).
Now, with Eq.~(\ref{xbftau1tau2tau3connection}) and the help of \textsc{Mathematica}~\cite{Mathematica12}, we finally find
\begin{eqnarray}\label{xBintermsofFkreisfunctions}
x_B&=&x_{B}^{\rm p}+3 F_{000}-\frac{B^2}{T}\left(
2 F_{100}+3 F_{010}\right)\nonumber\\
&&+B^2\left( 6 F_{200}+3 F_{020}+2 F_{110}-F_{101}
\right)+\mathcal{O}[B^3]\;.\nonumber\\
\label{eq:xBsmallB}
\end{eqnarray}
The remaining task is to determine $\Fkrei{j}(\tau_1,\tau_2,\tau_3)$, $j=1,2,3$, for $0 < B\ll T\ll 1$.

\subsection{Derivation of the functions \texorpdfstring{$\mathbf{\Fkreis{j}{lmn}}$}{Fkreisjlmn}}
\label{chapter:derivation-of-the-Ftau1tau2tau3functions}

In this subsection we list the formal expressions of the functions $\Fkreis{j}{lmn}$, $j=1,2,3$, 
for small magnetic fields at low temperatures, $0< B \ll T\ll1$. 

\subsubsection{Derivation of the functions \texorpdfstring{$\Fkreis{1}{lmn}$}{Fkreis1lmn}}
\label{chapter:derivation-of-the-Fonetau1tau2tau3functions}

We start from Eq.~(\ref{eq:Fkreis-introduction}) and~(\ref{eq:ghl1-Definitions}) ($\beta=1/T$, $\tau_1=\pm1$, $\tau_2=\pm1$, $\tau_3=\pm1$)
\begin{eqnarray}
\Fkrei{1}(\tau_1,\tau_2,\tau_3)&=&-\int_{-1}^1\rmd\omega\,\nonumber\\
&&\quad \times\gkrei{1}(\omega,\tau_2 B)\hkrei{1}(\omega,\tau_1 B)\lkrei{1}(\omega,\tau_3 B)\;.\nonumber\\
\end{eqnarray}
With Eq.~(\ref{eq:HTFunctionsForSmallB}) and~(\ref{eq:HTFunctionsForSmallB2}) we get for $B\ll1$
\begin{eqnarray}\label{eq:Fkreisonefunctions}
\Fkreis{1}{000}&=&-\int_{-1}^1\rmd\omega\,\frac{\rhotilde(\omega)}{1+e^{\beta\omega}}\left(H_T^{(0)}(\omega)\right)^2\;,\nonumber\\
\Fkreis{1}{100}&=&-\int_{-1}^{1}\rmd\omega\,\frac{\rhotilde(\omega)}{1+e^{\beta\omega}}H_T^{(0)}(\omega)H_T^{(1)}(\omega)\;,\nonumber\\
\Fkreis{1}{010}&=&-\int_{-1}^{1}\rmd\omega\,\frac{\rhotilde(\omega)\beta e^{\beta\omega}}{(1+e^{\beta\omega})^2}\left(H_T^{(0)}(\omega)\right)^2\;,\nonumber\\
\Fkreis{1}{001}&\equiv& \Fkreis{1}{100}\;,\nonumber\\
\Fkreis{1}{200}&=&-\int_{-1}^{1}\rmd\omega\,\frac{\rhotilde(\omega)}{1+e^{\beta\omega}}H_T^{(0)}(\omega)H_T^{(2)}(\omega)\;,\nonumber\\
\Fkreis{1}{020}&=&-\beta^2\int_{-1}^{1}\rmd\omega\,\rhotilde(\omega)\frac{e^{\beta\omega}\left(e^{\beta\omega}-1\right)}{2\left(e^{\beta\omega}+1\right)^3}\left(H_T^{(0)}(\omega)\right)^2\;,\nonumber\\
\Fkreis{1}{002}&\equiv&\Fkreis{1}{200}\;,\nonumber\\
\Fkreis{1}{110}&=&-\beta\int_{-1}^{1}\rmd\omega\,\frac{\rhotilde(\omega)e^{\beta\omega}}{\left(1+e^{\beta\omega}\right)^2}H_T^{(0)}(\omega)H_T^{(1)}(\omega)\;,\nonumber\\
\Fkreis{1}{101}&=&-\int_{-1}^{1}\rmd\omega\,\frac{\rhotilde(\omega)}{1+e^{\beta\omega}}\left(H_T^{(1)}(\omega)\right)^2\;,\nonumber\\
\Fkreis{1}{011}&\equiv&\Fkreis{1}{110}\;.
\end{eqnarray}
Here, the functions $H_T^{(i)}$ are given by
\begin{eqnarray}
H_T^{(0)}(\omega)&=&-\ln(1-\omega)+\ln(2\pi T)+\Rbar(\omega/T)\;,\nonumber\\
H_T^{(1)}(\omega)&=&\frac{1}{T}\Rbar^\prime(\omega/T)\;,\nonumber\\
H_T^{(2)}(\omega)&=&\frac{1}{2T^2}\Rbar^{\prime\prime}(\omega/T)\;,
\end{eqnarray}
see Sect.~\ref{chapter:properties-of-the-Rbar-function}.

\subsubsection{Derivation of the functions \texorpdfstring{$\Fkreis{2}{lmn}$}{Fkreis1lmn}}
\label{chapter:derivation-of-the-Ftwotau1tau2tau3functions}

We start from Eq.~(\ref{eq:Fkreistwo-introduction}) and~(\ref{eq:ghl2-Definitions})
\begin{eqnarray}
\Fkrei{2}(\tau_1,\tau_2,\tau_3)&=&-T\int_{-1}^1\rmd\omega\,\nonumber\\
&&\times\gkrei{2}(\omega,\tau_3 B)\hkrei{2}(\omega,\tau_1 B)(-\lkrei{2}(\omega,\tau_2 B)) \nonumber\\
\end{eqnarray}
and use
\begin{eqnarray}
(H_T^\prime)^{(0)}(\omega)&=&\frac{1}{1-\omega}+\frac{1}{T}\Rbar^\prime(\omega/T)\;,\nonumber\\
(H_T^\prime)^{(1)}(\omega)&=&\frac{1}{T^2}\Rbar^{\prime\prime}(\omega/T)\;,\nonumber\\
(H_T^{\prime})^{(2)}(\omega)&=&\frac{1}{2T^3}\Rbar^{\prime\prime\prime}(\omega/T)\;.
\end{eqnarray}
With the same procedure as in Sect.~\ref{chapter:derivation-of-the-Fonetau1tau2tau3functions} we find for $B\ll1$
\begin{eqnarray}\label{eq:Fkreistwofunctions1}
\Fkreis{2}{100}&=&T\int_{-1}^{1}\rmd\omega\,\frac{\rhotilde(\omega)}{1+e^{\beta\omega}}H_T^{(1)}(-\omega)(H_T^\prime)^{(0)}(\omega)\;,\nonumber\\
\Fkreis{2}{010}&=&T\int_{-1}^{1}\rmd\omega\,\frac{\rhotilde(\omega)}{1+e^{\beta\omega}}H_T^{(0)}(-\omega)(H_T^\prime)^{(1)}(\omega)\;,\nonumber\\
\Fkreis{2}{001}&=&T\beta\int_{-1}^{1}\rmd\omega\,\frac{\rhotilde(\omega)e^{\beta\omega}}{\left(1+e^{\beta\omega}\right)^2}H_T^{(0)}(-\omega)(H_T^\prime)^{(0)}(\omega)\;,\nonumber\\
\Fkreis{2}{200}&=&T\int_{-1}^{1}\rmd\omega\,\frac{\rhotilde(\omega)}{1+e^{\beta\omega}}H_T^{(2)}(-\omega)(H_T^\prime)^{(0)}(\omega)\;,\nonumber\\
\Fkreis{2}{020}&=&T\int_{-1}^{1}\rmd\omega\,\frac{\rhotilde(\omega)}{1+e^{\beta\omega}}H_T^{(0)}(-\omega)(H_T^\prime)^{(2)}(\omega)\;,\nonumber\\
\Fkreis{2}{002}&=&T\beta^2\int_{-1}^{1}\rmd\omega\,\rhotilde(\omega)\frac{e^{\beta\omega}\left(e^{\beta\omega}-1\right)}{2\left(1+e^{\beta\omega}\right)^3}\nonumber\\
&&\times H_T^{(0)}(-\omega)(H_T^\prime)^{(0)}(\omega)\;,\nonumber\\
\Fkreis{2}{110}&=&T\int_{-1}^{1}\rmd\omega\,\frac{\rhotilde(\omega)}{1+e^{\beta\omega}}H_T^{(1)}(-\omega)(H_T^\prime)^{(1)}(\omega)\;,\nonumber\\
\Fkreis{2}{101}&=&T\beta\int_{-1}^{1}\rmd\omega\,\frac{\rhotilde(\omega)e^{\beta\omega}}{\left(1+e^{\beta\omega}\right)^2}H_T^{(1)}(-\omega)(H_T^\prime)^{(0)}(\omega)\;,\nonumber\\
\Fkreis{2}{011}&=&T\beta\int_{-1}^{1}\rmd\omega\,\frac{\rhotilde(\omega)e^{\beta\omega}}{\left(1+e^{\beta\omega}\right)^2}H_T^{(0)}(-\omega)(H_T^\prime)^{(1)}(\omega)\;.\nonumber\\
\end{eqnarray}

\subsubsection{Derivation of the functions \texorpdfstring{$\Fbarkreis{lmn}$}{Fbar3kreislmn}}
\label{chapter:derivation-of-the-Fthreebartau1tau2tau3functions}

Starting from Eq.~(\ref{eq:Fkreisthree-introduction}) and~(\ref{eq:ghl3-Definitions}) we find
\begin{eqnarray}
\Fbarkrei(\tau_1,\tau_2,\tau_3)&=&T\int_{-1}^1\rmd\omega\,\gbarkrei(\omega,\tau_3 B)\nonumber\\
&&\quad\times\hbarkrei(\omega,\tau_1 B)(-\lbarkrei(\omega,\tau_2 B))\nonumber\; .\\
\end{eqnarray}
For $B\ll1$ we have
\begin{eqnarray}\label{eq:Fkreisthreebarfunctions}
\Fbarkreis{000}&=&-T\int_{-1}^1\rmd\omega\,\frac{\rhotilde(\omega)}{1+e^{\beta\omega}}H_T^{(0)}(\omega)(H_T^\prime)^{(0)}(\omega)\;,\nonumber\\
\Fbarkreis{100}&=&-T\int_{-1}^1\rmd\omega\,\frac{\rhotilde(\omega)}{1+e^{\beta\omega}}H_T^{(1)}(\omega)(H_T^\prime)^{(0)}(\omega)\;,\nonumber\\
\Fbarkreis{010}&=&T\int_{-1}^1\rmd\omega\,\frac{\rhotilde(\omega)}{1+e^{\beta\omega}}H_T^{(0)}(\omega)(H_T^\prime)^{(1)}(\omega)\;,\nonumber\\
\Fbarkreis{001}&=&T\beta\int_{-1}^1\rmd\omega\,\frac{\rhotilde(\omega)e^{\beta\omega}}{\left(1+e^{\beta\omega}\right)^2}H_T^{(0)}(\omega)(H_T^\prime)^{(0)}(\omega)\;,\nonumber\\
\Fbarkreis{200}&=&-T\int_{-1}^1\rmd\omega\,\frac{\rhotilde(\omega)}{1+e^{\beta\omega}}H_T^{(2)}(\omega)(H_T^\prime)^{(0)}(\omega)\;,\nonumber\\
\Fbarkreis{020}&=&-T\int_{-1}^1\rmd\omega\,\frac{\rhotilde(\omega)}{1+e^{\beta\omega}}H_T^{(0)}(\omega)(H_T^\prime)^{(2)}(\omega)\;,\nonumber\\
\end{eqnarray}
and
\begin{eqnarray}
\Fbarkreis{002}&=&-T\beta^2\int_{-1}^1\rmd\omega\,\rhotilde(\omega)\frac{e^{\beta\omega}\left(e^{\beta\omega}-1\right)}{2\left(1+e^{\beta\omega}\right)^3}\nonumber\\
&&\quad\times H_T^{(0)}(\omega)(H_T^\prime)^{(0)}(\omega)\;,\nonumber\\
\Fbarkreis{110}&=&T\int_{-1}^1\rmd\omega\,\frac{\rhotilde(\omega)}{1+e^{\beta\omega}}H_T^{(1)}(\omega)(H_T^\prime)^{(1)}(\omega)\;,\nonumber\\
\Fbarkreis{101}&=&T\beta\int_{-1}^1\rmd\omega\,\frac{\rhotilde(\omega)e^{\beta\omega}}{\left(1+e^{\beta\omega}\right)^2}H_T^{(1)}(\omega)(H_T^\prime)^{(0)}(\omega)\;,\nonumber\\
\Fbarkreis{011}&=&-T\beta\int_{-1}^1\rmd\omega\,\frac{\rhotilde(\omega)e^{\beta\omega}}{\left(1+e^{\beta\omega}\right)^2}H_T^{(0)}(\omega)(H_T^\prime)^{(1)}(\omega)\;.\nonumber\\
\end{eqnarray}

Finally, for the leading order, i.e., for the first line in Eq.~(\ref{FijkFormalExpansion}),
we need to address
\begin{eqnarray}\label{eq:Fkreistwo000plusFkreisthreebar000function}
\Fkreis{2}{000}+\Fkreis{3}{000}&=&T\int_{-1}^1\rmd\omega\,\frac{\rhotilde(\omega)}{1+e^{\beta\omega}}\nonumber\\
&&\times\left(H_T^{(0)}(-\omega)-H_T^{(0)}(\omega)\right)\left(H_T^{(0)}(\omega)\right)^\prime\;.\nonumber\\
\end{eqnarray}
Note that $\Fkreis{3}{000}=\Fbarkreis{000}$.
For convenience, we list again the functions $H_T^{(i)}$ and $(H_T^\prime)^{(i)}$,
\begin{eqnarray}
H_T^{(0)}(\omega)&=&-\ln(1-\omega)+\ln(2\pi T)+\Rbar(\omega/T)\;,\nonumber\\
H_T^{(1)}(\omega)&=&\frac{1}{T}\Rbar^\prime(\omega/T)\;,\nonumber\\
H_T^{(2)}(\omega)&=&\frac{1}{2T^2}\Rbar^{\prime\prime}(\omega/T)\;,\nonumber\\
(H_T^\prime)^{(0)}(\omega)&=&\frac{1}{1-\omega}+\frac{1}{T}\Rbar^\prime(\omega/T)\;,\nonumber\\
(H_T^\prime)^{(1)}(\omega)&=&\frac{1}{T^2}\Rbar^{\prime\prime}(\omega/T)\;,\nonumber\\
(H_T^{\prime})^{(2)}(\omega)&=&\frac{1}{2T^3}\Rbar^{\prime\prime\prime}(\omega/T) \; ,
\end{eqnarray}
expressed in terms of elementary functions and the function $\Rbar(x)$ and its derivatives.

\subsection{Calculation of the functions \texorpdfstring{$\mathbf{\Fkreis{j}{\boldsymbol{lmn} }}$}{Fkreisjlmn}}
\label{Fkreisjlmnderivations}

In this subsection we evaluate the functions $\Fkreis{1}{lmn}$, $\Fkreis{2}{lmn}$ and $\Fbarkreis{lmn}$ for small magnetic fields $B\ll1$ at low temperatures, $T\ll1 $. All occurring integrals are numerically verified and were calculated using \textsc{Mathematica}~\cite{Mathematica12}. 

\subsubsection{General procedure and useful relations}
\label{x3BGeneralProcedure}
The general procedure is as follows. We take the expressions $\Fkreis{j}{lmn}$ from Sect.~\ref{chapter:derivation-of-the-Fonetau1tau2tau3functions}, 
Sect.~\ref{chapter:derivation-of-the-Ftwotau1tau2tau3functions}, and Sect.~\ref{chapter:derivation-of-the-Fthreebartau1tau2tau3functions},
and simply insert the corresponding orders from the series expansion of the function $H_T(\omega,\tau B)$. 
Next, we substitute $x=\beta\omega=\omega/T$,
\begin{equation}
\int_{-1}^{1}\rmd\omega\,f(\beta\omega)= T\int_{-1/T}^{1/T}\rmd x\,f(x)\;.
\end{equation} 
In the following, we use one of the relations from Sect.~\ref{chapter:GeneralRelations} if not stated explicitly otherwise. 
In this way, the expressions reduce to explicitly calculable integrals ($\alpha_j$) and to those integrals in which the function $\Rbar(x)$ 
or its derivatives occur ($\mathcal{I}_i$). 
The integrals $\alpha_{j}$ and $\mathcal{I}_i$ are listed in Sect.~\ref{Integraltables}. 
The derivations of those integrals can be found in Sect.~\ref{chapter:derivation-of-the-integrals}.
Recall that the properties of the $\Rbar(x)$ function can be found in Sect.~\ref{chapter:properties-of-the-Rbar-function}.

\subsubsection{Evaluation of the function \texorpdfstring{$\Fkreis{1}{000}$}{Fkreis1000}}
Using the property of Fermi functions~(\ref{GIRFermi})
we have from Eq.~(\ref{eq:Fkreisonefunctions}) that
\begin{eqnarray}
\Fkreis{1}{000}&=&-\int_{-1}^1\rmd\omega\,\frac{1}{1+e^{\beta\omega}}\left(H_T^{(0)}(\omega)\right)^2\\
&=&-\int_{-1}^1\rmd\omega\,\frac{1}{1+e^{\beta\omega}}\nonumber\\
&&\times\left(-\ln(1-\omega)+\ln(2\pi T)+\Rbar(\omega/T)\right)^2\;,\nonumber\\
&=&\Akreis{1}{000}+\Bkreis{1}{000}
\end{eqnarray}
with
\begin{eqnarray}
\Akreis{1}{000}&=&-T\int_{0}^{1/T}\rmd x\,\left(\ln(2\pi T)-\ln(1+T x)+\Rbar(x)\right)^2\;,\nonumber\\
\Bkreis{1}{000}&=&-T\int_0^{1/T}\frac{\rmd x}{1+e^x}\nonumber\\
&&\quad\times\Bigl[\Big(\ln(2\pi T)-\ln(1-x T)+\Rbar(x)\Big)^2\nonumber\\
&&\quad\qquad-\Big(\ln(2\pi T)-\ln(1+x T)+\Rbar(x)\Big)^2\Bigr]\;.\nonumber \\
\end{eqnarray}
Note, in $\Akreis{1}{000}$ we have substituted $x\to -x$. 
We further write
\begin{eqnarray}
\Akreis{1}{000}&=&\Ckreis{1}{000}+\Dkreis{1}{000}+\Ekreis{1}{000}
\end{eqnarray}
with 
\begin{eqnarray}
\Ckreis{1}{000}&=&-T\alpha_1\;,\nonumber\\
\Dkreis{1}{000}&=&-2 T\left(\ln(2\pi T)\mathcal{I}_1-\mathcal{I}_2\right)\nonumber\;,\\
\Ekreis{1}{000}&=&-T \mathcal{I}_3\;.
\end{eqnarray}
This leads to
\begin{eqnarray}
\Akreis{1}{000}&=&-\frac{\pi^2}{6}-2\ln^2(2)+\frac{\pi^2}{2}T+\frac{\pi^2}{3}T^2\ln(2\pi T)\nonumber\\
&&+\pi^2 T^2\left(-\frac{1}{3}-4\ln \mathcal{A}+\ln 2\right)\;.
\end{eqnarray}

Using $\ln^2(1-Tx)-\ln^2(1+Tx)\sim \mathcal{O}[T^3 x^3]$ and $\ln(1+T x)-\ln(1-T x)\sim 2 T x$, we have
\begin{eqnarray}
\Bkreis{1}{000}&\approx&-4 T^2\left(\ln(2\pi T) \alpha_2+\mathcal{I}_4\right)\nonumber\\
&=&-\frac{\pi^2 T^2}{3}\left(1+\ln 2-12\ln \mathcal{A}  +\ln(2\pi T)\right)\;,\nonumber\\
\end{eqnarray}
where $\mathcal{A}\approx 1.28243$ is Glaisher's constant. Thus, finally, we arrive at
\begin{eqnarray}
\Fkreis{1}{000}&=&-\frac{\pi^2}{6}-2\ln^2(2)+\frac{\pi^2}{2}T\nonumber\\
&&+\frac{2\pi^2}{3}T^2\big(\ln 2 -1\big)+\mathcal{O}[T^3,T^3\ln T]\;.\nonumber\\
\end{eqnarray}

\subsubsection{Evaluation of the function \texorpdfstring{$\Fkreis{1}{100}$}{Fkreis1100}}
We start from Eq.~(\ref{eq:Fkreisonefunctions}) and use Eq.~(\ref{GIRFermi}) to find
\begin{eqnarray}
\Fkreis{1}{100}&=&-\int_{-1}^1\rmd\omega\,\frac{1}{1+e^{\beta\omega}}H_T^{(0)}(\omega)H_T^{(1)}(\omega)\nonumber\\
&=&-\int_{-1/T}^{1/T}\rmd x\,\frac{\Rbar^\prime(x)}{1+e^x}\nonumber\\
&&\quad\times\left(\ln(2\pi T)-\ln(1-T x)+\Rbar(x)\right)\nonumber\\
&=&\Akreis{1}{100}+\Bkreis{1}{100}+\mathcal{O}[T^2]
\end{eqnarray}
with 
\begin{eqnarray}
\Akreis{1}{100}&=&-\int_{-1/T}^0\rmd x\,\nonumber\\
&&\times \left(\ln(2\pi T)-\ln(1-T x)+\Rbar(x)\right)\Rbar^\prime(x)\nonumber\\
&=&-\ln(2\pi T) \mathcal{I}_5+\mathcal{I}_6-\mathcal{I}_7\nonumber\\
&=&\frac{1}{2}\bigg(
-\frac{\pi^2}{6}-(\EulerGamma+\ln 2)(\EulerGamma+3\ln 2)\nonumber\\
&&+\ln(\pi)\ln(4\pi)+2\ln(2\pi T)\left(\ln 2+\EulerGamma-\ln \pi \right)\nonumber\\
&&-\ln^2(T)
\bigg)-\frac{\pi^2 T^2}{6}\left(\ln 2 -12\ln \mathcal{A}+\ln(2\pi T)\right)\nonumber\\
\end{eqnarray}
and
\begin{eqnarray}
\Bkreis{1}{100}&=&-2\int_0^{1/T}\frac{\rmd x}{1+e^x}\Rbar^\prime(x)\left(\ln(2\pi T)+\Rbar(x)\right)\nonumber\\
&=&-2\left(\ln(2\pi T)\mathcal{I}_8+\mathcal{I}_9\right)\nonumber\\
&=&-1-\EulerGamma+\frac{\pi^2}{24}+2\ln 2\left(\EulerGamma+\ln 2 \right)\nonumber\\
&&+\ln(2\pi T)(1-2\ln 2)\;.
\end{eqnarray}

Thus, we arrive at
\begin{eqnarray}\label{Fkreis1-100}
\Fkreis{1}{100}&=&-1-\frac{\EulerGamma}{2}\left(2+\EulerGamma\right)-\frac{\pi^2}{24}+\left(1+\EulerGamma\right)\ln(2\pi T)\nonumber\\
&&-\frac{1}{2}\ln^2(2\pi T)+\mathcal{O}\left[T^2,T^2\ln T\right]\;.
\end{eqnarray}

\subsubsection{Evaluation of the function \texorpdfstring{$\Fkreis{1}{010}$}{Fkreis1010}}

We start from Eq.~(\ref{eq:Fkreisonefunctions}),
\begin{eqnarray}
\Fkreis{1}{010}&=&-\int_{-1}^1\rmd\omega\,\frac{\beta e^{\beta\omega}}{\left(1+e^{\beta\omega}\right)^2}\left(H_T^{(0)}(\omega)\right)^2\nonumber\\
&=&\frac{1}{1+e^{\beta\omega}}\left(H_T^{(0)}(\omega)\right)^2\bigg|_{-1}^1\nonumber\\
&&-2\int_{-1}^1\rmd\omega\,\frac{1}{1+e^{\beta\omega}}H_T^{(0)}(\omega)\left(H_T^{(0)}(\omega)\right)^\prime\;,\nonumber\\
\end{eqnarray}
where we integrated by parts. We single out $\Fkreis{1}{100}$, see Eq.~(\ref{Fkreis1-100}), and use the property of Fermi functions~(\ref{GIRFermi}),
\begin{eqnarray}
\Fkreis{1}{010}&=&-\ln^2(2)-2\int_{-1}^1\rmd\omega\,\nonumber\\
&&\times \frac{1}{1+e^{\beta\omega}}H_T^{(0)}(\omega)\left(\frac{1}{1-\omega}+\frac{1}{T}\Rbar^\prime(\omega/T)\right)\nonumber\\
&=&-\ln^2(2)+2\Fkreis{1}{100}\nonumber\\
&&-2\int_{-1}^1\rmd\omega\,\frac{1}{1+e^{\beta\omega}}H_T^{(0)}(\omega)\frac{1}{1-\omega}\nonumber\\
&=&-\ln^2(2)+2\Fkreis{1}{100}-2\left(\Akreis{1}{010}+\Bkreis{1}{010}\right)
\end{eqnarray}
with
\begin{eqnarray}
\Akreis{1}{010}&=&\int_{-1}^0\rmd \omega\, \frac{H_T^{(0)}(\omega)}{1-\omega}\;,\nonumber\\
\Bkreis{1}{010}&=&\int_0^1\rmd\omega\,\frac{1}{1+e^{\beta\omega}}\left(\frac{H_T^{(0)}(\omega)}{1-\omega}-\frac{H_T^{(0)}(-\omega)}{1+\omega}\right)\;.\nonumber\\
\end{eqnarray}
We find
\begin{eqnarray}
\Akreis{1}{010}&=&T\alpha_3+T \mathcal{I}_{10}\nonumber\\
&=&-\frac{\pi^2}{12}-\frac{\ln^2(2)}{2}-\ln 2\ln T+\frac{\ln 2 }{2}\ln(T^2)\nonumber\\
&&+\mathcal{O}\left(T^2\ln T \right)
\end{eqnarray}
and
\begin{eqnarray}
\Bkreis{1}{010}&=&\mathcal{I}_{11}=\mathcal{O}[T^2\ln T]\;.
\end{eqnarray}
Finally we obtain
\begin{equation}
\Fkreis{1}{010}=2\Fkreis{1}{100}+\frac{\pi^2}{6}+\mathcal{O}[T^2,T^2\ln T]\;.
\end{equation}

\subsubsection{Evaluation of the function \texorpdfstring{$\Fkreis{1}{001}$}{Fkreis1001}}
Due to symmetry, we have
\begin{equation}
\Fkreis{1}{001}=\Fkreis{1}{100}\;.
\end{equation}

\subsubsection{Evaluation of the function \texorpdfstring{$\Fkreis{1}{200}$}{Fkreis1200}}
With Eq.~(\ref{eq:Fkreisonefunctions}), we have
\begin{eqnarray}
\Fkreis{1}{200}&=&-\int_{-1}^1\rmd\omega\,\frac{1}{1+e^{\beta\omega}}H_T^{(0)}(\omega) H_T^{(2)}(\omega)\nonumber\\
&=&-\frac{1}{2T^2}\int_{-1}^{1}\rmd \omega\,\frac{\Rbar^{\prime\prime}(\omega/T)}{1+e^{\omega/T}}\nonumber\\
&&\quad\times\left(\ln(2\pi T)-\ln(1-\omega)+\Rbar(\omega/T)\right)\nonumber\\
&=&\Akreis{1}{200}+\Bkreis{1}{200}
\end{eqnarray}
with 
\begin{eqnarray}
\Akreis{1}{200}&=&-\int_{-1}^0\rmd\omega\,H_T^{(0)}(\omega)H_T^{(2)}(\omega)\;,\nonumber\\
\Bkreis{1}{200}&=&-\int_0^1\rmd\omega\,\frac{1}{1+e^{\beta\omega}}\nonumber\\
&&\quad \times\left(H_T^{(0)}(\omega)H_T^{(2)}(\omega)-H_T^{(0)}(-\omega)H_T^{(2)}(-\omega)\right)\;.\nonumber\\
\end{eqnarray}
We find
\begin{eqnarray}
\Akreis{1}{200}&=&-\frac{1}{2T}\left(\ln(2\pi T)\mathcal{I}_{12}-\mathcal{I}_{13}+\mathcal{I}_{14}\right)\nonumber\\
&=&\frac{\pi^2}{24 T}+\frac{1}{2}\left(\ln(2\pi T)-1-\EulerGamma\right)+\mathcal{O}[T^2\ln T]\nonumber\\
\end{eqnarray}
and
\begin{eqnarray}\label{eq:I15}
\Bkreis{1}{200}\approx&\mathcal{I}_{15}=\ln 2-\frac{3}{4}\;.
\end{eqnarray}
Finally,
\begin{eqnarray}
\Fkreis{1}{200}&=&-\frac{5}{4}-\frac{\EulerGamma}{2}+\ln 2+\frac{\pi^2}{24 T}+\frac{1}{2}\ln(2\pi T)\nonumber\\
&&+\mathcal{O}[T^2,T^2\ln T]\;.
\end{eqnarray}

\subsubsection{Evaluation of the function \texorpdfstring{$\Fkreis{1}{020}$}{Fkreis1020}}
We start from Eq.~(\ref{eq:Fkreisonefunctions}),
\begin{eqnarray}
\Fkreis{1}{020}&=&-\frac{\beta^2}{2}\int_{-1}^1\rmd\omega\,f(\omega)\left(H_T^{(0)}(\omega)\right)^2\;,\nonumber\\
f(\omega)&=&\frac{e^{\beta\omega}\left(e^{\beta\omega}-1\right)}{\left(1+e^{\beta\omega}\right)^3}=\frac{1}{4}\frac{\tanh(\beta\omega/2)}{\cosh^2(\beta\omega/2)}=-f(-\omega)\;.\nonumber\\
\end{eqnarray}
Since $f(\omega)$ is odd, we can use Eq.~(\ref{GIROdd}) to find
\begin{eqnarray}
\Fkreis{1}{020}&=&-\frac{1}{2T}\int_{0}^{1/T}\rmd x\, \frac{e^{x}\left(e^{x}-1\right)}{\left(1+e^{x}\right)^3}\nonumber\\
&&\quad\times\Bigl[\left(-\ln(1-T x)+\ln(2\pi T)+\Rbar(x)\right)^2\nonumber\\
&&\qquad-\left(-\ln(1+T x)+\ln(2\pi T)+\Rbar(x)\right)^2\Bigr]\;.\nonumber\\
\end{eqnarray}
At low temperatures, $T\ll1$, we further use 
\begin{eqnarray}
&&(H_T^{(0)}(x T))^2-(H_T^{(0)}(-xT))^2\nonumber\\
&&\quad\approx 4xT\left(\ln(2\pi T)+\Rbar(x)\right)+\mathcal{O}[T^2]
\end{eqnarray}
to find
\begin{eqnarray}
\Fkreis{1}{020}&=&-2\int_0^\infty\rmd x\,\frac{e^{x}\left(e^{x}-1\right)}{\left(1+e^{x}\right)^3}x\left(\ln(2\pi T)+\Rbar(x)\right)\nonumber\;,\\
&=&-2\left(\ln(2\pi T)\alpha_7+{I}_{30}\right)\;.
\end{eqnarray}
Thus,
\begin{equation}
\Fkreis{1}{020}=\frac{1}{2}+\EulerGamma-\ln(2\pi T)+\mathcal{O}[T^2,T^2\ln T]\;.
\end{equation}

\subsubsection{Evaluation of the function \texorpdfstring{$\Fkreis{1}{002}$}{Fkreis1002}}

Due to the symmetry, we have
\begin{equation}
\Fkreis{1}{002}=\Fkreis{1}{200}\;.
\end{equation}

\subsubsection{Evaluation of the function \texorpdfstring{$\Fkreis{1}{110}$}{Fkreis1110}}
We start from Eq.~(\ref{eq:Fkreisonefunctions}) and Eq.~(\ref{GIROdd}), since the exponential prefactor is even and $H_T^{(1)}$ is odd,
\begin{eqnarray}
\Fkreis{1}{110}&=&-\int_{-1}^1\rmd\omega\,\frac{\beta e^{\beta\omega}}{\left(1+e^{\beta\omega}\right)^2}H_T^{(0)}(\omega)H_T^{(1)}(\omega)\nonumber\\
&=&-\int_0^1\rmd\omega\,\frac{\beta e^{\beta\omega}}{\left(1+e^{\beta\omega}\right)^2}\nonumber\\
&&\quad \times\left(H_T^{(0)}(\omega)-H_T^{(0)}(-\omega)\right)\frac{\Rbar^\prime(\omega/T)}{T}\nonumber\\
&=&-\int_{0}^{1/T}\rmd x\,\frac{e^x}{\left(1+e^x\right)^2}\frac{\Rbar^\prime(x)}{T}\nonumber\\
&&\quad\times\left(-\ln(1- Tx)+\ln(1+T x)\right)\;.
\end{eqnarray}
At low temperatures $-\ln(1- Tx)+\ln(1+T x)\approx 2 x T$ and therefore
\begin{eqnarray}
\Fkreis{1}{110}&=&-2 \mathcal{I}_{17}=-\frac{1}{2}+\mathcal{O}[T^2,T^2\ln T]\;.
\end{eqnarray}

\subsubsection{Evaluation of the function \texorpdfstring{$\Fkreis{1}{101}$}{Fkreis1101}}
We start with Eq.~(\ref{eq:Fkreisonefunctions}) and use Eq.~(\ref{GIRFermi}) to find
\begin{eqnarray}
\Fkreis{1}{101}&=&-\int_{-1}^1\rmd\omega\,\frac{1}{1+e^{\beta\omega}}\left(H_T^{(1)}(\omega)\right)^2\nonumber\\
&=&-\int_{-1}^0\rmd\omega\,\left(H_T^{(1)}(\omega)\right)^2-\int_0^1\frac{\rmd \omega}{1+e^{\beta\omega}}\nonumber\\
&&\times\left(\left(H_T^{(1)}(\omega)\right)^2-\left(H_T^{(1)}(-\omega)\right)^2\right)\nonumber\\
&=&-\frac{1}{T}\mathcal{I}_{18}
\end{eqnarray}
because $\left[H_T^{(1)}(\omega)\right]^2-\left[H_T^{(1)}(-\omega)\right]^2=0$. We thus find
\begin{equation}
\Fkreis{1}{101}=1-\frac{\pi^2}{12 T}+\mathcal{O}[T^2,T^2\ln T]\;.
\end{equation}

\subsubsection{Evaluation of the function \texorpdfstring{$\Fkreis{1}{011}$}{Fkreis1011}}
Due to symmetry,
\begin{equation}
\Fkreis{1}{011}=\Fkreis{1}{110}\;.
\end{equation}

\subsubsection{Evaluation of the function \texorpdfstring{$\Fkreis{2}{100}$}{Fkreis2100}}
We start from Eq.~(\ref{eq:Fkreistwofunctions1}). Using the property of Fermi functions Eq.~(\ref{GIRFermi}), we evaluate
\begin{eqnarray}
\Fkreis{2}{100}&=&T\int_{-1}^{1}\rmd\omega\,\frac{\rhotilde(\omega)}{1+e^{\beta\omega}}H_T^{(1)}(-\omega)(H_T^\prime)^{(0)}(\omega)\nonumber\\
&=&-\Akreis{2}{100}-\Bkreis{2}{100}
\end{eqnarray}
with 
\begin{eqnarray}
\Akreis{2}{100}&=&\int_{-1/T}^0\rmd x\,\left(T\frac{\Rbar^\prime(x)}{1-T x}
+ \left(\Rbar^\prime(x)\right)^2\right)\;,\nonumber\\
\Bkreis{2}{100}&=&T\int_0^{1/T}\rmd x\,\frac{\Rbar^\prime(x)}{1+ e^x}\left(\frac{1}{1- Tx}+\frac{1}{1+Tx}\right)\nonumber\\
&\approx& 2T\int_0^\infty\rmd x\,\frac{\Rbar^\prime(x)}{1+e^x}
\end{eqnarray}
since $(1-Tx)^{-1}+(1+Tx)^{-1}\approx 2+\mathcal{O}[(xT)^2]$ at low temperatures, $T\ll1 $. We find 
\begin{eqnarray}
\Akreis{2}{100}&=&T \mathcal{I}_{19}+\mathcal{I}_{18}\nonumber\\
&=&\frac{\pi^2}{12}+T\left(-1-\EulerGamma+\ln(\pi T)\right)+\mathcal{O}[T^3\ln T]\nonumber\\
\end{eqnarray}
and 
\begin{eqnarray}
\Bkreis{2}{100}&=&2 T \mathcal{I}_{8}\nonumber\\
&=&T\left(2\ln 2-1\right)+\mathcal{O}[T^3]\;.
\end{eqnarray}
Thus, we have
\begin{eqnarray}
\Fkreis{2}{100}&=&-\frac{\pi^2}{12}+T\left(2+\EulerGamma-\ln2-\ln(2\pi T)\right)\nonumber\\
&&+\mathcal{O}[T^3,T^3\ln T]\;.
\end{eqnarray}

\subsubsection{Evaluation of the function \texorpdfstring{$\Fkreis{2}{010}$}{Fkreis2010}}
We start from Eq.~(\ref{eq:Fkreistwofunctions1}) and use again the property of Fermi functions Eq.~(\ref{GIRFermi}) to find
\begin{eqnarray}
\Fkreis{2}{010}&=&T\int_{-1}^{1}\rmd\omega\,\frac{\rhotilde(\omega)}{1+e^{\beta\omega}}H_T^{(0)}(-\omega)(H_T^\prime)^{(1)}(\omega)\nonumber\\
&=&\int_{-1/T}^{1/T}\rmd x\,\left(\ln(2\pi T)-\ln(1+Tx)+\Rbar(x)\right)\Rbar^{\prime\prime}(x)\nonumber\\
&=&\Akreis{2}{010}+\Bkreis{2}{010}
\end{eqnarray}
with 
\begin{eqnarray}
\Akreis{2}{010}&=&\int_{-1/T}^0\rmd x\,\nonumber\\
&&\quad \times\left(\ln(2\pi T)-\ln(1+T x)+\Rbar^\prime(x)\right)\Rbar^{\prime\prime}(x)\;,\nonumber\\
\Bkreis{2}{010}&=&\int_0^{1/T}\rmd x\,\frac{\Rbar^{\prime\prime}(x)}{1+e^x}\left(-\ln(1+Tx)+\ln(1-T x)\right)\nonumber\\
&\approx& -2 T\int_0^\infty\rmd x\,\frac{x\Rbar^{\prime\prime}(x)}{1+e^x}\;.
\end{eqnarray}
We further find
\begin{eqnarray}
\Akreis{2}{010}&=&\ln(2\pi T)\mathcal{I}_{12}-\mathcal{I}_{20}+\mathcal{I}_{14}\nonumber\\
&=&-\frac{\pi^2}{12}+T\left(1-\EulerGamma-\ln 2\right)+T\ln(\pi T)\;,\nonumber\\
\Bkreis{2}{010}&=&-2 T\mathcal{I}_{15}\nonumber\\
&=&2T\left(\ln 2-\frac{3}{4}\right)\;.
\end{eqnarray}
Thus, we arrive at
\begin{eqnarray}
\Fkreis{2}{010}&=&-\frac{\pi^2}{12}-\frac{T}{2}\left(1+2\EulerGamma\right)+T\ln(2\pi T)\nonumber\\
&&+\mathcal{O}[T^3,T^3\ln T]\;.
\end{eqnarray}

\subsubsection{Evaluation of the function \texorpdfstring{$\Fkreis{2}{001}$}{Fkreis2001}}
We start from Eq.~(\ref{eq:Fkreistwofunctions1}). Since $e^x/(1+e^x)^2$ is an even function in~$x$, we find with Eq.~(\ref{GIREven})
\begin{eqnarray} 
\Fkreis{2}{001}&=&T\beta\int_{-1}^{1}\rmd\omega\,\frac{\rhotilde(\omega)e^{\beta\omega}}{\left(1+e^{\beta\omega}\right)^2}H_T^{(0)}(-\omega)(H_T^\prime)^{(0)}(\omega)\nonumber\\
&=&T\int_{-1/T}^{1/T}\rmd x\,\frac{e^x}{(1+e^x)^2}\nonumber\\
&&\quad \times\left(\ln(2\pi T)-\ln(1+T x)+\Rbar(x)\right)\nonumber \\
&&\quad \times\left(\frac{1}{1-Tx}+\frac{\Rbar^\prime(x)}{T}\right)
\end{eqnarray}
and thus
\begin{eqnarray}
\Fkreis{2}{001}&=&T\int_0^{1/T}\rmd x\,\frac{e^x}{(1+e^x)^2}\left(\frac{1}{1+T x}+\frac{1}{1-T x}\right)\nonumber\\
&&\quad \times\left(\ln(2\pi T)+\Rbar(x)\right)\nonumber\\
&&-T\int_0^{1/T}\rmd x\,\frac{e^x}{(1+e^x)^2}\nonumber\\
&&\quad \times\left(\frac{\ln(1-T x)}{1+T x}+\frac{\ln(1+T x)}{1-T x}\right)\nonumber\\
&&+\int_0^{1/T}\rmd x\,\frac{e^x\Rbar^\prime(x)}{(1+e^x)^2}\left(\ln(1-T x)-\ln(1+T x)\right)\nonumber\\
\end{eqnarray}
for all temperatures. At $T\ll 1$, we can approximate
\begin{eqnarray}
\frac{1}{1+Tx}+\frac{1}{1-Tx}&\approx & 2+\mathcal{O}[(xT)^2]\;,\nonumber\\
\frac{\ln(1-T x)}{1+T x}+\frac{\ln(1+T x)}{1-Tx}&\approx &x^2 T^2+\mathcal{O}[(xT)^3]\;,\nonumber\\
\ln(1-T x)-\ln(1+T x)&\approx &-2xT+\mathcal{O}[(xT)^3]\;.\nonumber\\
\end{eqnarray}
Thus, we find
\begin{eqnarray}
\Fkreis{2}{001}&=&2 T\left(\ln(2\pi T)\alpha_5+\mathcal{I}_{16}\right)-2T\mathcal{I}_{17}+\mathcal{O}[T^3]\;.\nonumber\\
\end{eqnarray}
Finally, we arrive at
\begin{eqnarray}
\Fkreis{2}{001}&=&-\frac{T}{2}\left(3+2\EulerGamma-2\ln(2\pi T)\right)+\mathcal{O}[T^3,T^3\ln T]\;.\nonumber\\
\end{eqnarray}

\subsubsection{Evaluation of the function \texorpdfstring{$\Fkreis{2}{200}$}{Fkreis2200}}
We start with Eq.~(\ref{eq:Fkreistwofunctions1}) and use again Eq.~(\ref{GIRFermi}),
\begin{eqnarray}
\Fkreis{2}{200}&=&T\int_{-1}^{1}\rmd\omega\,\frac{\rhotilde(\omega)}{1+e^{\beta\omega}}H_T^{(2)}(-\omega)(H_T^\prime)^{(0)}(\omega)\nonumber\\
&=&\Akreis{2}{200}+\Bkreis{2}{200}
\end{eqnarray}
with 
\begin{eqnarray}
\Akreis{2}{200}&=&\frac{1}{2}\int_{-1/T}^0\rmd x\,\Rbar^{\prime\prime}(x)\left(\frac{1}{1-T x}+\frac{\Rbar^\prime(x)}{T}\right)\nonumber\\
\end{eqnarray}
and
\begin{eqnarray}
\Bkreis{2}{200}&=&\frac{1}{2}\int_0^{1/T}\rmd x\,\frac{\Rbar^{\prime\prime}(x)}{1+e^x}\nonumber\\
&&\quad\times\left(\frac{1}{1-T x}-\frac{1}{1+Tx}+2\frac{\Rbar^\prime(x)}{T}\right)\;,\nonumber\\
\end{eqnarray}
Furthermore, we find
\begin{eqnarray}
\Akreis{2}{200}&=&\frac{1}{2}\left(\mathcal{I}_{21}+\frac{\mathcal{I}_{22}}{T}\right)\nonumber\\
&=&\frac{T}{4}\left(2\EulerGamma-1-2\ln(\pi T)\right)
\end{eqnarray} 
and with $(1-Tx)^{-1}-(1+T x)^{-1}\approx 2 x T$ at $T\ll1 $
\begin{eqnarray}
\Bkreis{2}{200}&=&\int_0^\infty\rmd x\,\frac{\Rbar^{\prime\prime}(x)}{1+e^x}\left(x T+\frac{\Rbar^\prime(x)}{T}\right)\nonumber\\
&=&T\mathcal{I}_{15}+\frac{\mathcal{I}_{23}}{T}\nonumber\\
&=&\frac{\pi^2}{360 T}+T\left(\frac{3}{4}-\ln 2\right)\;.
\end{eqnarray}
Finally, we have
\begin{eqnarray}
\Fkreis{2}{200}&=&\frac{\pi^2}{360T}+\frac{T}{2}\left(1+\EulerGamma-\ln 2-\ln(2\pi T)\right)\nonumber\\
&&+\mathcal{O}[T^3,T^3\ln T]\;.
\end{eqnarray}

\subsubsection{Evaluation of the function \texorpdfstring{$\Fkreis{2}{020}$}{Fkreis2020}}
We start from Eq.~(\ref{eq:Fkreistwofunctions1}). 
With the property of Fermi functions Eq.~(\ref{GIRFermi}) we have
\begin{eqnarray}
\Fkreis{2}{020}&=&T\int_{-1}^{1}\rmd\omega\,\frac{\rhotilde(\omega)}{1+e^{\beta\omega}}H_T^{(0)}(-\omega)(H_T^\prime)^{(2)}(\omega)\nonumber\\
\end{eqnarray}
and therefore
\begin{eqnarray}
\Fkreis{2}{020}&=&\frac{1}{2T}\int_{-1/T}^0\rmd x\,\Rbar^{\prime\prime\prime}(x)\nonumber\\
&&\quad\times\left(\ln(2\pi T)-\ln(1+T x)+\Rbar(x)\right)\nonumber\\
&&+\frac{1}{T}\int_0^{1/T}\rmd x\,\frac{\Rbar^{\prime\prime\prime}(x)}{1+e^x}\nonumber\\
&&\quad\times\Bigl(\ln(2\pi T)+\Rbar(x)\nonumber\\
&&\qquad -\frac{\ln(1+ Tx)}{2}-\frac{\ln(1-Tx)}{2}\Bigr)\;.\nonumber\\
\end{eqnarray}
At low temperatures we approximate $\ln(1-Tx)+\ln(1+Tx)\approx -x^2T^2$ and find 
\begin{equation}
\Fkreis{2}{020}=\Akreis{2}{020}+\Bkreis{2}{020}
\end{equation}
with 
\begin{eqnarray}
\Akreis{2}{020}&=&\frac{1}{2T}\left(\ln(2\pi T)\mathcal{I}_{24}-\mathcal{I}_{25}+\mathcal{I}_{26}\right)\nonumber\\
&=&-\frac{7\zeta(3)}{4\pi^2 T}\left(\EulerGamma+2\ln 2-\ln(2\pi T)\right)\nonumber\\
&&+\frac{T}{4}\left(5-2\EulerGamma-4\ln 2+2\ln(2\pi T)\right)\;,\nonumber\\
\Bkreis{2}{020}&=&\frac{\ln(2\pi T)}{T}\mathcal{I}_{27}+\frac{1}{T}\mathcal{I}_{28}+\frac{T}{2}\mathcal{I}_{29}\nonumber\\
&=&\frac{T}{24}\left(2\left(12\ln 2-11\right)+3\zeta(3)\right)-\frac{\pi^2}{160 T}\nonumber\\
&&+\frac{\zeta(3)}{4\pi^2 T}\left(-3+4\EulerGamma+10\ln 2-4\ln(\pi T)\right)\nonumber\;.\\
\end{eqnarray}
Thus,
\begin{eqnarray}
\Fkreis{2}{020}&=&\frac{T}{2}\left(\frac{2}{3}-\EulerGamma+\ln(2\pi T)+\frac{\zeta(3)}{4}\right)\nonumber\\
&&+\frac{1}{T}\left(-\frac{\pi^2}{160}+\frac{3\zeta(3)}{4\pi^2}\left(-1-\EulerGamma+\ln(2\pi T)\right)\right)\nonumber\\
&&+\mathcal{O}[T^3,T^3\ln T]\;.
\end{eqnarray}

\subsubsection{Evaluation of the function \texorpdfstring{$\Fkreis{2}{002}$}{Fkreis2002}}
With Eq.~(\ref{eq:Fkreistwofunctions1}) we find that
\begin{eqnarray}
\Fkreis{2}{002}&=&T\beta^2\int_{-1}^{1}\rmd\omega\,\rhotilde(\omega)\frac{e^{\beta\omega}\left(e^{\beta\omega}-1\right)}{2\left(1+e^{\beta\omega}\right)^3}\nonumber\\
&&\quad\times H_T^{(0)}(-\omega)(H_T^\prime)^{(0)}(\omega)\;.
\end{eqnarray}
Since 
\begin{equation}
f(x)=\frac{e^x(e^x-1)}{2(1+e^x)^3}=\frac{\sinh^4(x/2)}{\sinh^3(x)}=\frac{1}{8}\frac{\tanh(x/2)}{\cosh(x/2)^2}
\end{equation}
is an odd function we can use Eq.~(\ref{GIROdd}). Furthermore, with 
\begin{eqnarray}
g(x)&=&\left(\ln(2\pi T)-\ln(1+Tx)+\Rbar(x)\right)\nonumber\\
&&\times\left(\frac{1}{1-T x}+\frac{\Rbar^\prime(x)}{T}\right)\;,
\end{eqnarray}
we can write
\begin{eqnarray}
\Fkreis{2}{002}&=&\frac{1}{T}\int_0^{1/T}\rmd x\,f(x)\left(g(x)-g(-x)\right)\nonumber\\
&=&\Akreis{2}{002}+\Bkreis{2}{002}+\Ckreis{2}{002}
\end{eqnarray}
with
\begin{eqnarray}
\Akreis{2}{002}&=&\int_0^{1/T}\rmd x\,\frac{e^x(e^x-1)}{2(1+e^x)^3}\nonumber\\
&&\quad\times\bigg(\frac{\ln(2\pi T)}{1-Tx}-\frac{\ln(2\pi T)}{1+Tx}\nonumber\\
&&\qquad +\frac{\ln(1-Tx)}{1+Tx}-\frac{\ln(1+T x)}{1-T x}\bigg)\;,\nonumber\\
\Bkreis{2}{002}&=&\int_0^{1/T}\rmd x\,\frac{e^x(e^x-1)}{2(1+e^x)^3}\Rbar(x)\nonumber\\
&&\quad\times\left(\frac{1}{1-T x}-\frac{1}{1+Tx}+\frac{2\Rbar^\prime(x)}{T}\right)\;,\nonumber\\
\end{eqnarray}
and
\begin{eqnarray}
\Ckreis{2}{002}&=&\int_0^{1/T}\rmd x\,\frac{e^x(e^x-1)}{2(1+e^x)^3}\Rbar^\prime(x)\nonumber\\
&&\quad\times\left(\frac{2\ln(2\pi T)}{T}-\frac{\ln(1-T x)}{T}-\frac{\ln(1+T x)}{T}\right)\;.\nonumber\\
\end{eqnarray}
At low temperatures we may approximate 
\begin{eqnarray}
k(x)&=&\frac{\ln(2\pi T)}{1-Tx}-\frac{\ln(2\pi T)}{1+Tx}+\frac{\ln(1-Tx)}{1+Tx}-\frac{\ln(1+T x)}{1-T x}\nonumber\\[3pt]
&\approx& 2 T x\left(\ln(2\pi T)-1\right)\;,
\end{eqnarray}
and
\begin{eqnarray}
\frac{1}{1-T x}-\frac{1}{1+T x}&\approx &2 T x\;,\nonumber\\[3pt]
-\frac{\ln(1- T x)}{T}-\frac{\ln(1+T x)}{T}&\approx & x^2 T\;.
\end{eqnarray}
Therefore,
\begin{eqnarray}
\Akreis{2}{002}&=&2 T\left(\ln(2\pi T)-1\right)\alpha_6=\frac{T}{2}\left(\ln(2\pi T)-1\right)\;,\nonumber\\
\Bkreis{2}{002}&=&T\mathcal{I}_{30}+\frac{\mathcal{I}_{31}}{T}\nonumber\\
&=&-\frac{T}{4}\left(1+2\EulerGamma\right)+\frac{\pi^2}{480 T}-\frac{3\zeta(3)}{4\pi^2 T}\left(1+\EulerGamma\right)\;,\nonumber\\
\Ckreis{2}{002}&=&\frac{\ln(2\pi T)}{T}\mathcal{I}_{32}+\frac{T}{2}\mathcal{I}_{33}\nonumber\\
&=&\frac{3\zeta(3)}{4\pi^2 T}\ln(2\pi T)+\frac{T}{2}\left(\frac{1}{6}+\frac{\zeta(3)}{4}\right)\;.
\end{eqnarray}
Finally, we arrive at
\begin{eqnarray}
\Fkreis{2}{002}&=&\frac{\pi^2}{480T}+\frac{3\zeta(3)}{4\pi^2 T}\left(-1-\EulerGamma+\ln(2\pi T)\right)\nonumber\\
&&+\frac{T}{2}\left(-\frac{4}{3}-\EulerGamma+\ln(2\pi T)+\frac{\zeta(3)}{4}\right)\nonumber\\
&&+\mathcal{O}[T^3,T^3\ln T]\;.
\end{eqnarray}

\subsubsection{Evaluation of the function \texorpdfstring{$\Fkreis{2}{110}$}{Fkreis2110}}
Using Eq.~(\ref{eq:Fkreistwofunctions1}) and the property of Fermi functions~(\ref{GIRFermi})
\begin{eqnarray}
\Fkreis{2}{110}&=&T\int_{-1}^{1}\rmd\omega\,\frac{\rhotilde(\omega)}{1+e^{\beta\omega}}H_T^{(1)}(-\omega)(H_T^\prime)^{(1)}(\omega)\;,\nonumber\\
&=&\Akreis{2}{110}+\Bkreis{2}{110}
\end{eqnarray}
with
\begin{eqnarray}
\Akreis{2}{110}&=&-\frac{1}{T}\mathcal{I}_{22}=\frac{T}{2}\;,\nonumber\\
\Bkreis{2}{110}&\approx&-\frac{1}{T}\int_0^\infty\rmd x\,\frac{1}{1+e^x}\nonumber\\
&&\qquad\times\left(\Rbar^\prime(x)\Rbar^{\prime\prime}(x)-\Rbar^\prime(-x)\Rbar^{\prime\prime}(-x)\right)\nonumber\\
&=&-\frac{2}{T}\mathcal{I}_{23}=-\frac{2}{T}\frac{\pi^2}{360}\;.
\end{eqnarray}

Thus,
\begin{equation}
\Fkreis{2}{110}=-\frac{\pi^2}{180 T}+\frac{T}{2}+\mathcal{O}[T^3,T^3\ln T]\;.
\end{equation}

\subsubsection{Evaluation of the function \texorpdfstring{$\Fkreis{2}{101}$}{Fkreis2101}}
We start from Eq.~(\ref{eq:Fkreistwofunctions1}) and find
\begin{eqnarray}
\Fkreis{2}{101}&=&T\beta\int_{-1}^{1}\rmd\omega\,\frac{\rhotilde(\omega)e^{\beta\omega}}{\left(1+e^{\beta\omega}\right)^2}H_T^{(1)}(-\omega)(H_T^\prime)^{(0)}(\omega)\;.\nonumber\\
\end{eqnarray}
Since $e^x(1+e^x)^{-2}\Rbar^{\prime\prime}(x)$ is even, we find with Eq.~(\ref{GIREven})
\begin{eqnarray}
\Fkreis{2}{101}&=&-\int_0^{1/T}\rmd x\,\frac{e^x}{(1+e^x)^2}\Rbar^\prime(x)\nonumber\\
&&\quad\times\left(\frac{1}{1-T x}-\frac{1}{1+T x}+2\frac{\Rbar^\prime(x)}{T}\right)\;.\nonumber\\
\end{eqnarray}
At low temperatures $(1-Tx)^{-1}-(1+Tx)^{-1}\approx 2 x T$, therefore
\begin{eqnarray}
\Fkreis{2}{101}&=&-2 \left(T\mathcal{I}_{17}+\frac{1}{T}\mathcal{I}_{34}\right)\;.
\end{eqnarray}
Thus,
\begin{equation}
\Fkreis{2}{101}=-\frac{\pi^2}{90T}-\frac{T}{2}+\mathcal{O}[T^3,T^3\ln T]\;.
\end{equation}

\subsubsection{Evaluation of the function \texorpdfstring{$\Fkreis{2}{011}$}{Fkreis2011}}
We start from Eq.~(\ref{eq:Fkreistwofunctions1}) and have
\begin{eqnarray}
\Fkreis{2}{011}&=&T\beta\int_{-1}^{1}\rmd\omega\,\frac{\rhotilde(\omega)e^{\beta\omega}}{\left(1+e^{\beta\omega}\right)^2}H_T^{(0)}(-\omega)(H_T^\prime)^{(1)}(-\omega)\nonumber\\
&=&\frac{1}{T}\int_{-1/T}^{1/T}\rmd x\,\frac{e^x}{(1+e^x)^2}\Rbar^{\prime\prime}(x)\nonumber\\
&&\quad\times\left(\ln(2\pi T)-\ln(1+T x)+\Rbar(x)\right)\;.
\end{eqnarray}
Since $e^x/(1+e^x)^2\Rbar^{\prime\prime}(x)$ is even we find 
\begin{eqnarray}
\Fkreis{2}{011}&=&\frac{1}{T}\int_0^{1/T}\rmd x\,\frac{e^x}{(1+e^x)^2}\Rbar^{\prime\prime}(x)\nonumber\\
&&\quad\times\Big(2\ln(2\pi T)+2\Rbar(x)\nonumber\\
&&\qquad-\ln(1+T x)-\ln(1-T x)\Big)\nonumber\\
&=&\frac{2}{T}\left(\ln(2\pi T)\mathcal{I}_{35}+\mathcal{I}_{36}+\frac{T^2}{2}\mathcal{I}_{37}\right)\;,\nonumber\\
\end{eqnarray}
where we approximate $-\ln(1+Tx)-\ln(1-Tx)\approx x^2 T^2$.
Therefore,
\begin{eqnarray}
\Fkreis{2}{011}&=&\frac{T}{12}\left(3\zeta(3)-4\right)-\frac{\pi^2}{144 T}\nonumber\\
&&+\frac{3\zeta(3)}{2\pi^2 T}\left(-1-\EulerGamma+\ln(2\pi T)\right)+\mathcal{O}[T^3,T^3\ln T]\;.\nonumber\\
\end{eqnarray}

\subsubsection{Evaluation of the  function \texorpdfstring{$\Fbarkreis{000}$}{F3bar000}}
Using the property of Fermi functions~(\ref{GIRFermi}), we start from Eq.~(\ref{eq:Fkreisthreebarfunctions})
\begin{eqnarray}
\Fbarkreis{000}&=&-T\int_{-1}^1\rmd\omega\,\frac{\rhotilde(\omega)}{1+e^{\beta\omega}}H_T^{(0)}(\omega)(H_T^\prime)^{(0)}(\omega)\;,\nonumber\\
&=&-T^2\int_{-1/T}^{1/T}\rmd x\,\frac{1}{1+e^x}\frac{1}{1-T x}\nonumber\\
&&\qquad\times\left(\ln(2\pi T)-\ln(1-T x)+\Rbar(x)\right)\nonumber\\
&&-T\int_{-1/T}^{1/T}\rmd x\,\frac{\Rbar^\prime(x)}{1+e^x}\nonumber\\
&&\qquad\times\left(\ln(2\pi T)-\ln(1-T x)+\Rbar(x)\right)\nonumber\\
&=&\Abarkreis{000}+\Bbarkreis{000}+\Cbarkreis{000}+\Dbarkreis{000}
\end{eqnarray}
with
\begin{eqnarray}
\Abarkreis{000}&=&-T^2\int_{-1/T}^0\rmd x\,\frac{1}{1-T x}\nonumber\\
&&\qquad\times\left(\ln(2\pi T)-\ln(1- Tx)+\Rbar(x)\right)\;,\nonumber\\
\end{eqnarray}
and
\begin{eqnarray}
\Bbarkreis{000}&=&-T^2\int_0^{1/T}\frac{\rmd x}{1+e^x}\nonumber\\
&&\qquad\times\biggl(\frac{\ln(2\pi T)-\ln(1-T x)+\Rbar(x)}{1-T x}\nonumber\\
&&\quad\qquad-\frac{\ln(2\pi T)-\ln(1+T x)+\Rbar(x)}{1+T x}\biggr)\;,\nonumber \\
\Cbarkreis{000}&=&-T \int_{-1/T}^0\rmd x\,\Rbar^\prime(x)\nonumber\\
&&\qquad\times\left(\ln(2\pi T)-\ln(1-T x)+\Rbar(x)\right)\;,\nonumber\\
\Dbarkreis{000}&=&-T\int_0^{1/T}\rmd x\,\frac{\Rbar^\prime(x)}{1+e^x}\nonumber\\
&&\qquad\times\Big(2\ln(2\pi T)+2\Rbar(x)\nonumber\\
&&\qquad\quad-\ln(1-T x)-\ln(1+T x)\Big)\;.\nonumber\\
\end{eqnarray}
For the first term we find
\begin{eqnarray}
\Abarkreis{000}&=&-T^2\left(\alpha_3+\mathcal{I}_{10}\right)\nonumber\\
&=&\frac{T}{12}\bigg(\pi^2+6\ln^2(2)+12\ln 2\ln T\nonumber\\
&&\qquad-6\ln2\ln(T^2)\bigg)\;.
\end{eqnarray}
At low temperatures, we have
\begin{eqnarray}
m(x)&=&\frac{\ln(2\pi T)-\ln(1-T x)}{1-T x}-\frac{\ln(2\pi T)-\ln(1+T x)}{1+T x}\nonumber\\
&\approx& Tx\left(1+\ln(2\pi T)\right)
\end{eqnarray}
and
\begin{equation}
\Rbar(x)\left(\frac{1}{1-T x}-\frac{1}{1+T x}\right)\approx 2 T\Rbar(x) x\;,
\end{equation}
so that
\begin{equation}
\Bbarkreis{000}\sim \mathcal{O}[T^3\ln(T)]
\end{equation}
in the limit $T\to 0$. 
For the third term, we have
\begin{eqnarray}
\Cbarkreis{000}&=&-T\left(\ln(2\pi T)\mathcal{I}_{5}-\mathcal{I}_{6}+\mathcal{I}_{7}\right)\nonumber\\
&=&\frac{T}{2}\biggl(
-\EulerGamma^2-\frac{\pi^2}{6}-4\EulerGamma\ln 2-\ln^2(2)\nonumber\\
&&\quad +2\EulerGamma\ln(2\pi)+2\ln 2\ln(\pi)-\ln^2(\pi)\nonumber\\
&&\quad+2\EulerGamma\ln T-4\ln(\pi)\ln T+2\ln(2\pi)\ln T\nonumber\\
&&\quad-\ln^2(T)\biggr) \;.
\end{eqnarray}
With $-\ln(1-T x)-\ln(1+T x)\approx x^2 T^2$, the last term reduces to
\begin{eqnarray}
\Dbarkreis{000}&=&-2 T\left(\ln(2\pi T)\mathcal{I}_{8}+\mathcal{O}[T^2]+\mathcal{I}_{9}\right)\nonumber\\
&=&-2 T\bigg(\frac{1}{2}-\frac{\pi^2}{48}+\EulerGamma\left(\frac{1}{2}-\ln 2\right)-\ln^2(2)\nonumber\\
&&\qquad+\left(\ln 2-\frac{1}{2}\right)\ln(2\pi T)\bigg)
\end{eqnarray}
at low temperatures. 
Finally, we arrive at
\begin{eqnarray}
\Fbarkreis{000}&=&
T\biggl[
-1-\EulerGamma-\frac{\EulerGamma^2}{2}+\frac{\pi^2}{24}-\ln2\ln(\pi)-\frac{\ln^2(\pi)}{2}\nonumber\\
&&\quad+\ln(2\pi T)\left(1+\EulerGamma-\ln T\right)+\frac{\ln^2(T)}{2}
\biggr]\nonumber\\
&&
+\mathcal{O}[T^3,T^3\ln T]\;.
\end{eqnarray}

\subsubsection{Evaluation of the  function \texorpdfstring{$\Fbarkreis{100}$}{F3bar100}}
With the property of Fermi functions Eq.~(\ref{GIRFermi}), we find from Eq.~(\ref{eq:Fkreisthreebarfunctions})
\begin{eqnarray}
\Fbarkreis{100}&=&-T\int_{-1}^1\rmd\omega\,\frac{\rhotilde(\omega)}{1+e^{\beta\omega}}H_T^{(1)}(\omega)(H_T^\prime)^{(0)}(\omega)\nonumber\\
&=&-T\int_{-1/T}^{1/T}\frac{\rmd x}{1+e^x}\frac{\Rbar^\prime(x)}{1-T x}-\int_{-1/T}^{1/T}\rmd x\,\frac{(\Rbar^\prime(x))^2}{1+e^x}\nonumber\\
&=&\Abarkreis{100}+\Bbarkreis{100}+\Cbarkreis{100}+\Dbarkreis{100}\;.
\end{eqnarray}
The first term becomes
\begin{eqnarray}
\Abarkreis{100}=-T\mathcal{I}_{19}=T\left(\EulerGamma-\ln(\pi T)\right)\;.
\end{eqnarray}
In the second term, we approximate $(1-Tx)^{-1}+(1+Tx)^{-1}\approx 2+\mathcal{O}[T^2]$
at low temperatures. Thus,
\begin{equation}
\Bbarkreis{100}=-2 T\mathcal{I}_{8}=-2T\left(\ln 2-\frac{1}{2}\right)\;.
\end{equation}
Moreover,
\begin{eqnarray}
\Cbarkreis{100}&=&-\mathcal{I}_{18}=T-\frac{\pi^2}{12}\;,\nonumber\\
\Dbarkreis{100}&=&0\;,
\end{eqnarray}
where we used $(\Rbar^\prime(x))^2-(\Rbar^\prime(-x))^2=0$ in the last term.
Finally, we arrive at
\begin{eqnarray}
\Fbarkreis{100}&=&-\frac{\pi^2}{12}+T\left(2+\EulerGamma-\ln2-\ln(2\pi T)\right)\nonumber\\
&&+\mathcal{O}[T^3,T^3\ln T]\;.
\end{eqnarray}

\subsubsection{Evaluation of the  function \texorpdfstring{$\Fbarkreis{010}$}{F3bar010}}
We start from Eq.~(\ref{eq:Fkreisthreebarfunctions}) and use the property of Fermi functions Eq.~(\ref{GIRFermi}),
\begin{eqnarray}
\Fbarkreis{010}&=&T\int_{-1}^1\rmd\omega\,\frac{\rhotilde(\omega)}{1+e^{\beta\omega}}H_T^{(0)}(\omega)(H_T^\prime)^{(1)}(\omega)\;,\nonumber\\
&=&\Abarkreis{010}+\Bbarkreis{010}
\end{eqnarray}
with 
\begin{eqnarray}
\Abarkreis{010}&=&\int_{-1/T}^0\rmd x\,\Rbar^{\prime\prime}(x)\nonumber\\
&&\quad\times\left(\ln(2\pi T)-\ln(1-T x)+\Rbar(x)\right)\nonumber\\
&=&\ln(2\pi T)\mathcal{I}_{12}-\mathcal{I}_{13}+\mathcal{I}_{14}\nonumber\\
&=&-\frac{\pi^2}{12}+T\left(1+\EulerGamma-\ln(2\pi T)\right)
\end{eqnarray}
and
\begin{eqnarray}
\Bbarkreis{010}&=&\int_0^{1/T}\rmd x\,\frac{\Rbar^{\prime\prime}(x)}{1+e^x}\left(-\ln(1-T x)+\ln(1+T x)\right)\nonumber\\
&=&2 T \mathcal{I}_{15}\nonumber\\
&=&T\left(\frac{3}{2}-2\ln 2\right)\;,
\end{eqnarray}
where we approximate $-\ln(1-Tx)+\ln(1+Tx)\approx 2xT$.
Finally, we find
\begin{eqnarray}
\Fbarkreis{010}&=&-\frac{\pi^2}{12}+T\left(\frac{5}{2}+\EulerGamma-2\ln 2-\ln(2\pi T)\right)\nonumber\\
&&+\mathcal{O}[T^3,T^3\ln T]\;.
\end{eqnarray}

\subsubsection{Evaluation of the  function \texorpdfstring{$\Fbarkreis{001}$}{F3bar001}}
We start from Eq.~(\ref{eq:Fkreisthreebarfunctions})
\begin{eqnarray}
\Fbarkreis{001}&=&T\beta\int_{-1}^1\rmd\omega\,\frac{\rhotilde(\omega)e^{\beta\omega}}{\left(1+e^{\beta\omega}\right)^2}H_T^{(0)}(\omega)(H_T^\prime)^{(0)}(\omega)\;,\nonumber\\
&=&\Abarkreis{001}+\Bbarkreis{001}
\end{eqnarray}
with 
\begin{eqnarray}
\Abarkreis{001}&=&T\int_{-1/T}^{1/T}\rmd x\,\frac{e^x}{(1+e^x)^2}\frac{1}{1-Tx}\nonumber\\
&&\quad\times\left(\ln(2\pi T)-\ln(1- Tx)+\Rbar(x)\right)\;,\nonumber\\
\Bbarkreis{001}&=&\int_{-1/T}^{1/T}\rmd x\,\frac{e^x}{(1+e^x)^2}\Rbar^\prime(x)\nonumber\\
&&\quad\times\left(\ln(2\pi T)-\ln(1-T x)+\Rbar(x)\right)\;.\nonumber\\
\end{eqnarray}
Since $e^x (1+e^x)^{-2}$ is even in~$x$, we can use Eq.~(\ref{GIREven}) to write
\begin{eqnarray}
\Abarkreis{001}&=&T\int_0^{1/T}\rmd x\,\frac{e^x}{(1+e^x)^2}\nonumber\\
&&\quad\times\bigg(\left(\ln(2\pi T)+\Rbar(x)\right)\left(\frac{1}{1-Tx}+\frac{1}{1+T x}\right)\nonumber\\
&&\qquad-\frac{\ln(1-Tx)}{1-T x}-\frac{\ln(1+T x)}{1+T x}\bigg)\nonumber\\
&=&2 T\ln(2\pi T)\alpha_5+2T \mathcal{I}_{16}\nonumber\\
&=&T\left(-1-\EulerGamma+\ln(2\pi T)\right)\;,
\end{eqnarray}
where we used $(1-Tx)^{-1}+(1+Tx)^{-1}\approx 2$ for $T\ll1$.
Moreover,
\begin{eqnarray}
\Bbarkreis{001}&=&\int_0^{1/T}\rmd x\,\frac{e^x}{(1+e^x)^2}\Rbar^\prime(x)\nonumber\\
&&\quad\times\left(-\ln(1-T x)+\ln(1+T x)\right)\;.
\end{eqnarray}
With $-\ln(1-T x)+\ln(1+T x)\approx 2x T$ we find
\begin{eqnarray}
\Bbarkreis{001}=2T\mathcal{I}_{17}=\frac{T}{2}\;.
\end{eqnarray}
Finally,
\begin{eqnarray}
\Fbarkreis{001}&=&-\frac{T}{2}\left(1+2\EulerGamma-2\ln(2\pi T)\right)\nonumber\\
&&+\mathcal{O}[T^3,T^3\ln T]\;.
\end{eqnarray}

\subsubsection{Evaluation of the  function \texorpdfstring{$\Fbarkreis{200}$}{F3bar200}}
We start from Eq.~(\ref{eq:Fkreisthreebarfunctions}) and use the property of Fermi functions~(\ref{GIRFermi}),
\begin{eqnarray}
\Fbarkreis{200}&=&-T\int_{-1}^1\rmd\omega\,\frac{\rhotilde(\omega)}{1+e^{\beta\omega}}H_T^{(2)}(\omega)(H_T^\prime)^{(0)}(\omega)\;,\nonumber\\
&=&-\frac{1}{2}\int_{-1/T}^{1/T}\rmd x\,\frac{1}{1+e^x}\frac{\Rbar^{\prime\prime}(x)}{1-T x}\nonumber\\
&&-\frac{1}{2T}\int_{-1/T}^{1/T}\rmd x\,\frac{\Rbar^\prime(x)\Rbar^{\prime\prime}(x)}{1+e^x}\nonumber\\
&=&\Abarkreis{200}+\Bbarkreis{200}+\Cbarkreis{200}+\Dbarkreis{200}
\end{eqnarray}
with 
\begin{eqnarray}
\Abarkreis{200}&=&-\frac{1}{2}\int_{-1/T}^0\rmd x\,\frac{\Rbar^{\prime\prime}(x)}{1-Tx}\;,\nonumber\\
\Bbarkreis{200}&=&-\frac{1}{2}\int_0^{1/T}\rmd x\,\frac{\Rbar^{\prime\prime}(x)}{1+e^x}\left(\frac{1}{1-T x}-\frac{1}{1+T x}\right)\;,\nonumber\\
\Cbarkreis{200}&=&-\frac{1}{2T}\int_{-1/T}^0\rmd x\,\Rbar^\prime(x)\Rbar^{\prime\prime}(x)\;,\nonumber\\
\Dbarkreis{200}&=&-\frac{1}{2T}\int_0^\infty\rmd x\,\frac{\Rbar^{\prime\prime}(x)}{1+e^x}\left(\Rbar^\prime(x)-\Rbar^\prime(-x)\right)\;.\nonumber\\
\end{eqnarray}
At $T\ll1 $ the expressions reduce to
\begin{eqnarray}
\Abarkreis{200}&=&-\frac{1}{2}\mathcal{I}_{21}=-\frac{T}{2}\left(\EulerGamma-\ln(\pi T)\right)\;,\nonumber\\
\Bbarkreis{200}&=&-T\mathcal{I}_{15}=T\left(\ln 2-\frac{3}{4}\right)\;,
\end{eqnarray}
where we used $(1-Tx)^{-1}-(1+Tx)^{-1}\approx 2 xT$ in the last step. Furthermore,
\begin{eqnarray}
\Cbarkreis{200}&=&-\frac{1}{2T}\mathcal{I}_{22}=\frac{T}{4}\;,\nonumber\\
\Dbarkreis{200}&=&-\frac{1}{T}\mathcal{I}_{23}=-\frac{1}{T}\frac{\pi^2}{360}\;.
\end{eqnarray}
Thus, we have
\begin{eqnarray}
\Fbarkreis{200}&=&-\frac{\pi^2}{360T}-\frac{T}{2}\left(1+\EulerGamma-\ln 2-\ln(2\pi T)\right)\nonumber\\
&&+\mathcal{O}[T^3,T^3\ln T]\;.
\end{eqnarray}

\subsubsection{Evaluation of the  function \texorpdfstring{$\Fbarkreis{020}$}{F3bar020}}
We start from Eq.~(\ref{eq:Fkreisthreebarfunctions}) and use the property of Fermi functions Eq.~(\ref{GIRFermi}),
\begin{eqnarray}
\Fbarkreis{020}&=&-T\int_{-1}^1\rmd\omega\,\frac{\rhotilde(\omega)}{1+e^{\beta\omega}}H_T^{(0)}(\omega)(H_T^\prime)^{(2)}(\omega)\;,\nonumber\\
&=&-\frac{1}{2T}\int_{-1/T}^{1/T}\rmd x\,\frac{\Rbar^{\prime\prime\prime}(x)}{1+e^x}\nonumber\\
&&\qquad\times\left(\ln(2\pi T)-\ln(1-T x)+\Rbar(x)\right)\nonumber\\
&=&\Abarkreis{020}+\Bbarkreis{020}
\end{eqnarray}
with
\begin{eqnarray}
\Abarkreis{020}&=&-\frac{1}{2T}\int_{-1/T}^0\rmd x\,\Rbar^{\prime\prime\prime}(x)\nonumber\\
&&\qquad\times\left(\ln(2\pi T)-\ln(1- Tx)+\Rbar(x)\right)\;,\nonumber\\
\Bbarkreis{020}&=&-\frac{1}{T}\int_0^{1/T}\rmd x\,\frac{\Rbar^{\prime\prime\prime}(x)}{1+e^x}\nonumber\\
&&\qquad\times\Bigl(\ln(2\pi T)+\Rbar(x)\nonumber\\
&&\qquad\quad-\frac{1}{2}\left(\ln(1-T x)+\ln(1+ Tx)\right)\Bigr)\;.\nonumber\\
\end{eqnarray}
We thus find
\begin{eqnarray}
\Abarkreis{020}&=&-\frac{1}{2T}\left(\ln(2\pi T)\mathcal{I}_{24}-\mathcal{I}_{38}+\mathcal{I}_{26}\right)\nonumber\\
&=&\frac{7\zeta(3)}{4\pi^2 T}\left(\EulerGamma+2\ln 2-\ln(2\pi T)\right)\nonumber\\
&&+\frac{T}{4}\left(-1+2\EulerGamma+4\ln 2-2\ln(2\pi T)\right)\nonumber\\
\end{eqnarray}
and, with $\ln(1-T x)+\ln(1+Tx)\approx -x^2T^2$,
\begin{eqnarray}
\Bbarkreis{020}&=&-\frac{1}{T}\left(\ln(2\pi T)\mathcal{I}_{27}+\mathcal{I}_{28}+\frac{T^2}{2}\mathcal{I}_{29}\right)\nonumber\\
&=&-\frac{T}{24}\left(-22+24\ln 2+3\zeta(3)\right)\nonumber\\
&&+\frac{1}{T}\biggl[\frac{\pi^2}{160}-\frac{\zeta(3)}{4\pi^2}\Big(-4\ln(2\pi T)\nonumber\\
&&\qquad\quad+4\EulerGamma-3+14\ln 2\Big)\biggr]\;.
\end{eqnarray}
Thus, we arrive at
\begin{eqnarray}
\Fbarkreis{020}&=&\frac{1}{T}\left(\frac{\pi^2}{160}+\frac{3\zeta(3)}{4\pi^2}\left(1+\EulerGamma-\ln(2\pi T)\right)\right)\nonumber\\
&&+\frac{T}{2}\left(\frac{4}{3}+\EulerGamma-\frac{\zeta(3)}{4}-\ln(2\pi T)\right)\nonumber\\
&&+
\mathcal{O}[T^3,T^3\ln T]\;.
\end{eqnarray}

\subsubsection{Evaluation of the  function \texorpdfstring{$\Fbarkreis{002}$}{F3bar002}}
We start from Eq.~(\ref{eq:Fkreisthreebarfunctions})
\begin{eqnarray}
\Fbarkreis{002}&=&-T\beta^2\int_{-1}^1\rmd\omega\,\rhotilde(\omega)\frac{e^{\beta\omega}\left(e^{\beta\omega}-1\right)}{2\left(1+e^{\beta\omega}\right)^3}\nonumber\\
&&\qquad\quad\times H_T^{(0)}(\omega)(H_T^\prime)^{(0)}(\omega)\nonumber\\
&=&\Abarkreis{002}+\Bbarkreis{002}
\end{eqnarray}
with 
\begin{eqnarray}
\Abarkreis{002}&=&-\frac{1}{2}\int_{-1/T}^{1/T}\rmd x\,\frac{e^x(e^x-1)}{(1+e^x)^3}\nonumber\\
&&\qquad\times \frac{\ln(2\pi T)-\ln(1-T x)+\Rbar(x)}{1-Tx}\;,\nonumber\\
\Bbarkreis{002}&=&-\frac{1}{2T}\int_{-1/T}^{1/T}\rmd x\,\frac{e^x(e^x-1)}{(1+e^x)^3}\Rbar^\prime(x)\nonumber\\
&&\qquad\times\left(\ln(2\pi T)-\ln(1-Tx)+\Rbar(x)\right)\;.\nonumber\\
\end{eqnarray}
Next, we use Eq.~(\ref{GIROdd}). At low temperatures
\begin{eqnarray}
\frac{1}{1-Tx}-\frac{1}{1+Tx}&\approx& 2 xT\;,\nonumber\\
-\frac{\ln(1-T x)}{1-T x}+\frac{\ln(1+T x)}{1+T x}&\approx& 2 xT
\end{eqnarray}
so that
\begin{eqnarray}
\Abarkreis{002}&\approx&-T\int_0^\infty\rmd x\,\frac{e^x(e^x-1)}{(1+e^x)^3}\nonumber\\
&&\qquad\times \left(x\left(1+\ln(2\pi T)\right)+\Rbar(x)x\right)\nonumber\\
&=&-T\left(\left(1+\ln(2\pi T)\right)\alpha_7+\mathcal{I}_{30}\right)\nonumber\\
&=&\frac{T}{4}\left(-1+2\EulerGamma-2\ln(2\pi T)\right)\;.
\end{eqnarray}
Next, using $-\ln(1-T x)-\ln(1+T x)\approx x^2T^2$,
\begin{eqnarray}
\Bbarkreis{002}&=&-\frac{1}{2T}\left(2\ln(2\pi T)\mathcal{I}_{32}+2\mathcal{I}_{31}+T^2\mathcal{I}_{33}\right)\nonumber\\
&=&-\frac{T}{24}\left(2+3\zeta(3)\right)-\frac{\pi^2}{480 T}\nonumber\\
&&-\frac{3\zeta(3)}{4\pi^2 T}\left(\ln(2\pi T)-1-\EulerGamma\right)\;.
\end{eqnarray}
Thus, we arrive at
\begin{eqnarray}
\Fbarkreis{002}&=&-\frac{\pi^2}{480 T}+\frac{3\zeta(3)}{4\pi^2 T}\left(1+\EulerGamma-\ln(2\pi T)\right)\nonumber\\
&&+T\left(-\frac{1}{3}+\frac{\EulerGamma}{2}-\frac{\zeta(3)}{8}\right)\nonumber\\
&-&\frac{T}{2}\ln(2\pi T)+\mathcal{O}[T^3,T^3\ln T]\;.
\end{eqnarray}

\subsubsection{Evaluation of the  function \texorpdfstring{$\Fbarkreis{110}$}{F3bar110}}
Equation~(\ref{eq:Fkreisthreebarfunctions}) gives
\begin{eqnarray}
\Fbarkreis{110}&=&T\int_{-1}^1\rmd\omega\,\frac{\rhotilde(\omega)}{1+e^{\beta\omega}}H_T^{(1)}(\omega)(H_T^\prime)^{(1)}(\omega)\;,\nonumber\\
&=&\frac{1}{T}\int_{-1/T}^{1/T}\rmd x\,\frac{\Rbar^\prime(x)\Rbar^{\prime\prime}(x)}{1+e^x}\nonumber\; ,\\
&=&\Abarkreis{110}+\Bbarkreis{110}
\end{eqnarray}
where we used the property of Fermi functions, see Eq.~(\ref{GIRFermi}). The remaining terms read
\begin{eqnarray}
\Abarkreis{110}&=&\frac{1}{T}\int_{-1/T}^0\rmd x\,\Rbar^\prime(x)\Rbar^{\prime\prime}(x)=\frac{1}{T}\mathcal{I}_{22}=-\frac{T}{2}\nonumber\\
\end{eqnarray}
and
\begin{eqnarray}
\Bbarkreis{110}&=&\frac{1}{T}\int_0^{1/T}\rmd x\,\frac{\Rbar^{\prime\prime}(x)}{1+e^x}\left(\Rbar^\prime(x)-\Rbar^\prime(-x)\right)\nonumber\\
&=&\frac{2}{T}\mathcal{I}_{23}\nonumber\\
&=&\frac{1}{T}\frac{\pi^2}{180}\;.
\end{eqnarray}
Finally,
\begin{equation}
\Fbarkreis{110}=\frac{\pi^2}{180 T}-\frac{T}{2}+\mathcal{O}[T^3,T^3\ln T]\;.
\end{equation}

\subsubsection{Evaluation of the  function \texorpdfstring{$\Fbarkreis{101}$}{F3bar101}}
We start from Eq.~(\ref{eq:Fkreisthreebarfunctions})
\begin{eqnarray}
\Fbarkreis{101}&=&T\beta\int_{-1}^1\rmd\omega\,\frac{\rhotilde(\omega)e^{\beta\omega}}{\left(1+e^{\beta\omega}\right)^2}H_T^{(1)}(\omega)(H_T^\prime)^{(0)}(\omega)\;,\nonumber\\
&=&\Abarkreis{101}+\Bbarkreis{101}\; ,
\end{eqnarray}
where we used Eq.~(\ref{GIREven}). The remaining terms read 
\begin{eqnarray}
\Abarkreis{101}&=&\int_{-1/T}^{1/T}\rmd x\,\frac{e^x}{(1+e^x)^2}\frac{\Rbar^\prime(x)}{1-Tx}\nonumber\;,\\
\Bbarkreis{101}&=&\frac{1}{T}\int_{-1/T}^{1/T}\rmd x\,\frac{e^x}{(1+e^x)^2}\left(\Rbar^\prime(x)\right)^2\;.
\end{eqnarray}
Using $(1-Tx)^{-1}-(1+Tx)^{-1}\approx 2x T$ at low temperatures, we arrive at
\begin{eqnarray}
\Abarkreis{101}&=&\int_0^{1/T}\rmd x\,\frac{e^x}{(1+e^x)^2}\Rbar^\prime(x)2xT\nonumber\\
&=&2 T\int_0^\infty\rmd x\,\frac{e^x}{(1+e^x)^2}\Rbar^\prime(x)x\nonumber\\
&=&2T\mathcal{I}_{17}=\frac{T}{2}
\end{eqnarray}
and
\begin{eqnarray}
\Bbarkreis{101}&=&\frac{2}{T}\int_{0}^{1/T}\rmd x\,\frac{e^x}{(1+e^x)^2}[\Rbar^\prime(x)]^2\nonumber\\
&=&\frac{2}{T}\mathcal{I}_{34}=\frac{1}{T}\frac{\pi^2}{90}\;.
\end{eqnarray}
Thus,
\begin{equation}
\Fbarkreis{101}=\frac{T}{2}+\frac{\pi^2}{90 T}+\mathcal{O}[T^3,T^3\ln T]
\;.
\end{equation}

\subsubsection{Evaluation of the  function \texorpdfstring{$\Fbarkreis{011}$}{F3bar011}}
We start from Eq.~(\ref{eq:Fkreisthreebarfunctions}) and use Eq.~(\ref{GIREven}),
\begin{eqnarray}
\Fbarkreis{011}&=&-T\beta\int_{-1}^1\rmd\omega\,\frac{\rhotilde(\omega)e^{\beta\omega}}{\left(1+e^{\beta\omega}\right)^2}H_T^{(0)}(\omega)(H_T^\prime)^{(1)}(\omega)\nonumber\\
&=&-\frac{1}{T}\int_{-1/T}^{1/T}\rmd x\,\frac{e^x}{(1+e^x)^2}\Rbar^{\prime\prime}(x)\nonumber\\
&&\qquad\times \left(\ln(2\pi T)-\ln(1-T x)+\Rbar(x)\right)\nonumber\\
&=&-\frac{1}{T}\int_0^{1/T}\rmd x\,\frac{e^x}{(1+e^x)^2}\Rbar^{\prime\prime}(x)\nonumber\\
&&\qquad\times \Big(2\ln(2\pi T)+2\Rbar(x)-\ln(1-T x)\nonumber\\
&&\qquad\quad-\ln(1+T x)\Big)\;.
\end{eqnarray}
With $-\ln(1-T x)-\ln(1+T x)\approx x^2 T^2$ for $T\ll1 $ we find
\begin{eqnarray}
\Fbarkreis{011}&=&-\frac{2}{T}\left(\ln(2\pi T)\mathcal{I}_{35}+\mathcal{I}_{36}+\frac{T^2}{2}\mathcal{I}_{37}\right)\nonumber\\
\end{eqnarray}
Finally,
\begin{eqnarray}
\Fbarkreis{011}&=&\frac{1}{T}\left(\frac{\pi^2}{144}+\frac{3\zeta(3)}{2\pi^2}\left(1+\EulerGamma-\ln(2\pi T)\right)\right)\nonumber\\
&&+\frac{T}{12}\left(4-3\zeta(3)\right)+\mathcal{O}[T^3,T^3\ln T]\;.
\end{eqnarray}

\subsubsection{Evaluation of the function \texorpdfstring{$\Fkreis{2}{000}+\Fbarkreis{000}$}{F2000plusF3bar000}}
We start from Eq.~(\ref{eq:Fkreistwo000plusFkreisthreebar000function}) and use the property of Fermi functions Eq.~(\ref{GIRFermi}),
\begin{eqnarray}
\Fkreis{2}{000}+\Fbarkreis{000}&=&T\int_{-1}^1\rmd\omega\,\frac{\rhotilde(\omega)}{1+e^{\beta\omega}}\nonumber\\
&&\quad\times\left(H_T^{(0)}(-\omega)-H_T^{(0)}(\omega)\right)(H_T^\prime)^{(0)}(\omega)\nonumber\\
&=&T^2\int_{-1/T}^{1/T}\rmd x\,\frac{1}{1+e^x}\nonumber\\
&&\qquad\times \frac{\ln(1-T x)-\ln(1+T x)}{1-T x}\nonumber\\
&&+T\int_{-1/T}^{1/T}\rmd x\,\frac{\Rbar^\prime(x)}{1+e^x}\nonumber\\
&&\qquad\times\left(\ln(1-T x)-\ln(1+T x)\right)\nonumber\\
&=&\Atwothreekreis+\Btwothreekreis+\Ctwothreekreis+\Dtwothreekreis\nonumber\\
\end{eqnarray}
with 
\begin{eqnarray}
\Atwothreekreis&=&T^2\alpha_{8}=T\frac{\pi^2}{12}\;,\nonumber\\
\Btwothreekreis&=&T^2\int_0^{1/T}\rmd x\,\frac{1}{1+e^x}\nonumber\\
&&\qquad\times\bigg(\frac{\ln(1-T x)-\ln(1+T x)}{1-T x}\nonumber\\
&&\qquad\quad-\frac{\ln(1+T x)-\ln(1-T x)}{1+Tx}\bigg)\nonumber\\
&\approx&-4T^3\alpha_{2}=-\frac{\pi^2}{3}T^3\;,
\end{eqnarray}
and
\begin{eqnarray}
\Ctwothreekreis&=&T\int_{-1/T}^0\rmd x\,\Rbar^\prime(x)\nonumber\\
&&\times\left(\ln(1-T x)-\ln(1+T x)\right)\;,\nonumber\\
&=&T\mathcal{I}_{6}-T\mathcal{I}_{39}\nonumber\\
&=&-\frac{\pi^2}{4}T+\frac{\pi^2}{3}T^3\;,\nonumber\\
\Dtwothreekreis&=&0
\end{eqnarray}
at low temperatures. 
Thus,
\begin{equation}
\Fkreis{2}{000}+\Fbarkreis{000}=-\frac{\pi^2}{6}T+\mathcal{O}[T^3,T^3\ln T]\;.
\end{equation}

\subsection{Summary of the term \texorpdfstring{$\mathbf{x_B}$}{xB}}

Now we are in the position to sum up the contributions to $x_B$. 
We combine all terms that we calculated in the previous subsections to evaluate the terms listed in Eq.~(\ref{FijkFormalExpansion}). 
This leads to
\begin{eqnarray}
F_{000}&=&-\frac{\pi^2}{6}-2\ln^2(2)+\frac{\pi^2}{6}T+\frac{2\pi^2}{3}T^2(\ln 2-1)\nonumber \\
&&+\mathcal{O}[T^4\ln T, T^4]\;,\nonumber\\
F_{100}&=&-2-\EulerGamma^2-\frac{\pi^2}{6}-\frac{\ln^2(2)}{2}-2\ln 2\ln\pi\nonumber\\
&&-\ln^2(\pi)-2\EulerGamma+2\ln(2\pi T)(1+\EulerGamma)\nonumber\\
&&-2\ln(2\pi)\ln T-\ln^2(T)-2T(-1+\ln 2)\nonumber\\
&&+\mathcal{O}[T^2,T^2\ln T]\nonumber\;,\\
F_{010}&=&-4-2\EulerGamma^2-\frac{\pi^2}{6}-\ln^2(2)+4\ln 2(1-\ln \pi)\nonumber\\
&&+4\ln\pi -2\ln^2(\pi)+4\EulerGamma(-1+\ln(2\pi T))\nonumber\\
&&+4\ln T(1-\ln(2\pi))-2\ln^2(T)\nonumber\\
&&-4T(-1+\ln 2)\nonumber \\
&& +\mathcal{O}[T^2,T^2\ln T]\;,\nonumber\\
F_{200}&=&\frac{1}{2T}\biggl[
-1-\EulerGamma-\frac{\EulerGamma^2}{2}+\frac{\pi^2}{8}-\ln 2\ln \pi-\frac{\ln^2(\pi)}{2}\nonumber\\
&&\quad -\ln T\ln(2\pi)+\ln(2\pi T)(1+\EulerGamma)-\frac{\ln^2(T)}{2}
\biggr]\nonumber\\
&&+\frac{1}{4}\left(-7-6\EulerGamma+4\ln 2+6\ln(2\pi T)\right)\nonumber \; ,\\
F_{020}&=&\frac{1}{T}\biggl[
-1-\EulerGamma-\frac{\EulerGamma^2}{2}-\frac{\pi^2}{8}-\ln 2\ln\pi-\frac{\ln^2(\pi)}{2}\nonumber\\
&&\quad-\ln T\ln(2\pi)+\ln(2\pi T)(1+\EulerGamma)-\frac{\ln^2(T)}{2}
\biggr]\nonumber\\
&&+\frac{1}{2}\left(11+6\EulerGamma-8\ln 2-6\ln(2\pi T)\right)\nonumber \; ,\\
F_{110}&=&\frac{1}{T}\biggl[
-1-\EulerGamma-\frac{\EulerGamma^2}{2}-\frac{\pi^2}{8}-\ln 2\ln\pi -\frac{\ln^2(\pi)}{2}\nonumber\\
&&\quad -\ln T \ln(2\pi)+\ln(2\pi T)(1+\EulerGamma)-\frac{\ln^2(T)}{2}
\biggr]\nonumber\\
&& +\frac{1}{2}\left(7+2\EulerGamma-6\ln 2-2\ln(2\pi T)\right)\nonumber \; , \\
F_{101}&=&-\frac{\pi^2}{4T}+5+2\EulerGamma-2\ln 2-2\ln(2\pi T) \; .
\end{eqnarray}
The corrections to the last four terms are of the order $\mathcal{O}[T\ln T,T]$.
We recall that $F_{100}=F_{001}$, $F_{200}=F_{002}$, and $F_{110}=F_{011}$ due to symmetries, 
see Eq.~(\ref{SymmetryOfFtau1tau2tau3}).

Therefore, with Eq.~(\ref{xBintermsofFkreisfunctions}), the final result of the term $x_B$ reads
\begin{equation}
\label{SummaryXB}
x_B(B,T)= x_{B,1}(T) +x_{B,2}B^2 +\frac{B^2}{T} x_{B,3}(T)  +\mathcal{O}[B^3]
\end{equation}
with
\begin{eqnarray}
   x_{B,1}(T) &=&    -\frac{\pi^2}{2}-6\ln^2(2)+2\pi^2 T+2\pi^2 T^2(\ln 2 -1)\nonumber \\
&& + \mathcal{O}[T^4,T^4\ln T] \; , 
\end{eqnarray}
and
\begin{eqnarray}
x_{B,2} &=&-8+6\ln2  + \mathcal{O}[T,T\ln T]\; , \nonumber \\
x_{B,3}(T) &=& 8+8\EulerGamma+4\EulerGamma^2-\frac{\pi^2}{6}+4\ln^2(2)\nonumber\\
&&+8\ln 2 \ln\pi+4\ln^2(\pi)-8\ln(2\pi T)(1+\EulerGamma)\nonumber\\
&&+8\ln(2\pi)\ln T +4\ln^2(T)\; .
\end{eqnarray}

\section{Approximation of the term \texorpdfstring{$\mathbf{x_A}$}{xA}}
\label{sm:xAterm-derivation}

In this section, we derive an analytic expression of the term $x_A$ that occurs in Sect.~\mainsectionRearrangement in the limit $0\ll B\ll T\ll 1$. 
This term can be evaluated numerically as well, see Sect.~\mainsectionnumericalcheckxA of the main paper.

\subsection{Evaluation of the \texorpdfstring{$\mathbf{x_A}$}{xA} term}
\label{chapter:evaluation-of-the-xA-term}

\subsubsection{Formal expansion of \texorpdfstring{${x_A}$}{xA}}

According to 
Eq.~(\mainxADefinitionthirdorder) 
and Eq.~(\mainxBDefinitionthirdorder) 
of the main paper, we have [$j_\perp=\rho_0(0)J_\perp$, $j_z=\rho_0(0)J_z$, $\beta=1/T$]
\begin{eqnarray}\label{x3Aeq}
x_A&=&T\tilde{M}_0(B)\int_0^\beta\rmd\lambda_3\,\int_0^{\lambda_3}\rmd\lambda_2\,\int_0^{\lambda_2}\rmd\lambda_1\,\nonumber\\
&&\times \Bigl(  W^+ \tilde{f}^2(\lambda_3-\lambda_1,B)- W^- \tilde{f}^2(\lambda_3-\lambda_1,-B)\nonumber\\
&&\quad-W^+\tilde{f}^2(\lambda_2-\lambda_1,B)+ W^-\tilde{f}^2(\lambda_2-\lambda_1,-B)\nonumber\\
&& \quad- W^+\tilde{f}^2(\lambda_3-\lambda_2,B)+ W^- \tilde{f}^2(\lambda_3-\lambda_2,-B)\Bigr)\nonumber\\
\end{eqnarray}
with $\rhotilde(0)=\rho_0(\epsilon)/\rho_0(0)$, and $W^{\pm}(B,T)$ and $f(\lambda,B)$ are defined in Eq.~(\ref{eq:Wplusminus-and-f-definition}). Note that
\begin{eqnarray}
\tilde{M}_0(B)=\frac{M_0(B)}{\rho_0(0)}\approx 2 B
\end{eqnarray}
for $B\ll1$ with $M_0(B)$ from 
Eq.~(\mainMzerodefinition) 
of the main paper.
We define ($\tau=\pm 1$ represents the sign of the magnetic field)
\begin{eqnarray}
x_A^{\alpha,\tau}&=&T\int_0^\beta\rmd\lambda_3\,\int_0^{\lambda_3}\rmd\lambda_2\,\int_0^{\lambda_2}\rmd\lambda_1\,\tilde{f}^2(\lambda_3-\lambda_1,\tau B)\;,\nonumber\\
x_A^{\beta,\tau}&=&T\int_0^\beta\rmd\lambda_3\,\int_0^{\lambda_3}\rmd\lambda_2\,\int_0^{\lambda_2}\rmd\lambda_1\,\tilde{f}^2(\lambda_2-\lambda_1,\tau B)\;,\nonumber\\
x_A^{\gamma,\tau}&=&T\int_0^\beta\rmd\lambda_3\,\int_0^{\lambda_3}\rmd\lambda_2\,\int_0^{\lambda_2}\rmd\lambda_1\,\tilde{f}^2(\lambda_3-\lambda_2,\tau B)\;,
\nonumber\\
\label{eq:xAabgoftau}
\end{eqnarray}
to find
\begin{eqnarray}\label{x_AapproxOrigin}
x_A&\approx& 2 B\Big(W^+x_A^{\alpha,+}-W^-x_A^{\alpha,-}-W^+x_A^{\beta,+}\nonumber\\
&&\qquad+ W^-x_A^{\beta,-}-W^+x_A^{\gamma,+}+W^- x_A^{\gamma,-}\Big)\;.\nonumber\\
\end{eqnarray}
Since $M_0(B)\sim B$ we only need the expansion of the remaining terms up to and including the linear order in $B$.

We define
\begin{eqnarray}
g_A^\alpha&=&\int_0^\beta\rmd\lambda_3\,\int_0^{\lambda_3}\rmd\lambda_2\,\int_0^{\lambda_2}\rmd\lambda_1\,e^{(\epsilon_1+\epsilon_2)(\lambda_3-\lambda_1)}\nonumber\\
g_A^\beta&=&\int_0^\beta\rmd\lambda_3\,\int_0^{\lambda_3}\rmd\lambda_2\,\int_0^{\lambda_2}\rmd\lambda_1\,e^{(\epsilon_1+\epsilon_2)(\lambda_2-\lambda_1)}\nonumber\\
g_A^\gamma&=&\int_0^\beta\rmd\lambda_3\,\int_0^{\lambda_3}\rmd\lambda_2\,\int_0^{\lambda_2}\rmd\lambda_1\,e^{(\epsilon_1+\epsilon_2)(\lambda_3-\lambda_2)}\;.\nonumber\\
\end{eqnarray}
This permits to rewrite the terms in Eq.~(\ref{eq:xAabgoftau}) in the form ($j=\alpha,\beta,\gamma$)
\begin{eqnarray}\label{xAjtau-Def}
x_A^{j,\tau}&=&T\int_{-1}^{1}\rmd\epsilon_1\,\int_{-1}^1\rmd\epsilon_2\\
&&\qquad\times \frac{\rhotilde(\epsilon_1)}{1+e^{\beta(\epsilon_1-\tau B)}}\frac{\rhotilde(\epsilon_2)}{1+e^{\beta(\epsilon_2-\tau B)}} g_A^j\;.\nonumber
\end{eqnarray}
Note that
\begin{eqnarray}
u(\lambda_3)&=&\int_0^{\lambda_3}\rmd\lambda_2\,\int_0^{\lambda_2}\rmd\lambda_1\,e^{(\epsilon_1+\epsilon_2)(\lambda_3-\lambda_2)}\nonumber\\
&=&\int_0^{\lambda_3}\rmd\lambda_2\,e^{(\epsilon_1+\epsilon_2)(\lambda_3-\lambda_2)}\lambda_2\nonumber\\
&=&-\frac{\lambda_3}{\epsilon_1+\epsilon_2}-\frac{1}{(\epsilon_1+\epsilon_2)^2}+\frac{e^{(\epsilon_1+\epsilon_2)\lambda_3}}{(\epsilon_1+\epsilon_2)^2}\nonumber\\
&=&\int_0^{\lambda_3}\rmd\lambda_2\,\left(-\frac{1}{\epsilon_1+\epsilon_2}+\frac{e^{(\epsilon_1+\epsilon_2)\lambda_2}}{\epsilon_1+\epsilon_2}\right)\nonumber\\
&=&\int_0^{\lambda_3}\rmd\lambda_2\,\int_0^{\lambda_2}\rmd\lambda_1\,e^{(\epsilon_1+\epsilon_2)(\lambda_2-\lambda_1)}\;,
\end{eqnarray}
which implies 
\begin{equation}
g_A^\beta\equiv g_A^\gamma
\end{equation}
and thus
\begin{equation}
x_A^{\beta,\tau}\equiv x_A^{\gamma,\tau}\;.
\end{equation}
Therefore, Eq.~(\ref{x_AapproxOrigin}) reduces to
\begin{eqnarray}\label{xAreduce}
x_A&\approx& 2 B\Big(W^+x_A^{\alpha,+}-W^-x_A^{\alpha,-}\nonumber\\
&&\qquad-2W^+x_A^{\beta,+}+2 W^-x_A^{\beta,-}\Big)
\end{eqnarray}
with $x_A^{\alpha,\tau}$ from Eq.~(\ref{xAjtau-Def}) and, with the help of \textsc{Mathematica}~\cite{Mathematica12},
\begin{eqnarray}
g_A^\alpha&=&\frac{1}{T}\frac{1}{(\epsilon_1+\epsilon_2)^2}+\frac{1}{T}\frac{e^{\beta(\epsilon_1+\epsilon_2)}}{(\epsilon_1+\epsilon_2)^2}\nonumber\\
&&-\frac{2e^{\beta(\epsilon_1+\epsilon_2)}}{(\epsilon_1+\epsilon_2)^3}+\frac{2}{(\epsilon_1+\epsilon_2)^3}\;,\nonumber\\
g_A^\beta&=&-\frac{1}{2T^2}\frac{1}{(\epsilon_1+\epsilon_2)}-\frac{1}{T}\frac{1}{(\epsilon_1+\epsilon_2)^2}\nonumber\\
&&-\frac{1}{(\epsilon_1+\epsilon_2)^3}+\frac{e^{\beta(\epsilon_1+\epsilon_2)}}{(\epsilon_1+\epsilon_2)^3} \; .
\end{eqnarray}
As in the previous section, we introduce the function $H_T$ whose behavior is well known at low temperatures,
\begin{eqnarray}
H_T(\omega,\tau B)&=&\int_{-1}^1\rmd\omega\,\frac{\rhotilde(\omega)}{1+e^{\beta(\omega-\tau B)}}\frac{1}{\omega+\epsilon}\nonumber\\
&=&-\ln(1-\omega)+\ln(2\pi T)+\Rbar\left(\frac{\omega+\tau B}{T}\right)\;,\nonumber\\
\end{eqnarray} 
see section~\ref{chapter:properties-of-the-Rbar-function}.
Thus,
\begin{eqnarray}
x_A^{\alpha,\tau}=\sum_{i=1}^4 R_i^\tau\;,\qquad x_A^{\beta,\tau}=\sum_{i=5}^8 R_i^\tau
\end{eqnarray}
with
\begin{eqnarray}
R_1^\tau&=&\int_{-1}^1\rmd\epsilon_1\int_{-1}^1\rmd\epsilon_2\,\frac{\rhotilde(\epsilon_1)}{1+e^{\beta(\epsilon_1-\tau B)}}\nonumber\\
&&\quad\times \frac{\rhotilde(\epsilon_2)}{1+e^{\beta(\epsilon_2-\tau B)}}\frac{1}{(\epsilon_1+\epsilon_2)^2}\nonumber\\
&=&\int_{-1}^1\rmd\omega\,\frac{\rhotilde(\omega)}{1+e^{\beta(\omega-\tau B)}}\left(- H_T^\prime(\omega,\tau B)\right)\;,\nonumber\\
R_2^\tau&=&\int_{-1}^1\rmd\epsilon_1\int_{-1}^1\rmd\epsilon_2\,\frac{\rhotilde(\epsilon_1)}{1+e^{\beta(\epsilon_1-\tau B)}}\nonumber\\
&&\quad\times \frac{\rhotilde(\epsilon_2)}{1+e^{\beta(\epsilon_2-\tau B)}}\frac{e^{\beta(\epsilon_1+\epsilon_2)}}{(\epsilon_1+\epsilon_2)^2}\nonumber\\
&=&e^{2\beta \tau B}\int_{-1}^1\rmd\epsilon_1\,\int_{-1}^1\rmd\epsilon_2\,\frac{\rhotilde(\epsilon_1)}{1+e^{\beta(\epsilon_1+\tau B)}}\nonumber\\
&&\qquad\quad\times \frac{\rhotilde(\epsilon_2)}{1+e^{\beta(\epsilon_2+\tau B)}}\frac{1}{(\epsilon_1+\epsilon_2)^2}\nonumber\\
&=&e^{2B\tau B}\int_{-1}^1\rmd\omega\,\frac{\rhotilde(\omega)}{1+e^{\beta(\omega+\tau B)}}\left(-H_T^\prime(\omega,-\tau B)\right)\;,\nonumber\\
\end{eqnarray}
where we substituted $\epsilon_i\to-\epsilon_i$, $i=1,2$. Recall, $\rhotilde(\epsilon)=\rhotilde(-\epsilon)$ is symmetric. Next, we have
\begin{eqnarray}
R_3^\tau&=&-2 T \int_{-1}^1\rmd\epsilon_1\int_{-1}^1\rmd\epsilon_2\,\frac{\rhotilde(\epsilon_1)}{1+e^{\beta(\epsilon_1-\tau B)}}\nonumber\\
&&\qquad\times \frac{\rhotilde(\epsilon_2)}{1+e^{\beta(\epsilon_2-\tau B)}}\frac{e^{\beta(\epsilon_1+\epsilon_2)}}{(\epsilon_1+\epsilon_2)^3}\nonumber\\
&=&Te^{2\beta\tau B}\int_{-1}^1\rmd \omega\,\frac{\rhotilde(\omega)}{1+e^{\beta(\omega+\tau B)}}\left(H_T^{\prime\prime}(\omega,-\tau B)\right)\nonumber\\
\end{eqnarray}
and 
\begin{eqnarray}
R_4^\tau&=&2T\int_{-1}^1\rmd\epsilon_1\int_{-1}^1\rmd\epsilon_2\,\frac{\rhotilde(\epsilon_1)}{1+e^{\beta(\epsilon_1-\tau B)}}\nonumber\\
&&\qquad\times \frac{\rhotilde(\epsilon_2)}{1+e^{\beta(\epsilon_2-\tau B)}}\frac{1}{(\epsilon_1+\epsilon_2)^3}\nonumber\\
&=&T\int_{-1}^1\rmd\omega\,\frac{\rhotilde(\omega)}{1+e^{\beta(\omega-\tau B)}}\left(H_T^{\prime\prime}(\omega,\tau B)\right)\;.\nonumber\\
\end{eqnarray}
Lastly,
\begin{eqnarray}
R_5^\tau&=&-\frac{1}{2T}\int_{-1}^1\rmd\epsilon_1\,\int_{-1}^1\rmd\epsilon_2\,\frac{\rhotilde(\epsilon_1)}{1+e^{\beta(\epsilon_1-\tau B)}}\nonumber\\
&&\qquad\times\frac{\rhotilde(\epsilon_2)}{1+e^{\beta(\epsilon_2-\tau B)}}\frac{1}{\epsilon_1+\epsilon_2}\\
&=&-\frac{1}{2T}\int_{-1}^1\rmd\omega\,\frac{\rhotilde( \omega)}{1+e^{\beta(\omega-\tau B)}}H_T(\omega,\tau B)\; .\nonumber
\end{eqnarray}
A closer inspection shows that
\begin{equation}
\label{x3AR6toR8}
R_6^\tau=-R_1^\tau\;, \quad R_7^\tau=-\frac{1}{2}R_4^\tau\;, \quad R_8^\tau=-\frac{1}{2}R^\tau_3\;.
\end{equation}
Therefore, the remaining task is the evaluation of the fives terms $R_i^\tau$, $i=1,\dots,5$. 

\subsection{Calculation of the functions \texorpdfstring{$\mathbf{R_j^{\boldsymbol{\tau}}}$}{Rjtau} of the term \texorpdfstring{$\mathbf{x_A}$}{xA}}
\label{CalcOfRj}

The following calculations are lengthy and we proceed analogously to the calculation of the $x_B$ term, see Sect.~\ref{x3BGeneralProcedure}. 
We first expand the function $H_T$ for small magnetic fields~$B$, see Sect.~\ref{HTforSmallFields}, 
and calculate the resulting integrals separately in Sect.~\ref{chapter:derivation-of-the-integrals}. 
Important properties of the function $\Rbar(x)$ can be found in Sect.~\ref{chapter:properties-of-the-Rbar-function}.

\subsubsection{Calculation of the \texorpdfstring{$R_1^{\tau}$}{R1tau} function}

We have
\begin{eqnarray}
R_1^\tau&=&\int_{-1}^1\rmd\omega\,\frac{\rhotilde(\omega)}{1+e^{\beta(\omega-\tau B)}}\left(-H_T^\prime(\omega,\tau B)\right)\nonumber\\
&=&R_{1,a}+R^\tau_{2,b}+R^\tau_{3,c}
\end{eqnarray}
with 
\begin{eqnarray}
R_{1,a}&=&-\int_{-1}^1\rmd \omega\,\frac{1}{1+e^{\omega/T}}\left(\frac{1}{1-\omega}+\frac{\Rbar^\prime(\omega/T)}{T}\right)\;,\nonumber\\
R_{1,b}^{\tau}&=&-\frac{\tau B}{T}\int_{-1}^1\rmd\omega\,\frac{e^{\omega/T}}{(1+e^{\omega/T})^2}\nonumber\\
&&\qquad\quad\times\left(\frac{1}{1-\omega}+\frac{\Rbar^\prime(\omega/T)}{T}\right)\;,\nonumber\\
R_{1,c}^{\tau}&=&-\tau B\int_{-1}^1\rmd\omega\,\frac{1}{1+e^{\omega/T}}\frac{\Rbar^{\prime\prime}(\omega/T)}{T^2}\;.
\end{eqnarray}
We thus find
\begin{eqnarray}
R_{1,a}&=&-(T\alpha_9+\mathcal{I}_5+2\mathcal{I}_8)\nonumber\\
&=&1+\EulerGamma-\ln2-\ln(2\pi T)-\frac{\pi^2 T^2}{3}+\mathcal{O}[T^4]\;,\nonumber\\
R_{1,b}^\tau&=&-\tau B \alpha_{10} = -\tau B\left(1+\frac{\pi^2 T^2}{3}\right)+\mathcal{O}[T^4]\;,\nonumber\\
R_{1,c}^\tau&=&-\tau B\frac{\mathcal{I}_{12}}{T}=-\tau B\left(1+\frac{\pi^2T^2}{3}\right)+\mathcal{O}[T^4]\;.\nonumber\\
\end{eqnarray}
In $R_{1,b}^\tau$ we used that $\Rbar^\prime(x)$ is odd in~$x$. In $R_{1,c}^\tau$ we used that $\Rbar^{\prime\prime}(x)$ is even in $x$. 
Thus, we have
\begin{eqnarray}
R_1^\tau&=&1+\EulerGamma-\ln 2-\ln(2\pi T)-\frac{\pi^2T^2}{3}\nonumber\\
&&-2\tau B\left(1+\frac{\pi^2T^2}{3}\right)+\mathcal{O}[T^4]+\mathcal{O}[B^2]\;.\nonumber\\
\end{eqnarray}

\subsubsection{Calculation of the \texorpdfstring{$R_2^{\tau}$}{R2tau} function}
We start from
\begin{eqnarray}
R_2^\tau&=&e^{2\beta \tau B}\int_{-1}^1\rmd\omega\,\frac{\rhotilde(\omega)}{1+e^{\beta(\omega+\tau B)}}\left(-H_T^\prime(\omega,-\tau B)\right)\nonumber\\
&=&R_{2,a}+R^\tau_{2,b}+R^\tau_{2,c}+R^\tau_{2,d}\;,
\end{eqnarray}
where
\begin{eqnarray}
R_{2,a}&=&-\int_{-1}^1\rmd\omega\,\frac{1}{1+e^{\omega/T}}\left(\frac{1}{1-\omega}+\frac{\Rbar^\prime(\omega/T)}{T}\right)\nonumber \\
&=&R_{1,a}\;,\nonumber\\
R_{2,b}^\tau&=&\frac{\tau B}{T}\int_{-1}^1\rmd\omega\,\frac{e^{\omega/T}}{(1+e^{\omega/T})^2}\left(\frac{1}{1-\omega}+\frac{\Rbar^\prime(\omega/T)}{T}\right)\nonumber\\
&=&-R_{1,b}^\tau\:,
\end{eqnarray}
and
\begin{eqnarray}
R_{2,c}^\tau&=&-\frac{2\tau B}{T}\int_{-1}^1\rmd\omega\,\frac{1}{1+e^{\omega/T}}\left(\frac{1}{1-\omega}+\frac{\Rbar^\prime(\omega/T)}{T}\right)\nonumber\\
&=&\frac{2\tau B}{T}R_{1,a}\;,\nonumber\\
R_{2,d}^\tau&=&\tau B\int_{-1}^1\rmd\omega\,\frac{1}{1+e^{\omega/T}}\frac{\Rbar^{\prime\prime}(\omega/T)}{T^2}\nonumber \\
&=&-R_{1,c}^\tau\;.
\end{eqnarray}
Thus,
\begin{eqnarray}
R_2^\tau&=&1+\EulerGamma-\ln2-\ln(2\pi T)-\frac{\pi^2 T^2}{3}\nonumber\\
&&+\tau B\left(\frac{2}{3}(3+\pi^2 T^2)\right)\nonumber\\
&&+\frac{2\tau B}{T}\left(1+\EulerGamma-\ln2-\ln(2\pi T)-\frac{\pi^2 T^2}{3}\right)\nonumber\\
&&+\mathcal{O}[T^4]\;.
\end{eqnarray}

\subsubsection{Calculation of the \texorpdfstring{$R_3^{\tau}$}{R3tau} function}
We have
\begin{eqnarray}
R_3^\tau&=&Te^{2\beta\tau B}\int_{-1}^1\rmd \omega\,\frac{\rhotilde(\omega)}{1+e^{\beta(\omega+\tau B)}}\left(H_T^{\prime\prime}(\omega,-\tau B)\right)\nonumber\\
&=&R_{3,a}+R_{3,b}^\tau+R_{3,c}^\tau+R_{3,d}^\tau
\end{eqnarray}
with 
\begin{eqnarray}
R_{3,a}&=&T\int_{-1}^1\rmd\omega\,\frac{1}{1+e^{\omega/T}}\left(\frac{1}{(1-\omega)^2}+\frac{\Rbar^{\prime\prime}(\omega/T)}{T^2}\right)\;,\nonumber\\
R_{3,b}^\tau&=&-\tau B\int_{-1}^1\rmd\omega\,\frac{e^{\omega/T}}{(1+e^{\omega/T})^2}\nonumber\\
&&\qquad\times\left(\frac{1}{(1-\omega)^2}+\frac{\Rbar^{\prime\prime}(\omega/T)}{T^2}\right)\;,\nonumber\\
R_{3,c}^\tau&=&2\tau B\int_{-1}^1\rmd\omega\,\frac{1}{1+e^{\omega/T}}\left(\frac{1}{(1-\omega)^2}+\frac{\Rbar^{\prime\prime}(\omega/T)}{T^2}\right)\nonumber\\
&=&\frac{2\tau B}{T}R_{3,a}\;,\nonumber\\
R_{3,d}^\tau&=&-\frac{\tau B}{T^2} \int_{-1}^1\rmd\omega\,\frac{1}{1+e^{\omega/T}}\Rbar^{\prime\prime\prime}(\omega/T)\;.
\end{eqnarray}
We find
\begin{eqnarray}
R_{3,a}&=&T^2\alpha_{11}+\mathcal{I}_{12}\nonumber\\
&=&\frac{3T}{2}+\frac{2\pi^2T^3}{3}+\mathcal{O}[T^5]\;,\nonumber\\
R_{3,b}^\tau&=&-\tau B(T\alpha_{12}+\frac{2}{T}\mathcal{I}_{35})\nonumber\\
&=&-\tau B\left(\frac{3\zeta(3)}{2\pi^2 T}+T+\pi^2 T^3\right)+\mathcal{O}[T^5]\;,\nonumber\\
R_{3,c}^\tau&=&\frac{2\tau B}{T}R_{3,a}\nonumber\\
&=&\tau B\left(3+\frac{4\pi^2T^2}{3}\right)+\mathcal{O}[T^4]\;,\nonumber\\
R_{3,d}^\tau&=&-\frac{B\tau}{T} (\mathcal{I}_{24}+2\mathcal{I}_{27})\nonumber\\
&=&-B\tau\left(\frac{3\zeta(3)}{2\pi^2 T}+T\right)+\mathcal{O}[T^3]\;.
\end{eqnarray}
Thus, we arrive at
\begin{equation}
R_3^\tau=\frac{3T}{2}+\tau B\left(-\frac{3\zeta(3)}{\pi^2 T}+3-2 T+\frac{4\pi^2T^2}{3}\right)+\mathcal{O}[T^3]\;.
\end{equation}

\subsubsection{Calculation of the \texorpdfstring{$R_4^{\tau}$}{R4tau} function}

We have
\begin{eqnarray}
R_4^\tau&=&T\int_{-1}^1\rmd\omega\,\frac{\rhotilde(\omega)}{1+e^{(\omega-\tau B)/T}}\left(H_T^{\prime\prime}(\omega,\tau B)\right)\nonumber\\
&=&R_{4,a}+R_{4,b}^\tau+R_{4,c}^\tau
\end{eqnarray}
with
\begin{eqnarray}
R_{4,a}&=&T\int_{-1}^1\rmd\omega\,\frac{1}{1+e^{\omega/T}}\left(\frac{1}{(1-\omega)^2}+\frac{\Rbar^{\prime\prime}(\omega/T)}{T^2}\right)\nonumber\\
&=&R_{3,a}\;,
\end{eqnarray}
and
\begin{eqnarray}
R_{4,b}^\tau&=&\tau B \int_{-1}^1\rmd\omega\,\frac{e^{\omega/T}}{(1+e^{\omega/T})^2}\nonumber\\
&&\qquad\times\left(\frac{1}{(1-\omega)^2}+\frac{\Rbar^{\prime\prime}(\omega/T)}{T^2}\right)\nonumber\\
&=&-R_{3,b}^\tau\;,\nonumber\\
R_{4,c}^\tau&=&\frac{\tau B}{T^2}\int_{-1}^1\rmd\omega\,\frac{1}{1+e^{\omega/T}}\Rbar^{\prime\prime\prime}(\omega/T)\nonumber\\
&=&-R_{3,d}^\tau\;.
\end{eqnarray}
Thus,
\begin{equation}
R_4^\tau=\frac{3T}{2}+\tau B\left(2T+\frac{3 \zeta(3)}{\pi^2 T}\right)+\mathcal{O}[T^3]\;.
\end{equation}

\subsubsection{Calculation of the \texorpdfstring{$R_5^{\tau}$}{R5tau} function}
We have
\begin{eqnarray}
R_5^\tau&=&-\frac{1}{2T}\int_{-1}^1\rmd\omega\,\frac{\rhotilde(\omega)}{1+e^{\beta(\omega-\tau B)}}H_T(\omega,\tau B)\nonumber\\
&=&R_{5,a}+R_{5,b}^\tau+R_{5,c}^\tau
\end{eqnarray}
with
\begin{eqnarray}
R_{5,a}&=&-\frac{1}{2T}\int_{-1}^1\rmd\omega\,\frac{1}{1+e^{\omega/T}}\nonumber\\
&&\quad\qquad\times\left(\ln(2\pi T)-\ln(1-\omega)+\Rbar(\omega/T)\right)\;,\nonumber\\
R_{5,b}^\tau&=&-\frac{\tau B}{2T^2}\int_{-1}^1\rmd\omega\,\frac{e^{\omega/T}}{(1+e^{\omega/T})^2}\nonumber\\
&&\quad\qquad\times\left(\ln(2\pi T)-\ln(1-\omega)+\Rbar(\omega/T)\right)\;,\nonumber\\
R_{5,c}^\tau&=&-\frac{\tau B}{2 T^2}\int_{-1}^1\rmd\omega\,\frac{1}{1+e^{\omega/T}}\Rbar^\prime(\omega/T)\,.
\end{eqnarray}
We find
\begin{eqnarray}
R_{5,a}&=&-\frac{1}{2}\left(\ln(2\pi T)\alpha_{13}-\alpha_{14}+\mathcal{I}_1\right)\nonumber\\
&=&\frac{\ln2}{T}-\frac{\pi^2 T}{6}+\mathcal{O}[T^3]\;,\nonumber\\
R_{5,b}^\tau&=&-\frac{\tau B}{2}\left(\frac{2\ln(2\pi T)}{T}\alpha_5-\frac{\alpha_{15}}{T}+\frac{2}{T}\mathcal{I}_{16}\right)\nonumber\\
&=&\tau B\left(-\frac{\pi^2 T}{12}+\frac{1+\EulerGamma-\ln(2\pi T)}{2T}\right)+\mathcal{O}[T^3]\;,\nonumber\\
R_{5,c}^\tau&=&-\frac{\tau B}{2T}\left(\mathcal{I}_5+2I_8\right)\nonumber\\
&=&\tau B\left(-\frac{\pi^2 T}{12}+\frac{1+\EulerGamma-\ln(2\pi T)}{2T}\right)+\mathcal{O}[T^3]\nonumber\\
&=&R_{5,b}^\tau\;.
\end{eqnarray}
Thus,
\begin{eqnarray}
R_5^\tau&=&\frac{\ln2}{T}-\frac{\pi^2 T}{6}\nonumber\\
&&+\tau B\left(\frac{1+\EulerGamma-\ln(2\pi T)}{T}-\frac{\pi^2T}{6}\right)+\mathcal{O}[T^3]\;.\nonumber\\
\end{eqnarray}

\subsection{Summary of the term \texorpdfstring{$\mathbf{x_A}$}{xA}}

We define
\begin{eqnarray}\label{xAalpha-and-xAbeta-Definition}
x_A^\alpha &=& W^+ x_A^{\alpha,+}- W^- x_A^{\alpha,-}\;,\nonumber\\
x_A^\beta&=&-2W^+x_A^{\beta,+}+2 W^-x_A^{\beta,-}
\end{eqnarray}
which leads to
\begin{equation}
x_A\approx 2 B(x_A^\alpha+x_A^\beta)
\end{equation}
for small magnetic fields~$B$.
Using the analytic expressions of the functions $R_i^\tau$ and Eq.~(\ref{x3AR6toR8}) we find
\begin{eqnarray}
x_A^{\alpha,\tau}&=&\sum_{i=1}^4R_i^\tau\nonumber\\
&=&2(1+\EulerGamma-\ln 2-\ln(2\pi T))\nonumber\\
&&+3T+\mathcal{O}[T^2]\nonumber\\
&&+\tau B\biggl(
\frac{2}{T}\left(1+\EulerGamma-\ln2-\ln(2\pi T)\right)\nonumber\\
&&\quad\qquad+3-\frac{2\pi^2}{3}T +\mathcal{O}[T^2]\biggr)+\mathcal{O}[B^3]\;,\nonumber\\
&& \\
x_A^{\beta,\tau}&=&\sum_{i=5}^8R_i^\tau\nonumber\\
&=&\frac{\ln 2}{T}-1-\EulerGamma+\ln 2+\ln(2\pi T)\nonumber\\
&&-\frac{T}{6}(9+\pi^2)+\frac{\pi^2 T^2}{6}+\mathcal{O}[T^3]\nonumber\\
&&+\tau B\biggl(\frac{1}{T}\left(1+\EulerGamma-\ln(2\pi T)\right)+\frac{1}{2}\nonumber\\
&&\quad\qquad-\frac{\pi^2}{6}T+\mathcal{O}[T^2]\biggr)+\mathcal{O}[B^2]\; ,
\end{eqnarray}
and
\begin{eqnarray}
x_A^\alpha &=&\mathcal{O}[T]\;, \label{eq:proveinmain}\\
x_A^\beta &=& B\left(\frac{2\ln 2 }{T^2}-\frac{1}{3}\left(12+\pi^2\right)\right)\nonumber\\
&&+\frac{B}{T}\bigl(-4-4\EulerGamma+4\ln(2\pi T)+2\ln 2 \bigr)\nonumber\\
&&+{\pi^2 T}+\mathcal{O}[B^2]\;.
\end{eqnarray}
The proof of Eq.~(\ref{eq:proveinmain}) can also be found in Appendix~\mainxAalphatermvanishes\ of the main paper.

Summarizing the contributions to the term $x_A$ term leads to 
\begin{eqnarray}
\label{eqSummaryXA}
x_A(B,T)&=&B^2\bigg(4\ln 2\frac{1}{T^2}\nonumber\\
&&+\frac{1}{T}\Big(8\big(\ln(2\pi T)-1-\EulerGamma\big)+4\ln2\Big)\nonumber\\
&&-\frac{2}{3}(12+\pi^2)+\mathcal{O}[T]\bigg)+\mathcal{O}[B^3]\;.\nonumber\\
\end{eqnarray}

\section{Properties of the functions \texorpdfstring{$\mathbf{H_T(\boldsymbol{\omega},\boldsymbol{\tau} B)}$}{HT-Func} and \texorpdfstring{$\mathbf{\overline{R}(x)}$}{Rbar(x)}}
\label{chapter:properties-of-the-Rbar-function}

In this section we collect the most important properties of the functions $H_T(\omega,\tau B)$ and $\overline{R}(x)$. 
In addition, we give the function $H_T$ for a general density of states in the last subsection. 
This is essential for the perturbation theory computations for a general density of states.

\subsection{Function \texorpdfstring{$\mathbf{H_T(\boldsymbol{\omega},\boldsymbol{\tau} B)}$}{Function HT}}
\subsubsection{Definition}

The function $H_T$ is defined as
\begin{eqnarray}
H_T(\omega,\tau B)&=&\int_{-1}^1\frac{\rmd \epsilon}{1+e^{\beta(\epsilon-\tau B)}}\frac{1}{\omega+\epsilon}\nonumber\\
&=&-\ln(1-\omega)+\ln(2\pi T)+\Rbar\left(\frac{\omega+\tau B}{T}\right)\;.\nonumber\\
\label{eq:HT-function-definition}
\end{eqnarray}
The function $\Rbar(x)$ will be discussed in Sect.~\ref{The-Rbar-function}.

\subsubsection{Derivation of the semi-analytical representation}

To derive Eq.~(\ref{eq:HT-function-definition}), we start from ($b=\tau B$)
\begin{equation}
H_T(\omega,b)=\int_{-1}^1\,\frac{\rmd\epsilon}{1+e^{\beta(\epsilon-b)}}\frac{1}{\omega+\epsilon}\;.
\end{equation}
We note that for $\eta=0^+$
\begin{eqnarray}
&&{\rm Re}\left((-\rmi)\int_0^\infty\rmd t\,e^{\rmi(\omega+\epsilon+\rmi\eta)t}\right)\nonumber\\
&&\quad={\rm Re}\left((-\rmi)\frac{1}{\rmi(\omega+\epsilon+\rmi\eta)}e^{\rmi(\omega+\epsilon+\rmi\eta)t}\right)\nonumber\\
&&\quad={\rm Re}\left(\frac{1}{\omega+\epsilon-\rmi\eta}\right) \nonumber\\
&&\quad=P\left(\frac{1}{\omega+\epsilon}\right)\;,
\end{eqnarray}
where $P$ denotes the Cauchy principal value. Thus,
\begin{eqnarray}
H_T(\omega,b)&=&{\rm Re}\bigg[(-\rmi)\int_0^\infty\rmd t\,e^{\rmi(\omega+\rmi\eta)t}\nonumber\\
&&\qquad\times\int_{-1}^1\rmd\epsilon\,\frac{e^{\rmi\epsilon t}}{1+e^{\beta(\epsilon-b)}}\bigg]\nonumber\\
&=&{\rm Re}\left[(-\rmi)\int_0^\infty\rmd t\,e^{\rmi(\omega+\rmi\eta)t}e^{\rmi b t}\,\mathcal{H}\right]
\end{eqnarray}
with 
\begin{eqnarray}
\mathcal{H}&=&T\int_{(-1-b)/T}^{(1-b)/T}\rmd u\,\frac{1}{1+e^u}e^{\rmi u Tt}\nonumber\\
&=&T\int_0^{(1-b)/T}\frac{\rmd u}{1+e^u}e^{\rmi u T t}+T\int_{(-1-b)/T}^0\rmd u\,e^{\rmi u T t}\nonumber\\
&&-T\int_{(-1-b)/T}^0\frac{\rmd u}{1+e^{-u}}e^{\rmi u T t}
\end{eqnarray}
and thus
\begin{eqnarray}
\mathcal{H}&=&\frac{T}{\rmi T t}\left(1-e^{-\rmi t(1+b)}\right)+T\int_0^{(1-b)/T}\frac{\rmd u}{1+e^u}e^{\rmi u T t}\nonumber\\
&&-T\int_0^{(1+b)/T}\frac{\rmd u}{1+e^u}e^{-\rmi u T t}\nonumber\\
&=&\frac{1}{\rmi t}\left(1-e^{-\rmi t/(1+b)}\right)+\rmesc\nonumber\\
&&+T\int_0^\infty\frac{\rmd u}{1+e^u}2\rmi\sin(uTt)
\end{eqnarray}
where $\rmesc$ means `exponential small corrections'. Thus,
\begin{eqnarray}
\mathcal{H}&=&\frac{1}{\rmi t}\left(1-e^{-\rmi t(1+b)}\right)+\frac{\rmi }{t}\left(1-\frac{\pi T t}{\sinh(\pi Tt)}\right)\;.\nonumber\\
\end{eqnarray}
With $\mathcal{H}$ we find [$H_T=H_T(\omega,b)$]
\begin{eqnarray}
H_T&=&{\rm Re}\left(\int_0^\infty\frac{\rmd t}{t}\left(e^{\rmi(\omega-1+\rmi\eta)t}-e^{\rmi(\omega+b+\rmi \eta)t}\right)\right)\nonumber\\
&&+{\rm Re}\left(\int_0^\infty\frac{\rmd t}{t}\left(1-\frac{\pi T t}{\sinh(\pi T t)}\right)e^{\rmi(\omega+b+\rmi \eta)t}\right)\nonumber\\
&=&\int_0^\infty\frac{\rmd t}{t}\Big(\cos\left((\omega-1)t\right)-\cos\left((\omega+b)t\right)\Big)\nonumber\\
&&+\int_0^\infty\frac{\rmd t}{t}\left(1-\frac{\pi T t}{\sinh(\pi T t)}\right)\cos\left((\omega+b)t\right)\nonumber\\
&=&\ln\left|\frac{\omega+b}{1-\omega}\right|\nonumber\\
&&+\int_0^\infty\frac{\rmd u}{u}\cos\left(\left(\frac{\omega+b}{T}\right)u\right)\left(1-\frac{\pi u}{\sinh(\pi u)}\right)\nonumber\\
\end{eqnarray}
with the help of \textsc{Mathematica}~\cite{Mathematica12}. Finally,
\begin{eqnarray}
H_T&=&\ln\left|\frac{\omega+b}{1-\omega}\right|+R\left(\frac{\omega+b}{T}\right)\nonumber \\
&=&-\ln|1-\omega|+\ln(2\pi T)+\Rbar\left(\frac{\omega+b}{T}\right)
\end{eqnarray}
with 
\begin{eqnarray}
R(x)&=&\ln\left(\frac{2\pi}{|x|}\right)+\Rbar(x)\;,\nonumber\\
\Rbar(x)&=&\frac{1}{2}\left(\psi\left(\frac{1}{2}+\frac{\rmi x}{2\pi}\right)+\psi\left(\frac{1}{2}-\frac{\rmi x}{2\pi}\right)\right)\;,\nonumber\\
&& \label{eq:defRbarfirsttime}
\end{eqnarray}
where $\psi(x)=\Gamma^\prime(x)/\Gamma(x)$ is the digamma function and $\Gamma(x)$ is the gamma function.
The functions $R(x)$ and $\Rbar(x)$ are further discussed in Sect.~\ref{The-Rbar-function}.

\subsubsection{Expansion for small magnetic fields}
\label{HTforSmallFields}
We want to evaluate the third order terms of the free energy up to and including the second order in $B$. Thus, assuming $-1<\omega<1$, we expand [$f'(\omega)=\partial f(\omega)/(\partial\omega)$]
\begin{eqnarray}\label{eq:HTFunctionsForSmallB}
H_T(\omega,\tau B)&=&H_T^{(0)}(\omega)+H_T^{(1)}(\omega)\tau B\nonumber\\
&&+H_T^{(2)}(\omega)(\tau B)^2 + \mathcal{O}[B^3]\;,\nonumber\\
H_T^\prime(\omega, \tau B)&=&(H_T^\prime)^{(0)}(\omega)+ (H_T^\prime)^{(1)}(\omega)\tau B \nonumber\\
&&+(H_T^\prime)^{(2)}(\tau B)^2+\mathcal{O}[B^3]\;,\nonumber\\
H_T^{\prime\prime}(\omega,\tau B)&=&(H_T^{\prime\prime})^{(0)}(\omega)+ (H_T^{\prime\prime})^{(1)}(\omega)\tau B\nonumber\\
&&+(H_T^{\prime\prime})^{(2)}(\tau B)^2+\mathcal{O}[B^3]
\end{eqnarray}
with 
\begin{eqnarray}\label{eq:HTFunctionsForSmallB2}
H_T^{(0)}(\omega)&=&-\ln(1-\omega)+\ln(2\pi T)+\Rbar(\omega/T)\;,\nonumber\\
H_T^{(1)}(\omega)&=&\frac{1}{T}\Rbar^\prime(\omega/T)\;,\nonumber\\
H_T^{(2)}(\omega)&=&\frac{1}{2T^2}\Rbar^{\prime\prime}(\omega/T)\;,\nonumber\\
(H_T^\prime)^{(0)}(\omega)&=&\frac{1}{1-\omega}+\frac{1}{T}\Rbar^\prime(\omega/T)\;,\nonumber\\
(H_T^\prime)^{(1)}(\omega)&=&\frac{1}{T^2}\Rbar^{\prime\prime}(\omega/T)\;,\nonumber\\
(H_T^{\prime})^{(2)}(\omega)&=&\frac{1}{2T^3}\Rbar^{\prime\prime\prime}(\omega/T)\;,\nonumber\\
(H_T^{\prime\prime})^{(0)}(\omega)&=&\frac{1}{(1-\omega)^2}+\frac{1}{T^2}\Rbar^{\prime\prime}(\omega)\;,\nonumber\\
(H_T^{\prime\prime})^{(1)}(\omega)&=&\frac{1}{T^3}\Rbar^{\prime\prime\prime}(\omega/T)\;,\nonumber\\
(H_T^{\prime\prime})^{(2)}(\omega)&=&\frac{1}{2 T^4}\Rbar^{(4)}(\omega/T)\;.
\end{eqnarray}
To calculate the function $H_T$, we only need to know the function $\Rbar(x)$.

\subsection{Function \texorpdfstring{$\mathbf{\Rbar(x)}$}{Rbar(x)}}
\label{The-Rbar-function}

\subsubsection{Definition}
We recall the definition of the function $\overline{R}(x)$ from Eq.~(\ref{eq:defRbarfirsttime}) 
\begin{equation}
\overline{R}(x)=\frac{1}{2}\left(\psi\left(\frac{1}{2}-\frac{\rmi x}{2\pi}\right)+\psi\left(\frac{1}{2}+\frac{\rmi x}{2\pi}\right)\right)\;,
\label{eq:Rbar-function-definition}
\end{equation}
where $\psi(x)=\Gamma^\prime(x)/\Gamma(x)$ is the digamma function and $\Gamma(x)$ is the gamma function. Obviously, $\overline{R}(x)$ is even, $\overline{R}(x)=\overline{R}(-x)$.

\subsubsection{Derivatives}
The $n$-th derivative of the function $\overline{R}(x)$ reads
\begin{eqnarray}
\overline{R}^{(n)}(x)&=&\frac{1}{2}\bigg(\left(\frac{\rmi}{2\pi}\right)^n\psi^{(n)}\left(\frac{1}{2}+\frac{\rmi x}{2\pi}\right)\nonumber\\
&&+\left(-\frac{\rmi}{2\pi}\right)^n\psi^{(n)}\left(\frac{1}{2}-\frac{\rmi x}{2\pi}\right)\bigg)\;,
\end{eqnarray}
where $\psi^{(n)}$ is the $n$-th derivative of the digamma function.

\subsubsection{Indefinite integrals}
\label{IndefiniteIntegrals}
The indefinite integral of the function $\Rbar(x)$ is the logarithm of the gamma function,
\begin{eqnarray}\label{IndefiniteIntegralOfRbar}
\mathcal{R}(x)&\equiv&\int\rmd x\,\Rbar(x)\nonumber\\
&=&\rmi \pi\left[\ln\left(\Gamma\left(\frac{1}{2}-\frac{\rmi x	}{2\pi}\right)\right)-\ln\left(\Gamma\left(\frac{1}{2}+\frac{\rmi x}{2\pi}\right)\right)\right]\;.\nonumber\\
\end{eqnarray}
The function $\mathcal{R}(x)$ has the properties
\begin{eqnarray}
\mathcal{R}(x)&=&-\mathcal{R}(-x)\;,\nonumber\\
\mathcal{R}(0)&=&0\;,\nonumber\\
\mathcal{R}(1/T)&=&-\left(\frac{1+\ln(2\pi T)}{T}\right)+\frac{\pi^2}{6}T+\frac{7\pi^4 T^3}{180}+\mathcal{O}(T^5)\nonumber\\
\label{eq:LogGamma-Function}
\end{eqnarray}
at low temperatures, $T\ll 1$.
For the indefinite integral over $\mathcal{R}(x)$ we find 
\begin{eqnarray}
\mathcal{S}(x)&=&\int\rmd x\,\mathcal{R}(x)\nonumber\\
&=&-2\pi^2\left(\Psi^{(2)}\left(\frac{\pi-\rmi x}{2\pi}\right)+\Psi^{(2)}\left(\frac{\pi+\rmi x}{2\pi}\right)\right)\;,\nonumber\\
\mathcal{S}(0)&=&-2\pi^2\left(3\ln\mathcal{A}+\frac{5\ln 2 }{12}+\frac{\ln \pi }{2}\right)\;,
\end{eqnarray}
where $\Psi^{(n)}(x)=\Gamma(-n,x)$ is the polygamma function at negative order. $\mathcal{A}$ is Glaisher's constant, $\mathcal{A}\approx 1.28243$. Moreover,
for $T\ll1$ we have
\begin{eqnarray}
\mathcal{S}(-1/T)&=&-\left(\frac{3+2\ln(2\pi T)}{4 T^2}\right)\nonumber\\
&&-\frac{\pi^2}{6}\left(24\ln \mathcal{A} +6\ln(2\pi)+\ln(2\pi T)\right)\nonumber\\
&&-\frac{7\pi^4T^2}{360}+\mathcal{O}[T^3]\;.
\end{eqnarray}

\subsubsection{Integral representations}
A crucial step for the computation of various integrals in this study is the use of integral representations of the function $\overline{R}(x)$,
\begin{eqnarray}
\overline{R}(x)&=&\int_0^\infty\frac{\rmd u}{u}\left(e^{-2\pi u}-\frac{\pi u}{\sinh(\pi u)}\cos(x u)\right)\;,\nonumber\\
\overline{R}(x)&=&-\ln(2\pi/|x|)\nonumber\\
&&+\int_0^\infty\frac{\rmd u}{u}\left(1-\frac{\pi u}{\sinh(\pi u)}\right)\cos(x u)\;,\nonumber\\
\overline{R}^\prime(x)&=&\int_0^\infty\rmd u\,\frac{\pi u}{\sinh(\pi u)}\sin(x u)\;,\nonumber\\
\overline{R}^{\prime\prime}(x)&=&\int_0^\infty\rmd u\,\frac{\pi u^2}{\sinh(\pi u)}\cos(x u)\;,\nonumber\\
\overline{R}^{\prime\prime\prime}(x)&=&-\int_0^\infty\rmd u\,\frac{\pi u^3}{\sinh(\pi u)}\sin(x u)\;.
\label{eq:integral-representations-of-the-rbar-function}
\end{eqnarray}
The second expression is obtained by using Eq.~(\ref{eq:Rbar-function-definition}) and the integral representation of the function $R(x)$,
\begin{equation}
R(x)=\int_0^\infty\frac{\rmd u}{u}\left(1-\frac{\pi u}{\sinh(\pi u)}\right)\cos(x u)\;.
\label{eq:integral-representations-of-the-r-function}
\end{equation}
Other useful relations are
\begin{eqnarray}
\frac{1}{1+e^x}&=&\frac{1}{\pi}\int_0^\infty\frac{\rmd u}{u}\sin(x u)\left(1-\frac{\pi u}{\sinh(\pi u)}\right)\;,\nonumber\\
x-\ln(1+e^x)&=&\frac{1}{\pi}\int_0^\infty\rmd u\,\frac{\cos(x u)}{u^2}\left(\frac{\pi u}{\sinh(\pi u)}-1\right)\;.\nonumber\\
\label{eq:other-useful-integral-representations}
\end{eqnarray}

\subsubsection{Series expansions and special function values}
We list some important function values of the $\Rbar(x)$ function at $x=0$,
\begin{eqnarray}
\Rbar(0)&=&-\EulerGamma-2\ln 2 =\Gamma(0,1/2)\;,\nonumber\\
\Rbar^\prime(0)&=&\Rbar^{\prime\prime\prime}(0)=0\;,\nonumber\\
\Rbar^{\prime\prime}(0)&=&\frac{7\zeta(3)}{2 \pi^2}=-\frac{\Gamma(2,1/2)}{4\pi^2}\;,\nonumber
\end{eqnarray}
where $\EulerGamma=0.577216$ is Euler's constant and $\zeta(x)$ is Riemann's zeta function. 

Moreover, for $T\ll 1$ we find
\begin{eqnarray}
\Rbar(1/T)&=&-\ln(2\pi T)-\frac{\pi^2}{6}T^2-\frac{7\pi^4}{60}T^4+\mathcal{O}[T^6]\;,\nonumber\\
\Rbar^\prime(1/T)&=&T+\frac{\pi^2}{3}T^3+\mathcal{O}(T^5)\;,\nonumber\\
\Rbar^{\prime\prime}(1/T)&=&-T^2+\mathcal{O}(T^4)\;,\nonumber\\
\Rbar^{\prime\prime\prime}(1/T)&=&2T^3+\mathcal{O}(T^5)\;.
\end{eqnarray}
Equivalently, for $x\to\infty$ we have
\begin{eqnarray}
\Rbar(x)&=&-\log\left(\frac{2\pi }{x}\right)-\frac{\pi^2}{6x^2}-\frac{7\pi^4}{60 x^4}+\mathcal{O}(x^{-6})\;,\nonumber\\
\Rbar^\prime(x)&=&\frac{1}{x}+\frac{\pi^2}{3x^3}+\mathcal{O}(x^{-5})\;,\nonumber\\
\Rbar^{\prime\prime}(x)&=&-\frac{1}{x^2}+\mathcal{O}(x^{-4})\;,\nonumber\\
\Rbar^{\prime\prime\prime}(x)&=&\frac{2}{x^3}+\mathcal{O}(x^{-5})\;.
\end{eqnarray}

\subsection{Function \texorpdfstring{$\mathbf{H_T}$}{HT} for a general density of states}
\label{htfunction-generaldos-finite_bfield}
For a general density of states, the function $H_T^{\rm g}(\omega,b)$, $b=\tau B$, can be expressed starting from the constant density of states,
\begin{eqnarray}
H_T^{\rm g}(\omega,b)&=&\int_{-1}^1\rmd\epsilon\,\frac{\rhotilde(\epsilon)}{1+e^{\beta(\epsilon-b)}}\frac{1}{\omega+\epsilon}\nonumber\\
&=&H_T(\omega,b)+\Delta H_T(\omega,b)
\end{eqnarray}
with
\begin{eqnarray}
H_T(\omega,b)&=&\int_{-1}^1\rmd\epsilon\,\frac{1}{1+e^{\beta(\epsilon-b)}}\frac{1}{\omega+\epsilon}\nonumber\\
&=&-\ln(1-\omega)+\ln(2\pi T)+\overline{R}\left(\frac{\omega+b}{T}\right)\;,\nonumber\\
\Delta H_T(\omega,b)&=&\int_{-1}^1\rmd\epsilon\,\frac{\rhotilde(\epsilon)-1}{1+e^{\beta(\epsilon-b)}}\frac{1}{\omega+\epsilon}
\end{eqnarray}
and 
\begin{equation}
\overline{R}(x)=\frac{1}{2}\left[\psi\left(\frac{1}{2}-\frac{\rmi x}{2\pi}\right)+\psi\left(\frac{1}{2}+\frac{\rmi x}{2\pi}\right)\right]=\overline{R}(-x)
\end{equation}
as before. For small $b$,
\begin{eqnarray}
\Delta H_T(\omega,b)&=&\Delta H_T^c(\omega)+bh^{(1)}_T(\omega)+
b^2h^{(2)}_T(\omega)\nonumber\\
\end{eqnarray}
with
\begin{eqnarray}
\Delta H_T(\omega)&=&\int_{-1}^1\rmd\epsilon\,\frac{\rhotilde(\epsilon)-1}{\omega+\epsilon}\frac{1}{1+e^{\beta\epsilon}}\;,\nonumber\\
h^{(1)}(\omega)&=&\frac{1}{T}\int_{-1}^1\rmd\epsilon\,\frac{\rhotilde(\epsilon)-1}{\omega+\epsilon}\frac{e^{\beta\epsilon}}{\left(1+ e^{\beta\epsilon}\right)^2}\;,\nonumber\\
h^{(2)}(\omega)&=&\frac{1}{2T^2}\int_{-1}^1\rmd\epsilon\,\frac{\rhotilde(\epsilon)-1}{\omega+\epsilon}\frac{e^{\beta\epsilon}}{(1+e^{\beta\epsilon})^2}\frac{e^{\beta\epsilon}-1}{e^{\beta\epsilon}+1}\;.\nonumber\\
\end{eqnarray}

\subsubsection{Calculation of \texorpdfstring{${h^{(1)}_T({\omega})}$}{h1T(w)}}

The substitution $x=\epsilon/T$ leads to ($\Omega=\omega/T$)
\begin{eqnarray}
h^{(1)}_T(\omega)&=&\frac{1}{T}\int_{-1/T}^{1/T}\rmd x\,\frac{\rhotilde(xT)-1}{\Omega+x}\frac{e^x}{\left(e^x+1\right)^2}\;.\nonumber\\
\end{eqnarray}
We find in the limit of low temperatures
\begin{eqnarray}
h_T^{(1)}(\omega)&=&\frac{\rhotilde^{\prime\prime}(0)}{2}\frac{1}{T}T^2\int_{-\infty}^\infty\rmd x\,\frac{x^2 e^x}{\left(e^x+1\right)^2}\frac{1}{\Omega+x}+\mathcal{O}(T^3)\nonumber\\
&=&\frac{\rhotilde^{\prime\prime}(0)}{2}T\,{\rm Re}\bigg[(-\rmi)\int_0^\infty\rmd t\,e^{\rmi(\Omega t+\rmi\eta)}\nonumber\\
&&\qquad\qquad\quad\times\int_{-\infty}^\infty\rmd x\,\frac{x^2 e^x}{\left(e^x+1\right)^2}e^{\rmi x t}\bigg]\;.
\end{eqnarray}
The first term of the second integrand is even in $x$, thus
\begin{eqnarray}
h_T^{(1)}(\omega)&=&\frac{\rhotilde^{\prime\prime}(0)}{2}T\,{\rm Re}\bigg[(-\rmi)\int_0^\infty\rmd t\,e^{\rmi(\Omega t+\rmi\eta)}\nonumber\\
&&\qquad\qquad\quad\times \int_{-\infty}^\infty\rmd x\,\frac{x^2 e^x}{\left(e^x+1\right)^2}\cos(x t)\bigg]\nonumber\\
&=&\frac{\rhotilde^{\prime\prime}(0)}{2}T\int_0^\infty\rmd t\,\sin(\Omega t)\left(-\frac{\pi^2}{2}\right)\left(\frac{1}{\sin(\pi t)}\right)^3\nonumber\\
&&\qquad\qquad\times\left[\pi t\left(3t\cosh(2\pi t)\right)-2\sinh(2\pi t)\right]\nonumber\\
&=&\frac{\rhotilde^{\prime\prime}(0)}{2}T\alpha_1\left(\frac{\omega}{T}\right)
\end{eqnarray}
with 
\begin{eqnarray}
\alpha_1(\omega)&=&-\Omega\frac{\Omega^8+4\pi^2\Omega^6+22 \Omega^4\pi^4-156\Omega^2\pi^6+81\pi^8}{\left(\Omega^4+10\Omega^2\pi^2+9\pi^4\right)^2}\nonumber\\
&&+\frac{\rmi\Omega^2}{4\pi}\bigg[\Gamma\left(1,-\frac{3}{2}+\frac{\rmi\Omega}{2\pi}\right)-\Gamma\left(1,-\frac{3}{2},-\frac{\rmi\Omega}{2\pi}\right)\bigg],\nonumber\\
\end{eqnarray}
where $\Gamma(n,x)$ is the $n$-th derivative of the digamma-function.

\subsubsection{Calculation of \texorpdfstring{${h_T^{(2)}({\omega})}$}{hT2(w)}}

Using the substitution $x=\epsilon/T$, for $T\ll1$ the second-order term becomes 
\begin{eqnarray}
h_T^{(2)}(\omega)&=&
\frac{1}{2 T^2}\int_{-1/T}^{1/T}\rmd x\,\frac{\rhotilde(x T)-1}{\Omega+x}\frac{e^x}{\left(e^x+1\right)^2}\frac{e^x-1}{e^x+1}\nonumber\\
&\approx& \left(-\frac{1}{2}\right)\frac{\rhotilde^{\prime\prime}(0)}{2}\int_{-\infty}^\infty\rmd x\,\frac{x^2}{\Omega+x}\frac{\rmd}{\rmd x}\frac{e^x}{\left(e^x+1\right)^2}\nonumber\\
&=&-\frac{\rhotilde^{\prime\prime}(0)}{4}\biggl(\int_{-\infty}^\infty\rmd x\,\left(x+\Omega-2\Omega+\frac{\Omega}{\Omega+x}\right)\nonumber\\
&&\qquad\qquad\times\frac{\rmd}{\rmd x}\frac{e^x}{\left(e^x+1\right)^2}\biggr)\; .
\end{eqnarray}
An integration by parts leads to
\begin{eqnarray}
   h_T^{(2)}(\omega) 
&=&-\frac{\rhotilde^{\prime\prime}(0)}{4}\bigg(-1+\Omega^2\bigg(\frac{e^x}{\left(e^x+1\right)^2}\frac{1}{\Omega+x}\bigg|_{-\infty}^\infty\nonumber\\
&&-\frac{\rmd}{\rmd\Omega}\int_{-\infty}^\infty\rmd x\,\frac{e^x}{\left(e^x+1\right)^2}\frac{1}{\Omega+x}\bigg)\bigg)\;.
\end{eqnarray}
Thus, we find ($\Omega=\omega/T$)
\begin{eqnarray}
h_T^{(2)}(\omega)&=&-\frac{\rhotilde^{\prime\prime}(0)}{4}\left(-1-\Omega^2\frac{\partial}{\partial \Omega} \tilde{h}_T^{(2)}(\Omega)\right)\nonumber\\
\tilde{h}_T^{(2)}(\Omega)&=&{\rm Re}\bigg[(-\rmi)\int_0^\infty\rmd t\,e^{\rmi(\Omega t+\rmi\eta)}\nonumber\\
&&\qquad\times\int_{-\infty}^\infty\rmd x\,\frac{e^x}{\left(e^x+1\right)^2}e^{\rmi x t}\bigg]\bigg)\;.
\end{eqnarray}
Again, the first term in the second integrand is even in $x$ and we find 
\begin{eqnarray}
h_T^{(2)}(\omega)&=&\frac{\rhotilde^{\prime\prime}(0)}{4}\left(1+\Omega^2\frac{\partial}{\partial\Omega}\int_0^\infty\rmd t\,\sin(\Omega t)\frac{\pi t}{\sinh(\pi t)}\right)\nonumber\\
&=&\frac{\rhotilde^{\prime\prime}(0)}{4}\left(1+\Omega^2\overline{R}^{\prime\prime}(\Omega)\right)+\mathcal{O}(T^2)
\end{eqnarray}
with ($\Omega=\omega/T$)
\begin{eqnarray}
\overline{R}(\Omega)&=&\frac{1}{2}\left(\Gamma\left(0,\frac{1}{2}+\frac{\rmi\Omega}{2\pi}\right)+\Gamma\left(0,\frac{1}{2}-\frac{\rmi\Omega}{2\pi}\right)\right)\;,\nonumber\\
\overline{R}^{\prime}(\Omega)&=&
\frac{\rmi}{4\pi}\left(\Gamma\left(1,\frac{1}{2}+\frac{\rmi\Omega}{2\pi}\right)-\Gamma\left(1,\frac{1}{2}-\frac{\rmi\Omega}{2\pi}\right)\right)\;,\nonumber\\
\overline{R}^{\prime\prime}(\Omega)&=&-\frac{1}{8\pi^2}\left(\Gamma\left(2,\frac{1}{2}+\frac{\rmi\Omega}{2\pi}\right)+\Gamma\left(2,\frac{1}{2}-\frac{\rmi\Omega}{2\pi}\right)\right)\;.\nonumber\\
\end{eqnarray}

\subsubsection{Calculation of \texorpdfstring{${\Delta H_T({\omega})}$}{DeltaHT(w)}}

The final term reads 
\begin{eqnarray}
\Delta H_T(\omega)&=&\int_{-1}^1\rmd\epsilon\,\frac{\rhotilde(\epsilon)-1}{\omega+\epsilon}\frac{1}{1+e^{\beta\epsilon}}\nonumber\\
&=&\int_{-1}^0\rmd\epsilon\,\frac{\rhotilde(\epsilon)-1}{\omega+\epsilon}\nonumber\\
&&+\int_0^1\rmd\epsilon\,\frac{\rhotilde(\epsilon)-1}{1+e^{\beta\epsilon}}\left(\frac{1}{\omega+\epsilon}-\frac{1}{\omega-\epsilon}\right)\nonumber\\
&=&\lambda_{<}^{\rm g}(\omega)-\lambda_<(\omega)\nonumber\\
&&+\int_0^{1/T}\rmd x\,\frac{\rhotilde(x T)-1}{1+e^x}\left(\frac{1}{\Omega+x}-\frac{1}{\Omega-x}\right)\nonumber\\
\end{eqnarray}
with $x=\epsilon/T$ in the last step. Here, we defined the half-Hilbert transformations
\begin{eqnarray}
    \lambda^{\rm g}_{<}(\omega)&=&\int_{-1}^0\rmd \epsilon \frac{\rhotilde(\epsilon)}{\omega+\epsilon}\;,\nonumber\\
    \lambda_{<}(\omega)&=&\int_{-1}^0\rmd \epsilon \frac{1}{\omega+\epsilon}\;.
\end{eqnarray}
We identify the remaining integral as
\begin{eqnarray}
h_T^{(0)}(\Omega)&=&T^2\int_0^\infty\rmd x\,\frac{\rhotilde^{\prime\prime}(0)/2}{1+e^x}x^2\left(\frac{1}{\Omega+x}-\frac{1}{\Omega-x}\right)\nonumber\\
&&+\mathcal{O}(T^2)\nonumber\\
&=&\frac{\rhotilde^{\prime\prime}(0)T^2}{2}\bigg(\int_0^\infty\rmd x\,\frac{2 x}{1+e^x}\nonumber\\
&&+\Omega^2\int_0^\infty\rmd x\,\frac{1}{1+e^x}\left(\frac{1}{\Omega+x}-\frac{1}{\Omega-x}\right)\bigg)\nonumber\\
&=&\frac{\rhotilde^{\prime\prime}(0)T^2}{2}\left(\frac{\pi^2}{6}+\Omega^2R(\Omega)\right)\nonumber\\
&=&\frac{\rhotilde^{\prime\prime}(0)}{2}\left(\frac{\pi^2T^2}{6}+\omega^2 R(\omega/T)\right)
\end{eqnarray}
with $R(x)=\ln(2\pi/|x|)+\overline{R}(x)$. Thus,
\begin{eqnarray}
\Delta H_T(\omega)&=&\lambda_{<}(\omega)-\lambda_{<}^{\rm c}(\omega)\\
&&+\frac{\rhotilde^{\prime\prime}(0)}{2}\left(\frac{\pi^2T^2}{6}+\omega^2R(\omega/T)\right)\;.\nonumber
\end{eqnarray}

\subsubsection{Summary}
We summarize
\begin{eqnarray}
H_T^{\rm g}(\omega,b)&=&H_T(\omega,b)+\lambda_{<}(\omega)-\lambda_{<}^c(\omega)\nonumber\\
&&+\frac{\rhotilde^{\prime\prime}(0)}{2}\left(\frac{\pi^2 T^2}{6}+\omega^2R(\omega/T)\right)\nonumber\\
&&+bh_T^{(1)}(\omega)+b^2h_T^{(2)}(\omega)\;.
\end{eqnarray}
At this point, the quantity $\lambda^{\rm g}_{<}(\omega)$ must be calculated individually for the chosen density of states. The half-Hilbert transformation
of the constant density of states, $\rhotilde(\omega)=1$, reads
 \begin{equation}
    \lambda_{<}(\omega)=
    \left\{\begin{array}{@{}lcl@{}}
        \displaystyle \ln\left(\frac{\omega}{\omega-1}\right)&, &\textnormal{ for }(|\omega|>1) \vee (-1<\omega< 0),\\[12pt]
        \displaystyle \ln\left(\frac{\omega}{1-\omega}\right)&,& \textnormal{ for }0<\omega<1\;.\\
        \end{array}
        \right.
\end{equation}

\section{Derivation of the integrals}
\label{subsec:DerviationOfAllIntegrals}

\subsection{Integrals of the second order contributions}
In this section we derive the integrals $\ell(B,T)$ and $l(B)$ that appear in the calculation of the second-order contributions to the impurity-induced free energy.

\subsubsection{Derivation of \texorpdfstring{$\ell(B,T)$}{ell(B,T)}}
\label{SM:subsubsec:derivation_ellBT}

The integral is defined in 
Eq.~(\mainellBTdef) 
of the main paper,
\begin{equation}
    \ell(B,T)=\int_0^\beta\rmd\lambda\,(\beta-\lambda)f(\lambda,B)f(\lambda,-B)
\end{equation}
We note [$\rho_0(-\epsilon)=\rho_0(\epsilon)$]
\begin{eqnarray}\label{eqappendixSternchen}
f(\beta-\lambda,B)&=&\int_{-\infty}^\infty\rmd\epsilon\,\rho_0(\epsilon)\frac{e^{\epsilon(\beta-\lambda)}}{1+e^{\beta(\epsilon-B)}}\nonumber\\
&=&\int_{-\infty}^\infty\rmd\epsilon\,\rho_0(\epsilon)\frac{e^{\epsilon\lambda}e^{-\epsilon\beta}}{1+e^{\beta(-\epsilon-B)}}\nonumber\\
&=&\int_{-\infty}^\infty\rmd\epsilon\,\rho_0(\epsilon)\frac{e^{\epsilon\lambda}e^{\beta B}}{e^{\beta(\epsilon+B)}+1}\nonumber\\
&=&e^{\beta B}f(\lambda,-B)\;,
\end{eqnarray}
and
\begin{equation}
f(\beta-\lambda,B)f(\beta-\lambda,-B)=f(\lambda,-B)f(\lambda,B)\;.
\end{equation}
Thus we find 
\begin{eqnarray}
\ell(B,T)\nonumber&=&\int_0^{\beta/2}\rmd\lambda\,(\beta-\lambda)f(\lambda,B)f(\lambda,-B)\nonumber\\
&&+\int_{\beta/2}^\beta\rmd\lambda\,(\beta-\lambda)f(\lambda,B)f(\lambda,-B)\nonumber\\
&=&\int_0^{\beta/2}\rmd\lambda\,(\beta-\lambda)f(\lambda,B)f(\lambda,-B)\nonumber\\
&&+\int_0^{\beta/2}\rmd\tilde{\lambda}\,\tilde{\lambda}f(\tilde{\lambda},B)f(\tilde{\lambda},-B)\;.
\end{eqnarray}
This leads to
\begin{eqnarray}
\ell(B,T)&=&\int_0^{\beta/2}\rmd\lambda\,\beta f(\lambda,B)f(\lambda,-B)\nonumber\\
&=&\beta\int_{-1}^1\frac{\rmd\epsilon_1}{1+e^{\beta(\epsilon_1-B)}}\int_{-1}^1\frac{\rmd\epsilon_2}{1+e^{\beta(\epsilon_2+B)}}\nonumber\\
&&\quad\times\int_0^{\beta/2}\rmd\lambda\,e^{\lambda(\epsilon_1+\epsilon_2)}\nonumber\\
&=&-\beta\int_{-1}^1\frac{\rmd\epsilon_1}{1+e^{\beta(\epsilon_1-B)}}\nonumber\\
&&\qquad\times\int_{-1}^1\frac{\rmd\epsilon_2}{1+e^{\beta(\epsilon_2+B)}}\frac{1}{\epsilon_1+\epsilon_2}+0\;,\nonumber \\
\end{eqnarray}
where we inserted the constant density of states, $\rho_0(\epsilon)=1/2$, in the second step and substituted $\epsilon_1\to -\epsilon_2,\;\epsilon_2\to-\epsilon_1$ in the third step. The zero occurs because of the symmetry. Thus,
\begin{eqnarray}
\ell(B,T)&=&-\beta\int_{-1-B}^{1-B}\frac{\rmd\epsilon_1}{1+e^{\beta\epsilon_1}}\int_{-1+B}^{1+B}\frac{\rmd\epsilon_2}{1+e^{\beta\epsilon_2}}\frac{1}{\epsilon_1+\epsilon_2}\nonumber\\
&=&-\beta\int_{-1-B}^1\frac{\rmd\epsilon_1}{1+e^{\beta\epsilon_1}}\int_{-1+B}^1\frac{\rmd\epsilon_2}{1+e^{\beta\epsilon_2}}\frac{1}{\epsilon_1+\epsilon_2}\;,\nonumber\\
\end{eqnarray}
where we used that the contribution around the upper integration limit is proportional to $e^{-1/T}$, $T\ll1$, in the last step. 
Then, using the property of Fermi functions from Eq.~(\ref{GIRFermi}), we find $[\ell\equiv \ell(B,T)]$
\begin{eqnarray}
\ell&=&-\beta\int_{-1-B}^0\rmd \epsilon_1\,\int_{-1+B}^0\rmd\epsilon_2\,\frac{1}{\epsilon_1+\epsilon_2}\nonumber\\
&&-\beta\int_{-1-B}^0\rmd\epsilon_1\,\int_0^1\frac{\rmd\epsilon_2}{1+e^{\beta\epsilon_2}}\left(\frac{1}{\epsilon_1+\epsilon_2}-\frac{1}{\epsilon_1-\epsilon_2}\right)\nonumber\\
&&-\beta\int_{-1+B}^0\rmd\epsilon_2\,\int_0^1\frac{\rmd\epsilon_1}{1+e^{\beta\epsilon_1}}\left(\frac{1}{\epsilon_1+\epsilon_2}-\frac{1}{\epsilon_2-\epsilon_1}\right)\nonumber\\
&&-\beta\int_0^1\frac{\rmd\epsilon_1}{1+e^{\beta\epsilon_1}}\int_0^1\frac{\rmd\epsilon_2}{1+e^{\beta\epsilon_2}}k(\epsilon_1,\epsilon_2)\;,
\end{eqnarray}
with
\begin{eqnarray}
k(\epsilon_1,\epsilon_2)&=&\frac{1}{\epsilon_1+\epsilon_2}-\frac{1}{\epsilon_1-\epsilon_2}-\frac{1}{\epsilon_2-\epsilon_1}+\frac{1}{-(\epsilon_1+\epsilon_2)}\nonumber\\
&=&0\;.
\end{eqnarray}
Thus, ($B\ll T\ll1$)
\begin{eqnarray}
\ell&=&\int_0^\beta\rmd\lambda\,(\beta-\lambda)f(\lambda,B)f(\lambda,-B)\nonumber\\
&=&-\beta\left(-2\ln2+B^2+\mathcal{O}[B^4]\right)\nonumber\\
&&-\beta\int_0^1\frac{\rmd\epsilon_2}{1+e^{\beta\epsilon_2}}\ln\left(\frac{1+B+\epsilon_2}{1+B-\epsilon_2}\right)\nonumber\\
&&-\beta\int_0^1\frac{\rmd\epsilon_1}{1+e^{\beta\epsilon_1}}\ln\left(\frac{1-B+\epsilon_1}{1-B-\epsilon_1}\right)\nonumber\\
&=&-\beta\left(-2\ln2+B^2\right)\nonumber\\
&=&-\beta T\int_0^\infty\frac{\rmd x}{1+e^x}\left(\frac{2xT}{1+B}+\frac{2xT}{1-B}\right)+\mathcal{O}[T^4]\nonumber\\
&=&-\beta\left(-2\ln2+B^2+\frac{T^2\pi^2}{6}\left(\frac{1}{1+B}+\frac{1}{1-B}\right)\right)\nonumber\\
&=&-\beta\left(-2\ln2+B^2+\frac{T^2\pi^2}{3}+\frac{\pi^2}{3}B^2T^2\right)+\mathcal{O}[B^3]\;.\nonumber\\
\end{eqnarray}

\subsubsection{Derivation of \texorpdfstring{$l(B)$}{l(B)}}
\label{SM:subsubsec:derivation_lb}

The integral is defined in 
Eq.~(\mainlBTdef) 
of the main paper,
\begin{equation}
l(B)=\int_{-1}^{1}\frac{\rmd\epsilon_1}{1+e^{\beta(\epsilon_1-B)}}\int_{-1}^1\frac{\rmd\epsilon_2}{1+e^{\beta(\epsilon_2-B)}}\frac{1}{\epsilon_1+\epsilon_2}
\end{equation}
Recall that we have
\begin{eqnarray}
H_T(\omega,B)&=&\int_{-1}^1\rmd\epsilon\,\frac{1}{1+e^{\beta(\epsilon-B)}}\frac{1}{\omega+\epsilon}\nonumber\\
&=&-\ln(1-\omega)+\ln(2\pi T)+\overline{R}\left(\frac{\omega+B}{T}\right)\;,\nonumber\\
\end{eqnarray}
where the properties of $H_T(\omega,B)$ and $\overline{R}(x)$ can be found in Appendix~\ref{chapter:properties-of-the-Rbar-function}. We find
\begin{eqnarray}\label{lbequalslapluslb}
l(B)&=&\int_{-1}^1\rmd\omega\,\frac{1}{1+e^{\beta(\omega-B)}}\nonumber\\
&&\quad\times\left(-\ln(1-\omega)+\ln(2\pi T)+\Rbar\left(\frac{\omega+B}{T}\right)\right)\nonumber\\
&=&l_a(B)+l_b(B)
\end{eqnarray}
with [$l_a=l_a(B)$, $l_b=l_b(B)$]
\begin{eqnarray}
l_a&=&\int_{-1}^B\rmd\omega\,\left(-\ln(1-\omega)+\ln(2\pi T)+\Rbar\left(\frac{\omega+B}{T}\right)\right)\nonumber\\
\end{eqnarray}
and
\begin{eqnarray}
l_b&=&\int_{-1}^B\rmd\omega\,\left(\frac{1}{1+e^{\beta(\omega-B)}}-1\right)\nonumber\\
&&\quad\times\left[-\ln(1-\omega)+\ln(2\pi T)+\Rbar\left(\frac{\omega+B}{T}\right)\right]\nonumber\\
&&+\int_B^1\rmd\omega\,\frac{1}{1+e^{\beta(\omega-B)}}\nonumber\\
&&\quad \times\left[-\ln(1-\omega)+\ln(2\pi T)+\Rbar\left(\frac{\omega+B}{T}\right)\right]\,.\nonumber \\
\end{eqnarray}
At $B/ T\ll1$ we find ($\beta=1/T$)
\begin{eqnarray}
l_a&\approx&\int_{-1}^B\rmd\omega\,\bigg(-\ln(1-\omega)+\ln(2\pi T)+\Rbar(\beta\omega)\nonumber\\
&&\qquad +\beta B\Rbar^\prime(\beta\omega)+\frac{(\beta B)^2}{2}\Rbar^{\prime\prime}(\beta\omega)\bigg)\nonumber\\
&\approx&\int_{-1}^0\rmd\omega\,\bigg(-\ln(1-\omega)+\ln(2\pi T)+\Rbar(\beta\omega)\nonumber\\
&&\qquad+\beta B\Rbar^\prime(\beta\omega)+\frac{(\beta B)^2}{2}\Rbar^{\prime\prime}(\beta\omega)\bigg)\nonumber\\
&&+B\left(-\ln(1-0)+\ln(2\pi T)+\Rbar(0)\right)\nonumber\\
&&+\frac{B^2}{2}\left(1+\beta\Rbar^\prime(0)\right)\nonumber\\
&=&1-2\ln 2+\ln(2\pi T)+T \mathcal{I}_1+\beta B T\mathcal{I}_5\nonumber\\
&&+\frac{(\beta B)^2}{2}T\mathcal{I}_{12}+B\left(\ln(2\pi T)+\Rbar(0)\right)+\frac{B^2}{2}\nonumber\\
\end{eqnarray}
with 
\begin{eqnarray}
\mathcal{I}_1&=&\int_0^{1/T}\rmd u\, \Rbar(u)\nonumber\\
&=&-\frac{\ln(2\pi T)}{T}-\frac{1}{T}+\frac{\pi^2T}{6}+\frac{7\pi^4T^3}{180}+\mathcal{O}[T^5]\;,\nonumber\\
\mathcal{I}_5&=&\int_{-1/T}^0\rmd u\,\Rbar^\prime(u)=-\EulerGamma+\ln\left(\frac{\pi T}{2}\right)+\mathcal{O}[T^2]\;,\nonumber\\
\mathcal{I}_{12}&=&\int_{-1/T}^0\rmd u\,\Rbar^{\prime\prime}(u)=T+\mathcal{O}[T^3]\;.
\end{eqnarray}
Note that the integrals $\mathcal{I}_j$ are usually quite complicated to approximate, so here we list their results only and shift their evaluation to chapter~\ref{chapter:derivation-of-the-integrals}. Thus, we find for the first contribution to $l(B)$
\begin{eqnarray}\label{la2ndorder}
l_a(B)&=&B^2+2B\left(-\EulerGamma-2\ln 2+\ln(2\pi T)\right)\nonumber\\
&&-2\ln 2+\frac{\pi^2T^2}{6}\;.
\end{eqnarray}
For the second term we find ($b=B/T\ll1$)
\begin{eqnarray}
l_b(B)&=&-T\int_{-b}^{1/T}\frac{\rmd x}{1+e^{x+b}}\nonumber\\
&&\qquad\times\left(-\ln(1+xT)+\ln(2\pi T)+\Rbar(x-b)\right)\nonumber\\
&&+T\int_b^{1/T}\frac{\rmd x}{1+e^{x-b}}\nonumber\\
&&\qquad\times\left(-\ln(1-xT)+\ln(2\pi T)+\Rbar(x+b)\right)\nonumber\\
&=&-T\int_{-b}^{1/T}\frac{\rmd x}{1+e^{x+b}}\bigg(-xT+\ln(2\pi T)\nonumber\\
&&\quad\qquad+\Rbar(x)-b\Rbar^\prime(x)+\frac{b^2}{2}\Rbar^{\prime\prime}(x)\bigg)\nonumber\\
&&+T\int_b^{1/T}\frac{\rmd x}{1+e^{x-b}}\bigg(xT+\ln(2\pi T)\nonumber\\
&&\quad\qquad+\Rbar(x)+b\Rbar^\prime(x)+\frac{b^2}{2}\Rbar^{\prime\prime}(x)\bigg)\;.
\end{eqnarray}
Since $T\ll1$ we can use $1/T\to\infty$ as integral limits in all terms where an exponential function appears, so that the corrections are only exponentially small,
\begin{eqnarray}
l_b(B)&=&T^2\int_{-b}^\infty\,\frac{\rmd x\, x}{1+e^{x+b}}+T^2\int_b^\infty\,\frac{\rmd x\,x}{1+e^{x-b}}\nonumber\\
&&-T\ln(2\pi T)\left(\int_{-b}^\infty\,\frac{\rmd x}{1+e^{x+b}}-\int_b^\infty\,\frac{\rmd x}{1+e^{x-b}}\right)\nonumber\\
&&-T\int_{-b}^\infty\,\frac{\rmd x}{1+e^{x+b}}\Rbar(x)+T\int_b^\infty\,\frac{\rmd x}{1+e^{x-b}}\Rbar(x)\nonumber\\
&&+B\int_{-b}^\infty\,\frac{\rmd x\,\Rbar^\prime(x)}{1+e^{x+b}}+B\int_b^\infty\,\frac{\rmd x\,\Rbar^\prime(x)}{1+e^{x-b}}\nonumber\\
&&-\frac{B^2}{2T}\int_0^\infty\,\frac{\rmd x}{1+e^x}\Rbar^{\prime\prime}(x)\nonumber\\
&&+\frac{B^2}{2T}\int_0^\infty\,\frac{\rmd x}{1+e^x}\Rbar^{\prime\prime}(x)\nonumber\\
&=&\frac{\pi^2T^2}{6}+0+T\int_0^\infty\rmd y\,\left(\Rbar(y+b)-\Rbar(y-b)\right)\nonumber\\
&&+B\int_0^\infty\,\frac{\rmd y}{1+e^y}\left(\Rbar^\prime(y+b)+\Rbar^\prime(y-b)\right)\nonumber\\
&=&\frac{\pi^2T^2}{6}+2b T\int_0^\infty\,\frac{\rmd y}{1+e^y}\Rbar^\prime(y)\nonumber\\
&&+2 B\int_0^\infty\,\frac{\rmd y}{1+e^y}\Rbar^\prime(y)\;.
\end{eqnarray}
With
\begin{equation}
\mathcal{I}_8=\int_0^\infty\,\rmd y\frac{\Rbar^\prime(y)}{1+e^y}=\ln 2-\frac{1}{2}
\end{equation}
we find for the second contribution to $l(B)$
\begin{equation}
l_b(B)=\frac{\pi^2T^2}{6}+4 B\left(\ln 2-\frac{1}{2}\right)
\end{equation}
and thus in total
\begin{eqnarray}
l(B)&=&-2\ln 2+\frac{\pi^2T^2}{3}\nonumber\\
&&+2B\left(-1-\EulerGamma+\ln(2\pi T)\right)+B^2 \; ,
\end{eqnarray}
using Eq.~(\ref{la2ndorder}).

\subsection{Derivation of the integrals \texorpdfstring{$\boldsymbol{{\mathcal{I}}}_{\mathbf{j}}$}{Ij}}
\label{chapter:derivation-of-the-integrals}

In this subsection we explicitly derive the integrals $\mathcal{I}_j$. 
We need the properties of the function $\Rbar(x)$ for each integral, so we will not always re-reference them, 
see section~\ref{The-Rbar-function} for details. 
Furthermore, we may use \textsc{Mathematica}~\cite{Mathematica12} for each integration and series expansion. 

For each integral we give the order of the omitted terms in the (asymptotic) low-temperature expansion. 
We do this semi-analytically: we subtract the analytical approximation from a numerical calculation of the integral so that only the correction term remains; then
we fit this correction term in a suitable power of $T$ and $T\ln(T)$ and thereby determine the order of the correction terms.
All integrals are verified numerically.

\subsubsection{Integral \texorpdfstring{$\mathcal{I}_1$}{I1}}
Since we know the indefinite integral of the $\Rbar(x)$ function,
we can determine the approximation of the integral at low temperatures simply as a series expansion,
\begin{eqnarray}
\mathcal{I}_1&=&\int_0^{1/T}\rmd x\,\Rbar(x) =\mathcal{R}(1/T)-\mathcal{R}(0)\nonumber\\
&\approx& -\frac{\ln(2\pi T)}{T}-\frac{1}{T}+\frac{\pi^2T}{6}+\frac{7\pi^4T^3}{180}+\mathcal{O}[T^5]\;.\nonumber\\
\end{eqnarray}

\subsubsection{Integral \texorpdfstring{$\mathcal{I}_2$}{I2}}
\label{DerivationI2}
We have to evaluate
\begin{eqnarray}
\mathcal{I}_2&=&\int_0^{1/T}\rmd x\,\ln(1+T x)\Rbar(x)\;.
\end{eqnarray}
In the integrand we add and subtract the leading order of the series expansion of $\Rbar(x)$ for $x\gg 1$ and the leading order of $\ln(1+T x)=Tx+\mathcal{O}[T^2]$ for $T\ll1$ and split the resulting integrals. Thus, we obtain analytically solvable integrals containing the required order at low temperatures and shift the non-analytical terms to higher order in $T$, which we can ignore. The more orders of the individual terms of the integrand we add and subtract again, the higher the order in $T$ that we can calculate analytically. 
Here, we find
\begin{eqnarray}
\mathcal{I}_2&=&-\int_0^{1/T}\rmd x\,\ln\left(\frac{2\pi}{x}\right)\ln(1+Tx)\nonumber\\
&&+\int_0^{1/T}\rmd x\,\left(\Rbar(x)+\ln\left(\frac{2\pi}{x}\right)\right)\nonumber\\
&&\quad\times \left(\ln(1+xT)-xT+xT\right)\nonumber\\
&=&-\frac{\ln T}{T}\left(2\ln 2-1\right)-\frac{\pi^2}{12T}\nonumber\\
&&-\frac{1}{T}\left(-2+\ln 2-\ln\pi+2\ln2\ln(2\pi )\right)\nonumber\\
&&+K_2^a+K_2^b+K_2^c
\end{eqnarray}
with 
\begin{eqnarray}
K_2^a&=&T\int_0^{1/T}\rmd x\,\left(\Rbar(x)+\ln\left(\frac{2\pi}{x}\right)\right)x\nonumber\\
&=&T\int_0^{1/T}\rmd x\,\ln\left(\frac{2\pi}{x}\right)x+T\mathcal{R}(x)x\Big|_0^{1/T}\nonumber\\
&&-T\int_0^{1/T}\rmd x\,\mathcal{R}(x)\nonumber\\
&=&\frac{\pi^2T}{6}\left(1+\ln 2-12\ln \mathcal{A}+\ln(2\pi T)\right)\;,
\end{eqnarray}
where $\mathcal{A}=1.28243$ is Glaisher's constant, and $\mathcal{R}(x)$ can be found in section~\ref{IndefiniteIntegrals}. 
In the second step we used a partial integration. Next, with 
\begin{eqnarray}
\Rbar(x)+\ln\left(\frac{2\pi}{x}\right)+\frac{\pi^2}{6x^2}&\sim& \frac{1}{x^4}\;\textnormal{ for }\;x\gg1\;,\nonumber\\
\ln(1+T x)-xT&\sim&\mathcal{O}[x^2]\;\textnormal{ for }\;x\to 0\nonumber \\
\end{eqnarray}
we obtain
\begin{eqnarray}
K_2^b&=&\int_0^{1/T}\rmd x\,\left(\Rbar(x)+\ln\left(\frac{2\pi}{x}\right)+\frac{\pi^2}{6x^2}\right)\nonumber\\
&&\quad \times\left(\ln(1+T x)-xT\right)\sim\mathcal{O}[T^3\ln T]\nonumber\\
\end{eqnarray}
and 
\begin{eqnarray}
K_2^c&=&-\frac{\pi^2}{6}\int_0^{1/T}\rmd x\,\frac{1}{x^2}\left(\ln(1+Tx)-xT\right)\nonumber\\
&=&\frac{\pi^2T}{6}\left(2\ln 2-1\right)\;.
\end{eqnarray}
Thus, we arrive at
\begin{eqnarray}
\mathcal{I}_2&=&-\frac{\ln T}{T}\left(2\ln 2-1\right)-\frac{\pi^2}{12 T}\nonumber\\
&&-\frac{1}{T}\left(-2+\ln 2-\ln \pi+2\ln 2\ln(2\pi )\right)\nonumber\\
&&+\frac{\pi^2 T}{6}\left(3\ln 2-12\ln \mathcal{A} +\ln(2\pi T)\right)+\mathcal{O}[T^3\ln T]\,.\nonumber\\
\end{eqnarray}

\subsubsection{Integral \texorpdfstring{$\mathcal{I}_3$}{I3}}
Using the same technique as before, we find
\begin{eqnarray}
\mathcal{I}_3&=&\int_0^{1/T}\rmd x\,\left(\Rbar(x)\right)^2\nonumber\\
&=&\int_0^{1/T}\rmd x\,\left(\Rbar(x)+\ln\left(\frac{2\pi}{x}\right)-\ln\left(\frac{2\pi}{x}\right)\right)^2\nonumber\\
&=&K_3^a+K_3^b+K_3^c\;,\nonumber\\
\end{eqnarray}
with
\begin{eqnarray}
K_3^a&=&\int_0^{1/T}\rmd x\,\ln^2\left(\frac{2\pi}{x}\right)\;,\nonumber\\
K_3^b&=&\int_0^{1/T}\rmd x\,\left(\Rbar(x)+\ln\left(\frac{2\pi}{x}\right)\right)^2\;,\nonumber\\
K_3^c&=&-2\int_0^{1/T}\rmd x\,\left(\Rbar(x)+\ln\left(\frac{2\pi}{x}\right)\right)\ln\left(\frac{2\pi}{x}\right)\;.\nonumber\\
\end{eqnarray}
The first integral gives
\begin{equation}
K_3^a=\frac{1+\left(\ln(2\pi T)+1\right)^2}{T}\;.
\end{equation}
The second integral is more difficult to analyze. Since $\Rbar(x)+\ln(2\pi/x)\sim x^{-2}$ we start from
\begin{eqnarray}
K_3^b&=&\int_0^\infty\rmd x\,\left(\Rbar(x)+\ln\left(\frac{2\pi}{x}\right)\right)^2+\mathcal{O}[T^3]\nonumber\\
&=&\int_0^\infty\,\frac{\rmd u_1}{u_1}\left(1-\frac{\pi u_1}{\sinh(\pi u_1)}\right)\nonumber\\
&&\quad\times\int_0^\infty\,\frac{\rmd u_2}{u_2}\left(1-\frac{\pi u_2}{\sinh(\pi u_2)}\right)\nonumber\\
&&\quad \times\frac{1}{2}\int_{-\infty}^\infty\rmd x\,\cos(xu_1)\cos(xu_2)+\mathcal{O}[T^3]\;,\nonumber\\
\end{eqnarray}
where we used the integral representation of $\Rbar(x)$.
The integral over~$x$ results in a Dirac delta distribution,
\begin{equation}
\int_{-\infty}^\infty\rmd x\,\cos(xu_1)\cos(xu_2)=\pi \delta(u_1-u_2)\;,
\end{equation}
thus ($\eta=0^+$)
\begin{eqnarray}
K_3^b&\approx&\frac{\pi}{2}\int_0^\infty\rmd u\,\frac{1}{u^2}\left(1-\frac{\pi u}{\sinh(\pi u)}\right)^2\nonumber\\
&=&\frac{\pi}{2}\bigg(\int_\eta^\infty\frac{\rmd u}{u^2}-2\int_\eta^\infty\frac{\rmd u}{u^2}\frac{\pi u}{\sinh(\pi u)}\nonumber\\
&&\qquad+\int_\eta^\infty\rmd u\,\frac{\pi^2}{\sinh(\pi u)^2}\bigg)\nonumber\\
&=&\frac{\pi }{2}\left(-\pi-2\int_0^\infty\frac{\rmd u}{u^2}\left(-1+\frac{\pi u}{\sinh(\pi u)}\right)\right)\nonumber\\
&=&-\frac{\pi^2}{2}-\pi^2\frac{1}{\pi}\int_0^\infty\frac{\rmd u}{u^2}\left(\frac{\pi u}{\sinh(\pi u)}-1\right)\nonumber\\
&=&\pi^2\left(\ln 2-\frac{1}{2}\right)\;.
\end{eqnarray}
The third integral becomes
\begin{eqnarray}
K_3^c&=&-2\int_0^\infty\frac{\rmd u}{u}\left(1-\frac{\pi u}{\sinh(\pi u)}\right)\nonumber\\
&&\qquad \times\int_0^{1/T}\rmd x\,\ln\left(\frac{2\pi }{x}\right)\cos(xu)\;,
\end{eqnarray}
where the integral over~$x$ results in
\begin{eqnarray}
\int_0^{1/T}\rmd x\,\ln\left(\frac{2\pi }{x}\right)\cos(xu)&=&\frac{1}{u}\ln(2\pi T)\sin\left(\frac{u}{T}\right)\nonumber\\
&&+\frac{1}{u}{\rm Si}\left(\frac{u}{T}\right)
\end{eqnarray}
with the sine integral
\begin{equation}
{\rm Si}(z)=\int_0^z\frac{\rmd t}{t}\sin(t)\;.
\end{equation}
Thus, ($\eta=0^+$, $u^{-2}(1-\pi u\sinh(\pi u)^{-1})\approx \pi^2/6$ at $u\ll1$)
\begin{eqnarray}
K_3^c&=&-2({\rm 1st}+{\rm 2nd})\;,\nonumber\\
{\rm 1st}&=&\int_0^\infty\frac{\rmd u}{u^2}\left(1-\frac{\pi u}{\sinh(\pi u)}\right)\sin\left(\frac{u}{T}\right)e^{-\eta u}\nonumber\\
&=&\int_0^\infty\rmd u\,\frac{\pi^2}{6}\sin\left(\frac{u}{T}\right)e^{-\eta u}\nonumber\\
&=&\frac{\pi^2T}{6}+\mathcal{O}[T^3]\;,\nonumber\\
{\rm 2nd}&=&\int_0^\infty\frac{\rmd u}{u^2}\left(1-\frac{\pi u}{\sinh(\pi u)}\right)\left({\rm Si}\left(\frac{u}{T}\right)-\frac{\pi}{2}\right)e^{-\eta u}\nonumber\\
&&+\frac{\pi}{2}\int_0^\infty\frac{\rmd u}{u^2}\left(1-\frac{\pi u}{\sinh(\pi u)}\right)\nonumber\\
&=&\frac{\pi^2}{2}\ln 2-\frac{\pi^2T}{6}+\mathcal{O}[T^3]\; ,
\end{eqnarray}
and we arrive at
\begin{eqnarray}
\mathcal{I}_3&=&\frac{1}{T}\big[1+\big[1+\ln(2\pi T)\big]^2\big]-\frac{\pi^2}{2}\nonumber\\
&&-\frac{\pi^2 T}{3}\big(\ln(2\pi T)-1\big)+\mathcal{O}[T^3\ln T]\;.\nonumber\\
\end{eqnarray}

\subsubsection{Integral \texorpdfstring{$\mathcal{I}_4$}{I4}}
With the integral representation of the function $\Rbar(x)$ we have
\begin{eqnarray}
\mathcal{I}_4&=&\int_0^\infty\rmd x\,\frac{x\Rbar(x)}{1+e^x}\nonumber\\
&=&\int_0^\infty\rmd x\,x\frac{\Rbar(x)+\ln(2\pi /x)}{1+e^x}-\int_0^\infty\rmd x\,x\frac{\ln(2\pi/x)}{1+e^x}\nonumber\\
&=&\int_0^\infty\frac{\rmd u}{u}\left(1-\frac{\pi u}{\sinh(\pi u)}\right)\int_0^\infty\rmd x\,x\frac{x\cos(xu)}{1+e^x}\nonumber\\
&&+\frac{\pi^2}{12}\left(1+\ln 2-12\ln \mathcal{A} \right)\nonumber\\
&=&\frac{\pi^2}{12}\left(1+\ln 2-12\ln \mathcal{A} \right)\;,
\end{eqnarray}
where $\mathcal{A}=1.28243$ is Glaisher's constant. Note, in the next to last step the first term with the two integrals adds up to zero.

\subsubsection{Integral \texorpdfstring{$\mathcal{I}_5$}{I5}}
Since we know the low temperature behavior of $\Rbar(x)$, we simply can write
\begin{eqnarray}
\mathcal{I}_5&=&\int_{-1/T}^0\rmd x\,\Rbar^\prime(x)\nonumber\\
&=&-\EulerGamma-2\ln 2+\ln(2\pi T)+\frac{\pi^2T^2}{6}+\mathcal{O}[T^4]\;.\nonumber\\
\end{eqnarray}

\subsubsection{Integral \texorpdfstring{$\mathcal{I}_6$}{I6}}
\label{derivationI6}
We need $\mathcal{I}_6$ up to order $\mathcal{O}[T^3\ln(T)]$, since $\mathcal{I}_{10}$ contains the term $\mathcal{I}_6/T$. Using the technique from section~\ref{DerivationI2}, we find
\begin{eqnarray}
\mathcal{I}_6&=&\int_{-1/T}^0\rmd x\,\Rbar^\prime(x)\ln(1-T x)\nonumber\\
&=&\int_{-1/T}^0\rmd x\,\left(\Rbar^\prime(x)-\frac{1}{x}-\frac{\pi^2}{3x^3}\right)\nonumber\\
&&\quad\times\left(\ln(1-Tx)+Tx+\frac{(Tx)^2}{2}\right)\nonumber\\
&&+\int_{-1/T}^0\rmd x\,\left(\frac{1}{x}+\frac{\pi^2}{3x^3}\right)\nonumber\\
&&\quad\times\left(\ln(1-Tx)+Tx+\frac{(Tx)^2}{2}\right)\nonumber\\
&&-\int_{-1/T}^0\rmd x\,\Rbar^\prime(x)\left(Tx+\frac{(Tx)^2}{2}\right)\, .\nonumber \\
\end{eqnarray}
The first integral is $\sim \mathcal{O}[T^4\ln(T)]$, and the second integral can be done by \textsc{Mathematica},
\begin{equation}
\int_{-1/T}^0\rmd x\,\left(\ln(1-Tx)+Tx+\frac{(Tx)^2}{2}\right)=\frac{3}{4}-\frac{\pi^2}{12}-\frac{\pi^2 T^2}{12}.
\end{equation}
The third integral can be approximated using some partial integrations,
\begin{eqnarray}
\mathcal{I}_6^c&=&\int_{-1/T}^0\rmd x\,\Rbar^\prime(x)\left(Tx+\frac{(Tx)^2}{2}\right)\nonumber\\
&=&T\int_{-1/T}^0\rmd x\,\Rbar^\prime(x)x+\frac{T^2}{2}\int_{-1/T}^0\rmd x\,\Rbar^\prime(x)x^2\nonumber\\
&=&T\Rbar(x)x\Big|_{-1/T}^0-T\int_{-1/T}^0\rmd x\,\Rbar(x)\nonumber\\
&&+\frac{T^2}{2}\Rbar(x)x^2\Big|_{-1/T}^0-T^2\mathcal{R}(x)x\Big|_{-1/T}^0\nonumber\\
&&+T^2\int_{-1/T}^0\rmd x\,\mathcal{R}(x)\;.
\end{eqnarray}
Recall that
\begin{eqnarray}
\int\rmd x\,\Rbar(x)&=&\mathcal{R}(x)+c\;,\nonumber\\
\int\rmd x\,\mathcal{R}(x)&=&\mathcal{S}(x)+c\;,
\end{eqnarray}
see section~\ref{IndefiniteIntegrals}. We thus find
\begin{eqnarray}
-T\Rbar(x)x\Big|_{-1/T}^0&\approx&\ln(2\pi T)+\frac{\pi^2 T^2}{6}\;,\nonumber\\
T\int_{-1/T}^0\rmd x\,\Rbar(x)&\approx&-\ln(2\pi T)-1+\frac{\pi^2 T^2}{6}\;,\nonumber\\
-\frac{T^2}{2}\Rbar(x)x^2\Big|_{-1/T}^0&\approx&-\frac{\ln(2\pi T)}{2}-\frac{\pi^2 T^2}{12}\;,\nonumber\\
T^2\mathcal{R}(x)x\Big|_{-1/T}^0&\approx&1+\ln(2\pi T)-\frac{\pi^2T^2}{6}\;,\nonumber\\
-T^2\int_{-1/T}^0\rmd x\,\mathcal{R}(x)&\approx&-\frac{3}{4}-\frac{\ln(2\pi T)}{2}\nonumber\\
&&-\frac{\pi^2T^2}{6}\bigl(\ln 2 +\ln(2\pi T)\nonumber\\
&&\qquad\qquad -12\ln \mathcal{A} \bigr)\;.
\end{eqnarray}
Again, $\mathcal{A}\approx 1.28243$ is Glaisher's constant. Thus, we finally arrive at
\begin{eqnarray}
\mathcal{I}_6&=&-\frac{\pi^2}{12}-\frac{\pi^2T^2}{6}\Bigl(\ln 2 +\ln(2\pi T)-12\ln \mathcal{A} \Bigr)\nonumber\\
&&+\mathcal{O}[T^4\ln T ]\;.
\end{eqnarray}

\subsubsection{Integral \texorpdfstring{$\mathcal{I}_7$}{I7}}
\label{derivationI7}
Using
\begin{equation}\label{RbarRbarprime-DerivativeRepresentation}
\Rbar(x)\Rbar'(x)=\frac{1}{2}\frac{\partial}{\partial x}\left(\Rbar(x)\right)^2 \; ,
\end{equation}
we obtain
\begin{eqnarray}
\mathcal{I}_7&=&\int_{-1/T}^0\rmd x\,\Rbar(x)\Rbar^\prime(x)\nonumber\\
&=&-\frac{1}{2}\int_0^{1/T}\rmd x\,\frac{\partial}{\partial x}\left(\Rbar(x)\right)^2=-\frac{1}{2}\left(\Rbar(x)\right)^2\Big|_0^{1/T}\nonumber\\
&=&\frac{1}{2}\left((\EulerGamma+2\ln 2)^2-\ln^2(2\pi T)\right)\nonumber\\
&&-\frac{\pi^2 T^2}{6}\ln(2\pi T)+\mathcal{O}[T^4\ln T ]\;.
\end{eqnarray}

\subsubsection{Integral \texorpdfstring{$\mathcal{I}_8$}{I8}}
Using the integral representation of $\Rbar^\prime(x)$ we find
\begin{eqnarray}
\mathcal{I}_8&=&\int_0^\infty\rmd x\,\frac{\Rbar^\prime(x)}{1+e^x}\nonumber\\
&=&\int_0^\infty\rmd u\,\frac{\pi u}{\sinh(\pi u)}\int_0^\infty\rmd x\,\frac{\sin(xu)}{1+e^x}\nonumber\\
&=&\int_0^\infty\rmd u\,\frac{\pi u}{\sinh(\pi u)}\frac{1}{2}\left(\frac{1}{u}-\frac{1}{\sinh(\pi u)}\right)\nonumber\\
&=&\ln 2 -\frac{1}{2}\;.
\end{eqnarray}

\subsubsection{Integral \texorpdfstring{$\mathcal{I}_9$}{I9}}
Using a partial integration and Eq.~(\ref{RbarRbarprime-DerivativeRepresentation}) we find
\begin{eqnarray}
\mathcal{I}_9&=&\int_0^\infty\,\frac{\rmd x}{1+e^x}\Rbar(x)\Rbar^\prime(x)\nonumber\\
&=&-\frac{1}{4}\left(\Rbar(0)\right)^2+\frac{1}{2}K_9^a\;,
\end{eqnarray}
because $(1+e^x)^{-1}\to 0$ for $x\to\infty$. Here, with the integral representation of $\Rbar(x)$ and of $\Rbar^\prime(x)$, the last term becomes
\begin{eqnarray}
K_9^a&=&\int_0^\infty\rmd x\,\frac{e^x}{(1+e^x)^2}\left(\Rbar(x)\right)^2\nonumber\\
&=&\int_0^\infty
\frac{\rmd u_1}{u_1}\int_0^\infty\frac{\rmd u_2}{u_2}\nonumber\\
&&\quad\times\int_0^\infty\rmd x\,\left(e^{-2\pi u_1}-\frac{\pi u_1\cos(x u_1)}{\sinh(\pi u_1)}\right)\nonumber\\
&&\quad\times \frac{e^x}{(1+e^x)^2}\left(e^{-2\pi u_2}-\frac{\pi u_2\cos(x u_2)}{\sinh(\pi u_2)}\right)\nonumber\\
&=&a_\eta+b_\eta+c_\eta\;.
\end{eqnarray}
The entire term is finite, whereas the individual integrals diverge. 
Therefore, we introduce the quantity $\eta\ll1$ which helps to avoid the divergences. 
As the last step, we perform the limit $\eta\to 0^+$. We have
\begin{eqnarray}
a_\eta&=&\int_\eta^\infty\frac{\rmd u_1}{u_1}\int_\eta^\infty\frac{\rmd u_2}{u_2}\left(\frac{1}{2}e^{-2\pi (u_1+u_2)}\right)\nonumber\\
&=&\frac{1}{2}\left(\EulerGamma+\ln(2\pi \eta)\right)^2\;,\nonumber\\
b_\eta&=&\int_\eta^\infty\frac{\rmd u_1}{u_1}\nonumber\\
&&\times\int_\eta^\infty\frac{\rmd u_2}{u_2}\,(-2)e^{-2\pi u_1}\frac{\pi u_2}{\sinh(\pi u_2)}\frac{\pi u_2}{\sinh2(\pi u_2)}\nonumber\\
&=&-[-1+\ln(2\pi \eta)][\EulerGamma+\ln(2\pi \eta)]\;.
\end{eqnarray}
For the third term we find
\begin{eqnarray}
c_\eta&=&\int_\eta^\infty\frac{\rmd u_1}{u_1}\int_\eta^\infty\frac{\rmd u_2}{u_2}\frac{\pi u_1}{\sinh(\pi u_1)}\frac{\pi u_2}{\sinh(\pi u_2)}\nonumber\\
&&\quad\times\frac{1}{4}\left(\frac{\pi(u_1-u_2)}{\sinh[\pi(u_1-u_2)]}+\frac{\pi(u_1+u_2)}{\sinh[\pi(u_1+u_2)]}\right)\nonumber\\
&=&\frac{\pi^2}{4}\int_\eta^\infty\rmd u_1\,\int_\eta^\infty\rmd u_2\,\nonumber\\
&&\quad \times\frac{2}{\cosh[\pi(u_1+u_2)]-\cosh[\pi(u_1-u_2)]}\nonumber\\
&&\quad \times \left(\frac{\pi(u_1-u_2)}{\sinh[\pi(u_1-u_2)]}+\frac{\pi(u_1+u_2)}{\sinh[\pi(u_1+u_2)]}\right)\,.\nonumber \\
\end{eqnarray}
We substitute $u_1=(x+y)/2$, $u_2=(x-y)/2$, $\rmd u_1\rmd u_2=\rmd x\rmd y/2$, $x=u_1+u_2$, $y=u_1-u_2$, to find
\begin{eqnarray}
c_\eta&=&\frac{\pi}{4}\int_{2\eta}^\infty\rmd x\,\int_{2\eta-x}^{x-2\eta}\rmd y\,\frac{1}{2}\frac{2}{\cosh(\pi x)-\cosh(\pi y)}\nonumber\\
&&\quad \times\left(\frac{\pi y}{\sinh(\pi y)}+\frac{\pi x}{\sinh(\pi x)}\right)\nonumber\\
&=&\frac{\pi^2}{2}\int_{2\eta}^\infty\frac{\rmd x\,\pi x}{\sinh(\pi x)}\int_0^{x-2\eta}\frac{\rmd y}{\cosh(\pi x)-\cosh(\pi y)}\nonumber\\
&&+\frac{\pi^2}{2}\int_0^\infty\frac{\rmd y\,\pi y}{\sinh(\pi y)}\int_{y+2\eta}^\infty\,\frac{\rmd x}{\cosh(\pi x)-\cosh(\pi y)}\;.\nonumber\\
\end{eqnarray}
Running the last integration in each case provides
\begin{eqnarray}
c_\eta&=&\alpha_\eta+\beta_\eta+\gamma_\eta
\end{eqnarray}
with
\begin{eqnarray}
\alpha_\eta&=&\frac{\pi^2}{2}\int_{2\eta}^\infty\rmd x\,\frac{x}{\sinh^2(\pi x)}\nonumber\\
&&\qquad\times\left(-\ln[\sinh(\pi\eta)]+\ln\left(\sinh[\pi(x-\eta)]\right)\right)\;,\nonumber\\
\beta_\eta&=&\frac{\pi^2}{2}\int_{2\eta}^\infty\rmd y\,\frac{y}{\sinh^2(\pi y)}\nonumber\\
&&\qquad\times\left(\ln[\sinh(\pi \eta)]-\pi y+\ln[\sinh(\pi \eta)]\right)\;,\nonumber\\
\gamma_\eta&=&\frac{\pi^2}{2}\int_0^{2\eta}\rmd y\,\frac{1}{\pi^2 y}\left(\ln[\pi(y+\eta)]-\ln(\pi\eta)\right)\;.
\end{eqnarray}
Moreover, with the Dilogarithm,
\begin{equation}
{\rm Li}_2(x)=-\int_0^1\frac{\rmd t}{t}\ln(1-x t)\;, \qquad\,x\leq 1\;,
\end{equation}
we find
\begin{eqnarray}
\gamma_\eta&=&\frac{\pi^2}{2}\int_0^2\rmd u\,\frac{1}{\pi^2 u}\left(\ln(u+1)\right)\nonumber\\
&=&\frac{\pi^2}{12}+\frac{\ln^2(2)}{4}+\frac{1}{2}{\rm Li}_2\left(-\frac{1}{2}\right)
\end{eqnarray}
and ($\eta\to 0^+$)
\begin{eqnarray}
\beta_{\eta}&=&\frac{\ln^2(\eta)}{2}+\ln(\eta)\left(\ln(2\pi)-1\right)-\frac{\pi^2}{12}\nonumber\\
&&+1-\ln(2\pi)+\frac{\ln^2(\pi)}{2}+\ln 2\ln(\pi)\;.
\end{eqnarray}
Furthermore, ($R\to \infty $)
\begin{eqnarray}
\alpha_\eta&=&\frac{\pi^2}{2}\int_2^{R/\eta}\rmd u\,\frac{\eta^2 u}{[\sinh(\pi u \eta)]^2}\nonumber\\
&&\qquad\times\ln\left(\frac{\sinh[\pi\eta(u-1)]\sinh[\pi \eta(u+1)]}{[\sinh(\pi\eta u)]^2}\right)\nonumber\\
&=&\frac{1}{2}\int_{2}^{R/\eta}\frac{\rmd u}{u}\Bigl[
-\ln 2+\pi\eta(u-1)\nonumber\\
&&\qquad +\ln\left(1-e^{-2\pi\eta(u-1)}\right)-\ln 2 +\pi\eta(u+1)\nonumber\\
&&\qquad+\ln\left(1-e^{-2\pi\eta(u+1)}\right)+2\ln 2\nonumber\\
&&\qquad +2\pi\eta u-2\ln\left(1-e^{-2\pi \eta u}\right)
\Bigr]\nonumber\\
&=&\frac{1}{2}\int_2^{\infty}\frac{\rmd u}{u}\left(\ln(u-1)+\ln(u+1)-2\ln u \right)\nonumber\\
&=&-\frac{1}{4}{\rm Li}_2\left(\frac{1}{4}\right)\;.
\end{eqnarray}
Thus,
\begin{eqnarray}
c_\eta&=&\frac{\pi^2}{12}+\frac{\ln^2(2)}{4}+\frac{1}{2}{\rm Li}_2\left(-\frac{1}{2}\right)+\frac{1}{2}\ln^2(\eta)\nonumber\\
&&+\ln \eta \left(\ln(2\pi)-1\right)-\frac{\pi^2}{12}+1-\ln(2\pi)\nonumber\\
&&+\frac{\ln^2(\pi)}{2}+\ln 2 \ln \pi -\frac{1}{4}{\rm Li}_2\left(\frac{1}{4}\right)\; .
\end{eqnarray}
Finally, summing all terms and letting $\eta\to 0^+$ gives
\begin{eqnarray}
\mathcal{I}_9&=&-\frac{1}{4}\left(-2\ln 2 -\EulerGamma\right)^2+\frac{1}{2}\left(a_\eta+b_\eta+c_\eta\right)\nonumber\\
&=&\frac{1}{2}-\frac{\pi^2}{48}+\EulerGamma\left(\frac{1}{2}-\ln 2 \right)-\ln^2(2) \;.
\end{eqnarray}

\subsubsection{Integral \texorpdfstring{$\mathcal{I}_{10}$}{I10}}
Using a partial integration we find
\begin{eqnarray}
\mathcal{I}_{10}&=&\int_{-1/T}^0\rmd x\,\frac{\Rbar(x)}{1-Tx}\nonumber\\
&=&-\frac{1}{T}\Rbar(x)\ln(1-T x)\Big|_{-1/T}^0\nonumber\\
&&+\int_{-1/T}^0\rmd x\,\Rbar^\prime(x)\ln(1-T x)\nonumber\\
&=&-\frac{\ln 2 }{T}\ln(2\pi T)-\ln 2 \frac{\pi^2T}{6}\nonumber\\
&&-\ln 2 \frac{7\pi^4T^3}{60}+\frac{1}{T}\mathcal{I}_6\nonumber\\
&=&-\frac{\pi^2}{12 T}-\frac{\ln 2 \ln(2\pi T)}{T}\nonumber\\
&&-\frac{\pi^2 T}{6}\left(2\ln 2+\ln(2\pi T)-12\ln \mathcal{A} \right)\nonumber\\
&&+\mathcal{O}[T^3\ln T ]\;.
\end{eqnarray}

\subsubsection{Integral \texorpdfstring{$\mathcal{I}_{11}$}{I11}}
We will show that $\mathcal{I}_{11}\sim \mathcal{O}[T^2]+\mathcal{O}[T^2\ln T ]$. We start from ($\rmesc$: exponential small corrections)
\begin{eqnarray}
\mathcal{I}_{11}&=&\int_0^1\frac{\rmd\omega}{1+e^{\omega/T}}\left(\frac{H_T^{(0)}(\omega)}{1-\omega}-\frac{H_T^{(0)}(-\omega)}{1+\omega}\right)\nonumber\\
&=&T\int_0^{1/T}\frac{\rmd x}{1+e^x}\bigg(\frac{-\ln(1-T x)+\ln(2\pi T)}{1-T x}\nonumber\\
&&\qquad-\frac{-\ln(1+T x)+\ln(2\pi T)}{1+T x}\bigg)\nonumber\\
&&+T\int_0^{1/T}\rmd x\,\frac{\Rbar(x)}{1+e^x}\left(\frac{1}{1-T x}-\frac{1}{1+T x}\right)\nonumber\\
&\approx &\rmesc+2T^2\left(1+\ln(2\pi T)\right)\int_0^\infty\rmd x\,\frac{x}{1+e^x}
\nonumber\\
&&+2T^2\int_0^\infty\rmd x\,\frac{\Rbar(x)x}{1+e^x}\nonumber\\
&=&\rmesc+2T^2\left(1+\ln(2\pi T)\right)\alpha_2+2T^2\mathcal{I}_4\nonumber\\
&=&\frac{\pi^2 T^2}{6}\left(1+\ln(2\pi T)\right)\nonumber\\
&&+\frac{\pi^2 T^2}{6}\left(1+\ln 2 -12\ln \mathcal{A} \right)+\mathcal{O}[T^4\ln T ]\;.\nonumber\\
\end{eqnarray}
Note, in the third step we used
\begin{eqnarray}
p(x,T)&=&\frac{-\ln(1-T x)+\ln(2\pi T)}{1-T x} \nonumber \\
&& -\frac{-\ln(1+T x)+\ln(2\pi T)}{1+T x}\nonumber \\
&\approx& 2x T\left(1+\ln(2\pi T)\right)
\end{eqnarray}
and
\begin{eqnarray}
\frac{1}{1-T x}-\frac{1}{1+T x}\approx 2xT
\end{eqnarray}
at low temperatures.

\subsubsection{Integral \texorpdfstring{$\mathcal{I}_{12}$}{I12}}
Since we do know the behavior from $\Rbar^\prime(x)$ at low temperatures, we can write
\begin{eqnarray}
\mathcal{I}_{12}&=&\int_{-1/T}^0\rmd x\,\Rbar^{\prime\prime}(x)=T+\frac{\pi^2 T^3}{3}+\mathcal{O}[T^5]\;.\nonumber\\
\end{eqnarray}

\subsubsection{Integral \texorpdfstring{$\mathcal{I}_{13}$}{I13}}
We need $\mathcal{I}_{13}$ up to and including order $T^3\ln(T)$ since the integral $\mathcal{I}_{19}$ involves the term $\mathcal{I}_{13}/T$. 
For this purpose we use an analogous approach as in Sect.~\ref{DerivationI2} to find
\begin{eqnarray}
\mathcal{I}_{13}&=&\int_{-1/T}^0\rmd x\,\Rbar^{\prime\prime}(x)\ln(1-T x)\nonumber\\
&=&\int_{-1/T}^0\rmd x\,\left(\Rbar^{\prime\prime}(x)+\frac{1}{x^2}+\frac{\pi^2}{x^4}\right)\nonumber\\
&&\quad \times\left(\ln(1-T x)+xT+\frac{(xT)^2}{2}+\frac{(xT)^3}{3}\right)\nonumber\\
&&+K_{13}^a+K_{13}^b\;.
\end{eqnarray}
The first term is $\sim T^5\ln(T)$ and thus can be ignored. The second term reads
\begin{eqnarray}
K_{13}^a
&=&-\int_{-1/T}^0\rmd x\,\left(\frac{1}{x^2}+\frac{\pi^2}{x^4}\right)\nonumber\\
&&\quad\times\left(\ln(1-T x)+xT+\frac{(xT)^2}{2}+\frac{(xT)^3}{3}\right)\nonumber\\
&=&T\left(-\frac{4}{3}+2\ln 2 \right)+\frac{\pi^2 T^3}{3}\left(2\ln 2 -\frac{5}{6}\right)\;.\nonumber\\
\end{eqnarray}
For the last term we use partial integration,
\begin{eqnarray}
K_{13}^b&=&-\int_{-1/T}^0\rmd x\,\Rbar^{\prime\prime}(x)\left(xT+\frac{(xT)^2}{2}+\frac{(xT)^3}{3}\right)\nonumber\\
&=&K_{13}^c+K_{13}^d+K_{13}^e
\end{eqnarray}
with 
\begin{eqnarray}
K_{13}^c&=&-T\int_{-1/T}^0\rmd x\,x\Rbar^{\prime\prime}(x)\;,\nonumber\\
K_{13}^d&=&-\frac{T^2}{2}\int_{-1/T}^0\rmd x\,x^2\Rbar^{\prime\prime}(x)\;,\nonumber\\
K_{13}^e&=&-\frac{T^3}{3}\int_{-1/T}^0\rmd x\,x^3\Rbar^{\prime\prime}(x)\;.
\end{eqnarray}
We find
\begin{eqnarray}
K_{13}^c&=&-T x\Rbar^\prime(x)\Big|_{-1/T}^0+T\int_{-1/T}^0\Rbar^\prime(x)\nonumber\\
&=&T\left(1-\EulerGamma-2\ln 2 +\ln(2\pi T)\right)+\frac{\pi^2T^3}{2}\;,\nonumber\\
K_{13}^d&=&-\frac{T^2}{2}x^2\Rbar^\prime(x)\Big|_{-1/T}^0+T^2 x\Rbar(x)\Big|_{-1/T}^0\nonumber\\
&&-T^2\int_{-1/T}^0\Rbar(x)\nonumber\\
&=&\frac{T}{2}-\frac{\pi^2T^3}{2}\;,\nonumber\\
K_{13}^e&=&-\frac{T^3}{3}x^3\Rbar^\prime(x)\Big|_{-1/T}^0+T^3x^2\Rbar(x)\Big|_{-1/T}^0\nonumber\\
&&-2T^3 x\mathcal{R}(x)\Big|_{-1/T}^0+2T^3\int_{-1/T}^0\rmd x\,\mathcal{R}(x)\nonumber\\
&=&-\frac{T}{6}\nonumber\\
&&+\frac{\pi^2T^3}{18}\left(11+6\ln 2-72\ln \mathcal{A}+6\ln(2\pi T)\right)\;.\nonumber\\
\end{eqnarray}
In total, we arrive at
\begin{eqnarray}
\mathcal{I}_{13}&=&T\left(-\EulerGamma+\ln(2\pi T)\right)\nonumber\\
&&+\frac{\pi^2 T^3}{3}\left(1+3\ln 2 -12\ln \mathcal{A} +\ln(2\pi T)\right)\nonumber\\
&&+\mathcal{O}[T^5\ln T ]\;.
\end{eqnarray}

\subsubsection{Integral \texorpdfstring{$\mathcal{I}_{14}$}{I14}}
With a partial integration we find
\begin{eqnarray}
\mathcal{I}_{14}&=&\int_0^{1/T}\rmd x\,\Rbar(x)\Rbar^{\prime\prime}(x)\nonumber\\
&=&\Rbar(x)\Rbar^\prime(x)\Big|_0^{1/T}-\int_0^\infty\rmd x\,\left(\Rbar^\prime(x)\right)^2\nonumber\\
&&+\int_{1/T}^\infty\rmd x\,\left(\Rbar^\prime(x)\right)^2\nonumber\\
&=&-T\ln(2\pi T)+K_{14}^a+K_{14}^b
\end{eqnarray}
with
\begin{eqnarray}
K_{14}^b&=&\int_{1/T}^\infty\rmd x\,\left(\Rbar^\prime(x)\right)^2\nonumber\\
&\approx&\int_{1/T}^\infty\rmd x\,\frac{1}{x^2}\nonumber\\
&=&T\;.
\end{eqnarray}
and
\begin{eqnarray}
K_{14}^a&=&-\int_0^\infty\rmd x\,\left(\Rbar^\prime(x)\right)^2\nonumber\\
&=&-\int_0^\infty\rmd u_1\,\frac{\pi u_1}{\sinh(\pi u_1)}\int_0^\infty\rmd u_2\,\frac{\pi u_2}{\sinh(\pi u_2)}\nonumber\\
&&\qquad\frac{1}{2}\int_{-\infty}^\infty\rmd x\,\sin(xu_1)\sin(xu_2)\nonumber\\
&=&-\int_0^\infty\rmd u_1\,\frac{\pi u_1}{\sinh(\pi u_1)}\int_0^\infty\rmd u_2\,\frac{\pi u_2}{\sinh(\pi u_2)}\nonumber\\
&&\qquad\times \frac{1}{2}\frac{1}{4}4\pi\delta(u_1-u_2)\nonumber\\
&=&-\frac{\pi}{2}\int_0^\infty\rmd u\,\left(\frac{\pi u}{\sinh(\pi u)}\right)^2=-\frac{\pi^2}{12}\;.
\end{eqnarray}
Thus, we arrive at the final result
\begin{equation}
\mathcal{I}_{14}=-T\ln(2\pi T)+T-\frac{\pi^2}{12}+\mathcal{O}[T^3\ln T]\;.
\end{equation}

\subsubsection{Integral \texorpdfstring{$\mathcal{I}_{15}$}{I15}}
Using the integral representation of $\Rbar^{\prime\prime}(x)$ we get
\begin{eqnarray}
\mathcal{I}_{15}&=&\int_0^\infty\rmd x\,\frac{x\Rbar^{\prime\prime}(x)}{1+e^x}\nonumber\\
&=&\int_0^\infty\rmd x\,\frac{\pi u^2}{\sinh(\pi u)}\int_0^\infty\rmd x\,\frac{x\cos(xu)}{1+e^x}\nonumber\\
&=&\frac{3}{4}-\ln 2\;.
\end{eqnarray}

\subsubsection{Integral \texorpdfstring{$\mathcal{I}_{16}$}{I16}}
Using a partial integration we find
\begin{eqnarray}
\mathcal{I}_{16}&=&\int_0^\infty\rmd x\,\frac{e^x}{(1+e^x)^2}\Rbar(x)\nonumber\\
&=&-\frac{\Rbar(x)}{1+e^x}\Big|_0^\infty +\int_0^\infty\rmd x\,\frac{\Rbar^\prime(x)}{1+e^x}\nonumber\\
&=&-\frac{1}{2}\left(\EulerGamma+2\ln 2\right)+\mathcal{I}_{8}\nonumber\\
&=&-\frac{1}{2}\left(1+\EulerGamma\right)\;.
\end{eqnarray}

\subsubsection{Integral \texorpdfstring{$\mathcal{I}_{17}$}{I17}}
Using the integral representation of $\Rbar^\prime(x)$ we find
\begin{eqnarray}
\mathcal{I}_{17}&=&\int_0^\infty\rmd x\,\frac{xe^x}{(1+e^x)^2}\Rbar^\prime(x)\nonumber\\
&=&\int_0^\infty\rmd u\,\frac{\pi u}{\sinh(\pi u)}\int_0^\infty\rmd x\,\frac{x e^x\sin(xu)}{(1+e^x)^2}\nonumber\\
&=&\frac{1}{4}\;.
\end{eqnarray}

\subsubsection{Integral \texorpdfstring{$\mathcal{I}_{18}$}{I18}}
Using the integral representation of $\Rbar^\prime(x)$ we obtain
\begin{eqnarray}
\mathcal{I}_{18}&=&
\int_0^\infty\rmd x\,\left(\Rbar^\prime(x)\right)^2-\int_{1/T}^\infty\rmd x\,\frac{1}{x^2}\nonumber\\
&=&-T+\frac{1}{2}\int_0^\infty\rmd u_1\,\frac{\pi u_1}{\sinh(\pi u_1)}\int_0^\infty\rmd u_2\,\frac{\pi u_2}{\sinh(\pi u_2)}\nonumber\\
&&\hphantom{-T+\frac{1}{2}\int_0^\infty} \times\int_{-\infty}^\infty\rmd x\,\sin(xu_1)\sin(xu_2)\nonumber\\
&=&-T+\frac{\pi}{2}\int_0^\infty\rmd u\,\left(\frac{\pi u}{\sinh(\pi u)}\right)^2\nonumber\\
&=&-T+\frac{\pi^2}{12}+\mathcal{O}[T^3]\;.
\end{eqnarray}

\subsubsection{Integral \texorpdfstring{$\mathcal{I}_{19}$}{I19}}
\label{derivationI19}
We use a partial integration to find
\begin{eqnarray}
\mathcal{I}_{19}&=&\int_{-1/T}^0\rmd x\,\frac{\Rbar^\prime(x)}{1-T x}\nonumber\\
&=&-\Rbar^\prime(x)\frac{\ln(1-T x)}{T}\Big|_{-1/T}^0\nonumber\\
&&+\frac{1}{T}\int_{-1/T}^0\rmd x\,\Rbar^{\prime\prime}(x)\ln(1-T x)\nonumber\\
&=&-\ln 2-\frac{\pi^2 T^2}{3}\ln 2 +\mathcal{O}[T^4]+\frac{\mathcal{I}_{13}}{T}\nonumber\\
&=&-\EulerGamma+\ln(\pi T)\nonumber\\
&&+\frac{\pi^2 T^2}{3}\left(1+2\ln 2 -12\ln\mathcal{A}+\ln(2\pi T)\right)\nonumber\\
&&+\mathcal{O}[T^4\ln T ]\;.
\end{eqnarray}

\subsubsection{Integral \texorpdfstring{$\mathcal{I}_{20}$}{I20}}
\label{DerivationI20}
We follow the approach in Sect.~\ref{DerivationI2} to find 
\begin{eqnarray}
\mathcal{I}_{20}&=&\int_{-1/T}^0\rmd x\,\ln(1+T x)\Rbar^{\prime\prime}(x)\nonumber\\
&=&\int_{-1/T}^0\rmd x\,\left(\ln(1+Tx)-T x\right)\Rbar^{\prime\prime}(x)\nonumber\\
&&+T\int_{-1/T}^0\rmd x\,\Rbar^{\prime\prime}(x) x\nonumber\\
&=&\int_{-1/T}^0\rmd x\,\left(\ln(1+T x)-Tx\right)\left(\Rbar^{\prime\prime}(x)+\frac{1}{x^2}\right)
\nonumber\\
&&-\int_{-1/T}^0\rmd x\,\left(\ln(1+Tx)-T x\right)\frac{1}{x^2}\nonumber\\
&&+T x\Rbar^\prime(x)\Big|_{-1/T}^0-T\int_{-1/T}^0\rmd x\,\Rbar^\prime(x)\nonumber\\
&=&T\left(\EulerGamma+\ln 2-\ln(\pi T)\right)+\mathcal{O}[T^3\ln T]\;,
\end{eqnarray}
where we used a partial integration in the next to last step.

\subsubsection{Integral \texorpdfstring{$\mathcal{I}_{21}$}{I21}}
We use an analogous approach as for the integral $\mathcal{I}_{20}$ 
and find
\begin{eqnarray}
\mathcal{I}_{21}&=&\int_{-1/T}^0\rmd x\,\frac{\Rbar^{\prime\prime}(x)}{1-T x}\nonumber\\
&=&\int_0^{1/T}\rmd x\,\left(\frac{1}{1+T x}-1+Tx\right)\left(\Rbar^{\prime\prime}(x)+\frac{1}{x^2}\right)\nonumber\\
&&+\Rbar^\prime(x)\Big|_0^{1/T}-T x\Rbar^\prime(x)\Big|_0^{1/T}+ T\Rbar(x)\Big|_0^{1/T}\nonumber\\
&&-\int_0^{1/T}\rmd x\,\left(\frac{1}{1+Tx}-1+T x\right)\frac{1}{x^2}\nonumber\\
&=&T\left(\EulerGamma-\ln(\pi T)\right)+\mathcal{O}[T^3\ln T]\;.
\end{eqnarray}

\subsubsection{Integral \texorpdfstring{$\mathcal{I}_{22}$}{I22}}
We use an analogous approach  as for the integral $\mathcal{I}_{7}$ 
to obtain
\begin{eqnarray}
\mathcal{I}_{22}&=&\int_{-1/T}^0\rmd x\,\Rbar^\prime(x)\Rbar^{\prime\prime}(x)\nonumber\\
&=&\frac{1}{2}\int_{-1/T}^0\rmd x\,\frac{\partial}{\partial x}\left(\Rbar^\prime(x)\right)^2\nonumber\\
&=&-\frac{T}{2}+\mathcal{O}[T^4]\;.
\end{eqnarray}

\subsubsection{Integral \texorpdfstring{$\mathcal{I}_{23}$}{I23}}
Using the integral presentations of $\Rbar^\prime(x)$ and $\Rbar^{\prime\prime}(x)$ we find
\begin{eqnarray}
\mathcal{I}_{23}&=&\int_0^\infty\rmd x\,\frac{1}{1+e^x}\Rbar^\prime(x)\Rbar^{\prime\prime}(x)\nonumber\\
&=&\frac{1}{2}\frac{\big[\Rbar^\prime(x)\big]^2}{1+e^x}\Big|_0^\infty+\frac{1}{2}\int_0^\infty\rmd x\,\frac{\big[\Rbar^\prime(x)\big]^2e^x}{(1+e^x)^2}\nonumber\\
&=&\frac{1}{2}\int_0^\infty\rmd u_1\,\frac{\pi u_1}{\sinh(\pi u_1)}\int_0^\infty\rmd u_2\,\frac{\pi u_2}{\sinh(\pi u_2)}K_{23}^a\nonumber\\
\end{eqnarray}
with 
\begin{eqnarray}
\frac{K_{23}^a}{\pi}&=&\frac{1}{\pi}\int_0^\infty\rmd x\,\frac{e^x}{(1+e^x)^2}\sin(xu_1)\sin(xu_2)\nonumber\\
&=&\frac{1}{4}\left(\frac{u_1-u_2}{\sinh[\pi(u_1-u_2)]}-\frac{u_1+u_2}{\sinh[\pi(u_1+u_2)]}\right)\nonumber\\
&=&\frac{u_1\cosh\pi u_1\sinh\pi u_2-u_2\cosh\pi u_2\sinh\pi u_1}{\cosh2\pi u_1-\cosh2\pi u_2}\;.\nonumber\\
\end{eqnarray}
Thus,
\begin{eqnarray}
\mathcal{I}_{23}&=&-\frac{\pi}{2}\int_0^\infty\rmd u_1\,\frac{\pi u_1^2\cosh(\pi u_1)}{\sinh(\pi u_1)}K_{23}^b
\end{eqnarray}
with
\begin{eqnarray}
K_{23}^b&=&\int_0^\infty\rmd u_2\,\frac{2\pi u_2}{\cosh(2\pi u_2)-\cosh(2\pi u_1)}\nonumber\\
&=&\int_1^\infty\rmd \lambda\,\frac{1}{2\pi \lambda}\frac{\ln\lambda}{(\lambda+1/\lambda)/2-A}
\end{eqnarray}
where $\lambda=e^{2\pi u_2}$, $\ln(\lambda)=2\pi u_2$, and $A=\cosh(2\pi u_1)>1$. Therefore,
\begin{eqnarray}
K_{23}^b&=&\int_1^\infty\,\frac{\rmd \lambda}{\pi}\frac{\ln \lambda }{e^{2\pi u_1}-e^{-2\pi u_1}}\nonumber\\
&&\quad \times\left(\frac{1}{\lambda-e^{2\pi u_1}}-\frac{1}{\lambda-e^{-2\pi u_1}}\right)\nonumber\\
&=&\frac{1}{2\pi \sinh(2\pi u_1)}\left(\frac{\pi^2}{3}-2\pi^2u_1^2-2{\rm Li}_2(e^{-2\pi u_1})\right)\;,\nonumber\\
\end{eqnarray}
where ${\rm Li}_2(x)$ is the polylogarithm with $n=2$ (dilogarithm).
This leads to
\begin{eqnarray}
\mathcal{I}_{23}&=&K_{23}^c+K_{23}^d
\end{eqnarray}
with
\begin{eqnarray}
K_{23}^c&=&-\frac{1}{8\pi}\int_0^\infty\rmd u_1\,\left(\frac{\pi u_1}{\sinh(\pi u_1)}\right)^2\left(\frac{\pi^2}{3}-2\pi u_1^2\right)\nonumber\\
&=&\frac{\pi^2}{720}\;.
\end{eqnarray}
In the remaining term,
\begin{eqnarray}
K_{23}^d&=&-\frac{1}{4\pi}\int_0^\infty\rmd u\,\left(\frac{\pi u}{\sinh(\pi u)}\right)^2\nonumber\\
&&\qquad\quad\times\int_0^1\frac{\rmd t}{t}\ln\left(1-e^{-2\pi u}t\right)\;,
\end{eqnarray}
we substitute $\mu=e^{-2\pi u}t$, $\rmd \mu=e^{-2\pi u}\rmd t$ to write
\begin{eqnarray}
K_{23}^d&=&-\frac{1}{4\pi}\int_0^1\frac{\rmd \mu}{\mu}\ln(1-\mu)\nonumber\\
&&\qquad\quad\times\int_0^{\ln\left((\mu^{-1})/(2\pi)\right)}\rmd u\,\left(\frac{\pi u}{\sinh(\pi u)}\right)^2\nonumber\\
&=&-\frac{1}{4\pi}\int_0^1\frac{\rmd \mu}{\mu}\ln(1-\mu)f(\mu)
\end{eqnarray}
with 
\begin{equation}
f(\mu)=\frac{1}{6\pi}\bigg(\pi^2-6\ln(1-\mu)\ln \mu +\frac{3\mu\ln^2(\mu)}{\mu-1}-6{\rm Li}_2(\mu)\bigg)\;.
\end{equation}
The $\mu$ integral can be done with \textsc{Mathematica} to give 
\begin{equation}
K_{23}^d=\frac{\pi^2}{720}
\end{equation}
and finally
\begin{equation}
\mathcal{I}_{23}=\frac{\pi^2}{360}\;.
\end{equation}

\subsubsection{Integral \texorpdfstring{$\mathcal{I}_{24}$}{I24}}
Since we know the behavior of $\Rbar^{\prime\prime}(x)$ at low temperatures, we readily find
\begin{equation}
\mathcal{I}_{24}=\int_{-1/T}^0\rmd x\,\Rbar^{\prime\prime\prime}(x)
=\frac{7\zeta(3)}{2\pi^2}+T^2+\mathcal{O}[T^4]\;.
\end{equation}

\subsubsection{Integral \texorpdfstring{$\mathcal{I}_{25}$}{I25}}
We use the procedure as for the integral $\mathcal{I}_{2}$ 
to find
\begin{eqnarray}
\mathcal{I}_{25}&=&\int_{-1/T}^0\rmd x\,\ln(1+T x)\Rbar^{\prime\prime\prime}(x)\nonumber\\
&=&\int_{-1/T}^0\rmd x\,\left(\ln(1+T x)-xT+\frac{x^2T^2}{2}\right)\nonumber\\
&&\quad \times\left(\Rbar^{\prime\prime\prime}(x)-\frac{2}{x^3}\right)\nonumber\\
&&+\int_{-1/T}^0\rmd x\,\left(\ln(1+T x)-xT+\frac{x^2T^2}{3}\right)\frac{2}{x^3}\nonumber\\
&&+T\int_{-1/T}^0\rmd x\,\Rbar^{\prime\prime\prime}(x)x-\frac{T^2}{2}\int_{-1/T}^0\rmd x\,\Rbar^{\prime\prime\prime}(x)x^2\nonumber\\
&=&\int_{-1/T}^0\rmd x\,\left(\ln(1+T x)-xT+\frac{x^2T^2}{2}\right)\nonumber\\
&&\quad \times\left(\Rbar^{\prime\prime\prime}(x)-\frac{2}{x^3}\right)\nonumber \\ +\frac{3T^2}{2}
&&+\frac{3T^2}{2} +T\Rbar^{\prime\prime}(x)x\Big|_{-1/T}^0-T\Rbar^\prime(x)\Big|_{-1/T}^0\nonumber\\
&&-\frac{T^2}{2}\Rbar^{\prime\prime}(x)x^2\Big|_{-1/T}^0+T^2\Rbar^\prime(x)x\Big|_{-1/T}^0\nonumber\\
&&-T^2\Rbar(x)\Big|_{-1/T}^0\nonumber\\
&=&T^2\left(\EulerGamma-2+\ln 2 -\ln(\pi T)\right)+\mathcal{O}[T^4\ln T ]\;.\nonumber\\
\end{eqnarray}

\subsubsection{Integral \texorpdfstring{$\mathcal{I}_{26}$}{I26}}
To approximate $\mathcal{I}_{26}$ we use $K_{28}^a$ from $\mathcal{I}_{28}$ to find
\begin{eqnarray}
\mathcal{I}_{26}&=&\int_{-1/T}^0\rmd x\,\Rbar(x)\Rbar^{\prime\prime\prime}(x)\nonumber\\
&=&-\int_0^\infty\rmd x\,\Rbar(x)\Rbar^{\prime\prime\prime}(x)+\int_{1/T}^\infty\rmd x\,\Rbar(x)\Rbar^{\prime\prime\prime}(x)\nonumber\\
&\approx&-K_{28}^a-\int_{1/T}^\infty\rmd x\,\ln\left(\frac{2\pi }{x}\right)\frac{2}{x^3}\nonumber\\
&=&-\frac{7\zeta(3)}{2\pi^2}\left(\EulerGamma+2\ln 2 \right)\nonumber\\
&&+\frac{T^2}{2}\left(1-2\ln(2\pi T)\right)+\mathcal{O}[T^4\ln T]
\end{eqnarray}
with
\begin{equation}
K_{28}^a=\frac{7\zeta(3)}{2\pi^2}\left(\EulerGamma+2\ln2\right)\;.
\end{equation}

\subsubsection{Integral \texorpdfstring{$\mathcal{I}_{27}$}{I27}}
We use the integral representation of $\Rbar^{\prime\prime\prime}(x)$ to find
\begin{eqnarray}
\mathcal{I}_{27}&=&\int_0^\infty\rmd x\,\frac{\Rbar^{\prime\prime\prime}(x)}{1+e^x}\nonumber\\
&=&-\int_0^\infty\rmd u\,\frac{\pi u^2}{\sinh(\pi u)}\int_0^\infty\rmd x\,\frac{\sin(xu)}{1+e^x}\nonumber\\
&=&-\frac{\zeta(3)}{\pi^2}\;.
\end{eqnarray}

\subsubsection{Integral \texorpdfstring{$\mathcal{I}_{28}$}{I28}}
\label{integral28}
To derive
\begin{eqnarray}
\mathcal{I}_{28}&=&\int_0^\infty\rmd x\,\frac{\Rbar(x)\Rbar^{\prime\prime\prime}(x)}{1+e^x}
\end{eqnarray}
we define 
\begin{eqnarray}
K_{28}^a&=&\int_0^\infty\rmd x\,\Rbar(x)\Rbar^{\prime\prime\prime}(x)\nonumber\\
K_{28}^b&=&\int_{-\infty}^\infty\rmd x\,\frac{\Rbar(x)\Rbar^{\prime\prime\prime}(x)}{1+e^x}\nonumber\\
&=&-K_{28}^a\nonumber\\
&&+\int_0^\infty\frac{\rmd x}{1+e^x}\left(\Rbar(x)\Rbar^{\prime\prime\prime}(x)-\Rbar(-x)\Rbar^{\prime\prime\prime}(-x)\right)\nonumber\\
&=&-K_{28}^a+2\mathcal{I}_{28}\;,
\end{eqnarray}
where we used Eq.~(\ref{GIRFermi}). Thus, from those two functions we can implicitly calculate $\mathcal{I}_{28}$,
\begin{equation}
\mathcal{I}_{28}=\frac{1}{2}\left(K_{28}^a+K_{28}^b\right)\;.
\end{equation}
We perform a partial integration to find
\begin{eqnarray}
K_{28}^a&=&\Rbar^{\prime\prime}(x)\Rbar(x)\Big|_0^{\infty}-\int_0^\infty\rmd x\,\Rbar^{\prime\prime}(x)\Rbar^\prime(x)\nonumber\\
&=&-\Rbar^{\prime\prime}(0)\Rbar(0)-\frac{1}{2}\left(\Rbar^\prime(x)\right)_0^\infty\nonumber\\
&=&\left(2\ln 2 +\EulerGamma\right)\frac{7\zeta(3)}{2\pi^2}\;.
\end{eqnarray}
With the integral representation and $\eta\ll1$ (later we perform $\eta\to 0^+$) the second term reduces to
\begin{eqnarray}
K_{28}^b&=&-\int_0^\infty\rmd u_1\int_0^\infty\rmd u_2\int_{-\infty}^\infty\rmd x\,\frac{e^{\eta x}}{1+e^x}\nonumber\\
&&\qquad \times\left(\frac{e^{-2\pi u_1}}{u_1}-\frac{\pi \cos(x u_1)}{\sinh(\pi u_1)}\right)\frac{\pi u_2^3 \sin(xu_2)}{\sinh(\pi u_2)}\nonumber\\
&=&\alpha_\eta^a+\alpha_\eta^b
\end{eqnarray}
with 
\begin{eqnarray}
\alpha_\eta^a&=&-\frac{\pi}{2}\int_\eta^\infty\rmd u_1\int_0^\infty \rmd u_2
\frac{-2e^{-2\pi u_1}}{u_1\sinh(\pi u_2)}\frac{\pi u_2^3}{\sinh(\pi u_2)}\;,\nonumber\\[3pt]
\alpha_\eta^b&=&-\frac{\pi}{2}\int_\eta^\infty\rmd u_1\int_0^\infty \rmd u_2
\frac{\pi}{\sinh(\pi u_1)}\frac{\pi u_2^3}{\sinh(\pi u_2))}\nonumber\\
&&\qquad \times\left(\frac{1}{\sinh[\pi(u_1+u_2)]}-\frac{1}{\sinh[\pi(u_1-u_2)]}\right)\;.\nonumber \\
\end{eqnarray}
We find ($\eta=0^+$)
\begin{eqnarray}
\alpha_\eta^a&=&\pi\int_\eta^\infty\rmd u_1\,\frac{e^{-2\pi u_1}}{u_1}\int_0^\infty\rmd u_2\,\frac{\pi u_2^3}{\sinh^2(\pi u_2)}\nonumber\\
&\approx&-\frac{3\zeta(3)}{2\pi^2}\left(\ln \eta +\EulerGamma+\ln(2\pi)\right)\;,\nonumber\\[3pt]
\alpha_\eta^b&=&-\frac{\pi}{2}\int_0^\infty\rmd u_2\,\frac{\pi u_2^3}{\sinh(\pi u_2)}\int_\eta^\infty\rmd u_1\,\frac{\pi}{\sinh(\pi u_1)}\nonumber\\
&&\qquad \times\left(\frac{1}{\sinh[\pi(u_1+u_2)]}-\frac{1}{\sinh[\pi(u_1-u_2)]}\right)\nonumber\\
&=&-\frac{\pi}{2}\int_0^\infty\rmd u_2\,\frac{\pi u_2^3}{\sinh(\pi u_2)}\left(\frac{-2}{\sinh(\pi u_2)}\right)\nonumber\\
&&\qquad\times\left(\ln \eta +\ln \pi-\ln(\sinh(\pi u_2))\right)\nonumber\\
&=&\gamma_\eta^a+\gamma_\eta^b+\gamma_\eta^c
\end{eqnarray}
with 
\begin{eqnarray}
\gamma_\eta^a&=&\ln \eta \pi\int_0^\infty\rmd u_2\,\frac{\pi u_2^3}{\sinh^2(\pi u_2)}=\ln(\eta)\frac{3\zeta(3)}{2\pi^2}\;,\nonumber\\
\gamma_\eta^b&=&\pi\int_0^\infty\rmd u_2\,\frac{\pi u_2^3}{\sinh^2(\pi u_2)}\left(\ln(2\pi)-u_2\pi\right)\nonumber\\
&=&\ln(2\pi )\frac{3\zeta(3)}{2\pi^2}-\frac{\pi^2}{30}\;,\nonumber\\
\gamma_\eta^c&=&-\pi\int_0^\infty\rmd u_2\,\frac{\pi u_2^3}{\sinh^2(\pi u_2)}\ln(1-e^{-2\pi u_2})\nonumber\\
&=&\sum_{n=1}^\infty\frac{1}{n}\int_0^\infty\rmd u_2\,\frac{\pi^2u_2^3}{\sinh^2(\pi u_2)}e^{-2\pi u_2n}\nonumber\\
&=&\sum_{n=1}^\infty\frac{1}{n}\frac{3}{2\pi^2}\left(\zeta(3,n+1)-n\zeta(4,n+1)\right)\;.\nonumber \\
\end{eqnarray}
For the zeta-functions we use
\begin{eqnarray}
\zeta(3,n+1)=\sum_{l=0}^\infty(l+n+1)^{-3}\;,\nonumber\\
\zeta(4,n+1)=\sum_{l=0}^\infty(l+n+1)^{-4}\;,
\end{eqnarray}
to write
\begin{eqnarray}
\gamma_\eta^c&=&\frac{3}{2\pi^2}\sum_{l=0}^\infty\nonumber\\
&&\qquad\times\left[\sum_{n=1}^\infty\left[\frac{1}{l+n+1}\right]^3\frac{1}{n}-\sum_{n=1}^\infty\left[\frac{1}{l+n+1}\right]^4\right]\nonumber\\
&=&\frac{3}{2\pi^2}\sum_{l=0}^\infty\bigg(\frac{\EulerGamma+\Gamma(0,l+2)}{(l+1)^3}-\frac{\Gamma(1,l+2)}{(l+1)^2}\nonumber\\
&&\qquad\qquad+\frac{1}{2}\frac{\Gamma(2,l+2)}{l+1}-\frac{\Gamma(3,l+2)}{6}\bigg)\nonumber\\
&=&\frac{3}{2\pi^2}\left(\frac{\pi^4}{72}-\zeta(3)\right)\;.
\end{eqnarray}
In total,
\begin{equation}
\alpha_\eta^b\approx\frac{3\zeta(3)}{2\pi^2}\left(\ln \eta +\ln(2\pi)-1\right)-\frac{\pi^2}{30}+\frac{\pi^2}{48}\; ,
\end{equation}
and, therefore,
\begin{eqnarray}
K_{28}^b=-\frac{\pi^2}{80}-\frac{3\zeta(3)}{2\pi^2}\left(1+\EulerGamma\right)\;.
\end{eqnarray}
Finally,
\begin{eqnarray}
\mathcal{I}_{28}&=&\frac{1}{2}\left(K_{28}^a+K_{28}^b\right)\nonumber\\
&=&-\frac{\pi^2}{160}+\frac{\zeta(3)}{4\pi^2}\left(-3+4\EulerGamma+14\ln 2\right)\;.
\end{eqnarray}
Note, a contour integral 
\begin{equation}
\int_{-\infty}^\infty\rmd x\,\frac{\Rbar(x)\Rbar^{\prime\prime\prime}(x)}{1+e^x}\neq \oint\rmd z\,\frac{\Rbar(z)\Rbar^{\prime\prime\prime}(z)}{1+e^z}
\end{equation}
does \emph{not} give the correct result.

\subsubsection{Integral \texorpdfstring{$\mathcal{I}_{29}$}{I29}}
With the integral representation of $\Rbar^{\prime\prime\prime}(x)$ we find
\begin{eqnarray}
\mathcal{I}_{29}&=&\int_0^\infty\rmd x\,\frac{\Rbar^{\prime\prime\prime}(x)x^2}{1+e^x}\nonumber\\
&=&-\int_0^\infty\rmd u\,\frac{\pi u^2}{\sinh(\pi u)}\int_0^\infty\rmd x\,\frac{\sin(xu)x^2}{1+e^x}\nonumber\\
&=&-\frac{11}{6}+2\ln 2+\frac{\zeta(3)}{3}\;.
\end{eqnarray}

\subsubsection{Integral \texorpdfstring{$\mathcal{I}_{30}$}{I30}}
Using a partial integration we find
\begin{eqnarray}
\mathcal{I}_{30}&=&\int_0^\infty\rmd x\,\frac{e^x(e^x-1)x}{(1+e^x)^3}\Rbar(x)\nonumber\\
&=&-\frac{e^x}{(1+e^x)^2}x\Rbar(x)\Big|_0^\infty\nonumber\\
&&+\int_0^\infty\rmd x\,\frac{e^x}{(1+e^x)^2}\left(x\Rbar^\prime(x)+\Rbar(x)\right)\nonumber\\
&=&\mathcal{I}_{17}+\mathcal{I}_{16}\nonumber\\
&=&-\frac{1}{4}(1+2\EulerGamma)\;.
\end{eqnarray}

\subsubsection{Integral \texorpdfstring{$\mathcal{I}_{31}$}{I31}}
Using partial integrations we get
\begin{eqnarray}
\mathcal{I}_{31}&=&\int_0^\infty\rmd x\,\frac{e^x(e^x-1)}{(1+e^x)^3}\Rbar(x)\Rbar^\prime(x)\nonumber\\
&=&-\frac{e^x}{(1+e^x)^2}\Rbar(x)\Rbar^\prime(x)\Big|_0^\infty\nonumber\\
&&+\int_0^\infty\rmd x\,\frac{e^x}{(1+e^x)^2}\left(\Rbar(x)\Rbar^{\prime\prime}(x)+(\Rbar^\prime(x))^2\right)\nonumber\\
&=&-\frac{1}{1+e^x}\left(\Rbar(x)+\Rbar^{\prime\prime}(x)+[\Rbar^\prime(x)]^2\right)_0^\infty\nonumber\\
&&+\int_0^\infty\frac{\rmd x}{1+e^x} \bigg(\Rbar^\prime(x)\Rbar^{\prime\prime}(x)+\Rbar(x)\Rbar^{\prime\prime}(x)\nonumber\\
&&\qquad\qquad\qquad+2\Rbar^\prime(x)\Rbar^{\prime\prime}(x)\bigg)\nonumber\\
&=&-\frac{7\zeta(3)}{4\pi^2}\left(\EulerGamma+2\ln 2\right)+3\int_0^\infty\rmd x\,\frac{\Rbar^\prime(x)\Rbar^{\prime\prime}(x)}{1+e^x}\nonumber\\
&&+\int_0^\infty\rmd x\,\frac{\Rbar(x)\Rbar^{\prime\prime\prime}(x)}{1+e^x}\nonumber\\
&=&-\frac{7\zeta(3)}{4\pi^2}\left(\EulerGamma+2\ln 2\right)+3\mathcal{I}_{26}+\mathcal{I}_{31}\nonumber\\
&=&\frac{\pi^2}{480}-\frac{3\zeta(3)}{4\pi^2}\left(1+\EulerGamma\right)\;.
\end{eqnarray}

\subsubsection{Integral \texorpdfstring{$\mathcal{I}_{32}$}{I32}}
Using the integral representation of $\Rbar^\prime(x)$ leads to
\begin{eqnarray}
\mathcal{I}_{32}&=&\int_0^\infty\rmd x\,\frac{e^x(e^x-1)}{(1+e^x)^3}\Rbar^\prime(x)\nonumber\\
&=&\int_0^\infty\rmd u\,\frac{\pi u}{\sinh(\pi u)}\int_0^\infty\rmd x\,\frac{e^x(e^x-1)}{(1+e^x)^3}\sin(xu)\nonumber\\
&=&\frac{3\zeta(3)}{4\pi^2}\;.
\end{eqnarray}

For the derivation of the 
integrals $\mathcal{I}_{33}$, $\mathcal{I}_{34}$, $\mathcal{I}_{35}$, $\mathcal{I}_{36}$, $\mathcal{I}_{37}$, 
and $\mathcal{I}_{38}$
we use partial integration.

\subsubsection{Integral \texorpdfstring{$\mathcal{I}_{33}$}{I33}}
Using a partial integration we find
\begin{eqnarray}
\mathcal{I}_{33}&=&\int_0^\infty\rmd x\,\frac{e^x(e^x-1)}{(1+e^x)^3}\Rbar^\prime(x)x^2\nonumber\\
&=&
\int_0^\infty\rmd x\,\frac{e^x}{(1+e^x)^2}\left(\Rbar^{\prime\prime}(x)x^2+2\Rbar^\prime(x) x\right)\nonumber\\
&=&-\frac{1}{1+e^x}\left.\left(\Rbar^{\prime\prime}(x)x^2+2\Rbar^\prime(x) x\right)\right|_0^\infty\nonumber\\
&&+\int_0^\infty\rmd x\,\frac{\Rbar^{\prime\prime\prime}(x)x^2+4\Rbar^{\prime\prime}(x)x+2\Rbar^\prime(x)}{1+e^x}\nonumber\\
&=&\int_0^\infty\rmd x\,\frac{\Rbar^{\prime\prime\prime}(x)x^2+4\Rbar^{\prime\prime}(x)x+2\Rbar^\prime(x)}{1+e^x}\nonumber\\
&=&\mathcal{I}_{32}+4\mathcal{I}_{15}+2\mathcal{I}_{8}
=\frac{1}{6}+\frac{\zeta(3)}{3}\;.
\end{eqnarray}

\subsubsection{Integral \texorpdfstring{$\mathcal{I}_{34}$}{I34}}
We have
\begin{eqnarray}
\mathcal{I}_{34}&=&\int_0^\infty\rmd x\,\frac{e^x}{(1+e^x)^2}\left(\Rbar^\prime(x)\right)^2\nonumber\\
&=&-\frac{1}{1+e^x}\left(\Rbar^\prime(x)\right)^2\Big|_0^\infty+2\int_0^\infty\rmd x\,\frac{\Rbar^\prime(x)\Rbar^{\prime\prime}(x)}{1+e^x}\nonumber\\
&=&2\mathcal{I}_{26}
= \frac{\pi^2}{180}\;.
\end{eqnarray}

\subsubsection{Integral \texorpdfstring{$\mathcal{I}_{35}$}{I35}}
Using a partial integration again we find
\begin{eqnarray}
\mathcal{I}_{35}&=&\int_0^\infty\rmd x\,\frac{e^x}{(1+e^x)^2}\Rbar^{\prime\prime}(x)\nonumber\\
&=&-\frac{1}{1+e^x}\Rbar^{\prime\prime}(x)\Big|_0^\infty+\int_0^\infty\rmd x\,\frac{\Rbar^{\prime\prime\prime}(x)}{1+e^x}\nonumber\\
&=&\frac{7\zeta(3)}{4\pi^2}+\mathcal{I}_{30}
= \frac{3\zeta(3)}{4\pi^2}\; .
\end{eqnarray}

\subsubsection{Integral \texorpdfstring{$\mathcal{I}_{36}$}{I36}}
We have
\begin{eqnarray}
\mathcal{I}_{36}&=&\int_0^\infty\rmd x\,\frac{e^x}{(1+e^x)^2}\Rbar(x)\Rbar^{\prime\prime}(x)\nonumber\\
&=&-\frac{1}{1+e^x}\Rbar(x)\Rbar^{\prime\prime}(x)\Big|_0^\infty\nonumber\\
&&+\int_0^\infty\rmd x\,\frac{1}{1+e^x}\left(\Rbar(x)\Rbar^{\prime\prime\prime}(x)+\Rbar^\prime(x)\Rbar^{\prime\prime}(x)\right)\nonumber\\
&=&-\frac{7\zeta(3)}{4\pi^2}\left(\EulerGamma+2\ln 2\right)+\mathcal{I}_{28}+
\mathcal{I}_{23}\nonumber\\
&=&-\frac{\pi^2}{288}-\frac{3\zeta(3)}{4\pi^2}(1+\EulerGamma)\;.
\end{eqnarray}

\subsubsection{Integral \texorpdfstring{$\mathcal{I}_{37}$}{I37}}
With a partial integration we find
\begin{eqnarray}
\mathcal{I}_{37}&=&\int_0^\infty\rmd x\,\frac{x^2e^x}{(1+e^x)^2}\Rbar^{\prime\prime}(x^2)\nonumber\\
&=&-\frac{1}{1+e^x}\Rbar^{\prime\prime}(x)x^2\Big|_0^\infty\nonumber\\
&&+\int_0^\infty\frac{\rmd x}{1+e^x}\left(\Rbar^{\prime\prime\prime}(x)x^2+2x\Rbar^{\prime\prime}(x)\right)\nonumber\\
&=&\mathcal{I}_{32}+2\mathcal{I}_{15}=-\frac{1}{3}+\frac{\zeta(3)}{4}\;.
\end{eqnarray}

\subsubsection{Integral \texorpdfstring{$\mathcal{I}_{38}$}{I38}}
Using a partial integration leads to
\begin{eqnarray}
\mathcal{I}_{38}&=&\int_{-1/T}^0\rmd x\,\Rbar^{\prime\prime\prime}(x)\ln(1-T x)\nonumber\\
&=&\Rbar^{\prime\prime}(x)\ln(1-T x)\Big|_{-1/T}^0\nonumber\\
&&+T\int_{-1/T}^0\rmd x\,\frac{\Rbar^{\prime\prime}(x)}{1-T x}\nonumber\\
&=&\ln 2T^2+T\mathcal{I}_{24}\nonumber\\
&=&{T^2}\left({\EulerGamma+2\ln 2 -\ln(2\pi T)}\right)\nonumber\\
&&+\mathcal{O}[T^4\ln T ]\;.
\end{eqnarray}

\subsubsection{Integral \texorpdfstring{$\mathcal{I}_{39}$}{I39}}
Following the derivation of the integral~$\mathcal{I}_2$ we find
\begin{eqnarray}
\mathcal{I}_{39}&=&\int_{-1/T}^0\rmd x\,\Rbar^\prime(x)\ln(1+T x)\nonumber\\
&=&\int_{-1/T}^0\rmd x\,\left(\Rbar^\prime(x)-\frac{1}{x}-\frac{\pi^2}{3 x^3}\right)\nonumber\\
&&\qquad \times\left(\ln(1+T x)-T x+\frac{(Tx)^2}{2}\right)\nonumber\\
&&+\int_{-1/T}^0\rmd x\,\left(\frac{1}{x}+\frac{\pi^2}{3x^3}\right)\nonumber\\
&&\qquad\times\left(\ln(1+T x)-T x+\frac{(Tx)^2}{2}\right)\nonumber\\
&&-\int_{-1/T}^0\rmd x\,\Rbar^\prime(x)\left(-T x+\frac{(T x)^2}{2}\right)\;.\nonumber\\
\end{eqnarray}
The first term is $\sim T^4\ln(T)$ and the second term gives
\begin{eqnarray}
\mathcal{I}_{39}^b&=&\int_{-1/T}^0\rmd x\left[\frac{1}{x}+\frac{\pi^2}{3x^3}\right]\left[\ln(1+T x)-T x+\frac{T^2x^2}{2}\right]\nonumber\\
&=&-\frac{5}{4}+\frac{\pi^2}{6}+\frac{\pi^2 T^2}{4}+\mathcal{O}[T^4]\;.
\end{eqnarray}
With some partial integrations we find for the last term
\begin{eqnarray}
\mathcal{I}_{39}^c&=&-\int_{-1/T}^0\rmd x\,\Rbar^\prime(x)\left(-T x+\frac{(Tx)^2}{2}\right)\nonumber\\
&=&T x\Rbar(x)\Big|_{-1/T}^0-T\mathcal{I}_1-\frac{T^2}{2}x^2\Rbar(x)\Big|_{-1/T}^0\nonumber\\
&&+T^2x\mathcal{R}(x)\Big|_{-1/T}^0-T^2\int_{-1/T}^0\rmd x\,\mathcal{R}(x)\nonumber\\
&=&\frac{5}{4}-\frac{\pi^2 T^2}{12}\left(7+\ln 16 -24\ln\mathcal{A}+2\ln(\pi T)\right)\;.\nonumber\\
\end{eqnarray}
Thus, we arrive at
\begin{eqnarray}
\mathcal{I}_{39}&=&\frac{\pi^2}{6}-\frac{\pi^2T^2}{6}\left(2+\ln 2 +\ln(2\pi T)-12\ln \mathcal{A}\right)\nonumber\\
&&+\mathcal{O}[T^4\ln T]\;.
\end{eqnarray}

\section{Mathematical details}
\label{mathematicaldetails}

In this section we list some important relations.

\subsection{Useful relations}
\label{chapter:GeneralRelations}

Let $c\in\mathbb{R}$. Let $f_1(x)=f_1(-x)$ be an even function, $f_2(x)=-f_2(-x)$ be an odd function, and $f_3(x)$ be an arbitrary function. Then
\begin{eqnarray}
\int_{-c}^{c}\rmd x\,\frac{f_3(x)}{1+e^x}&=&\int_{-c}^0\rmd x f_3(x)\nonumber\\
&&+\int_0^{c}\rmd x\,\frac{1}{1+e^x}\left(f_3(x)-f_3(-x)\right),\label{GIRFermi}\nonumber\\
&&\phantom{blubb}\\
\int_{-c}^c\rmd x\,\frac{f_1(x)}{1+e^x}&=&\int_{-c}^0\rmd x\,f_1(x)\;,\nonumber\\
\int_{-c}^c\rmd x\,\frac{f_2(x)}{1+e^x}&=&\int_{-c}^0\rmd x\,f_2(x)\nonumber\\
&&+2\int_0^c\rmd x\,\frac{1}{1+e^x}f_2(x)\;,\nonumber\\
\int_{-c}^{c}\rmd x\,f_1(x)f_3(x)&=&\int_0^c\rmd x\,f_1(x)\left(f_3(x)+f_3(-x)\right)\;,\nonumber\\
&&\label{GIREven}\\
\int_{-c}^{c}\rmd x\,f_2(x)f_3(x)&=&\int_0^c\rmd x\,f_2(x)\left(f_3(x)-f_3(-x)\right)\;.\nonumber\\
\label{GIROdd}
\end{eqnarray}
In our work, we have $T\ll 1$ and $c=1/T\gg 1$. Therefore, we can replace the upper integration limit $c \to \infty$ in those integrals where an exponential term is present 
in the denominator of the integrand because the resulting corrections are exponentially small. This will be done henceforth, ignoring exponentially small corrections.

\subsection{Substitution of the \texorpdfstring{$\mathbf{\lambda}$}{lambda}-integrals}
\label{Appendix:SubstitutionOfTheLambdaIntegrals}
The handling of the multi-dimensional $\lambda$ integrals in the second and especially in the third order is not trivial. 
Therefore, we explicitly perform the proper substitutions in this subsection.

\subsubsection{Double integral}
\label{subsec:DoubleIntegral}
We face the following type of integral,
\begin{equation}
I_2=\int_0^\beta\rmd\lambda_1\int_0^{\lambda_1}\rmd\lambda_2\,F(\lambda_1-\lambda_2)\;,
\end{equation}
see Appendix~VI in the supplemental material of~\cite{PhysRevB.101.075132}.

Let $\lambda=\lambda_1+\lambda_2$ and $u=\lambda_1-\lambda_2$, thus 
\begin{eqnarray}
\lambda_1&=&\frac{1}{2}(\lambda+u)\;,\nonumber\\
\lambda_2&=&\frac{1}{2}(\lambda-u)\;.
\end{eqnarray}
Therefore, we obtain the determinant of the Jacobian matrix,
\begin{eqnarray}
\rmd\lambda_1\rmd\lambda_2&=&\left|\frac{\partial(\lambda_1,\lambda_2)}{\lambda,u}\right|\rmd u\,\rmd\lambda=\frac{1}{2}\rmd u\,\rmd\lambda\;.
\end{eqnarray}
With
\begin{eqnarray}
{\rm (i)}&0\leq\lambda_1\leq \beta\;\;&\Rightarrow\;\;0\leq\frac{1}{2}(u+\lambda)\leq \beta\;,\nonumber\\
{\rm (ii)}&0\leq\lambda_2\leq\lambda_1\;\;&\Rightarrow\;\;0\leq\frac{1}{2}(\lambda-u)\leq \frac{1}{2}(\lambda+u)\nonumber\; ,\\
\end{eqnarray}
we find for the limits of the integrals $-u\leq\lambda\leq 2\beta-u$, $-u\leq u$ so that
$u\geq 0$ and $u\leq \lambda$ must hold. Thus, we arrive at
\begin{equation}
I_2=\frac{1}{2}\int_0^\beta\rmd u\,\int_u^{2\beta-u}\rmd \lambda F(u)=\int_0^\beta\rmd u F(u)(\beta-u)\;.
\end{equation}

\subsubsection{Triple integral}
\label{subsec:TripleIntegral}
For the triple integral,
\begin{eqnarray}
I_3&=&\int_0^\beta\rmd\lambda_1\,\int_0^{\lambda_1}\rmd\lambda_2\,\int_0^{\lambda_2}\rmd\lambda_3\\
&&\quad\times F(\Lambda_1-\lambda_2)G(\lambda_2-\lambda_3)H(\lambda_1-\lambda_3)\nonumber
\end{eqnarray}
we denote $\lambda=\lambda_1+\lambda_2+\lambda_3$, $u=\lambda_1-\lambda_3$ and $v=\lambda_1-\lambda_2$, so $u-v=\lambda_2-\lambda_3$ and find
\begin{eqnarray}
\lambda_1&=&\frac{1}{3}(\lambda+u+v)\;,\nonumber\\
\lambda_2&=&\frac{1}{3}(\lambda+u-2 v)\;,\nonumber\\
\lambda_3&=&\frac{1}{3}(\lambda-2u+v)\;.
\end{eqnarray}
The corresponding determinant of the Jacobian matrix becomes
\begin{eqnarray}
\rmd\lambda_1\,\rmd\lambda_2\,\rmd\lambda_3=\displaystyle\left|\frac{\partial(\lambda_1,\lambda_2,\lambda_3)}{\partial(\lambda,u,v)}\right|=\frac{1}{3}\rmd\lambda\,\rmd u\,\rmd v\;.\nonumber\\
\end{eqnarray}
For the integration limits we obtain the three conditions
\begin{eqnarray}
{\rm (i)}&:&0\leq\frac{1}{3}(\lambda+u+v)\leq\beta\;,\nonumber\\
{\rm (ii)}&:&\frac{1}{3}(\lambda+u-2v)\leq \frac{1}{3}(\lambda+u+v)\;,\nonumber\\
{\rm (iii)}&:&0\leq \frac{1}{3}(\lambda-2u+v)\leq \frac{1}{3}(\lambda+u-2v)\;.
\end{eqnarray}
They are equivalent to
\begin{eqnarray}
{\rm (i)}&:&\lambda\geq -(u+v)\; \wedge\; \lambda\leq 3\beta-(u+v)\;\;\nonumber\\
&&\quad \Rightarrow\;\;-(u+v)\leq \lambda\leq 3\beta-(u+v)\;,\nonumber\\
{\rm( ii)}&:&\lambda\geq 2v-u\;\wedge\; v\geq 0\;,\nonumber\\
{\rm (iii)}&:&\lambda \geq 2u-v\;\wedge \;v\leq u\;,
\end{eqnarray}
so that 
\begin{eqnarray}
{\rm (ii)}+{\rm (iii)}&:& 0\leq v\leq u\;.
\end{eqnarray}
Since $-(u+v)\leq 2u-v$ we find
\begin{equation}
2u-v\leq \lambda\leq 3\beta-(u+v)
\end{equation}
if $2u-v\leq 3\beta-(u+v)$ or 
\begin{equation}
u\leq \beta\;.
\end{equation}
Thus,
\begin{eqnarray}
I_3&=&\frac{1}{3}\int_0^\beta\rmd u\,\int_0^u\rmd v\,\int_{2u-v}^{3\beta-(u+v)}\rmd\lambda\,\nonumber\\
&&\quad\times F(v)G(u-v)H(u)\nonumber\\
&=&\int_0^\beta\rmd u\,(\beta-u)H(u)\int_0^u\rmd v\,F(v)G(u-v)\;.\nonumber\\
\end{eqnarray}

\subsection{Derivative of the Hilbert transform}
\label{appendix:Lambda0PrimeVon0}

To calculate the free energy of the Ising-Kondo model we need the derivative of the Hilbert transform at the origin, $\omega=0$, see Eq.~(\ref{lambda0primeoccur}).

The Hilbert transform of the density of states is defined by a principle value integral as long as $|\omega|\leq 1$, 
see Eq.~(\ref{Hilberttransformed}). With $\eta=0^+$ we can represent the Hilbert transform in the form
\begin{equation}
\Lambda_0(\omega)=\int_{-1}^{\omega-\eta}\rmd\epsilon\,\frac{\rho_0(\epsilon) }{\omega-\epsilon}+\int_{\omega+\eta}^1\rmd\epsilon\,\frac{\rho_0(\epsilon)}{\omega-\epsilon}\;.
\end{equation}
Therefore we find the derivative with respect to $\omega$,
\begin{eqnarray}
 \Lambda_0^\prime(\omega)&=&\frac{\rho_0(\omega-\eta)  }{\omega-(\omega-\eta)}+\frac{\rho_0(\omega+\eta)}{\omega-(\omega+\eta)}(-1)\nonumber\\
 &&-\int_{-1}^{\omega-\eta}\rmd\epsilon\,\frac{\rho_0(\epsilon)}{(\omega-\epsilon)^2}-\int_{\omega+\eta}^1\rmd\epsilon\,\frac{\rho_0(\epsilon)}{(\omega-\epsilon)^2}\;.\nonumber\\
\end{eqnarray}
For $\omega=0$ we thus find
\begin{eqnarray}
\Lambda_0^\prime(0)&=&\frac{2\rho_0(0)}{\eta}-\int_{-1}^{-\eta} \rmd\epsilon\,\frac{\rho_0(\epsilon)-\rho_0(0)}{\epsilon^2 }\nonumber\\
&&-\rho_0(0)\int_{-1}^{-\eta}\rmd\epsilon\,\frac{1 }{\epsilon^2}-\int_{\eta}^1\rmd\epsilon\,\frac{\rho_0(\epsilon)-\rho_0(0)}{\epsilon^2}\nonumber\\
&&-\rho_0(0)\int_\eta^1\rmd\epsilon\,\frac{1}{\epsilon^2}\nonumber\\
&=&2\rho_0(0)-\int_{-1}^1\rmd\epsilon\,\frac{\rho_0(\epsilon)-\rho_0(0)}{\epsilon^2}\:.
\end{eqnarray}
Thus, we know $\Lambda_0^\prime(0)$ without having to calculate $\Lambda_0(\omega)$ for all $\omega$. For the constant density of states, $\rho_0(\epsilon)=1/2$, we have
\begin{equation}
    \Lambda_0^\prime(0)=1\;.
\end{equation}

\section{Integral tables}
\label{Integraltables}
\setcounter{equation}{0}
\renewcommand{\theequation}{I-\arabic{equation}}
In this section, we list all definitions and analytical approximations of the integrals  $\mathcal{I}_i$ and $\alpha_j$.

\subsection{Integrals \texorpdfstring{$\boldsymbol{\mathcal{I}_i}$}{Ii}}
In this subsection we collect all 39 integrals $\mathcal{I}_i$.
\begin{eqnarray}
\mathcal{I}_1&=&\int_0^{1/T}\rmd x\, \Rbar(x)\\
&=&-\frac{\ln(2\pi T)}{T}-\frac{1}{T}+\frac{\pi^2T}{6}+\frac{7\pi^4T^3}{180}+\mathcal{O}[T^5]\,,\nonumber 
\end{eqnarray}
\begin{eqnarray}
\mathcal{I}_2&=&\int_0^{1/T}\rmd x\,\ln(1+T x)\Rbar(x)\\
&=&-\frac{\ln T}{T}\left(2\ln 2-1\right)-\frac{\pi^2}{12 T}\nonumber\\
&&-\frac{1}{T}\left(-2+\ln 2-\ln \pi+2\ln 2\ln(2\pi )\right)\nonumber\\
&&+\frac{\pi^2 T}{6}\left(3\ln 2-12\ln \mathcal{A}+\ln(2\pi T)\right)+\mathcal{O}[T^3\ln T]\,,\nonumber
\end{eqnarray}
\begin{eqnarray}
\mathcal{I}_3&=&\int_0^{1/T}\rmd x\,\left(\Rbar(x)\right)^2\nonumber\\
&=&\frac{1}{T}\big(1+\big(1+\ln(2\pi T)\big)^2\big)\nonumber\\
&&-\frac{\pi^2}{2}-\frac{\pi^2 T}{3}\big(\ln(2\pi T)-1\big)+\mathcal{O}[T^3\ln T]\,,
\end{eqnarray}
\begin{equation}
\mathcal{I}_4=\int_0^\infty\rmd x\,\frac{x \Rbar(x)}{1+e^x}=\frac{\pi^2}{12}\left(1+\ln 2-12\ln\mathcal{A}\right)\, ,
\end{equation}
\begin{eqnarray}
\mathcal{I}_5&=&\int_{-1/T}^0\rmd x\,\Rbar^\prime(x)\nonumber\\
&=&-\EulerGamma-2\ln 2+\ln(2\pi T)+\frac{\pi^2T^2}{6}+\mathcal{O}[T^4]\, ,
\end{eqnarray}
\begin{eqnarray}
\mathcal{I}_6&=&\int_{-1/T}^0\rmd x\,\Rbar^\prime(x)\ln(1-T x)\nonumber\\
&=&-\frac{\pi^2}{12}-\frac{\pi^2T^2}{6}\left(\ln 2+\ln(2\pi T)-12\ln\mathcal{A}\right)\nonumber\\
&&+\mathcal{O}[T^4\ln T]\,
\end{eqnarray}
\begin{eqnarray}
\mathcal{I}_7&=&\int_{-1/T}^0\rmd x\, \Rbar(x)\Rbar^\prime(x)\nonumber\\
&=&\frac{1}{2}\left(\left(\EulerGamma+2\ln 2\right)^2-\ln^2(2\pi T)\right)\nonumber\\
&&-\frac{\pi^2T^2}{6}\ln(2\pi T)+\mathcal{O}\left(T^4\ln T\right)\,,
\end{eqnarray}
\begin{equation}
    \mathcal{I}_8=\int_0^\infty\rmd x\,\frac{\Rbar^\prime(x)}{1+e^x}=\ln 2-\frac{1}{2}\,,
\end{equation}
\begin{eqnarray}
\mathcal{I}_9&=&\int_0^\infty\rmd x\,\frac{\Rbar(x)\Rbar^\prime(x)}{1+e^x}\nonumber\\
&=&\frac{1}{2}-\frac{\pi^2}{48}-\EulerGamma\left(\ln 2-\frac{1}{2}\right)-\ln^2(2) \, ,
\end{eqnarray}
\begin{eqnarray}
\mathcal{I}_{10}&=&\int_{-1/T}^0\rmd x\,\frac{\Rbar(x)}{1-T x}\nonumber\\
&=&-\frac{\pi^2}{12 T}-\frac{\ln 2\ln(2\pi T)}{T}\nonumber\\
&&-\frac{T \pi^2}{6}\left(2\ln 2+\ln(2\pi T)-12\ln \mathcal{A}\right)\nonumber\\
&&+{O}[T^3\ln T]\,,
\end{eqnarray}
\begin{eqnarray}
    \mathcal{I}_{11}&=&\int_0^1\rmd\omega\,\frac{1}{1+e^{\beta\omega}}\left(\frac{H_T^{(0)}(\omega)}{1-\omega}-\frac{H_T^{(0)}(-\omega)}{1+\omega}\right)\nonumber\\
&=&\frac{\pi^2 T^2}{6}\left(1+\ln(2\pi T)\right)\nonumber\\
&&+\frac{\pi^2 T^2}{6}\left(1+\ln 2-12\ln\mathcal{A}\right)\nonumber\\
&&+\mathcal{O}[T^4\ln T]\,,
\end{eqnarray}
\begin{equation}
\mathcal{I}_{12}=\int_{-1/T}^0\rmd x\,\Rbar^{\prime\prime}(x)=T+\frac{\pi^2 T^3}{3}+\mathcal{O}[T^5]\,,
\end{equation}
\begin{eqnarray}
\mathcal{I}_{13}&=&\int_{-1/T}^0\rmd x\,\Rbar^{\prime\prime}(x)\ln(1-T x)\nonumber\\
&=&T\left(-\EulerGamma+\ln(2\pi T)\right)\nonumber\\
&&+\frac{\pi^2 T^3}{3}\left(1+3\ln
 2-12\ln\mathcal{A}+\ln(2\pi T)\right)\nonumber\\
 &&+\mathcal{O}[T^5\ln T]\,,
 \end{eqnarray}
 \begin{eqnarray}
\mathcal{I}_{14}&=&\int_{-1/T}^0\rmd x\,\Rbar(x)\Rbar^{\prime\prime}(x)\nonumber\\
&=&-\frac{\pi^2}{12}+T-T\ln(2\pi T)+\mathcal{O}[T^3\ln T]\,,
\end{eqnarray}
\begin{equation}
\mathcal{I}_{15}=\int_{0}^\infty\rmd x\,\frac{x \Rbar^{\prime\prime}(x)}{1+e^x}=\frac{3}{4}-\ln 2\,,
\end{equation}
\begin{equation}
    \mathcal{I}_{16}=\int_0^\infty\rmd x\,\frac{e^x}{\left(1+e^x\right)^2}\Rbar(x)=-\frac{1}{2}\left(1+\EulerGamma\right)\,,
    \end{equation}
\begin{equation}
    \mathcal{I}_{17}=\int_0^\infty\rmd x\,\frac{e^x}{\left(1+e^x\right)^2}x\Rbar^\prime(x)=\frac{1}{4}\,,
    \end{equation}
\begin{equation}
\mathcal{I}_{18}=\int_{-1/T}^0\rmd x\,\left(\Rbar^\prime(x)\right)^2=\frac{\pi^2}{12}-T+\mathcal{O}[T^3]\,,
\end{equation}
\begin{eqnarray}
\mathcal{I}_{19}&=&\int_{-1/T}^0\rmd x\,\frac{\Rbar^\prime(x)}{1-T x}\nonumber\\
&=&-\EulerGamma+\ln(\pi T)\nonumber\\
&&+\frac{\pi^2 T^2}{3}\left(1+2\ln 2-12\ln\mathcal{A}+\ln(2\pi T)\right)\nonumber\\
&&+\mathcal{O}[T^4\ln T]\,,
\end{eqnarray}
\begin{eqnarray}
\mathcal{I}_{20}&=&\int_{-1/T}^0\rmd x\,\ln(1+T x)\Rbar^{\prime\prime}(x)\nonumber\\
&=&T\left(\EulerGamma+2\ln 2-\ln(2\pi T)\right)+\mathcal{O}[T^3\ln T]\,,\nonumber\\
\end{eqnarray}
\begin{eqnarray}
\mathcal{I}_{21}&=&\int_{-1/T}^0\rmd x\,\frac{\Rbar^{\prime\prime}(x)}{1-Tx}\nonumber\\
&=&T\left(\EulerGamma-\ln(\pi T)\right)+\mathcal{O}[T^3\ln(T)] \, ,
\end{eqnarray}
\begin{equation}
\mathcal{I}_{22}=\int_{-1/T}^0\rmd x\,\Rbar^\prime(x)\Rbar^{\prime\prime}(x)=-\frac{T^2}{2}+\mathcal{O}[T^4]\,,
\end{equation}
\begin{equation}
\mathcal{I}_{23}=\int_0^\infty\rmd x\,\frac{\Rbar^\prime(x)\Rbar^{\prime\prime}(x)}{1+e^x}=\frac{\pi^2}{360}\,,
\end{equation}
\begin{equation}
    \mathcal{I}_{24}=\int_{-1/T}^0\rmd x\,\Rbar^{\prime\prime\prime}(x)=\frac{7\zeta(3)}{2\pi^2}+T^2+\mathcal{O}[T^4]\,,
    \end{equation}
\begin{eqnarray}
\mathcal{I}_{25}&=&\int_{-1/T}^0\rmd x\,\ln(1+T x)\Rbar^{\prime\prime\prime}(x)\\
&=&T^2\left(\EulerGamma-2+2\ln 2-\ln(2\pi T)\right)+\mathcal{O}[T^4\ln T]\,,\nonumber
\end{eqnarray}
\begin{eqnarray}
    \mathcal{I}_{26}&=&\int_{-1/T}^0\rmd x\,\Rbar(x)\Rbar^{\prime\prime\prime}(x)\nonumber\\
&=&-\frac{7\zeta(3)}{2\pi^2}\left(\EulerGamma+2\ln 2\right)\nonumber\\
&&+\frac{T^2}{2}\left(1-2\ln(2\pi T)\right)+\mathcal{O}[T^4\ln T]\,,
\end{eqnarray}
\begin{equation}
    \mathcal{I}_{27}=\int_0^\infty\rmd x\,\frac{\Rbar^{\prime\prime\prime}(x)}{1+e^x}=-\frac{\zeta(3)}{\pi^2}\,,
    \end{equation}
    \begin{eqnarray}
\mathcal{I}_{28}&=&\int_0^\infty\rmd x\,\frac{\Rbar(x)\Rbar^{\prime\prime\prime}(x)}{1+e^x}\\
&=&-\frac{\pi^2}{160}+\frac{\zeta(3)}{4\pi^2}\left(-3+4\EulerGamma+14\ln 2\right)\,,\nonumber
\end{eqnarray}
\begin{equation}
    \mathcal{I}_{29}=\int_0^\infty\rmd x\,\frac{\Rbar^{\prime\prime\prime}(x)x^2}{1+e^x}=-\frac{11}{6}+2\ln 2+\frac{\zeta(3)}{4}\,,
    \end{equation}
\begin{equation}
    \mathcal{I}_{30}=\int_0^\infty\rmd x\,\frac{e^x(e^x-1)}{(1+e^x)^3}\Rbar(x)x=-\frac{1}{4}\left(1+2\EulerGamma\right)\,,
    \end{equation}
\begin{eqnarray}
    \mathcal{I}_{31}&=&\int_0^\infty\rmd x\,\frac{e^x(e^x-1)}{(1+e^x)^3}\Rbar(x)\Rbar^\prime(x)\nonumber\\
&=&\frac{\pi^2}{480}-\frac{3 \zeta(3)}{4\pi^2}(1+\EulerGamma)\,,
\end{eqnarray}
\begin{equation}
    \mathcal{I}_{32}=\int_0^\infty\rmd x\,\frac{e^x(e^x-1)}{(1+e^x)^3}\Rbar^\prime(x)=\frac{3\zeta(3)}{4\pi^2}\,,
    \end{equation}
\begin{equation}
\mathcal{I}_{33}=\int_0^\infty\rmd x\,\frac{e^x(e^x-1)}{(1+e^x)^3}\Rbar^\prime(x)x^2=\frac{1}{6}+\frac{\zeta(3)}{4}\,,
\end{equation}
\begin{equation}
\mathcal{I}_{34}=\int_0^\infty\rmd x\,\frac{e^x}{(1+e^x)^2}\left(\Rbar^\prime(x)\right)^2=\frac{\pi^2}{180}\,,
\end{equation}
\begin{equation}
\mathcal{I}_{35}=\int_0^\infty\rmd x\,\frac{e^x}{(1+e^x)^2}\Rbar^{\prime\prime}(x)=\frac{3\zeta(3)}{4\pi^2}\,,
\end{equation}
\begin{eqnarray}
\mathcal{I}_{36}&=&\int_0^\infty\rmd x\,\frac{e^x}{(1+e^x)^2}\Rbar^{\prime\prime}(x)\Rbar(x)\nonumber\\
&=&-\frac{\pi^2}{288}-\frac{3\zeta(3)}{4\pi^2}(1+\EulerGamma)\,,
\end{eqnarray}
\begin{equation}
    \mathcal{I}_{37}=\int_0^\infty\rmd x\,\frac{e^x}{(1+e^x)^2}\Rbar^{\prime\prime}(x)x^2=-\frac{1}{3}+\frac{\zeta(3)}{4}\,,
    \end{equation}
\begin{eqnarray}
    \mathcal{I}_{38}&=&\int_{-1/T}^0\rmd x\,\Rbar^{\prime\prime\prime}(x)\ln(1-T x)\\
&=&T^2\left(\EulerGamma+2\ln 2-\ln(2\pi T)\right)+\mathcal{O}[T^4\ln T] \, ,\nonumber
\end{eqnarray}
and, finally,
\begin{eqnarray}
\mathcal{I}_{39}&=&\int_{-1/T}^0\rmd x\,\Rbar^\prime(x)\ln(1+T x)\nonumber\\
&=&\frac{\pi^2}{6}-\frac{\pi^2T^2}{6}\left(2+\ln2+\ln(2\pi T)-12\ln\mathcal{A}\right)\nonumber\\
&&+\mathcal{O}[T^4\ln T]\,.
\end{eqnarray}

\subsection{\texorpdfstring{$\boldsymbol{\alpha}_{\mathbf{j}}$}{alphaj} integrals}
\setcounter{equation}{0}
\renewcommand{\theequation}{A-\arabic{equation}}

Now we list the integrals $\alpha_i$. 
Note that any integral in which the integral limit goes to infinity has exponentially small corrections, so our expansions are asymptotic in these cases.
\begin{eqnarray}
\alpha_1&=&\int_0^{1/T}\rmd x\,\left(\ln(2\pi T)-\ln(1+T x)\right)^2\nonumber\\
&=&\frac{1}{T}\Big(2-\ln^2(2)+\ln^2(\pi)\nonumber\\
&&\qquad-2\ln2\ln\pi+2(\ln\pi-\ln2)\nonumber\\
&&\qquad +\ln T\left(2+2\ln \pi-2\ln2+\ln T\right)\Big)\;,\nonumber \\
\end{eqnarray}

\begin{equation}
\alpha_2=\int_0^\infty\rmd x\,\frac{x}{1+e^x}=\frac{\pi^2}{12}\;,
\end{equation}

\begin{eqnarray}
\alpha_3&=&\int_{-1/T}^0\rmd x\,\frac{\ln(2\pi T)-\ln(1-T x)}{1-T x}\nonumber\\
&=&\frac{\ln 2}{2 T}\ln(2\pi^2 T^2)\;,
\end{eqnarray}

\begin{equation}
\alpha_4=\int_0^\infty\rmd x\,x\frac{\tanh(x/2)}{\cosh^2(x/2)}=2\;,
\end{equation}

\begin{equation}
    \alpha_5=\int_0^\infty\rmd x\,\frac{e^x}{(1+e^x)^2}=\frac{1}{2}\;,
    \end{equation}
    
    \begin{equation}
\alpha_6=\int_0^\infty\rmd x\,x\frac{e^x(e^x-1)}{2(1+e^x)^3}=\frac{1}{4}\;,
\end{equation}

\begin{equation}
    \alpha_7=\int_0^\infty\rmd x\,\frac{xe^x(e^x-1)}{(1+e^x)^3}=\frac{1}{2}\;,
    \end{equation}
    
\begin{equation}
    \alpha_{8}=\int_{-1/T}^0\rmd x\,\frac{\ln(1-T x)-\ln(1+T x)}{1-T x}=\frac{\pi^2}{12T}\;,
    \end{equation}


\begin{equation}
\alpha_{9}=\int_{-1/T}^{1/T}\rmd x\,\frac{1}{1+e^{x}}\frac{1}{1-T x}
=\frac{\ln 2}{T}+\frac{\pi^2T}{6}+\mathcal{O}[T^3]\;,
\end{equation}


\begin{equation}
\alpha_{10}=\int_{-\infty}^\infty\rmd x\,\frac{e^{x}}{(1+e^{x})^2}\frac{1}{1-T x}
=1+\frac{\pi^2T^2}{3}+\mathcal{O}[T^3]\; ,
\end{equation}


\begin{equation}
\alpha_{11}=\int_{-1/T}^{1/T}\rmd x\,\frac{1}{1+e^{x}}\frac{1}{(1-T x)^2}
=\frac{1}{2T}+\frac{\pi^2 T}{3}+\mathcal{O}[T^3]\;,
\end{equation}

\pagebreak[4]


\begin{equation}
    \alpha_{12}=\int_{-1/T}^{1/T}\rmd x\,\frac{e^{x}}{(1+e^{x})^2}\frac{1}{(1-T x)^2}
=1+\pi^2 T^2+\mathcal{O}[T^3]\;,
\end{equation}


\begin{equation}
\alpha_{13}=\int_{-1/T}^{1/T}\rmd x\,\frac{1}{1+e^{x}}=\frac{1}{T}\;,
\end{equation}

\begin{eqnarray}
\alpha_{14}&=&\int_{-1/T}^{1/T}\rmd x\,\frac{1}{1+e^{x}}\ln(1-T x)\nonumber\\
&=&\frac{-1+2\ln2}{T}-\frac{\pi^2 T}{6}+\mathcal{O}[T^3]\; ,
\end{eqnarray}


\begin{equation}
\alpha_{15}=\int_{-1/T}^{1/T}\rmd x\,\frac{e^{x}}{(1+e^{x})^2}\ln(1-T x)
=-\frac{\pi^2T^2}{6}+\mathcal{O}[T^3]\;.
\end{equation}

\bibliography{kondo}